\documentclass{report}
\usepackage[utf8]{inputenc}
\usepackage{graphicx}
\usepackage{url}
\usepackage{amsmath}
\usepackage{amssymb}
\usepackage[english]{babel}
\DeclareMathOperator{\e}{e}
\usepackage[table]{xcolor}
\usepackage{supertabular}
\usepackage[
backend=biber,
style=apa,backref,backrefstyle=none
]{biblatex}
\usepackage{bbold}
\usepackage[pdfborder={0 0 0}]{hyperref}
\usepackage{comment}
\usepackage[T1]{fontenc}
\usepackage{physics}
\usepackage[toc,page]{appendix}
\usepackage{array, tabularx}
\usepackage{gensymb}
\usepackage{doi}
\usepackage{tikz}
\usepackage[nottoc,numbib]{tocbibind}
\usepackage{multirow}
\usetikzlibrary{quantikz2}
\DeclareUnicodeCharacter{2009}{} 
\definecolor{darkblue}{rgb}{0, 0, 0.8}

\newcommand{\beq}{\begin{equation}}
\newcommand{\eeq}{\end{equation}}

\newcommand\pdfmath[1]{\texorpdfstring{$#1$}{#1}}

\newcommand{\be}{\begin{equation}}
\newcommand{\ee}{\end{equation}}
\newcommand{\bea}{\begin{eqnarray}}
\newcommand{\eea}{\end{eqnarray}}
\newcommand{\ba}{\begin{align}\begin{split}}
\newcommand{\ea}{\end{split}\end{align}}
\title{Quantum computing for energy industry}

\begin{document}

\begin{titlepage}
  \begin{sffamily}
  \begin{center}

    
    \textsc{\LARGE Laboratoire Charles Fabry, EDF R\&D}\\[2cm]

    
    \textsc{PhD manuscript}\\[1.5cm]

    { \huge \bfseries Quantum simulation for strongly interacting fermions with neutral atoms array : towards the simulation of materials of interest\\[0.4cm] }

    \vfill

    \begin{minipage}{0.4\textwidth}
      \begin{flushleft} \large
        Antoine \textsc{Michel}\\
      \end{flushleft}
    \end{minipage}
    \begin{minipage}{0.4\textwidth}
      \begin{flushleft} \large
        \emph{Supervisors:} Antoine \textsc{Browaeys} et Christophe \textsc{Domain} \\
        \emph{Co-supervisor:} Thierry \textsc{Lahaye}
      \end{flushleft}
    \end{minipage}

    \vfill


  \end{center}
  \end{sffamily}
\end{titlepage}

\textbf{\Large{List of acronyms}} \\
AQS : Analog Quantum Simulation \\
AT : Adiabatic Theorem \\
BP : Barren Plateau \\
CDMFT : Cluster Dynamical Mean-Field Theory \\
CMFT : Cluster Mean Field Theory \\
DFT : Density Functional Theory \\
DFPT : Density Functional Perturbation Theory \\
DMFT : Dynamical Mean-Field Theory\\
DQS : Digital Quantum Simulation \\
EDF : Electricité de France \\
GGA : Generalized Gradient Approximation \\
HF : Hartree Fock \\
HPC : High Performance Computing \\
IASCC : Irradiated Assisted Stress Corrosion Cracking \\
LDA : Local Density Approximation  \\
LTO : Long Term Operator \\
MF : Mean-Field \\
MIS : Maximum Independent Set \\
MOT : Magneto-Optical Trap \\
NISQ : Noisy Intermediate Scale Quantum \\
NPP : Nuclear Power Plant \\
PKA : Primary Knocked-on Atoms \\
PWR : Pressurized Water Reactor \\
PQC : Parametrized Quantum Circuit \\
QA : Quantum Annealing \\
QAOA : Quantum Approximate Optimization Algorithm \\
QC : Quantum Computing \\
QEC : Quantum Error Correction \\
QFT : Quantum Fourier Transform \\
QMC : Quantum Monte Carlo \\
QPU : Quantum Processor Unit \\
QuAltOA : Quantum Alternating Operator Ansatz \\
QUBO : Quadratic Unconstrained Binary Optimization\\
RQP : Rydberg Quantum Processor \\
SLM : Spatial Light Modulator \\
SSMF : Slave Spin Mean Field \\
STO : Slater Type Orbital \\
VQA : Variational Quantum Algorithm \\
VQE : Variational Quantum Eigensolver \\

\chapter*{Remerciements}
C'est peut-être la partie la plus difficile à écrire pour moi, car je sais que c'est celle qui va être lue en premier, mais toute cette aventure n'aurait été possible sans la participation de beaucoup de personnes que je souhaite, par conséquent, remercier ici. 
Je souhaite tout d'abord remercier mon jury de thèse, Silke Biermann, Benoît Vermersch, Guido Pupillo, David Clément, Bruno Senjean et Thomas Ayral. Leur lecture approfondie de mon manuscrit ainsi que leurs retours et leurs commentaires sur mon travail m'ont permis de mettre en perspective ma recherche et sont des outils précieux pour la poursuivre.

Cette thèse étant CIFRE, je tiens à remercier la direction du laboratoire Charles Fabry pour son accueil, et notamment Patrick Georges, mais aussi EDF R\&D et plus particulièrement Stéphane Taunier, Julien Stodolna et Marion Gorce pour m'avoir donné la chance d'évoluer au sein du groupe T27 du département MMC. Je tiens à remercier également Valérie Kervargant pour son aide très précieuse sur l'organisation de mes voyages en France et à l'étranger.

Je souhaite aussi remercier grandement Antoine Browaeys et Thierry Lahaye, mes deux directeurs de thèse, de m'avoir fait confiance pour ce projet ambitieux et d'avoir relevé le défi d'encadrer pour la première fois une thèse CIFRE, mais aussi une thèse théorique. Le mode d'encadrement était plus distendu qu'à l'accoutumé et la pandémie n'a pas aidé, mais vous avez toujours fait l'effort de garder le contact et un œil attentif sur mon travail. Vous avez toujours su être présent dans les moments charnières de la thèse et vos conseils et remarques ainsi que votre immense culture et expérience de la recherche m’ont été très utiles pour mener à bien mon travail et me suivront, j'en suis sûr, jusqu'à la fin de ma carrière. Je tiens aussi à remercier chaleureusement Christophe Domain, mon co-superviseur, avec qui j'ai passé la plupart de mon temps durant ces trois ans. Malgré la nouveauté du sujet, tu t'es investi avec moi pour essayer de comprendre et de relier les enjeux d'EDF sur le vieillissement des matériaux et la simulation quantique. Ta bienveillance et ta patience sont des qualités rares, surtout pour un doctorant à qui l'on doit expliquer encore au bout de trois ans comment fonctionne (et vieillit) une centrale nucléaire. Ta culture du numérique (je suis toujours friand de tes anecdotes sur le début du calcul HPC !) et de la physique m'a permis de prendre énormément de recul sur mon sujet et de comprendre les enjeux du calcul quantique. Merci de m'avoir introduit de la meilleure des manières à ce tout nouveau monde pour moi et je suis ravi de continuer avec toi sur ce sujet passionnant !

Cette thèse n'aurait tout simplement pas vu le jour sans le travail de Marc Porcheron (que je n'ai pas eu la chance de côtoyer longtemps), Stéphane Tanguy et Etienne Décossin qui ont lancé le programme quantique à EDF. Merci à vous d'avoir été aussi visionnaire et de croire en la physique~! Je tiens aussi à remercier tout particulièrement Joseph Mikael. Tu m'as tout de suite accueilli à bras ouverts dans le groupe même après un an de COVID et je m'y suis senti immédiatement dans mon élément. Je ne pense pas qu'il existe beaucoup de personnes aussi enthousiastes pour le monde du calcul quantique, et l'énergie que tu déploies pour faire vivre le projet m'impressionne un peu plus chaque jour. 
Je souhaite aussi remercier toute l'équipe qui gravite autour du quantique~: Paulin, Pascale, Cyril, Cyril $\times 2$, Quentin, Rodolphe, Mohamed, Ulysse, Naomi (et j'en oublie certainement), nos échanges et diversités de compétences initiales furent un moteur pour ma thèse.

Durant toute la durée de ma thèse, j'ai pu compter sur le soutien de PASQAL et tout particulièrement de Loïc Henriet qui a suivi et participé à mon travail, merci pour la confiance !

Je souhaite aussi remercier grandement Sebastian Grijalva et Thomas Ayral. Vous avez officieusement participé activement à l'encadrement de ma thèse et avez pris énormément de votre temps pour travailler avec un jeune doctorant que vous ne connaissiez pas et pour cela, je vous en serai éternellement reconnaissant.

Je tiens aussi à remercier Julien Villmejane pour m'avoir fait confiance en me confiant des TP et des TD d'électronique en première année de l'IOGS malgré mes souvenirs qui étaient lointains.

Pendant les trois années de thèse (en fait les deux dernières), j'ai pu partager mes galères avec Clément, mon coloc' de bureau, ainsi qu'avec Clément bis, Lucie, Maxime et je les remercie pour cela. j'ai aussi fait de super rencontres à l'école des Houches que ce soit Andréas, Marion, Julien (j'espère que tu bats tout ton labo au ping maintenant !), Alexandre, Gabriel et d'autres que j'oublie !

En-dehors du monde quantique, le soutien de mes amis a été particulièrement important et je souhaite tout d'abord remercier Alexis, Sylvain et Anis qui ont fait le déplacement pour ma soutenance. C'est une amitié qui commence (la sauce) à bien durer dans le temps et j'espère que cela perdurera jusqu'à nos 80 ans quand on ouvrira une bonne bouteille... Mais j'ai aussi pu compter sur Laury, le chaleureux Antoine, Ambre, Antho, Thomas, Seddik (qui est le seul à s'être engagé dans la même galère que moi, merci pour la balade jusqu'au lac où on ne peut pas se baigner et pour les memes échangés), Alexandre et Amélie. Je voudrais aussi remercier Kamil pour m'avoir carry sur Ash (contrairement à Anis et Sylvain).

Parce qu'il est important de vider sa tête des codes Python et des équations, je souhaite remercier le club de tennis de table de Villemomble et tout particulièrement Nico, Jérèm, Loris, Lad, Cédric, Vibol, Yves et tout ceux que j'oublie pour m'avoir pris dans leurs équipes ! Promis les gars, on remonte en R3 bientôt.

Un grand merci aussi à ma famille (et ma belle-famille !) pour le soutien sans faille et surtout mes parents.

Enfin, je souhaite finir par toi qui me suis depuis maintenant 7 ans. De ma déprime d'école d'ingénieur à aujourd'hui, tu ne m'as jamais lâché. Ces trois années ont été très intenses avec de précieux moments de joie, mais aussi des moments difficiles. Néanmoins, nous sommes toujours restés soudés et pour rien au monde, je n'aurais souhaité vivre ces années avec quelqu'un d'autre. Tu es le dénominateur commun de toutes mes réussites, que tu le veuilles ou non. J'espère que l'on regardera The Big Bang Theory encore ensemble quand on sera vieux.

\tableofcontents

\chapter{General introduction}
\section{Industrial material aging context}
\begin{figure}[!h]
    \centering
    \includegraphics[width=0.3 \textwidth]{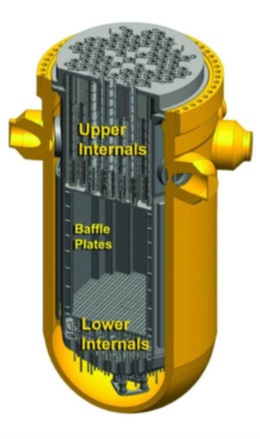}
    \caption{View of a vessel in PWR.}
    \label{fig:interne}
\end{figure}

The energy industry faces major challenges today with the threat of global warming, growth of raw material prices and political instabilities. Regarding this context, it is crucial for industries like Electricité de France EDF to deliver a sustainable source of electricity. In France, this electric power is mainly provided by nuclear energy (63\% of the total production in 2022 \footnote{RTE bilan électrique 2022 \url{https://assets.rte-france.com/prod/public/2023-02/Bilan-electrique-2022-synthese.pdf}}). This proportion dropped in 2022 due to the discovery of stress corrosion in Nuclear Power Plant (NPP). Global warming is expected to have a huge impact on our society \parencite{RN15}; therefore, reaching a zero carbon emission energy industry in 2050 is a major goal. From this standpoint, extending the lifetime of the nuclear power plant over 60 years (Long Term Operation-LTO) has been raised to be an important step \footnote{RTE futures energétiques 2050 \url{https://assets.rte-france.com/prod/public/2021-12/Futurs-Energetiques-2050-principaux-resultats.pdf}}. All currently operating NPPs in France use the pressurized water reactor technology (PWR). One reactor is made of a pressure vessel containing the core internals supporting the fuel assemblies in which the fissile materials are, Fig. \ref{fig:interne}. 

When operating, internal structures undergo harsh stresses: temperature locally up to 350\degree C, mechanical constraints, intense neutron irradiation up to 120 dpa (displacement per atom), corrosion. The combination of these factors induces creep, a potential risk of swelling (for which VVER or austenitic materials such as fuel cladding are concerned) but no evidence of this phenomena have been found in PWR reactors today, and thus changes the properties of materials inside the vessels. It has been shown that irradiation increases the formation of cracks in PWR components such as bolts \parencite{christiaen_modelisation_2018}. These phenomena are known as the Irradiation Assisted Stress Corrosion Cracking (IASCC). To anticipate materials aging of NPPs, it is needed to understand these phenomena at the atomic scale, where the formation of defects appears. Several works have been done to study defects evolution in austenitic steels \parencite{pare_previsions_2022} with \textit{ab initio} method such as the Density Functional Theory (DFT) \parencite{piochaud} or Monte-Carlo algorithms (kinetic Monte-Carlo or rate theory) \parencite{fokt_modelling_2021} with encouraging results. Nevertheless, simulating complex phenomena where quantum correlations are important, such as para-magnetism, is still difficult with these tools. Austenitic steels are in the para-magnetic phase in power plants conditions of temperature and pressure, and it is well known that magnetic phases play an important role in energies of mitigation and defects \parencite{ekholm_influence_2010, alling_effect_2010}. Therefore, being able to simulate accurately strong quantum correlations could help, \textit{in fine}, to increase NPP lifetime. Indeed, DFT is in a framework multiscale modelling approach and output data of atomic scale simulation can be used as input data of macroscopic scale simulations.

In addition, the aging of lithium-ion batteries is also a major issue to renew the car fleet or for static energy storage, which is crucial to reduce greenhouse gas emissions. Oxides are very difficult to simulate because most of these materials have strong-correlated electrons that state-of-the-art methods struggle to design \parencite{birkl}. We are surrounded by batteries in our everyday life, and understanding how batteries age is decisive if an all electrical world is considered. Another field where simulations are important is solar cells. The difficulty is to reliably simulate excited states to understand the efficiency degradation. \textit{Ab initio} methods reach their limits when it comes to study excited states of a material. 
The energy industry therefore needs new methods of simulation to understand how materials of interest age and to be able to avoid enormous costs like changing vessel interns after a cracking or an explosion of a battery following a great loss of the capacity. First principles computations have encountered great successes \parencite{PhysRevB.65.024103,PhysRevB.84.140411,piochaud} but they are limited in great significance area and need to be improved or replaced because of quantum correlations being too important. \\

\begin{figure}
    \centering
    \includegraphics[width=1.\textwidth]{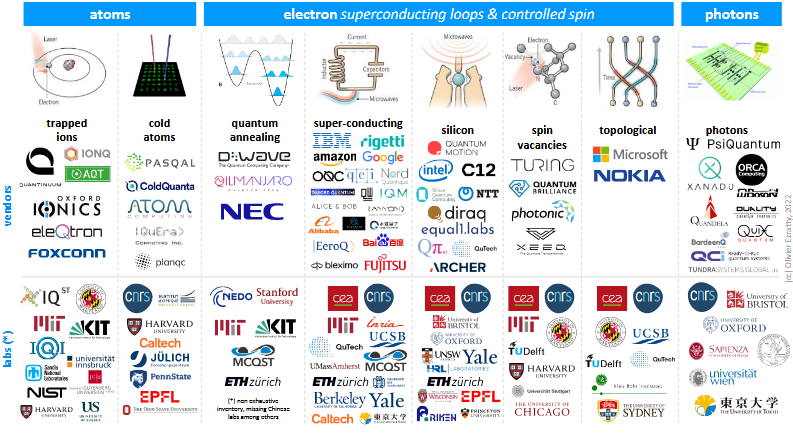}
    \caption{Overview of quantum computers builders in the world with the three main qubit carriers: the atoms, the electrons and the photons (figure taken from \parencite{ezratty2022understanding}).}
    \label{fig:technologies}
\end{figure}

\section{Quantum computing context}

In 1982, the physicist Richard Feynman proposed to simulate quantum phenomena with a quantum computer \parencite{Feynman1982}. This is the birth of quantum computing. The principle is to manipulate qubits, instead of classical bits, which lie in vectorial space. Thanks to the superposition of quantum states, it is possible to manipulate the coefficients $a$ and $b$ in a state $a\ket{0} + b \ket{1}$ where $|a|^2 + |b|^2 = 1$. We are not dealing only with the bit $0$ or $1$ but with both at the same time.  In 1994, Peter Shor proposed an algorithm based on quantum computing and Quantum Fourier Transform (QFT) to factor any integer $N$ in a polynomial time \parencite{shor}, threatening to break public-key cryptography schemes. In other words, if a quantum computer with enough noiseless qubits is built, it could break most of the internet security and the world internet network would be endangered. Thenceforth, a race for the universal quantum computer has begun and is still ongoing today. The number of applications is great: cryptography, optimization, differential partial equations, combinatorial problems and of course materials and chemistry.  Several technologies are being tested with their own pros and cons all over the world (examples in Fig. \ref{fig:technologies}). Researchers also focus on designing original quantum algorithms because it is impossible to convert a classical algorithm into a quantum algorithm \parencite{montanaro_quantum_2016}. Nevertheless, quantum computing is still a promise as no real quantum advantage has been shown up to date. Nowadays, we are in the Noisy Intermediate Scale Quantum Computer (NISQ) era, \parencite{preskill_quantum_2018} where noisy quantum simulators with a few qubits are available. Therefore, original algorithms considering noise have to be designed and a great amount of effort is put in the emulation of quantum computing. This means that High Performance Computing (HPCs) are used to numerically simulate the behavior of a quantum computer, considering noise and limitations. This very important step in the creation of an algorithm allows testing it with a few qubits to anticipate and to mitigate the effect of noise. Moreover, quantum algorithms are often in two parts:
\begin{itemize}
    \item A quantum part which is supposed to be solved by a real quantum computer.
    \item A classical part that is tackled by a classical quantum computer.
\end{itemize}
The two parts are self-correlated and aim at minimizing an energy and finding the groundstate energy of a system through the variational principle, for instance.

The work presented in this manuscript lies between the energy industry challenges described above and the breakthroughs in quantum computing. We show how we can study strong-correlated electrons on realistic quantum simulators, having in mind the issue of aging materials for energy industry. 

\section{EDF projects}

My PhD is part of a quantum project and contributing to a material modelling project: 
\begin{itemize}
    \item a multiscale modelling approach to understand and anticipate defects and develop aging models of materials in PWR (ULTIMATE project);
    \item a quantum project (SI quantique) which aims at studying potential applications of quantum computing for EDF.
\end{itemize}

\subsection{Multiscale modelling}

\textit{Ab initio} atomic simulations are a part of a wide modelling program at EDF, starting from simulating neutron irradiation to finite elements method. The idea of this multiscale modelling is to use data and results of atomic scale simulation to help modelling higher orders of magnitude problems, and so on until the macroscopic scale is reached. This encompasses simulating the energy distribution of the Primary Knocked-on Atoms (PKA) from the neutron spectrum, \textit{ab initio} methods such as DFT, Molecular Dynamics to characterize elementary mechanisms at the atomic scale and also to simulate displacement cascades and dislocation-defect interactions,  Monte-Carlo and rate theory to simulate microstructure evolution under irradiation, Dislocation Dynamics for plasticity modelling and so on \parencite{adjanor_overview_2010,becquart_modeling_2011}.

\begin{figure}
    \centering
    \includegraphics[width = 1 \linewidth]{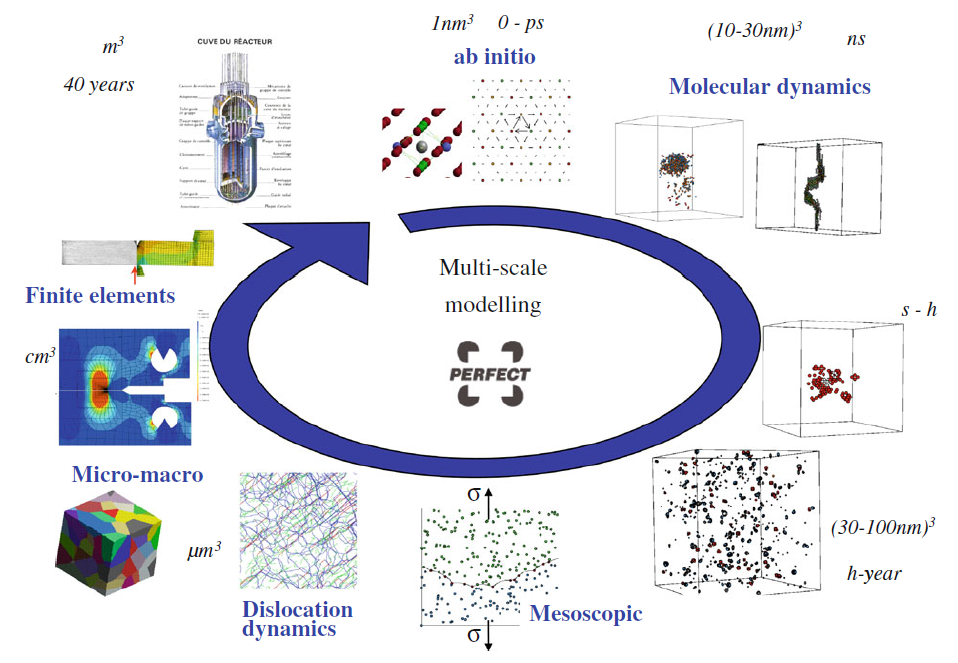}
    \caption{Multiscale modelling applied in the PERFECT project to the pressure vessel steels. Taken from \parencite{becquart_modeling_2011}.}
    \label{fig:multiscale}
\end{figure}

A schematic representation of multiscale modelling is shown in Fig.~\ref{fig:multiscale}.

Therefore, quantum computing could help to improve results of all the loop of multiscale modelling by enhancing DFT materials modelling.

\subsection{The quantum project}
The Research and Development section of EDF is the biggest in Europe. Several crucial issues are studied there to deliver sustainable source of electricity in France (and other countries) and at the same time, keeping a balance in economy. Quantum computing holds the promise to solve numbers of relatable problems. The advent of QC could change everything in computing and simulation, and there is no mapping between classical simulations and the quantum world: one has to rewrite all algorithms to benefit QC.
That was the starting point of the quantum project at EDF in 2018: being active in this field and starting to design our own algorithms for our own use case. Several sector and direct applications have been identified such as partial differential equation (with the HHL algorithm see Sec.~\ref{sec:QPA}), post-quantum cryptography (Shor algorithm Sec.~\ref{sec:QPA}) the smart charging optimization problem \parencite{dalyac_qualifying_2021} and, of course, material simulation. All use case can be seen in Tab.~\ref{tab:use_case}. The quantum group at EDF collaborated with a wide variety of actors from French start-ups to IBM in the United States of America.
\begin{table}[!h]
    \centering
    \begin{tabular}{|c|c|}
    \hline
    \rowcolor{lightgray}
    R\&D sectors & Direct application \\
    \hline
    \multirow{4}{*}{Production} & Materials simulation (internals, batteries) \\  & Probabilistic Safety Assessment studies \\ & Partial differential equations \\ & Quantum metrology networks \\
    \hline
     Network & Optimal power flow \\
    \hline
    \multirow{2}{*}{Energy management} & Optimisation/smart charging  \\ & Machine learning \\
    \hline 
    Information Technologies & Post-quantum cryptography \\
     \hline
    \end{tabular}
    \caption{Sectors and direct applications of quantum computing for EDF}
    \label{tab:use_case}
\end{table}


\section{Contributions of this work}
The contributions of this work are the following:

First, I present the first hybrid digital-analog variational algorithm on the Rydberg Quantum Processor (RQP), a NISQ device, which aims at finding the groundstate of molecules. This algorithm encompasses all specificities from the device. I show a numerical implementation of this algorithm and prove that it can be run on an actual device in a reasonable amount of time. The efficiency of this algorithm is shown on molecules H$_\text{2}$, LiH and BeH$_\text{2}$. All steps of the algorithm are discussed. The transformation from a molecular algorithm to a qubits algorithm, the state preparation, the measurement process with the derandomization method are studied \parencite{PhysRevA.107.042602}. We propose a roadmap for bigger molecules and an experimental implementation. This work results from a strong collaboration with the start-up PASQAL.

Secondly, I introduce a new hybrid algorithm to observe the Mott transition on 2D-Hubbard model on a RQP. The slave-spin method is used to transfer the complexity of the strongly-correlated electrons of Hubbard model on Ising-like Hamiltonian coupled to a free electron's system via a mean-field mapping. We study the feasibility of this algorithm on an experimental device by exploring the impact of all sources of error known up to date. I show that the Mott transition can be observed on a RQP. I also demonstrate that we can observe the effect of a quantum quench on the quasi-particle weight of the system on RQP. This algorithm is the first step toward simulation of bulk materials on a noisy quantum simulator. This work results from a strong collaboration with the EVIDEN/ATOS company.

\section{Manuscript organization}
This work is structured as follows:\\

In chapter \ref{Bibliography}, I give an overview of state-of-the-art methods to simulate electrons and magnetism "classically". Successes and limitations of these tools are presented. Then, after an introduction of theoretical quantum principles, I present some of the most recent NISQ quantum algorithms and their applications with a focus on methods to simulate fermionic systems.\\

In chapter \ref{Rydberg}, I present the Rydberg platform, the physics behind it and how to perform quantum computing with it. The digital and the analog approaches are presented.\\

In chapter \ref{chemistry}, I give the method and results of a digital-analog quantum eigensolver designed for the RQP. After a presentation of the molecular to qubits Hamiltonian mapping, we present an analog simulation implementation to find the groundstate of the H$_\text{2}$ molecule, which is inspired from the Unitary coupled-cluster ansatz. Finally, I present the protocol for bigger molecules and the results we obtain numerically.\\

In chapter \ref{hubbard}, I introduce the slave-spin method and I show how it can be employed to predict the Mott metallic-insulator  transition and the out of equilibrium behavior in the 2D-Hubbard model on a RQP. All the protocol and theory behind the algorithm are described. The numerical simulation of a realistic implementation on a noisy quantum simulator is shown and discussed.\\

Chapter \ref{conclusion} presents the conclusion and outlook of this work.

\chapter{Simulating correlated matter: from classical to quantum}\label{Bibliography}
\section{Forewords}

The famous citation of Richard Feynman "Nature isn't classical, dammit, and if you want to make a simulation of nature, you'd better make it quantum mechanical, and by golly it's a wonderful problem, because it doesn't look so easy" reminds us that simulating nature with a quantum computer seems to be a good idea but most importantly, it is not straightforward. Indeed, there are $\approx 10^{23}$ interacting electrons in all pieces ($\approx 1 $ cm$^3$) of all the matter surrounding us. Even if only one-half spin degree of freedom is considered, it means that $\approx 2^{10^{23}}$ states are allowed for a small piece of matter, which is more than the number of particles in the universe. Therefore, in front of this unreachable complexity, humans had to be creative. The first step is attributable to Erwin Schrödinger and Werner Heisenberg who first described the behavior of non-relativist quantum particles with the Schrödinger equation (and matrix calculation):
\begin{equation}\label{eq:schrodinger}
    i\hbar \frac{\partial \psi(t,\Vec{r})}{\partial t} = -\frac{\hbar^2}{2m}\Delta\psi(t,\Vec{r}) + V(t,\Vec{r})\psi(t,\Vec{r})
\end{equation}
with $\hbar$ the Planck constant divided by $2\pi$, $\psi(t,\Vec{r})$ the wavefunction of the particle, $m$ its mass, and $ V(t,\Vec{r})$ all external potentials the particle undergoes. This equation combined with periodic boundaries condition (PBC) has led to the first quantum revolution with the band theory \parencite{kittel_introduction_2018} which is at the heart of all today's electronic devices. Nevertheless, approximations can not stand when considering certain class of materials. For instance, complex alloys can not be described with this theory. With the democratization of computers, the idea of solving the Schrödinger equation with an algorithm to describe and understand complex phenomena has been more and more studied. Indeed, as Richard Feynman said, the world is quantum and if we want to improve our lives and our society, the quantum world has to reveal all its secrets. Since particles are in interaction and the complexity of materials is huge, approximations have to be made to hope for an advancement. In this section, I review some of the most ubiquitous methods to simulate strongly correlated electrons and spins. They all consider approximations, but they have allowed to understand a large part of complex materials. These methods are used today at EDF for instance to improve and anticipate material aging. I discuss successes but also limitations of these methods. These limitations can be really bounding in some fields where quantum correlations are very strong \parencite{abrikosov_recent_2016, ho2018promise}. Quantum computing (QC) holds the promise to solve really hard materials' problem unreachable today. In the second part of this chapter, I give an overview of the theoretical bases of QC and how to simulate interacting electrons on a quantum computer. The last part of this bibliography is dedicated to modern methods to perform quantum computing on NISQ device and how to simulate fermionic problems on them.   

\section{\textit{Ab initio} or first principles methods}\label{sec:dft}

The challenge of simulating materials relies on solving the Schrödinger equation (Eq. \ref{eq:schrodinger}). It is only possible in specific cases (the harmonic oscillator for instance with $\hat{H} = \frac{\hat{P}^2}{2m} + \frac{1}{2}k\hat{X}^2$). In general, approximations are necessary to reach observable values. In this section, I describe the Hartree-Fock method and the Density Functional Theory (DFT), electronic structure-calculation methods widely used all around the world and state-of-the-art methods in a large part of chemistry and material simulation at the atomic level.
\subsubsection{The Born-Oppenheimer approximation}
Let us consider a periodic material. It is composed of electrons and nuclei in interactions. The Hamiltonian of such a system can be written:
\begin{equation}\label{eq:Hmat}
    H_\text{mat} = \sum_{k=1}^M \nabla_{\hat{\vec{R}}_\text{k}}^2 - \sum_{i=1}^N \nabla_{\vec{\text{r}}_\text{i}}^2 + \frac{1}{2} \sum_{i_1 \neq i_2 = 1}^N \frac{e^2}{|\vec{r}_{i_1} - \vec{r}_{i_2}|} + \frac{1}{2} \sum_{k_1 \neq k_2 = 1}^M \frac{Z_{k_1}Z_{k_2}}{|\vec{R}_{k_1} - \vec{R}_{k_2}|} - \sum_i^N \sum_k^M \frac{Z_k e}{|\vec{r}_{i}-\vec{R}_{k}|}
\end{equation}
with $M$ and $N$ being the number of nucleus and electrons respectively, $\vec{R}_k$ and $\vec{r}_i$ the coordinates in 3D-space of nucleus and electrons respectively. $e$ is the electric charge and $Z_k$ the number of protons inside the nucleus. The first two terms are the kinetic energy of protons and electrons, and the other terms are the Coulomb interaction energy. The first approximation to simplify this equation was proposed by Max Born and Robert Oppenheimer in 1927. It relies on the ratio of mass between nucleus and electrons being equal to $\approx 2000$. It is therefore possible to consider that electrons see the nucleus moving adiabatically. The kinetic energy of the nucleus is then neglected, and the Coulomb interaction between protons is a constant. From electron's point of view, they undergo the following Hamiltonian:

\begin{equation}\label{eq:He}
    H_\text{e} = E + U + V
\end{equation}
where $E$ is the kinetic energy of electrons, $U$ is the interaction between electrons and $V$ the interaction of electrons with the stationary external potential (protons interaction).
The purpose now is to find the groundstate of such a Hamiltonian for a well-defined $V$. It can actually be solved for small systems in small basis set but obviously, trying to solve this equation with the vectorstate $\ket{\psi_0}$ for a material is out of reach for classical computer and will always be.

\subsection{Hartree-Fock method}
The Hartree-Fock (HF) method \parencite{echenique_mathematical_2007} is based on finding a good approximation of the wave-function of the system to compute the ground state energy. The method often assumes that the N-body wave-function describing the ground state of a system is what we called a Slater determinant:

\begin{equation}\label{eq:Slater}
    \Psi(\textbf{r}_1,\textbf{r}_2,\dots,\textbf{r}_N) = \frac{1}{\sqrt{N!}}\begin{vmatrix} \psi_1(\textbf{r}_1) & \psi_2(\textbf{r}_1)  & \dots & \psi_N(\textbf{r}_1)\\
    \psi_1(\textbf{r}_2) & \psi_2(\textbf{r}_2) & \dots & \psi_N(\textbf{r}_2) \\
    \vdots & \vdots & \ddots & \vdots \\
    \psi_1(\textbf{r}_N) & \psi_2(\textbf{r}_N) & \dots & \psi_N(\textbf{r}_N)
    \end{vmatrix}
\end{equation}
with orthogonality relations:
\begin{equation}
    \delta_{i,j} = \int \psi^{\dagger}_i(\textbf{x}) \psi_j(\textbf{x}) d\textbf{x}
\end{equation}
where $\textbf{x} = (\textbf{r},\sigma)$.
The next approximation is to find how to express these spatial-spin wave-functions. In general, linear combinations of Gaussian functions (STO-nG) are chosen as basis set. This leads to good approximations without too much complexity \parencite{stewart_small_2003}.
The variational principle is then used to solve the HF equations.

This principle states that all approximations of ground state will lead to an energy greater or equal to the exact ground state energy. The Rayleigh Ritz wave-function variational principle can be expressed as:

\begin{equation}\label{eq:variationalprinciples}
    E_0 = \underset{\ket{\psi}}{\text{min}} \bra{\Psi}{H_\text{e}}\ket{\Psi}
\end{equation}
where $\bra{\psi}\ket{\psi} = 1$.

The HF method is at the center of many quantum chemistry numerical simulations today \parencite{helgaker_molecular_2001}.
\subsection{Fundamentals of density functional theory}
Finding a good wave-function for the system can be complicated and can lead to huge approximations. Another approach is to focus on the one-particle density of the system. Indeed, this observable gives a lot of information of the system and can be easier to compute than the wave-function. This is the very first step of the Density Functional Theory (DFT).

The variational theorem can be reformulated in terms of one-electron density defined as:
\begin{equation}
    n(\textbf{r}) = N \int \dots \int |\ket{\psi(\textbf{x}_0 \textbf{x}_1 \dots \textbf{x}_N)}|^2 d\textbf{x}_0 d\textbf{x}_1 \dots d\textbf{x}_N.
\end{equation}
Indeed, the Hohenberg-Kohn theorem states that:
\begin{itemize}
    \item the external potential is a unique functional of the electronic density up to a constant; \footnote{Proof: let's consider two external potentials $V_1(\textbf{r}) $ and $V_2(\textbf{r})$ such as $V_1(\textbf{r}) \neq V_2(\textbf{r})  + \mathrm{cst}$.  The two corresponding Hamiltonians $H_1(V_1)$ and $H_2(V_2)$ have therefore different groundstates $\ket{\psi_1}$ and $\ket{\psi_2}$ and groundstate energies $E_1$ and $E_2$ but let's suppose the same one-electron density n(\textbf{r}). We have $E_1 = \bra{\psi_1} H_1 \ket{\psi_1} = \bra{\psi_1} H_2 + V_1 - V_2 \ket{\psi_1} > E_2 + \bra{\psi_1} V_1-V_2 \ket{\psi_1} = E_2 + \int (V_1(r) -V_2(r)) n(r) dr$ and $E_2 = \bra{\psi_2} H_2 \ket{\psi_2} = \bra{\psi_2} H_1 + V_2 - V_1 \ket{\psi_2} > E_1 + \int (V_2(r) -V_1(r)) n(r) dr$. This leads to $E_1 + E_2 > E_1 +E_2$ which is absurd. So, the one-electron density has to be different for the two Hamiltonians.}
    \item the functional that gives the energy of the system gives the ground state energy if and only if the density is the ground-state density.
\end{itemize}
This is the foundation of the density functional theory. If we define the universal density functional, 
\begin{equation}\label{eq:unive_func}
F[n] = \bra{\psi[n]} {E} + {U} \ket{\psi[n]},
\end{equation}
The variational principle can be written:
\begin{equation}\label{eq:var_dens}
E_0 = \underset{\text{n}}{\text{min}} \Big ( F[n] + \int V(\textbf{r}) n(\textbf{r}) d\textbf{r} \Big ).
\end{equation}

The final step is to map this interacting functional of energy onto $N$ non-interacting one-electron systems with an effective external potential $V_\text{eff}(\textbf{r})$ generating the same density n(\textbf{r}) as the real system. The state of such a system can be described as a Slater determinant of one-body wave functions $\phi_i(\textbf{r})$. Therefore, for the free-electrons (or Kohn-Sham) system, the density is $n(\textbf{r}) = \sum_i |\phi_i(\textbf{r})|^2 $ where $i$ browses all occupied orbitals and the ground-state energy is $E_0 = \underset{\phi}{\mathrm{min}} \Big ( \bra{\phi} T_0 \ket{\phi} + \bra{\phi} V_\text{eff} \ket{\phi} \Big ) $ With $T_0$ the kinetic energy of the free-electrons system. By decomposing $F[n]$ as $\bra{\phi} T_0 \ket{\phi}+ E_\text{Hxc}[n]$ with $E_\text{Hxc}[n] = \frac{1}{2}\int \int \frac{n(\textbf{r}_1) n(\textbf{r}_2)}{|\textbf{r}_1 - \textbf{r}_2 |}d\textbf{r}_1\textbf{r}_2 + E_\text{xc}[n]$, one can obtain:
\begin{equation}
   E[n] = T_0[n] + \frac{1}{2}\int \int \frac{n(\textbf{r}_1) n(\textbf{r}_2)}{|\textbf{r}_1 - \textbf{r}_2 |}d\textbf{r}_1\textbf{r}_2 + E_\text{xc}[n] + \int V(\textbf{r}) n(\textbf{r}) d\textbf{r}
\end{equation}
with $T_0[n] = \sum_i \int \phi_i^{*}(\textbf{r}) (\frac{-\nabla^2}{2})\phi_i(\textbf{r})d\textbf{r} $. As a result, $V_\text{eff}(\textbf{r}) = V(\textbf{r}) + \int \frac{n(\textbf{r'})}{|\textbf{r'} - \textbf{r} |}d\textbf{r}' + \frac{\delta E_\text{xc}[n]}{\delta n(\textbf{r}) }$.

\begin{figure}
    \centering
    \includegraphics[width = 0.6 \textwidth]{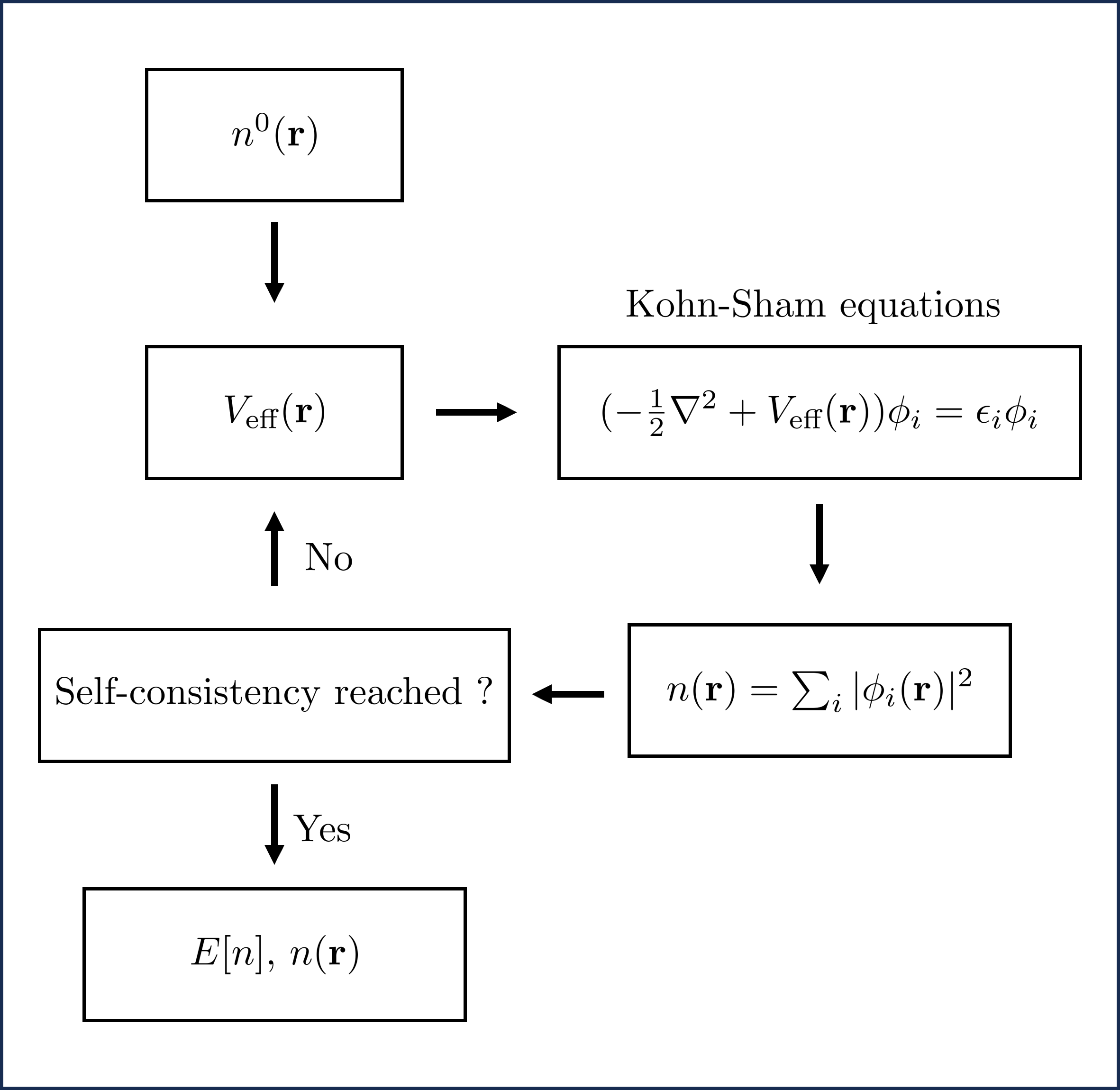}
    \caption{Self-consistency loop to solve Kohn-Sham equations.}
    \label{fig:DFT}
\end{figure}

We obtain the Kohn-Sham equations:
\begin{equation}\label{eq:KS}
\Big (-\frac{1}{2} \nabla^2 + V(\textbf{r}) + \int \frac{n(\textbf{r'})}{|\textbf{r'} - \textbf{r} |}d\textbf{r}'+ \frac{\delta E_\text{xc}[n]}{\delta n(\textbf{r}) } \Big )\phi_i(\textbf{r)} = \epsilon_i \phi_i(\mathbf{r}).
\end{equation}
This equation can be solved by a self-consistent loop described in Fig.~\ref{fig:DFT}. One starts from a first guess for the density $n^{0}(\textbf{r})$. From this density, one can calculate $V_\text{eff}(\textbf{r})$ and solve the Kohn-Sham equation, Eq.~\ref{eq:KS}. We then recover $n^{k}(\textbf{r})$ from wave-functions $\phi_i(\textbf{r})$ and we compare it to the previous density $n^{k-1}(\textbf{r})$. The loop goes on until self-consistency is reached. Everything in the DFT is exact except that $E_{xc}$ and $\frac{\delta E_{xc}[n]}{\delta n(\textbf{r}) }$ are unknown and must be approximated. I will give a few examples of common approximations used in this theory.

\subsection{The Local Density Approximation (LDA)}
A strong but simple approximation is to consider that $E_{xc}$ depends on the density the same way as an electron gas:
\begin{equation}
    E_\text{xc}[n] = \int \epsilon_\textbf{xc}(n(\textbf{r}))n(\textbf{r})d\textbf{r}
\end{equation}
and 
\begin{equation}
    \frac{\delta E_\text{xc}[n]}{\delta n(\textbf{r}) } = \epsilon_\text{xc} + n(\textbf{r}) \frac{\delta \epsilon^{\text{LDA}}_\text{xc}(n(\textbf{r}))}{\delta n(\textbf{r})}
\end{equation}
where $\epsilon_{xc}$ is the exchange-correlation term of a uniform electron gas of density $n(\textbf{r})$. This approximation relies on empirical results and on powerful transferability and universality properties of interaction, leading to successful results during the last decades.
It is easy to consider spinfull model with the Local Spin Density Approximations where densities of spins up $n_{\uparrow}(\textbf{r})$ and spins down $n_{\downarrow}(\textbf{r})$ are added in the term:
\begin{equation}
    E_\text{xc}^\text{LSDA}[n_\uparrow, n_\downarrow] = \int \varepsilon_\text{XC}(n_\uparrow, n_\downarrow) n(\mathbf r) \,\mathrm d \mathbf r
\end{equation}
\subsection{The generalized-gradient approximation (GGA)}
Another approach consists in considering gradients $\nabla n(\textbf{r})$, $\nabla^2 n(\textbf{r})$ and so on, of the density. whereas results can be worse than LDA in some cases (like describing gas-phase reaction barriers), it is used to describe metals and molecule–metal surface reactions \parencite{lazzeri_impact_2008, gerrits_density_2020}. In certain cases, LDA can have better results than GGA \parencite{barrera_lda_2005}.

\subsection{The LDA + U method}
The LDA method is blind to the orbital dependency of the Coulomb and exchange interaction. In \parencite{anisimov_first-principles_1997}, the authors proposed to add an interaction term, for localized electrons, driven by an interaction $U$ just as proposed by P.W Anderson \parencite{anderson_localized_1961}. This method gives a correct description of the Mott insulators (see Sec.~\ref{sec:mott}) and oxides such as NiO \parencite{anisimov_first-principles_1997,
bengone2000implementation} and U$\mathrm{O}_2$ \parencite{yu2009first,
shi2010optical,
dorado2013advances} which are materials of interest for EDF.

\subsection{Hybrid/metaGGA approach}
metaGGA methods includes a part in the exchange-correlation function which depends on the kinetic energy, and higher order of gradients of the density. In addition, one, can mix a Hartree-Fock (HF) and a Kohn-Sham exchange to obtain better results \parencite{toulouse_new_2002}.

\subsection{Beyond DFT}
In many DFT works, paramagnetic phases are modeled as non-magnetic which can lead to erroneous conclusions \parencite{abrikosov_recent_2016}. A realistic treatment of magnetism in materials can be crucial to perform a predictive description of some properties, like defect's energies in austenitic steels. Two orthogonal pictures were first proposed to simulate magnetism: 
\begin{itemize}
    \item the itinerant model,
    \item the local magnetic moment model.
\end{itemize}
The first one is based on the band theory of electrons whereas the latter assumes localized electrons on atoms which carry therefore a local moment. The answer lies in mixing these two approaches \parencite{abrikosov_recent_2016}. For instance, the spin dynamics is based on the equivalence between a strongly interacted system with onsite interactions and an electron living in a system with fluctuating charge and spin fields.
For instance, the Disordered Local Moment (DLM) model considers that the full electrons model does not cover its phase space in time. Indeed, each of the spins can flip after a time denotes the spin-flip time and points in more or less random direction.
An example of magnetic interactions for this type of model is the longitudinal spin-fluctuation Hamiltonian $H_\mathrm{lsf}$:
\begin{equation}
    H_\mathrm{lsf} =  N J^{(0)}(\overline{M}) + \sum_i J^{(1)}(\overline{M},M_i) - \sum_{i,j \neq i} J_{i,j}(\overline{M},M_i,M_j)) \mathbf{M}_i \mathbf{M}_j
\end{equation}
with $\mathbf{M}_i$ being the magnetization vector and $M_i$ its norm, $J^{(0)}(\overline{M})$ is the energy of the magnetically disordered system where the mean value of the local magnetic moment is $\overline{M} = 1/N \sum_i M_i$.
$J^{(1)}(\overline{M},M_i)$ is the energy needed to alter the value of the spin $i$ into $M_i$ from the imposed average value $\overline{M}$. Finally, $J_{i,j}$ is the interaction energy between two spins where all other local moments are disordered.

Many other approaches exist, \parencite{abrikosov_recent_2016}, and choosing the good magnetic model strongly depends on the problem.

\subsection{Successes and limitations}
The potency of the DFT relies on its simplicity and its polynomial complexity. When it first came out, the method exceeded all previous methods in terms of efficiency and duration of computation. Despite being quite simple, the LDA approximations have shown his efficiency and has confirmed a lot of experimental results \parencite{burke_perspective_2012}. Several improvements have been done such as Density Functional Perturbation Theory (DFPT) to study phonons in crystals and periodic materials \parencite{baroni_phonons_2001}.
However, several complex systems remain unreachable for DFT.
\begin{itemize}
    \item In practice, computing excited energies is still difficult with this method as it relies on variational principle which is efficient to find the ground  state. Therefore, computing gap in semiconductor or photovoltaic materials are still out of reach \parencite{burke_perspective_2012}.
    \item In batteries, complex phenomena can happen, especially for oxides (for instance Li-ion batteries or NiO materials \parencite{rohrbach_molecular_2005}) and \textit{ab initio} simulations struggle to recover good results corresponding to experimental data \parencite{birkl,ma_computer_2018}. Simple and empirical approximations of the exchange-correlation terms are not enough to encompass all phenomena observed experimentally, especially when electrons tend to be delocalized \parencite{burke_perspective_2012}.
    \item In magnetic materials, spins have a quantum behavior and the simulation of a paramagnetic material suffers from too many approximations \parencite{abrikosov_recent_2016}. An example is the austenitic steel in nuclear power plant: it has been shown \parencite{piochaud} that simulating paramagnetism with DFT (and molecular dynamics as DFT works at $0$ K) leads to find the antiferromagnetic phase the most stable one to study defects, whereas it is known the paramagnetic phase is the most stable. In that specific case, one must go further to properly study magnetic phases in materials.
\end{itemize}

\section{Quantum approaches}

\subsection{The Hubbard model}
Starting from first principles methods is not the only way to describe the quantum world.
Another approach to study strongly correlated electrons system is to describe it with quantum mechanics formalism. By considering second quantization \parencite{coleman_introduction_2015}, one can express states and observables in the Fock state, which is much more convenient when the number of particle is big (such as in materials).
Electrons are described by quantum mechanics and then tend to be delocalized to minimize their energy (wave behavior) but as charged particles, they repel one another and
thus try to be as localized as possible in order to avoid paying this potential energy price. The minimal way of describing this strive is the famous example of the toy model called “the Hubbard model”. The system is simple: we consider fermions (or bosons) on a lattice where particles have a probability to jump on nearest-neighbors sites (tight-binding) and interact locally when they see each other on the same site \parencite{hubbard}:

\begin{align}
    \begin{split}
    H_\text{Hubbard} =& \sum_{i,j,\sigma} t_{i,j} d_{i\sigma}^{\dagger} d_{j\sigma} + U \sum_{i} n_{i,\uparrow}n_{i,\downarrow}  \\& + \mu \sum_i (n_{i,\uparrow} + n_{i,\downarrow})
      \label{eq:Hubbard}
      \end{split}
\end{align}
where $t_{i,j}$ is the hopping term between sites $i$ and $j$, $d_{i,\sigma}^{\dagger}$ ($d_{i,\sigma}$) the fermionic creation (annihilation) operator following anticommutation rule $\{d_{i,\sigma}^{\dagger},d_{j,\sigma'}\} = \delta_{i,j,\sigma,\sigma'}$, $n_{i,\sigma} = d_{i,\sigma}^{\dagger}d_{i,\sigma}$, $U$ the onsite interaction (describing the Coulomb interaction) and $\mu$ the chemical potential.
This is the single-band Hubbard model (only one orbital per site). This model in 2 dimensions is conjectured to be describing (reasonably) high-temperature superconductors \parencite{anderson_resonating_1987}. The doping and several orbital cases are still active fields of research in 2D and 3D. Solving this model could lead to major discoveries and breakthroughs in fields of experimental and theoretical physics \parencite{noauthor_hubbard_2013}.

The ratio $U/t$ controls the importance of interaction energy over kinetic energy. If $U/t \rightarrow 0$, this means that we are dealing with free electrons and the Hamiltonian can be diagonalized in the Fourier space by setting $c_{k,\sigma}^{\dagger} = \sum_{i}e^{i.k.R_i}c_{i,\sigma}^{\dagger} $ such that $H_\mathrm{Hubbard} = \sum_{k} \epsilon_k c_{k}^{\dagger}c_{k} $. The many-body ground
state is a Slater determinant made up of these modes of energy $<\epsilon_f$ (Fermi energy) propagating with the dispersion relation $\epsilon(k)$.
On the other hand, if $U/t \rightarrow \infty$, the Hamiltonian is diagonal in the real space and eigenstates are of the form $\ket{n_{(0,\sigma)},\dots, n_{(i,\sigma)},\dots, n_{(N,\sigma)}}$ where $n_{i,\sigma} = 0$ or  $1$. In the case of half-filling, each site is populated with one particle and the first excited state (so one site double-occupied) costs an energy of $U$.
We can therefore observe two regimes: one where electrons can move freely such as in a metal and the other where the kinetic energy is negligible, and the system is an insulator.  
This describes what we call the Mott physics.

\subsection{The Mott physics: starting point of highly correlated electrons}\label{sec:mott}
Back in the early 1900s, band's theory was one of the greatest breakthroughs discovered thanks to the quantum theory. It is the starting point of transistors discovery which has led to the huge establishment of computers in our lives (for instance the one on which I am typing these lines) and the Moore's law \parencite{moore_cramming_2006}. Still, while band theory provided a good classification for many solids at the time, Verwey and de Boer \parencite{verwey_cation_1936}
discovered in 1936 that some materials such a nickel oxide (NiO), an oxide
with 3D valence electrons, were very poor conductors. From this postulate, Mott and Peierls \parencite{mott_discussion_1937} proposed in 1937 to hypothesize 
that in these materials, “the electrostatic interaction between the electrons prevents them from moving at all”,
explaining their insulating behavior. Mott’s paper was the opening remark of the field of strongly correlated
materials.

\subsection{The Dynamical Mean-Field Theory (DMFT)}
\paragraph{The mean-field approximation}
Let's consider the toy model of statistical physics, the Ising model:
\begin{equation}\label{eq:class_ising}
    H_\mathrm{Ising} = J \sum_{i,j} \sigma_i \sigma_j + h\sum_j \sigma_j\end{equation}
with $\sigma = \{-1;+1\}$ and $z$ the coordination number of the lattice. This model is exactly solvable in 1D and 2D \parencite{onsager_crystal_1944}. Nevertheless, a first approach to solve such a model is the single-site Mean-Field (MF) method \parencite{weiss_hypothese_1907}. This means that we only consider one site interacting with a MF obtained by a self-consistent loop. Mathematically, $\sigma_i \sigma_j \approx \langle \sigma_i \rangle \sigma_j + \langle \sigma_j \rangle \sigma_i - \langle \sigma_i \rangle \langle \sigma_j \rangle $. We can define $m = \langle \sigma \rangle$ and as a result, Eq.~\ref{eq:class_ising} can be written (neglecting constant terms):
\begin{equation}
    H_\mathrm{Ising} \approx (mJz+h) \sum_i \sigma_i.
\end{equation}

$h_\mathrm{eff} = mJz+h $ is called the Weiss field and must be found by the self-consistency condition $m = \mathrm{tanh}(\beta mJz+h)$. $m$ is the relevant degree of freedom, as known as the order parameter.
The dynamical mean-field theory is based on the same reasoning.
In this section, I describe the basis of the DMFT following \parencite{dmftgeorges}.
\paragraph{The Green function}
Treating interacting particles is very hard, and the complexity of such a system grows exponentially with the number of particle (or size of the system in the Fock space). We therefore need some general way of examining
the change of the system in response to external effects (potential, temperature…) even though we can not diagonalize the Hamiltonian.
One of the mathematical tools often used in many-body physics is the Green's function. It is defined as:
\begin{equation}\label{eq:green-function}
    \mathcal{G}(\textbf{r},\textbf{r}',t-t') = -i \bra{\phi} T d_{i,\sigma}(\textbf{r},t) d_{j,\sigma'}^{\dagger}(\textbf{r}',t')\ket{\phi}
\end{equation}
where $\ket{\phi}$ is the many-body groundstate and $\textbf{r}$, $t$ are space coordinates, time coordinates respectively. $T$ is the ordering operator, such as:
\begin{align}
    \begin{split}
         T d_{i,\sigma} (\textbf{r},t) d_{j,\sigma'}^{\dagger}(\textbf{r}',t') &= d_{i,\sigma} (\textbf{r},t) d_{j,\sigma'}^{\dagger}(\textbf{r}',t') \, (t > t')\\
          T d_{i,\sigma} (\textbf{r},t) d_{j,\sigma'}^{\dagger}(\textbf{r}',t') &= \pm d_{j,\sigma'}^{\dagger} (\textbf{r}',t')d_{i,\sigma}(\textbf{r},t) \, (t < t').
    \end{split}
\end{align}
In the case of periodic system, $\mathcal{G}$ only depends on $(\textbf{r}-\textbf{r}')$. In the Hubbard model, fermionic operators only depends on time and site number. We can define the \textit{local} Green's function at a given lattice site:
\begin{equation}
    \mathcal{G}^{\sigma}(\tau-\tau') = -\langle T d_{i,\sigma}(\tau) d_{i,\sigma'}(\tau')\ket{\phi}
\end{equation}
where $\tau$ is the imaginary time defined as $\tau = -it$.
In the classical mean-field theory, the local magnetization $m_i$ is represented as a single spin on site $i$ coupled to an effective Weiss field $h_\mathrm{eff}$. The reasoning is rigorously analog in the DMFT \parencite{dmftgeorges}. Let's consider the Hamiltonian of an Anderson impurity model, $H_\mathrm{And} = H_\mathrm{atom} + H_\mathrm{bath} + H_\mathrm{coupling}$ in which:
\begin{align}
    \begin{split}
        H_\mathrm{atom} &= U n_{\uparrow}^cn_{\downarrow}^c -\mu(n_{\uparrow}^c + n_{\downarrow}^c)\\
        H_\mathrm{bath} &= \sum_{l,\sigma} \epsilon_l a_{l,\sigma}^{\dagger}a_{l,\sigma}\\
        H_\mathrm{coupling} &= \sum_{l,\sigma} V_l (a_{l,\sigma}^{\dagger}c_{\sigma} + c_{\sigma}^{\dagger}a_{l,\sigma}).
    \end{split}
\end{align}
with $\mu$ the chemical potential of the impurity/atom.
This Hamiltonian is describing free fermions (a bath described by the $a_l^{\dagger}$'s) coupled (via the interaction $V_l$) to a single-site (the “impurity”). The idea is to consider that the Green's function of this model coincides with the local Green's function of the
lattice Hubbard model under consideration. The parameters $V_l$ and $\epsilon_l$ are only taken into account in the hybridization function 
\begin{equation}
    \Delta(i\omega_n) = \sum_l \frac{|V_l|^2}{i\omega_n - \epsilon_l}.
\end{equation}
The Weiss dynamical field (as known as the self-energy of the impurity model) of this method is then defined as:
\begin{align}
\begin{split}
    \hat{\sum}_{imp}(i\omega_n) &= \hat{\mathcal{G}}_{0,imp}^{-1} - \hat{G}_{imp}^{-1}(i\omega_n) \\
    &= i\omega_n + \mu - \Delta(i\omega_n) - \hat{G}_{imp}^{-1}(i \omega_n)
\end{split}
\end{align}
with $\mathcal{G}_0$ being the Green's function of the impurity model when $U=0$ and $G$ the interacting Green's function. Let's now consider the Green's function of the original lattice model (the Hubbard model):
\begin{equation}
    \hat{G}(\textbf{k},i\omega_n) = \frac{1}{i\omega_n + \mu -\epsilon_{\textbf{k}}-\hat{\sum}(\textbf{k},i\omega_n)}.
\end{equation}
The approximation relies now on considering that the lattice self-energy coincides with the impurity. This means that:
\begin{equation}
    \hat{\sum}_{i,i} \approx \hat{\sum}_{imp} \, \,\,, \, \hat{\sum}_{i\neq j} \approx 0.
\end{equation}
As a result, we obtain the self-consistency relation:
\begin{equation}
    \sum_{\textbf{k}} \frac{1}{\Delta(i\omega_n) + \hat{G}(i\omega_n)^{-1} - \epsilon_\textbf{k}} = \hat{G}(i\omega_n).
\end{equation}
\begin{figure}
    \raggedleft
    \includegraphics[width=1 \linewidth,trim={0.5cm 0.5cm 0cm 0cm},clip]{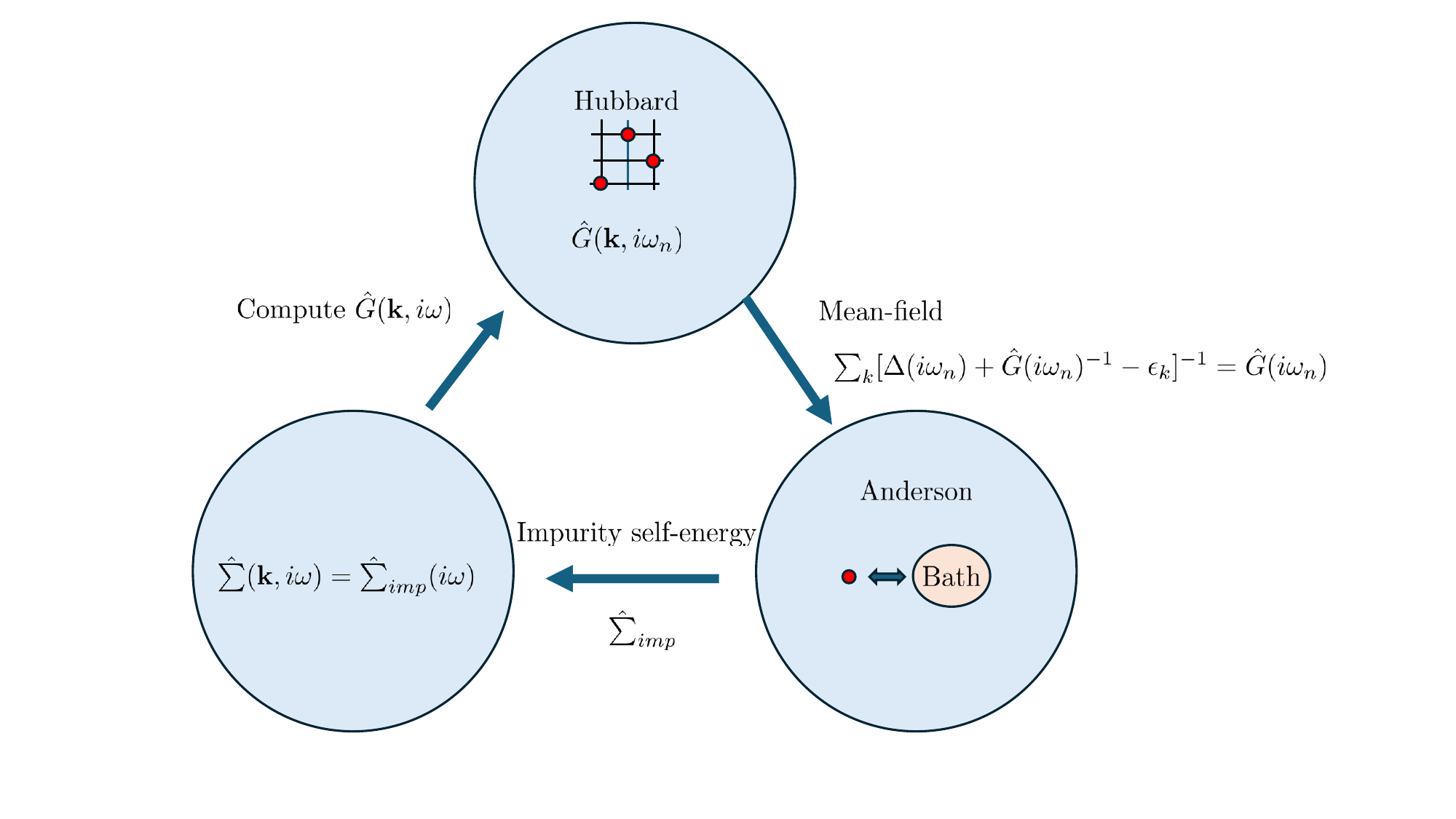}
    \caption{The DMFT loop (inspire from \parencite{ayral_nonlocal_2015})}
    \label{fig:DMFT}
\end{figure}
A schematic representation of DMFT loops is shown in Fig.~\ref{fig:DMFT}.
The DMFT is exact in the limit of infinite coordination lattices. This is also true for the mean-field in classical mechanics. It is also true in the extreme cases $U=0$ and $t_{i,j}=0$.
The DMFT has encountered many successes in the calculation of Mott transition (see \parencite{georges_dynamical_1996} for more details).
\subsection{Extensions of these methods}
\begin{figure}
    \centering
    \includegraphics[width=0.8 \linewidth]{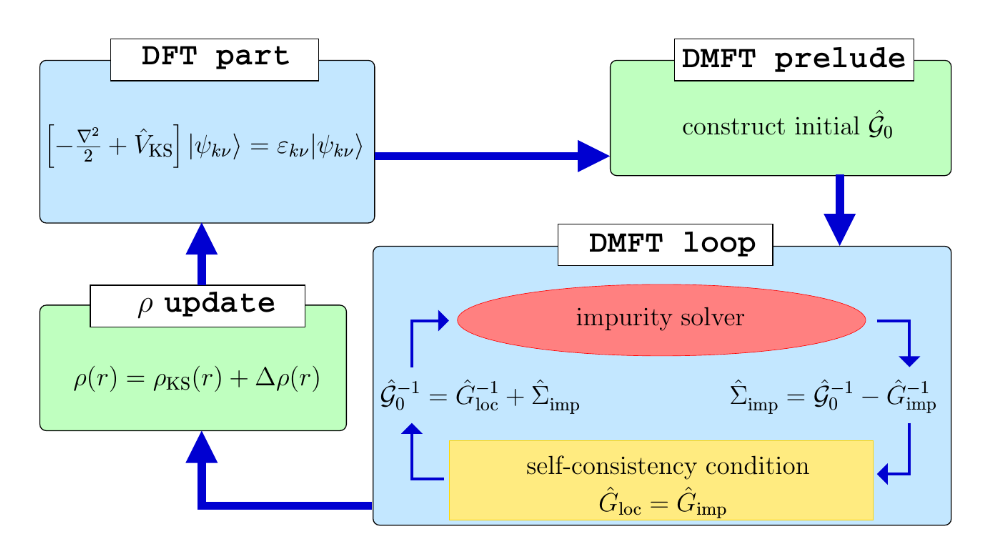}
    \caption{Double loop in the DFT+DMFT method. Taken from \parencite{lechermann_oxide_2018}.}
    \label{fig:dft+dmft}
\end{figure}

The major weakness of the DMFT is the lack of correlations due to considering a single site. The Cluster DMFT (CDMFT) has been developed to consider a cluster of sites coupled to a bath resulting in a better comprehension of Mott transitions\parencite{park_cluster_2008,misra_chapter_2008}. In addition, the DFT and DMFT can be combined  \parencite{biermann_dynamical_2005,haule_free_2015,park_computing_2014,doi:10.1146/annurev-matsci-070218-121825,lechermann_oxide_2018,sim2023intraatomic} which enables an efficient calculation of the total energy of realistic correlated electron systems (see Fig.~\ref{fig:dft+dmft}).

\subsection{Quantum Monte Carlo}
To end this section, I will talk a bit about Quantum Monte Carlo (QMC) methods. This approach aims at solving complex many-body problems by estimating explicitly the groundstate wave-function of the system. To this aim, a stochastic numerical integration is performed with the famous Metropolis algorithm \parencite{wessel20137,metropolis_equation_2004}. These algorithms have a polynomial complexity ($O(N^3)-O(N^4)$) and are widely used to solve and simulate spins model \parencite{sandvik_quantum_1991, stapmanns_thermal_2018}. Some works also study materials \parencite{wagner_transition_2007,esler_accelerating_2012}. A mix between \textit{Ab initio} approaches and QMC is possible \parencite{wagner_quantum_2014}. However, the famous sign problem arises for fermionic systems \parencite{pan_sign_2022} but also for frustrated spin systems \parencite{henelius_sign_2000, alet_sign-problem-free_2016}. QMC approaches are methods used for decades with a lot of successes but some strong limitations exist.

\subsection{Conclusion}
In this section, I have briefly described first principles methods, Green function methods and quantum Monte-Carlo approaches to many-body systems. These approaches are at the center of modern numerical methods to understand properties of materials at the atomic scale. During the last decades, these methods have contributed to understand and design complex phenomena in materials in academic and industry research. Investigations on improving them are still ongoing.
Nevertheless, describing correlations in strong correlated systems seems to be out of reach without approximations. A spark in the dark has appeared with the promise of quantum computing.

\section{Overview of quantum computing}
Whether quantum theory is exact or holds hidden variables has been a thorny debate among physicists in the early 1900s. Albert Einstein said that “gods don't play dice” and there was room for doubt. Nevertheless, the quantum theory was confirmed with many experiments during all the century until the final hit, the Aspect's experiment that has violated Bell's inequalities in 1982 \parencite{PhysRevLett.49.91}. Since this major breakthrough awarded by the Nobel Prize in 2022, researchers and industrials invest a huge amount of money and energy to build what we call a quantum computer. In this section, I describe the foundation of quantum computing principles and some advanced methods to use it. I also describe the impact of this research and the promise of quantum computing for materials and chemistry, which could change our society.

In "classical" computing, the smallest unit of information is stored in Bits "0" or "1". Thanks to transistors and their reduced size ($\approx$ few nm), it is possible to apply millions of operations per second in modern computers, such as the 'AND' or the 'XOR' operators. In the quantum world, the information is stored in vector states. The combination of such states allows performing this kind of operation to bits '0' and '1' at the same time. We are not dealing with bits anymore, but with qubits.
These two states are distinguishable, and we will use the Dirac notation in the following $\ket{0} =  \begin{pmatrix}
1\\
0
\end{pmatrix}$  and $\ket{1} = 
\begin{pmatrix}
0\\
1
\end{pmatrix}$ 
(equivalent to $\ket{\uparrow}$ and $\ket{\downarrow}$ or $\ket{g}$ and $\ket{r}$ in our notations). These states can be created via photons, excited states in atoms or superconducting devices. One of the main properties of these states is their dynamical behavior, being driven by the Schrödinger equation. It is then possible to apply unitary operators to evolve the system through time. We have therefore access to a new algebraic set of "quantum gates"  replacing the "classical gates" of qubit. In the following, $\hbar = 1$. Most of this section is inspired from \parencite{chuang}.

\subsection{Basis}
\begin{figure}
    \centering
    \includegraphics[width=0.6 \linewidth]{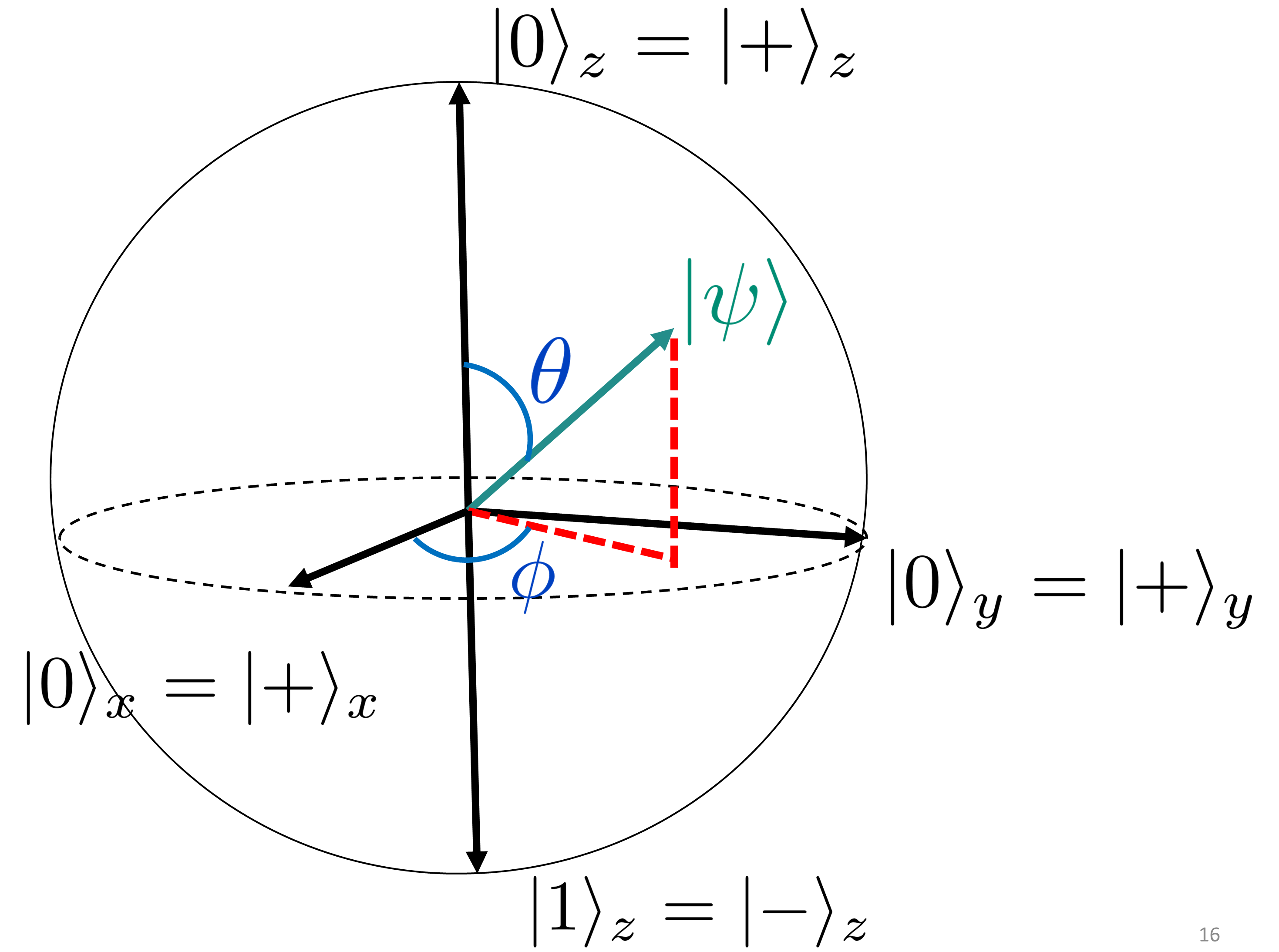}
    \caption{The Bloch sphere is a graphical tool to represent a qubit. A state $\psi$ lies on the surface of the sphere and can entirely be described by the two angles $\theta$ and $\phi$.}
    \label{fig:bloc_sphere}
\end{figure}

Let's start with a pure state. One can write the state of a qubit as:
\begin{equation}
    \ket{\psi} = a\ket{0} + b\ket{1}
\end{equation}
where $a$ and $b$ are complex coefficients and $|a|^2+|b|^2 = 1$. A qubit is always written in the $z$ basis. A qubit can evolve under the application of operators. A very important set of such operators are the Pauli matrices, 
\begin{equation}
   \mathcal{I} =   \begin{pmatrix}
    1 & 0 \\
    0 & 1 
    \end{pmatrix}, \quad
    S^x = \begin{pmatrix}
    0 & 1\\
    1 & 0
    \end{pmatrix}, \quad
    S^y = \begin{pmatrix}
    0 & -i\\
    i & 0
    \end{pmatrix},\quad
    S^z = \begin{pmatrix}
    1 & 0\\
    0 & -1
    \end{pmatrix}.
\end{equation}
It is a basis for the $2\times 2$ Hermitian matrices real vector. This means that any hermitian operator $U$ ($U^{\dagger} = U$) is a linear combination of these 4 matrices. They are also unitary ($U^{\dagger}U = \mathcal{I}$). 
Usually, a qubit is represented on a Bloch sphere, Fig.~\ref{fig:bloc_sphere}. Indeed, in the $z$ basis, any state can be written $\cos(\frac{\theta}{2})\ket{0}_z + e^{i\phi} \sin(\frac{\theta}{2})\ket{1}_z$ with $\phi \in [0, 2\pi]$ and $\theta \in [0, \pi]$ up to a global phase. As an example,  $\ket{0}_x$ (the eigenstate of $S^x$ with the corresponding eigenvalue $1$) can be read on the Bloch sphere
\begin{equation}
    \cos(\frac{\pi}{4})\ket{0} +  e^{0} \sin(\frac{\pi}{4})\ket{1} = \frac{1}{\sqrt{2}}(\ket{0}+\ket{1}) = \ket{0}_x
\end{equation}
but also $\ket{1}_y$ (the eigenstate of $S^y$ with the corresponding eigenvalue $-1$)
\begin{equation}
     \cos(\frac{\pi}{4})\ket{0} +  e^{-i\frac{\pi}{2}} \sin(\frac{\pi}{4})\ket{1} = \frac{1}{\sqrt{2}}(\ket{0}-i \ket{1})  = \ket{1}_y.
\end{equation}

\subsection{Many qubits system and entangled states}
Considering only one qubit is not very useful but it is possible to deal with several qubits (let's note $N$ qubits) by considering the tensor product of states. For instance, if we suppose that a system $A$ is in the state $\psi_A = a_{A} \ket{0} + b_{A}\ket{1}$ and a system $B$ is in the state $\psi_B = a_{B} \ket{0} + b_{B} \ket{1} $, we can describe the full system $AB$ by the separable state  \begin{equation}
\ket{\psi_A}\otimes\ket{\psi_B} = a_{A}a_{B}\ket{00} + a_{A}b_{B}\ket{01} + b_{A}a_{B}\ket{10} + b_{A}b_{B}\ket{11}.
\end{equation}
Separable means that the state of the whole system can be written as a tensor product of the state of the two sub-systems. 
One can generalize with $\ket{\Psi}_\mathrm{tot} = \bigotimes_{i=1}^{N} \ket{\psi}_{i}$ but also for any operator $\hat{O}_\mathrm{tot} = \bigotimes_{i=1}^{N} \hat{O_i}$.
What makes quantum physics distinct from other physics fields is the strange property of entanglement. If we go back to the two qubits system, it can be, for instance, in what are called Bell states:
\begin{equation}
    \frac{\ket{00}+\ket{11}}{\sqrt{2}} \,\,\,\, \text{or} \,\,\,\,  \frac{\ket{01}+\ket{10}}{\sqrt{2}}.
\end{equation}
These states can not be written as a tensor product $\ket{\psi_A}\bigotimes \ket{\psi_B}$. In addition, performing a measurement on one system imposes the state of the other system instantly.  This counter-intuitive concept is described by Albert Einstein as “a spooky action at a distance” and is at the center of the promises of quantum computing. One can remark that the amount of information to store to be able to fully describe a quantum state grows exponentially ($2^N$) with the number of qubit.

\subsection{Density matrix and mixed states}
The most general way of representing the state of a quantum system is with the density matrix representation:
\begin{equation}
    \rho = \sum_i p_i \ket{\psi}_i\bra{\psi}_i.  
\end{equation}
In this representation, we consider a statistical ensemble of $N_s$ system, each one being in a pure state $\ket{\psi}_i$. $p_i$ is the probability of finding the quantum system in a pure state $\ket{\psi_i}$ and $\sum_i p_i = 1$. This state is called a mixed state. Using density matrices and mixed states is useful when the state prepared is not fully known. For instance, if we consider two ensembles of one qubit with one ensemble in the state $\ket{0}$ and the other in the state $\ket{1}$, the corresponding density matrix will be:
\begin{equation}
    \rho = \begin{pmatrix}
        \frac{1}{2} & 0 \\
        0 & \frac{1}{2}
    \end{pmatrix}.
\end{equation}
The state described here is a fully mixed state.
Non-zero off-diagonal elements in the density matrix identify the presence of quantum coherence in a system. One interesting value to compute is the purity of the system, \textit{i.e} $P = Tr(\rho^2)$. If $P = 1$ the system is in a pure state and, on the other hand, if $P = \frac{1}{2^N}$, we have a fully mixed state. The density matrix representation is very useful to describe open system, as it is very difficult to know the state of the whole system: $(\mathrm{system}\bigotimes \mathrm{environment})$.

\subsection{Close system and quantum logic operators}\label{sec:close}
For a closed system, the evolution of a quantum state is driven by the Schrödinger equation, leading to:
\begin{equation}
    i \hbar \frac{\partial \ket{\psi(\tau)}}{\partial \tau} = H(\tau) \ket{\psi(\tau)}
\end{equation}
where $H(\tau)$  is the time dependent Hamiltonian of the system. A direct consequence is the unitary nature of any evolution operator acting on a close system (a unitary operator means $U^{\dagger}U = \mathcal{I}$). Therefore, only unitary operators can be used to evolve a quantum state. This can be interpreted as the norm preservation of a quantum state.
To know the state after the time-evolution, one must measure the state (so the operator $S^z$ in the $z$ basis) several times. Indeed, if the system is in the state $\frac{\ket{0} + \ket{1}}{\sqrt{2}}$, performing on measurement of $S^z$ one time will give $-1$ with probability $|a|^2 = 1/2$ and $1$ with probability $|b|^2 = 1/2$. One has to perform many measurement repetitions to obtain these probabilities. The mean value of $S^z$ is obtained by probability of mean value $-\frac{N_0}{N_\mathrm{m}}+ \frac{N_1}{N_\mathrm{m}} $ and a standard-deviation proportional to $1/ \sqrt{N_\mathrm{m}}$ where $N_0$ is the number of $1$ obtained after measurement, $N_1$ is the number of $-1$ obtained after measurement and $N_\mathrm{m} = N_0 + N_1$ is the total number of measurements. This is a direct consequence of the quantum projection principle (or collapse of the wave-function), and it is called the quantum projection noise as it brings errors to the computation of an observable (this will be called shot noise in the following).

The quantum computing is based on applying unitary operators or quantum logical gates (abbreviated as gates) on the system, just as in classical computing. The procedure to perform quantum computing is the following:
\begin{itemize}
    \item start with a register of $N $ qubits in a well-known initial state;
    \item a quantum circuit which carries all time-ordered gates to apply to the register is defined;
    \item at the end of the quantum circuit, a measurement is performed to obtain a set of $N$ classical bits (due to the quantum projection noise).
\end{itemize}

\begin{figure}
    \centering
    \begin{quantikz}[font=\large]
   \lstick{$\ket{0}$}&\gate{U_0}&\ctrl{1} &  &\\
\lstick{$\ket{0}$}&& \targ{} & \gate[2]{U_{1,2}} & \\
\lstick{$\ket{0}$}& & & &
\end{quantikz}
\caption{Example of a quantum circuit with three qubits. The initial state is $\ket{000}$}
    \label{fig:circuit}
\end{figure}
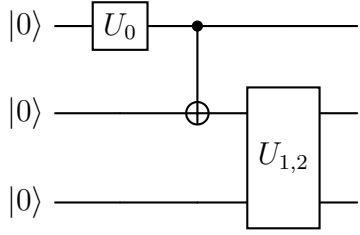

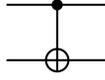
\begin{figure}
    \centering
   \begin{quantikz}[font=\large]
&\ctrl{1} & \\
& \targ{} &
\end{quantikz}
    \caption{Graphical, a representation of a CNOT gate in a quantum circuit}
    \label{fig:Cnot}
\end{figure}

An example of a quantum circuit is shown in Fig.~\ref{fig:circuit}.
A quantum circuit allows manipulating $2^N$ coefficients, whereas only $N$ bits are controlled at a time in classical computing. Single qubit operators $U = \begin{pmatrix}
    U_{11} & U_{12} \\
    U_{21} & U_{22}
\end{pmatrix}$ are then $2\times 2$ unitary matrices. It changes coefficient $a$ and $b$ in $\ket{\psi} = a\ket{0} + b\ket{1}$ into $a'$ and $b'$ such as:
\begin{equation}
    U\ket{\psi} = \begin{pmatrix}
    U_{11} & U_{12} \\
    U_{21} & U_{22}
\end{pmatrix} \begin{pmatrix}
    a \\ b
\end{pmatrix} = \begin{pmatrix}
    a' \\ b'
\end{pmatrix}, \,\,\,\, |a'|^2 + |b'|^2 = 1
\end{equation}
This means that any quantum evolution on  a quantum circuit is reversible (contrary to classical circuit).
More generally, applying a unitary matrix to a qubit is like moving the quantum state on the surface of the Bloch sphere. One performs rotations in all directions while conserving the norm. Thus, any arbitrary single qubit unitary operator can be written in the form:
\begin{equation}
U = e^{i \alpha} R_{\textbf{n}}(\theta)
\end{equation}
with $R_{\textbf{n}}(\theta) = exp(-i \theta (\textbf{n} \cdot \boldsymbol{S})/ 2) = \cos{\frac{\theta}{2}}\mathcal{I} - i\sin{\frac{\theta}{2}}(n_xS^x + n_yS^y + n_zS^z)$ and $\alpha$ a real number $\in [0, 2\pi]$. Here, $\textbf{n} = (n_x, n_y, n_z)$ is a real unit vector in three dimensions and $\boldsymbol{S} = (S^x, S^y, S^z)$.
Some very useful gates can be pointed out, such as the Hadamard gates (denoted $H$) or  the phase shift gate (denoted $R_{\phi}$): 
\begin{equation}
    H = \frac{1}{\sqrt{2}}\begin{pmatrix}
    1 & 1\\
    1 & -1
    \end{pmatrix} = i e^{-i \frac{\pi}{2}(S^x-S^z)/\sqrt{2}}, \,\,\,\,\ 
    R_{\phi} = \begin{pmatrix}
    1 & 0\\
    0 & e^{i\phi}
    \end{pmatrix}  =  e^{i \frac{\phi}{2}}e^{-i \phi S^z}
\end{equation}
Still, considering only one-qubit operators is not very useful as we can not create entangled states. Two-qubit gates can do the job. The idea is to make two qubits interact. For instance, the CNOT gate considers two input qubits: one is the control qubit and the other is the target qubit such as if the control qubit is in the state $\ket{1}$, the other qubit "flips" $\ket{0} \rightarrow \ket{1}$ and $\ket{1} \rightarrow \ket{0}$, otherwise it remains the same. The matrix representation of the CNOT gate is 
\begin{equation}
\mathrm{CNOT} = \begin{pmatrix}
    1 & 0 & 0 & 0 \\
    0 & 1 & 0 & 0 \\
    0 & 0 & 0 & 1 \\
    0 & 0 & 1 & 0
\end{pmatrix}
\end{equation}
and it is graphically represented in Fig.~\ref{fig:Cnot}.
Combining 1-qubit and 2-qubit gates can bring entanglement in the system. For instance, if the target qubit is in state $\ket{0}$ and the control qubit is in state $\ket{1}$ at the start of the circuit, applying a Hadamard gate to the control qubit and then a CNOT creates a Bell state for the target qubit $\frac{\ket{00}-\ket{11}}{\sqrt{2}}$. Some other 2-qubit gates are widely used, such as the $\textit{SWAP}$ gate 
\begin{equation}
    \textit{SWAP} = \begin{pmatrix}
       1 & 0 & 0 & 0 \\
       0 & 0 & 1 & 0 \\
       0 & 1 & 0 & 0 \\
       0 & 0 & 0 & 1
    \end{pmatrix}.
\end{equation}
We can also think of N-qubit gates, but they are very difficult to implement in practice. Hopefully, universal quantum gates sets exist, which means that only with specific finite size sets of 1-qubit and 2-qubit gates, one can approximately reproduce any unitary operation. The error on the approximation reduces to $0$ when the number of gates is infinite. For example, the rotation operator $R_{\textbf{n}}(\theta)$, the phase shift gate $R_{\phi}$ and the CNOT gate are commonly used to form a universal quantum gate set.

The quantum computing is then based on the evolution through time of a state of a closed system thanks to 1-qubit and 2-qubits unitary operators. In practice, two very different approaches exist \parencite{georgescu_quantum_2014}: the digital and the analog approaches.
\subsection{Digital approach}
Following the fact that any unitary operator can be (approximately) decomposed in product of 1-qubit and 2-qubit gates, it is possible to emulate the evolution of any Hamiltonian with a quantum circuit. This is the Digital Quantum Simulation (DQS). In the following, this Hamiltonian is called the \textbf{target} Hamiltonian. 
In quantum computing, the ansatz is a way to prepare a specified quantum state with a circuit. It encompasses the initial state preparation, but also the circuit itself. The purpose of a good ansatz is to reach the desire quantum state with the fewer gates or measurements as possible.

Preparing the initial state of an ansatz can be very difficult if some specific properties are needed. In the case of fermionic computation, it can be very useful to start with a state that respects antisymmetries in the quantum state. Usually, the state $\ket{0}^{\bigotimes N}$ is the easiest to prepare as it is the native state of the device, and it has been proven that one can prepare a fermionic state with all possible permutations with polynomial resources \parencite{abrams_simulation_1997}. Ancilla qubit can also be useful to prepare the initial state \parencite{wang_quantum_2011}. An ancilla qubit is a qubit acting in the circuit but which is not a qubit describing the original problem. It is added to the quantum circuit only to help the algorithm to run faster. It does not contain properties of the wanted state, but it helps in the quantum computation.

DQS is not easy because sometimes the number of gates needed grows exponentially with the size of the system but for most Hamiltonians with local terms, this number grows in a polynomial way \parencite{chuang}. Let us consider that the target Hamiltonian is a sum of constant local interaction terms
$H = \sum_l H_l$. Therefore, mimicking the evolution of $H$ in time means that the circuit must be as close as possible of $U(t) = \e^{-i\sum_l H_l t}$.
Two cases are possible:
\begin{itemize}
    \item Either, for all indices $l$ and $l'$, $[H_l,H_{l'}] = 0$ and $U = \prod_l e^{-iH_lt}$ and the decomposition in gates is direct.
    \item Either, and that is the vast majority of cases, $[H_l,H_{l'}] \neq 0$ and one need to use the Trotter decomposition. 
\end{itemize}
The idea is to cut the total time of evolution in infinitesimal slices, such as $U(t) \approx (e^{-i\sum_l H_l \Delta t})^{\frac{t}{\Delta t}}$.
Therefore, we obtain \parencite{chuang}
\begin{equation}\label{eq:Trotter}
    U(t) \approx \prod_{N_t = \frac{t}{\Delta t}} (\prod_l e^{-iH_l\Delta t})^{N_t} + O((\Delta t)^2)
\end{equation}
and when $\Delta t \rightarrow 0 $ we recover a product of local terms. Similarly, one can decompose $e^{i(A+B)\Delta t}$, where $[A,B] \neq 0$, into:
\begin{equation}
    e^{i(A+B)\Delta t } = e^{i A \Delta t /2} e^{i B \Delta t } e^{i A \Delta t / 2} + O((\Delta t)^3).
\end{equation}
In some cases, the goal is to reach a specific state with specific properties instead of mimicking the evolution of a specific Hamiltonian. The Shor algorithm \parencite{shor} is a perfect example of this. Details about this algorithm will be provided in Sec.\ref{sec:method}. For this type of algorithm, the QFT is very useful.
The idea is the same as for the classical discrete Fourier transform, where we take a vector of complex numbers $(x_0,x_1, \dots, x_{N-1})$ as inputs, and it outputs the transformed data as a vector of complex numbers $(y_0, y_1, \dots, y_{N-1})$ defined as:
\begin{equation}
    y_k = \frac{1}{\sqrt{N}} \sum_{j=0}^{N-1} x_j e^{2i\pi jk/N}
\end{equation}
The QFT is exactly the same transformation applying on an orthonormal basis. We define states $\ket{0}, \dots, \ket{n-1}$ with $\ket{0}=\ket{0}^{\bigotimes N}, \ket{1} = \ket{0}^{\bigotimes (N-1)} \bigotimes \ket{1}, \dots, \ket{n-1} = \ket{1}^{\bigotimes N}$ and $n = 2^N$. Considering a state $\ket{\psi} = \sum_{j=0}^{n-1} \alpha_j \ket{j}$
The QFT is the defined as:
\begin{align}
\begin{split}
    \ket{\psi} &\rightarrow \sum_{k=0}^{n-1} \beta_k \ket{k} \\
    \beta_k &= \frac{1}{\sqrt{n}}\sum_{j=0}^{n-1} e^{2i\pi jk/n} \alpha_j \ket{j}.
\end{split}
\end{align}
The digital approach is based on the fact that "anything" can be simulated with only a universal set of gates. The gates are applied with in an ordered-time sequence. The last step is the measurement of the state at the output of the circuit.

As described in previous sections (Sec.~\ref{sec:close}), measuring a quantum state means getting the eigenvalue $\lambda_k$ of a hermitian operator $\hat{O}$ (measurement tool) corresponding to the eigenstate $\ket{k}$ with the probability $|\bra{\psi}\ket{k}|^2$ where $\ket{\psi}$ is the state just before the measurement. If $\ket{\psi}$ is not an eigenstate of $\hat{O}$ but a linear combination of its eigenstates, $N_{m}$ measurements are needed to reconstruct statistically the value of $\langle \hat{O} \rangle$ with an error of $\approx 1/\sqrt{N_m}$. In general, Quantum State Tomography (QST) can be used to characterize a quantum state but it requires resources that grow
exponentially with the size of the system, making it inefficient for large quantum systems \parencite{dariano_quantum_2003}. The measurement can also be performed in the quantum circuit and not only at the end, that the case for protective measurements \parencite{choi_quantum_2020}, to perform Quantum Error Correction (QEC). The QEC for digital quantum computing is out of the scope of this thesis but it is a crucial part of digital quantum computating today \parencite{roffe_quantum_2019}.

The measurement procedure is therefore a major step of quantum computing, and a good measurement procedure is needed to reduce the cost of computation.

In practice, we do not have access to perfect qubits but to what is called \textbf{physical} qubits \textit{i.e.} qubits with errors. In classical computing, the system is repeated a great amount of time in parallel to avoid errors but in the quantum world, the no-cloning theorem \parencite{wootters_single_1982} prevents this method. Two types of error can appear in qubits:
\begin{itemize}
    \item the bit-flip error $\ket{0} \rightarrow \ket{1}$ and $\ket{1} \rightarrow \ket{0}$
    \item the phase-flip error $\frac{\ket{0}+\ket{1}}{\sqrt{2}} \rightarrow \frac{\ket{0}-\ket{1}}{\sqrt{2}}$ and $\frac{\ket{0}-\ket{1}}{\sqrt{2}} \rightarrow \frac{\ket{0}+\ket{1}}{\sqrt{2}}$.
\end{itemize}
\textit{A contrario}, \textbf{logical} qubits are noiseless qubits and can be directly used for computation. QEC allows to count the number of effective logical qubits that are available for an algorithm from the number of physical qubits available in the device. Another source of error is the gate fidelity and most importantly the fidelity of entangling gates. The fidelity $\mathcal{F}$ of a gate is a number between $0$ and $1$ which measures the capacity of an experimental gate to reproduce the theoretical expected state. If the fidelity is $\mathcal{F} = 0$, it always reproduces an orthogonal state whereas if $\mathcal{F} = 1$ (impossible in practice), it reproduces exactly the target state at each application.

\subsection{Analog approach}
\begin{figure}
    \centering
    \includegraphics[width=0.5 \linewidth]{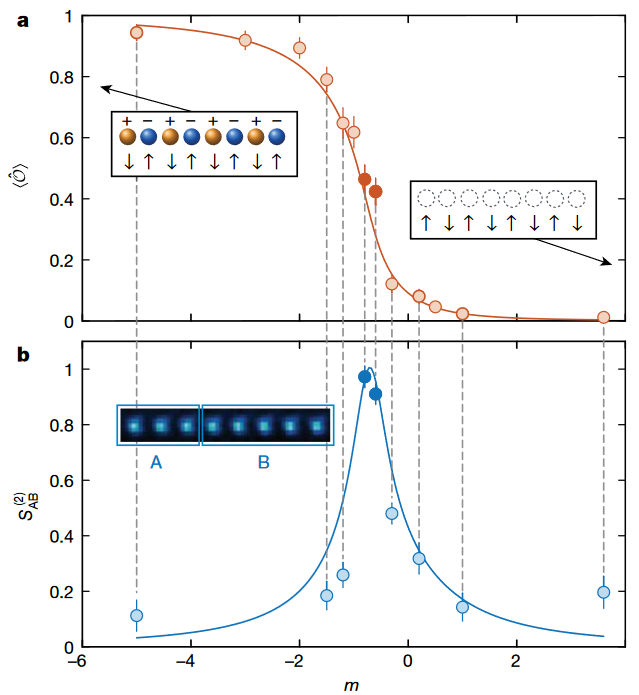}
    \caption{Quantum phase transition observed on trapped-ions device with Analog Quantum Simulation. (\textbf{a}) is the order parameter of the phase transition and (\textbf{b}) is the second-order Rényi entanglement entropy. Experimental data are represented by circles, and solid lines are theoretical results from exact diagonalization. The error bars are standard errors of the mean $\langle \hat{\mathcal{O}} \rangle$ and $S^{(2)}_\text{AB} =$ tr($\rho^2_\text{A})$ (with the A–B
bipartition shown in the inset), respectively. Taken from \parencite{kokail_self-verifying_2019} }
    \label{fig:VQS_kokail}
\end{figure}

That being said, a "perfect" quantum computer or simulator (without noise or with very efficient QEC) does not exist today and the number of qubits or gates is limited. As an example, the Shor algorithm would need several thousands logical qubits which means millions of physical qubits... which is far from being reached. John Preskill \parencite{preskill_quantum_2018} has defined this era as the Noisy Intermediate Scale Quantum (NISQ) era where only few number of noisy qubits are available and, unfortunately, DQS boils down to the number of logical qubits and the fidelity of gates. Nevertheless, some quantum devices can simulate quantum mechanics by quantum means. This means that the intrinsic physics of the device follows a quantum behavior and, therefore, can mimic another system of interest. The Hamiltonian of the system is called the \textit{resource} Hamiltonian and the Hamiltonian to mimic is still the \textit{target} Hamiltonian. Here, the challenge is to map the resource Hamiltonian onto the target Hamiltonian. It is not always possible, and in most cases, only a partial mapping can be done with, for instance, symmetries in common. An important advantage of Analog Quantum Simulation (AQS) is the fidelity of the simulation of the resource Hamiltonian. Usually, this type of device can control more qubits with a better fidelity than for digital quantum computing. In addition, even in the presence of noise, one can recover physical properties of interest. For instance, when looking at a quantum phase transition, the qualitative behavior with not all quantitative quantities can be enough. At first glance, AQS seems simpler than DQS, but it comes with major inconveniences. First, the mapping between the resource and the target Hamiltonian is not always direct. For instance, qubits are not fermions and therefore looking at properties of a fermionic system with a driven qubits Hamiltonian needs several steps. Second, AQS is way less flexible than DQS and not anything can be simulated. In most cases, states reached by AQS are not known and problems which are hard to describe with a physical Hamiltonian (such as the Shor algorithm) can not be solved with analog simulators.
An example of the success of AQS has been done on trapped-ions simulator \parencite{kokail_self-verifying_2019}. They have investigated the physics of the Schwinger model — a toy problem for lattice quantum electrodynamics by leveraging the similarity between the symmetries of a 20-ions quantum
simulator and those of the Schwinger model. The resource Hamiltonian is the 'XY' Hamiltonian
\begin{equation}
    H_\mathrm{R} = \sum_{i=1}^{N-1} \sum_{j=i+1}^{N} J_{i,j}(S_i^+S_j^-+S_i^-S_j^+) + B \sum_{i=1}^N S_i^z
\end{equation}
with $J_{i,j}$ the long-range antiferromagnetic interaction term. This term is controllable and $J_{i,j} \approx J_0/|i-j|^{\alpha}$ with $0<\alpha<3$. $B$ is an effective, uniform magnetic field. $S_i^+ = (S_i^x + iS_i^y)/2$ and $S_i^- = (S_i^+)^{\dagger} =  (S_i^x - iS_i^y)/2$.
On the other hand, the target Hamiltonian (Schwinger model) is 
\begin{equation}
    H_\mathrm{T} = w\sum_{j=1}^{N-1} (S_j^+ S_{j+1}^- + S_{j+1}^- S_{j}^+ )+ \frac{m}{2}\sum_{j=1}^N (-1)^j S_j^z + \overline{g}\sum_{j=1}^N L_j^2
\end{equation}
where $L_j$ depends on $S^z$. The crucial point, here, is the shared symmetries of the two Hamiltonians: in both cases, $ [H, S^z_\mathrm{tot}] = 0$. This helps to reduce the complexity of the "quantum circuit" and the number of parameters in the quantum computation (see Sec.~ \ref{sec:H2}). The result is the observation of a quantum phase transition up to 20 qubits despite all the noise and errors in the system (Fig.~\ref{fig:VQS_kokail}) with a divergent derivative for one observable and a divergence for the other. Because of the finite size system, these divergences do not go to infinity. Nevertheless, the behavior of the target Hamiltonian is reproduced at least qualitatively.

\subsection{Conclusion} Quantum computing can be performed in two different ways: 
\begin{itemize}
    \item A digital approach with the possibility to simulate any Hamiltonian at a cost of a high number of gates and qubits needed to reach an advantage.
    \item An analog approach leading to simulate with a better precision specific Hamiltonian feasible in practice.
\end{itemize}
The promises could be world changing or at least improve drastically the search of solutions for certain classes of problems and, hence, overcome classical limitations. Despite many public and private investments, it is far from a forgone conclusion that quantum computing will be inevitable one day because of noise and errors. A new field has therefore emerged: the hybrid algorithms. The quantum algorithms are divided in two parts: one quantum part where the quantum simulator or computer solves a very specific quantum problem and one classical part which helps the quantum part to reach his target. More globally, many hybrid and fully quantum algorithms are developed today to mitigate errors and being implemented on today's devices.
In the next section, I will present some of these methods with their success and also the improvements needed to reach the famous "quantum supremacy" or, more realistically, "quantum advantage".

\section{Methods}\label{sec:method}
In this section, I describe some of state-of-the-art methods of quantum computing. I also depict the slave-spin theory and how to implement it on a Rydberg Quantum Processor (RQP) leading to an hybrid quantum-classical algorithm.
\subsection{The quantum phase algorithm}\label{sec:QPA}
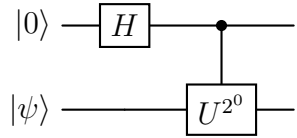
\begin{figure}[!h]
    \centering
   \begin{quantikz}[font=\large]
\lstick{$\ket{0}$}&\gate{H}&\ctrl{1} & \\
\lstick{$\ket{\psi}$}& & \gate{U^{2^0}} &
\end{quantikz}
    \caption{Quantum circuit for the quantum phase estimation at first order.}
    \label{fig:QPE_0}
\end{figure}

In many quantum problems, the goal is to find the eigenstates of an operator $U$. When $N$ is too big, classical computer struggles to diagonalize a $2^N \times 2^N$ matrix.
The quantum phase algorithm proposes to overcome this. The first thing to notice is that an eigenvector of a unitary operator is only defined by its phase as its modulus square is $1$. If $\lambda$ is an eigenvalue corresponding to the eigenstate $\ket{\psi}$ of a unitary operator, $\bra{\psi} U^{\dagger} U \ket{\psi} = \bra{\psi}\ket{\psi} \lambda^* \lambda = |\lambda|^2 $ and because $U^{\dagger}U = \mathcal{I}$, $ |\lambda|^2 = 1$. Therefore, any eigenstate is of the form $\lambda = e^{2i\pi \phi}$ with $\phi \in [0,1]$.
As $\phi \in [0,1]$, one can write it in binary notation $\phi = 0.\phi_1 \phi_2 \dots \phi_N = \phi_1\times 2^{-1} + \phi_2\times 2^{-2} + \dots \phi_N\times 2^{-N} $. Let's consider the circuit, Fig.~\ref{fig:QPE_0}. After the Hadamard gate, the state is $(\ket{0}\ket{\psi}+\ket{1}\ket{\psi})/\sqrt{2}$. Then the $C_{U^{2^j}}$ gate is applied: this means that if the control qubit is $\ket{1}$, the gate ${U^{2^j}}$, with $j$ an integer, is applied to $\ket{\psi}$, otherwise it is not. The power of $2^j$ will be explained later. Here, $j=0$ and hence, the state at the end of the circuit is $(\ket{0}\ket{\psi}+e^{2i\pi \phi}\ket{1}\ket{\psi})/\sqrt{2}$.

Let's now consider Fig.~\ref{fig:QPE_1}. It is straightforward to see that $U^2 \ket{\psi} = e^{2\pi(2 \phi)}$ and more generally, $U^{2^j} \ket{\psi} = e^{2\pi(2^j \phi)}$. In addition, $\phi = 0.\phi_1 \phi_2 = \phi_1\times2^{-1} + \phi_2\times2^{-2}$, $2\phi = \phi_1.\phi_2$ and therefore, $e^{i 2\pi (2\phi)} = e^{i 2\pi (\phi_1)}e^{i 2\pi (0.\phi_2)} = e^{i 2\pi (0.\phi_2)}$ because $\phi_1$ is an integer. More generally, $e^{i 2\pi (2^j\phi)} = e^{i 2\pi 0.\phi_{j+1}\phi_{j+2} \dots}$. Going back to the case $j=1$, the output state is $\Big ((\ket{0}+e^{2i\pi 0.\phi_1\phi_2}\ket{1}(\ket{0}+e^{2i\pi 0.\phi_2}\ket{1})\ket{\psi} \Big )/2$ if $\phi = 0.\phi_1 \phi_2$ is assumed. For an arbitrary $n$, the output state is 
\begin{equation}\label{eq:after_cu}
    \frac{1}{\sqrt{2^n}} \sum_{k \in \{0,1\}^n}e^{2i\pi(\phi\times k)}\ket{k} \ket{\psi}
\end{equation}
with $k$ a binary number.
 The final step is to consider the inverse Fourier transform ($\mathrm{QFT^{-1}}$) in the canonical basis $\ket{x}, \, \,\,\, x \in \{0,1\}^n$.
 The $\mathrm{QFT}$ is defined as 
 \begin{equation}
     \ket{x} \rightarrow \ket{x_\mathrm{QTF}} = \frac{1}{\sqrt{2^n}}\sum_{k \in \{0,1\}^n} e^{2i\pi (x \times k) /2^n} \ket{k}
 \end{equation}
and looks very much like Eq.~\ref{eq:after_cu}. Finally, applying the inverse Fourier transform leads to the state $\ket{\phi} = \ket{\phi_n}\ket{\phi_{n-1}}\dots \ket{\phi_1}$ which can be directly measured (ignoring the state $\ket{\psi}$ which remains the same). Thanks to this method, one can estimate the $n$ first digits of the phase and therefore the eigenvalue of any unitary operator. A strong asset of this method is the state $\ket{\psi}$ can be unknown. Indeed, one can decompose the state $\ket{\psi}$ in the basis of eigenvalue of $U$ and thus recover the result described above for all eigenstates.

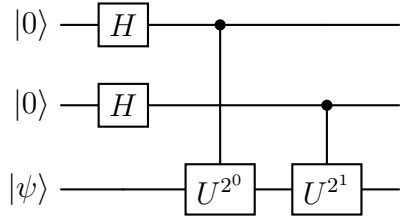
\begin{figure}[!h]
    \centering
   \begin{quantikz}[font=\large]
   \lstick{$\ket{0}$}&\gate{H}&\ctrl{2} &  &\\
\lstick{$\ket{0}$}&\gate{H}& &\ctrl{1} & \\
\lstick{$\ket{\psi}$}& & \gate{U^{2^0}}&\gate{U^{2^1}} &
\end{quantikz}
    \caption{Quantum circuit for the quantum phase estimation at second order.}
    \label{fig:QPE_1}
\end{figure}

\begin{figure}[!h]
    \centering
   \begin{quantikz}[font=\large]
   \lstick{$\ket{0}$}&\gate{H}&\ctrl{4} && & \ \ldots\ &  & & \gate[4]{QFT^{-1}} &\\
   \lstick{$\ket{0}$}&\gate{H}&&\ctrl{3} &  & \ \ldots\ & & &&\\ 
   \lstick{\vdots}  && \ \vdots\ & \ \vdots\ & \ \vdots\ &  \ \vdots \ &  \ \vdots \ &  & & \\
\lstick{$\ket{0}$}&  \gate{H} & && & \ \ldots\ & \ctrl{1} && &\\
\lstick{$\ket{\psi}$}& & \gate{U^{2^0}}&\gate{U^{2^1}} &  & \ \ldots\ & \gate{U^{2^j}} && &
\end{quantikz}
    \caption{Complete quantum circuit for the quantum phase estimation.}
    \label{fig:QPE_1}
\end{figure}
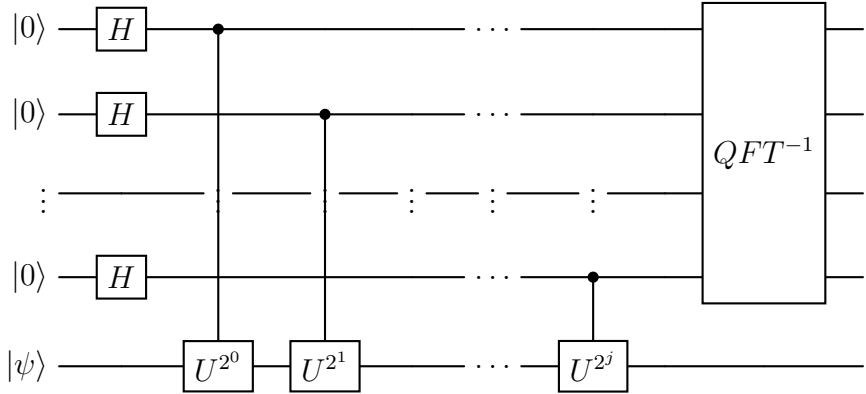

This algorithm and the QFT are at the heart of several algorithms which are candidate to show quantum supremacy : the Shor algorithm to decompose any number into prime numbers \parencite{shor} with polynomial resources, the HHL (stands for Harrow, Hassidim and Lloyd, the authors of the original article) \parencite{harrow_quantum_2009} algorithm which solves linear equations with a complexity of $O(\mathrm{log}(N))$ where $N$ is the number of variables in the system and the Grover algorithm \parencite{grover_fast_1996} for unstructured search that finds with high probability the unique input to a black box function that produces a particular output value. For instance, consider a function $f \{0,1, \dots, N-1\} \rightarrow \{0,1\}$ and you want to find out the relation between inputs and outputs of the function. Classically, the fastest way to do so is to test all possibilities, which means $O(N)$ measurement. The Groover algorithm proposes to find the solution with a probability higher than $1/2$ with only $O( \sqrt{N})$ measurements.

Despite being very promising, these algorithms need a lot of qubits and high fidelity gates that are not accessible today, and we do not know when they will be. The challenge is very hard and it is a long journey before being able to show a quantum advantage over classical algorithms but the reward would be to crack most communication encryption of our modern world, so maybe it is worth it. But it remains a bet. Only the number $15$ was decomposed into prime numbers thanks to the Shor algorithm \parencite{monz_realization_2016}. Because no one wants to wait several decades, some other methods have been explored.

\subsection{The Variational Quantum  Algorithm (VQA)}

\begin{figure}[!h]
    \centering
    \includegraphics[width=0.8 \textwidth]{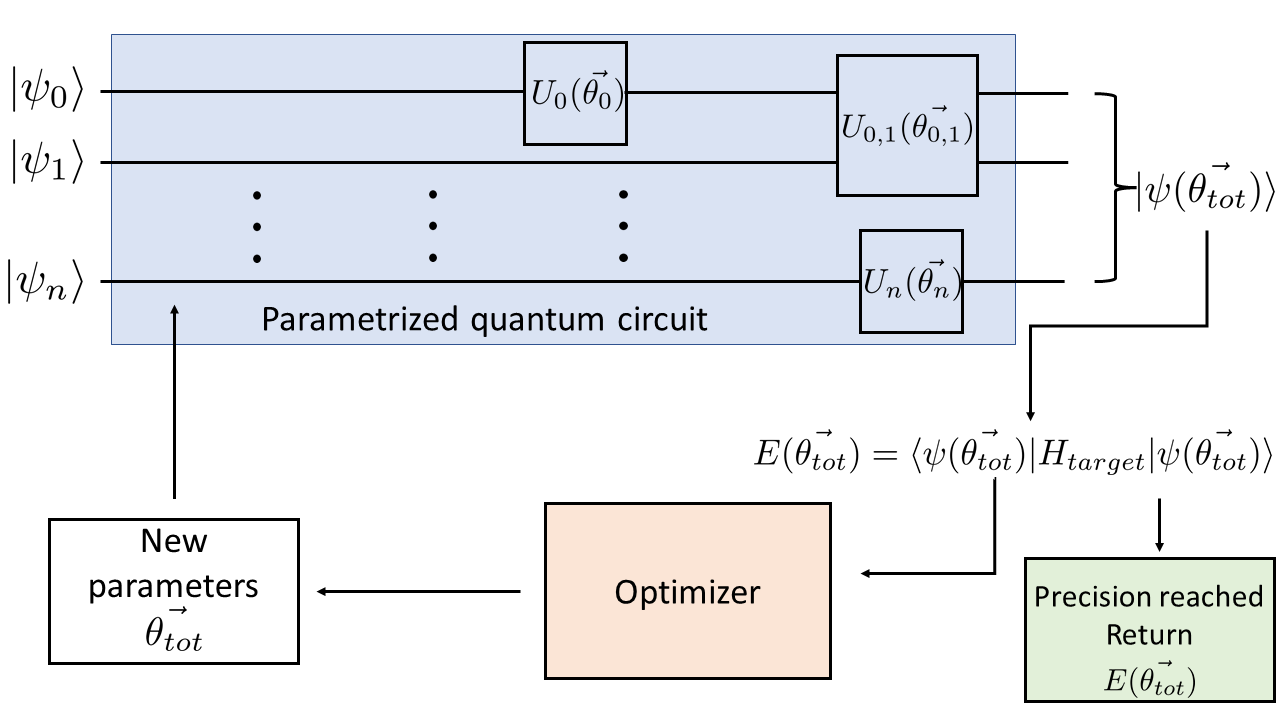}
    \caption{The Variational Quantum Algorithms loop. An initial state is prepared as an input. Then time-ordered parametrized (the parameters are regrouped in a vector $\mathbf{\theta}$) gates are applied to this state. The output is a vector state depending on parameters. This state is measure to obtain a bitstring and the target Hamiltonian is measured in this bitstring state. This procedure is repeated $N$ times to estimate the mean value $E(\mathbf{\theta}_\mathrm{tot}) = \langle H_\mathrm{T} \rangle$. Then a classical optimizer looks for new parameters to minimize this mean value. The procedure is done again with the new parameters. These loops are performed until a criteria is reached on the value of $\langle H_\mathrm{target} \rangle$ is reached or a fixed number of loops is exceeded. The final value of $\langle H_\mathrm{T} \rangle$ is expected to be very close to the real ground-state energy.}
    \label{fig:VQE}
\end{figure}

 As described in Sec.~\ref{sec:dft}, computing the groundstate of strongly-correlated many-body Hamiltonians would help researchers but also companies in energy, chemistry, health, drugs and therefore change the society. The good news is that, it often relies on finding the groundstate of such a Hamiltonian. 
 In all minds, quantum computing aims at exceeding classical computing. However, classical algorithms benefit from several decades of development, and they are far from being obsolete. The idea to combine both quantum and classical worlds was first proposed by Alberto Peruzzo $\textit{et al.}$ in \parencite{Peruzzo2014}. Such an algorithm which combines quantum and classical realms is called a hybrid quantum-classical algorithm. More specifically, one specific algorithm is well suited to find the groundstate of an operator: the Variational Quantum Algorithm (VQA). The principle is to prepare a state with a gate-parametrized quantum circuit. This means that parameters can be manually adjusted in gates of the circuit (such as angle $\theta$ in rotation matrix $R_{\textbf{n}}(\theta)$) to change the output as a function of parameters. The aim is to be able to span the most of Hilbert space possible with the different parameters. One starts with an input state and a unitary parametrized gate (called Parametrized Quantum Circuit PQC) $U(\mathbf{\theta})$ is applied to it. The state is measured at the end of the circuit, and the target operator (often the target Hamiltonian, noted as $H_\mathrm{T}$ in the following) is measured in this state by statistical repetition of the measurement. When the goal of VQA is to find the lower eigen value of an operator, it is often called the Variatonal Quantum Eigensolver (VQE). The key point is that now a classical optimizer looks for the parameters to minimize the measured value. Then the computation is done again with the new parameters and the loop goes on until a criterion is reached. At the end, the energy measured in the state obtained should be close to the real ground state energy of the target operator. A graphical example of the VQA algorithm is shown in Fig.~\ref{fig:VQE}. In the following, I describe each step of this algorithm and how to leverage them to increase the performance of VQA.

One of the main advantages of VQA methods is that they provide a general framework
that can be used to solve a wide array of problems. Indeed, the basic elements are always the same, whereas the structure can be different from a problem to another. VQA algorithm is an active field of research \parencite{mcardle_quantum_2020, Cao2019, Bharti2021, cerezo_variational_2021, mcclean_theory_2016} and have been performed on a lot of architectures \parencite{Peruzzo2014, kandala_hardware-efficient_2017, omalley_scalable_2016, nam_ground-state_2020, colless_computation_2018}. It is at the center of quantum computing today and the most promising way to show a quantum advantage in the next few years.

The first step of VQA is to encode our problem as a cost function to minimize $C$. This cost function is usually the mean value of a Hamiltonian:
\begin{equation}\label{eq:cost_function}
    \bra{\psi_\mathrm{init}}U(\mathbf{\theta})^{\dagger} H_\mathrm{T}  U(\mathbf{\theta})\ket{\psi_\mathrm{init}} = \langle H_\mathrm{T} \rangle_{\mathbf{\theta}}
\end{equation} 
where $\ket{\psi_\mathrm{init}}$ is the initial state and $U(\mathbf{\theta})$ the PQC describing the circuit.
Other cost functions can be chosen, such as the Gibbs objective function \parencite{li_quantum_2020}
\begin{equation}
    G = -\mathrm{ln}\langle e^{-\eta H_\mathrm{T}}\rangle
\end{equation}
where the parameter $\eta$ has to be tuned. But in the end, when $\eta$ is small, it reduces to the problem Eq.~\ref{eq:cost_function}. Another target of VQA can be a specific state, $\ket{\Psi}$ and then this algorithm is used to test the fidelity of a circuit. The cost function is then 
\begin{equation}
     \bra{\psi_\mathrm{init}}U(\mathbf{\theta})^{\dagger} \ket{\Psi}\bra{\Psi} U(\mathbf{\theta})\ket{\psi_\mathrm{init}} = | \bra{\Psi}U(\mathbf{\theta})\ket{\psi_\mathrm{init}}|^2.
\end{equation}
That is the case in \parencite{cerezo_cost_2021, havlicek_supervised_2019,
barison_efficient_2021} for specific problems. However, one needs to know the target state to use this cost function, and it is often not the case in materials where the groundstate is obviously not known.
The main goal of VQA is to find the set of parameters $\theta$ to reach:
\begin{equation}
    \mathrm{min}_{\theta} (C(\mathbf{\theta}))
\end{equation}
and therefore the ground-state energy of the target Hamiltonian through the variational principle (see \ref{eq:variationalprinciples}) in the specific case of a VQE algorithm for instance.

One has to ensure that the target Hamiltonian can be measured in the computational basis, \textit{i.e.} it is a sum of Pauli strings
\begin{equation}
  H_\mathrm{T} = \sum_{s=1}^{N_S} {\bf c}_s \bigg( \bigotimes_{j=1}^N S_j^{(s)}\bigg)
\end{equation}
with $N_S$ the number of Pauli strings and $N$ the number of qubits. $S_j$ are the Pauli matrices. If $H_\mathrm{T}$ is expressed this way, the measurement in the computational basis is straightforward. Unfortunately, the original problem is not always described with qubits and therefore, one must find a way to map the initial Hilbert space into the qubit Hilbert space. Depending on the problem considered, this step can be easy or can need a mapping (see Sec.~\ref{sec:chem}). Then an expectation value of $H_\mathrm{T}$ is the sum of expectation values of Pauli operators
\begin{equation}
     \langle H_\mathrm{T} \rangle_{\mathbf{\theta}} = \sum_{s=1}^{N_S} {\bf c}_s \bigg( \langle\bigotimes_{j=1}^N S_j^{(s)} \rangle_{\mathbf{\theta}}\bigg).
\end{equation}

The next essential building block is of course the quantum circuit or the ansatz. The common definition of all VQA algorithm is the state obtained at the end of the circuit:
\begin{equation}
    \ket{\psi(\mathbf{\theta})} = U(\mathbf{\theta}) \ket{\psi_\mathrm{init}}
\end{equation}
where $\ket{\psi_\mathrm{init}}$ is the initial state of the circuit. The initial state preparation strongly relies on the problem and the circuit. A good start offers better performance with for instance an initial state that follows problem symmetries or is easy to prepare on the quantum hardware. The choice of $U(\mathbf{\theta})$ will greatly impact the success of the VQA. From the perspective of the problem, the \textit{ansatz} influences both the convergence speed and the closeness of the final state to a state that optimally solves
the problem. On the other hand, one can have to take into account the hardware on which the VQA is performed: gates, parameters, errors, and even the method (digital or analog) can differ a lot between architectures. For instance, some entangling gates such as CNOT can have a great fidelity in some hardwares, but it would be better to choose the $\mathrm{SWAP}$ gate on another ones. In practice, most of the ansatz developed can be classified  as problem-inspired or hardware efficient, depending on their structure and
application.
\subparagraph{Problem-inspired ansatz}
The idea here is to draw from the target Hamiltonian to get the circuit. One example is the Trotter decomposition, Eq.~\ref{eq:Trotter}. Chemists have developed their own ansatz way before the advent of the VQA. It is called the \textit{Unitary coupled cluster ansatz} \parencite{bartlett_alternative_1989} and will be developer deeper in Sec.~\ref{sec:app} and Sec.~\ref{chemistry}. In a simpler manner, the variational Hamiltonian ansatz aims at reducing the number of parameters and accelerate the convergence \parencite{mcclean_theory_2016, progress_vqe}
by considering terms of the fermionic target Hamiltonian itself. For this purpose, we consider that $H_\mathrm{T} = \sum_l H_l$ and the ansatz is then
\begin{equation}
    U = \prod_{l=1} e^{i\theta_l H_l}.
\end{equation}
A very promising algorithm of the NISQ era is the Quantum Approximate Optimization Algorithm (QAOA) \parencite{farhi_quantum_2014}. It aims at approximate solutions to combinatorial optimization problems. The cost function is designed to encode a combinatorial problem by means of bit strings that form the computational basis. A key point of this algorithm is a theoretical guarantee of convergence when the depth of the quantum circuit increases. It consists of applying $p$ times two layers of non commutating operators $C$ (the operator of the cost function) and $M$ the mixing operator. Mathematically, the prepared wave function is of the form (taking notations from \parencite{dalyac_qualifying_2021}):
\begin{equation}
    \ket{Z_{\mathbf{\gamma},\mathbf{\beta}}} = e^{-i\beta_pM}e^{-i\gamma_p C}\dots e^{-i\beta_0 M}e^{-i\gamma_0 C}\ket{\psi_\mathrm{init}}.
\end{equation}
$(\mathbf{\gamma},\mathbf{\beta})$ are sets of parameters to optimize and $p$ is called the QAOA level or depth. The performance of the QAOA algorithm improves with this value in a perfect case (in the absence of noise). For satisfying constraints of the original problem or experimental device limitations, penalties can be added to the cost function. Nevertheless, it is often not sufficient and thinking on how to encode the constraints directly in the ansatz seems to be the most promising way \parencite{nguyen_quantum_2023}. This approach was implemented on a quantum emulator considering Rydberg atoms device on Maximum Independent Set (MIS) and max-k-cut problems with promising results \parencite{dalyac_qualifying_2021}. The two problems are drawn from the rapidly growing sector of smart-charging of electrical vehicles, in which EDF is strongly involved. The QAOA method can be compared with a Trotterisation of the annealing method (see \ref{sec:annealing}). One can generalize this method with layers of two unitary operators $U(\gamma)$ (phase-separation) and $V(\beta)$ (mixing) which do not necessarily emerge from the time evolution of a specific Hamiltonian. This is called the Quantum Alternating Operator Ansatz \parencite{hadfield_measurements_2020} (also QAOA in the literature). Following \parencite{Bharti2021}, we will abbreviate this ansatz as QuAltOA. In my work, I have tested this ansatz for chemistry problems with Rydberg atoms device and compared it with a new hardware-efficient digital-analog ansatz and I have shown that in this case, hardware-ansatz is more efficient.

\paragraph{Hardware-efficient ansatz}

Problem-inspired ansatz seems great theoretically with good symmetries between the problem and the unitary operator, reduced parameters number, better efficiency... But when it comes to implement it on a real quantum computer, reality is catching up to us. Indeed, each architecture can implement gates with more or less efficiency (some gates are impossible to implement on some devices) and errors, decoherence, limited fidelity impose to use hardware-efficient circuits. That is hardware-efficient ansatz. Even if the circuit is not well suited for the original problem, choosing gates that can be well implemented on a specific device can lead to better results than with a problem-inspired device. The first example of this is in \parencite{kandala_hardware-efficient_2017} where IBM chose to build a parametrized circuit with specific gates that can be well implemented on a superconductor device. Back then, it was one of the most impressive results of quantum computing for chemistry and spin system. Often, hardware efficient ansatz relies on only one or two entangling gates and just a few single-qubit gates. For the specific case of superconducting qubit, the lack of connectivity between qubits leads to choose specific gates that are not linked with the original problem. All the difficulty of studying VQA in The NISQ era is to balance the two ansatz. As explained in the previous section, it is possible to map a circuit with specific gates into another circuit with another gates, with the same results. Most of the studies take the problem-inspired ansatz and implement it with a hardware efficient circuit \parencite{nam_ground-state_2020,hempel_quantum_2018}. In addition, one can choose a specific architecture to solve a problem as  symmetries of implementable gates match with symmetries of the problem \parencite{kokail_self-verifying_2019}. In fact, it seems that choosing the architecture with respect to problem symmetries to reduce the complexity and length of the circuit is one of the major challenges of the NISQ era (and maybe beyond).

\paragraph{Classical optimization}
The quantum part of VQA is very important and challenging, but the optimization of parameters is also an active field of research. Thus far, the effect of noise and errors have only been considered for the quantum part, but the impact on the optimizer is huge. Even natural laws of physics themselves limit the efficiency of the optimization: one have to perform many measurements to obtain the mean value of an observable with a great precision. Thus, a good optimizer should try to minimize the number of measurements or function evaluations. Last but not least, it should be resilient to noisy values coming from limited fidelity, quantum decoherence and so on. In other words, the perfect optimizer should converge toward the minimum value of an observable with only few noisy sets of data which is not an easy task. Therefore, choosing a good optimizer is crucial \parencite{lavrijsen_classical_2020}. 
The first idea is to use gradient descent algorithm, which is a local based research algorithm. It is based on the postulate that one can have access to the derivative of the cost function $\frac{\partial C(\mathbf{\theta})}{\partial \theta_i}$. This value indicates the direction in which the objective function shows the greatest change \parencite{piskor_using_2022}. More sophisticated or free-gradient algorithms have been tried such as the genetic algorithm \parencite{wakaura_evaluation_2021,PhysRevA.107.042602} which is a global optimizer (it aims at finding a global minimum) at a cost of more function evaluations. Finding the good optimizer for different problems is still an active field of research \parencite{gacon_simultaneous_2021,bonet-monroig_performance_2023} that I will not develop here, but useful insights can be found in \parencite{progress_vqe, Bharti2021,cerezo_variational_2021}.

\paragraph{Measurements}

Once the state is obtained at the output of the circuit, one must measure the state in the most efficient way to facilitate the work of the quantum optimizer and reduced the number of total measurements. The most direct approach is to decompose the output state in the basis of the target operator $O$. To proceed, a unitary operation can be applied to project the state into the diagonal basis of the observable. Let's consider an example: we have a state $\ket{\psi(\mathbf{\theta})}_z$  of $N$ qubits at the output of the circuit, diagonal in the $z$ basis, and we want to know the mean value of the operator $\bigotimes^N Y$. One needs to perform a rotation at the end of the circuit to project the state in the $y$ basis:

\begin{equation}
    Y = R^{\dagger}_x(\pi/2) Z R_x(\pi/2)
\end{equation}

and 
\begin{equation}
    \langle Y \rangle = \bra{\psi(\mathbf{\theta})}R^{\dagger}_x(\pi/2) Z R_x(\pi/2) \ket{\psi(\mathbf{\theta})}.
\end{equation}

Therefore, the rotation $R_x(\pi/2)$ must be applied to all qubits at the end of the circuit. This example is quite simple, but let's now consider a general target Hamiltonian $H_\mathrm{T} = \sum_{s=1}^{N_S} {\bf c}_s \bigg( \bigotimes_{j=1}^N S_j^{(s)}\bigg)$ where $S$ can be either $S^z$, $S^x$ or $S^y$. For each Pauli string, a specific tensor product of rotations is needed. This means that the number of gates to apply to each qubit is proportional to the number of different Pauli strings in the Hamiltonian. In many-body and chemistry target Hamiltonian, the number of terms can grow rapidly and therefore the number of measurement can become considerable (keeping in mind that for each Pauli strings, $N_s$ measurements are needed to have an error of $1/\sqrt{N_s}$ on the mean value). For one Pauli string, the number of measurement samples needed to estimate $\bigotimes^N_j S_j $ with an additive error of at most $\epsilon$ with a failure probability of at most $\delta$ is bounded by Hoeffding's inequality \parencite{huang2019nearterm}
\begin{equation}
    N_s \geq \frac{2}{\epsilon^2}\mathrm{log}(\frac{2}{\delta}).
\end{equation}

The first idea is to group Pauli strings that can be measured simultaneously to minimize the number of measurements \parencite{kandala_hardware-efficient_2017,mcclean_theory_2016} but this can be a NP-hard problem. 

One of the most powerful methods  is the classical shadow method with randomized \parencite{huang_predicting_2020} and derandomized \parencite{huang_efficient_2021} measurement (details about this method is shown in \ref{derand}).

One advantage of VQA algorithm is the possibility to implement it with AQS \parencite{kokail_self-verifying_2019, PhysRevA.107.042602} and DQS \parencite{kandala_hardware-efficient_2017,nam_ground-state_2020}.

\paragraph{Limitations and challenges}

VQA algorithm offers a good alternative to full quantum algorithm and many promising results have yet been shown theoretically and experimentally in this paradigm. Yet, theoretical limitations (so not considering device limitations and errors) have been demonstrated recently. The first one is the Barren Plateau (BP) \parencite{mcclean_barren_2018}. In this article, it has been shown that the expectation value of the gradient of the cost function decays exponentially to zero as a function of the number of qubits. This is true for Randomized Parameterized Quantum Circuit (RPQC) but also for a wide class of reasonable parameterized quantum circuits. As a result, the parameters landscape is essentially flat. Hence, in a BP, one needs an exponentially large precision to resolve against finite sampling noise and determine a cost-minimizing direction. This phenomenon is true for free-gradient methods \parencite{Arrasmith_2021} and of course gradient-based methods. Even noise has been shown to be a source of BP regardless of the ansatz employed \parencite{wang_noise-induced_2021}.  This could annihilate any hope on the much anticipated quantum advantage, and methods to overcome this issue have to be found.

Other challenges remain, such as the expressibility of the ansatz (the capability of the parametrized quantum gates to explore all vector states of the Hilbert space) or the reachability (whether it is easy to find a quantum state that minimize the cost function). Even the effects of noise are still studied today \parencite{kubler_adaptive_2020, gentini_noise-resilient_2020,  fontana_non-trivial_2022}.

VQA algorithms pave the way to reach a quantum advantage. Combining High-performance computing (HPC) and Quantum Processor Unit (QPU) is a hot topic, and this could help to overcome difficulties described above. One can also think of parallelization of quantum device, just as the start-up Welinq proposes \parencite{welinq}.

Many challenges remain but in the NISQ era, variational algorithms seem to be the most promising way to show that quantum computing can surpass most advanced classical methods in specific problems.

\subsection{Quantum Annealing}\label{sec:annealing}
Another possible approach with NISQ is the Quantum Annealing (QA) \parencite{finnila_quantum_1994, kadowaki_quantum_1998, de_falco_numerical_1988}. The idea is to use quantum fluctuation to reach a specific state of a Hamiltonian. In practice, the Adiabatic Theorem (AT) is often used \parencite{albash_adiabatic_2018}. 
Starting from a well-known eigenstate of a realizable Hamiltonian $H_0$, one can slowly tune the parameters of the device to reach a more complicated Hamiltonian (the target Hamiltonian $H_\mathrm{T}$). The adiabatic theorem stipulates that the state of the system at the end of annealing is (approximately) the corresponding eigenstate of the target Hamiltonian (the eigenstate with the same level of excited energy). For instance, if the system is in the groundstate of $H_0$ at the beginning, it will be in the groundstate of $H_\mathrm{T}$ at the end. If it starts in the first excited state of $H_0$, it will end in the first excited state of $H_\mathrm{T}$ and so on for all eigenstates to the most excited state. Mathematically, a time-dependent parameter $s(t)$ is slowly tuned from $0$ to $1$
\begin{equation}
    H(s) = s(t) H_\mathrm{T} + (1-s(t))H_0 \,\,\,\, 
\end{equation}
where $s(t=0) = 0$ and $s(t=T_\mathrm{final}) = 1$ and $T_\mathrm{final}$ is the final time of the procedure.
 QA has been performed with DQS on the D-Wave machine \parencite{kairys_simulating_2020}. It attempts to solve problems in a particular form called Quadratic Unconstrained Binary Optimization
(QUBO) with the Ising Hamiltonian \parencite{lucas_ising_2014}. However, no quantum advantage or speed-up over classical simulations have yet been found \parencite{cho_quantum_2014,amin_searching_2015}.
AT is well suited for AQS if the problem can be translated into finding the groundstate of the resource Hamiltonian. A recent success of quantum annealing is with Rydberg Quantum Processor where an anti-ferromagnetic has been built up to 196 spins (or qubits) \parencite{scholl_quantum_2021} by dynamically tuning the parameters of the Hamiltonian quasi-adiabatically whereas numerical simulations struggle at calculating ground state of a quantum Ising Hamiltonian for more than 32 spins.

 The adiabatic theorem is not really flexible in terms of problems that can be tackled, but if a solvable problem with this method is found, it can provide good results and a potential path towards a quantum advantage (see Sec.~\ref{hubbard}).

\paragraph{Conclusion on methods}
We have seen the limitations of full-quantum algorithm such as QPE on today's device. Reaching a quantum advantage with the Shor algorithm is only conceivable in a distant future, hence, many other tools have been developed to run algorithms on NISQ computers.
One of them are hybrid algorithms, which combine a quantum part with few gates and qubits and a classical part helping the quantum part to reach its target. Despite having theoretical challenges that remain, hybrid algorithms are the most promising way to solve real-world problems. Materials and chemistry simulation could help to find room temperature superconductor, new drugs, new materials are to anticipate materials aging. In the next section, I present modern methods to apply quantum computing to solve many-body physics problems.


\section{Quantum computing for chemistry and many-body physics}\label{sec:app}
\subsection{From qubits to fermions}\label{sec:transform}
In atoms, electrons are assumed to be in a fixed potential created by nucleus (Born-Oppenheimer approximation). Therefore, we can describe them with a kinetic potential and the Coulomb interaction between electrons. In the second quantization, the electronic Hamiltonian is used to rewrite Eq.~\ref{eq:Hmat} as: 
\be
\label{eq:2nd-quant-ham}
 H = \sum_{p,q}h_{pq}a_p^{\dagger}a_q + \frac{1}{2}\sum_{p,q,r,s}h_{pqrs}a_p^{\dagger}a_q^{\dagger}a_ra_s.
\ee
where the coefficients $h_{pq}$ and $h_{pqrs}$ encode the spatial and spin configuration of each of the electrons and depend on the inter-nuclear and inter-electron distances $\mathbf R, \mathbf r$: 
\be\label{eq:2nd-quant-terms_main}
\begin{aligned}
    h_{pq}&=\int d \mathbf x \phi_p^* (\mathbf x)\left(-\frac{\nabla^2}{2}-\sum_i\frac{\mathcal Z_i}{|\mathbf R_i- \mathbf r|}\right)\phi_q(\mathbf x) \\
    h_{pqrs}&=\int d\mathbf x_1 d\mathbf x_2\frac{\phi_p^*(\mathbf x_1)\phi_q^*(\mathbf x_2)\phi_r(\mathbf x_1)\phi_s(\mathbf x_2)}{|\mathbf r_1- \mathbf r_2|}.
\end{aligned}
\ee

Next, we map the fermionic $a^{\dagger}$ operators acting on Fock states of $n$ orbitals to a Hilbert space of operators acting on spin states of $N$ qubits. This corresponds to the quantum processors' effective interaction Hamiltonians, quantum gates and measurement basis.
In my work, I have studied VQA algorithms applied to electrons system and so on, fermions. As it is well-known, fermions follow the Pauli principle and fermionic wave-functions have to be antisymmetrized to respect this fundamental principle. Qubits (or spin) do not have the same statistics, and the corresponding states are not naturally antisymmetrized. Hopefully, several methods have been developed to map qubits to fermions and vice versa. Here, I describe the ones I have used during my PhD: the Jordan-Wigner transform \parencite{JordanWigner1928} and the Bravyi-Kitaev transform \parencite{Bravyi2002}.
Let's consider a many-body electrons system. The annihilation $c$ and creation $c^{\dagger}$ operators follow an anticommutation rule $\{c_i,c_j^{\dagger}\} = \delta_{i,j}$. The system's Hamiltonian is written with this type of operator. For Pauli operators $\{S^i,S^j\} = 2 \delta_{i,j} \mathcal{I}$ where $i$ and $j$ $\in \{0,x,y,z\}$ and, therefore, it is not easy to describe the target Hamiltonian in terms of Pauli operators. To alleviate this issue, Pascual Jordan and Eugene Wigner proposed the following transformation \parencite{Seeley_2012}
\
\begin{align}
\begin{split}
    c_1^{\dagger} &= (S^-)\otimes I \otimes I \dots \otimes I\\
    c_2^{\dagger} &= \sigma^z\otimes(S^-)\otimes I \otimes I \dots \otimes I\\
    \vdots\\
     c_n^{\dagger} &= \sigma^z \otimes \sigma^z \otimes \dots \otimes \sigma^z \otimes (S^-).\\
     \end{split}
\end{align}
where $S^- = \frac{S^x+iS^y}{2}$ and ${S^-}^{\dagger}= S^+ = \frac{S^x-iS^y}{2}$.
Therefore, the relations 
\begin{equation}
    \{c_i,c_j^{\dagger}\} = \delta_{i,j}, \quad \{c_i,c_j\}=0, \quad \{c_i^{\dagger}, c_j^{\dagger}\}=0
\end{equation}
are well fulfilled. The main advantage of this mapping is its simplicity, but as a result, any fermionic operator becomes a Pauli string of $N$ Pauli matrices, and we lost locality properties of the original Hamiltonian. It is therefore really hard to emulate such a Hamiltonian as $H_\mathrm{T} \neq \sum_l H_l$. In addition, this transformation needs $O (N)$ operations on qubits to perform one fermionic operation.
In JW transform, the occupation stays local but parity of the wave-function is delocalized.

The \textbf{parity} mapping is the exact inverse \parencite{Seeley_2012}. In this mapping, the parity is stored locally. The idea is to consider the occupation number basis state (fermionic state) $\ket{f_n,f_{n-1},\dots,f_0}$ (in the following, all sums of binary variables are taken modulo 2) and apply the transformation:
\begin{equation}
    p_i = \sum_j[\pi_n]_{i,j} f_j
\end{equation}

where $n$ is the number of orbitals and $\pi_n$ is upper triangular matrix:
\begin{equation}
    [\pi_n]_{i,j} = 1 \, \,\, \mathrm{if} \,\,\, i<j \,\,\, , \,\,\, [\pi_n]_{i,j} = 0  \,\,\, \mathrm{otherwise}.
\end{equation}

Unlike the JW transform, one can not represent the creation and annihilation operators of a particle in an orbital $j$ simply by applying $S^{\pm}_j$ because one qubit does not store the occupation of an orbital.  It is needed to look at qubit $(j-1)$. If the qubit $(j-1)$ is in the state $\ket{0}$, then it represents accurately the occupation orbital $j$ and then one needs to apply $S^+$ to the parity basis to simulate $a^{\dagger}$ one the fermionic basis. On the other hand, if the qubit $(j-1)$ is in the state $\ket{1}$, the qubit $j$ inverts parity and the operator $S^+$ needs to be applied. The same reasoning applied for the annihilation operator. 
The equivalent of $S^{\pm}$ in the parity basis is:

\begin{equation}
    P_j^{\pm} = S_j^{\pm} \otimes \ket{0}\bra{0}_{j-1} - S_j^{\mp}\otimes\ket{1}\bra{1}_{j-1} = \frac{1}{2}(S^x_j \otimes Z_{j-1} \mp i S^y_j)
\end{equation}

Furthermore, creating or annihilating a particle in orbital $j$ implies a change in the parity data and we must update the cumulative sums for all $k > j$ by applying the operator $S^x$. It corresponds to add $1$ to all qubits $k > j$ modulo 2.

At the end of the day, the creation and annihilation operators in the parity mapping are written:

\begin{align}
\begin{split}
    a^{\dagger} &= S^x_n \otimes S^x_{n-1} \otimes \dots \otimes S^x_{j+1} \otimes P_j^{+}\\
    a &= S^x_n \otimes S^x_{n-1} \otimes \dots \otimes S^x_{j+1} \otimes P_j^{-}.
\end{split}
\end{align}
As it is clearly seen, we just replaced the loss of locality with matrices $Z$ in the JW transform with the loss of locality with matrices $X$ in the parity basis. Thus, the total operation to apply on each qubit is still $O(N)$.

In the middle of both transforms, there is the Bravyi-Kitaev mapping (BK) \parencite{Bravyi2002}. 
The idea is to store the parity and occupation non-locally. A qubit $j$ stores the occupation only if $j$ is even. If $j$ is odd, the corresponding qubit holds a partial sum of the occupation set of orbitals less than index $j$. The transformation is quite complicated and I refer to \parencite{Seeley_2012} for more details. What is important is the BK transform only needs $O(\mathrm{log}_2(N))$ transformation to go from fermions to qubits and therefore is the simplest transform from a digital circuit complexity point of view.

\subsection{Unitary coupled cluster ansatz}
The simplest approximation for a complete basis of Fock space for fermions is the set of Slater Determinants (Eq.~\ref{eq:Slater}). In other words, we approximate the eigenstates of the target Hamiltonian to be of the form:
\begin{equation}
    \ket{\psi} = \prod_j (a^{\dagger}_j)^{\gamma_j} \ket{00\dots 0}
\end{equation}
where $\gamma_j = 0,1$ and $a^{\dagger}_j$ creates a particle in the orbital $j$. The corresponding state with the JW transform is
\begin{equation}
    \ket{\psi}_\mathrm{JW} = \prod_j (S^x_j)^{\gamma_j} \ket{00\dots 0}
\end{equation}

Usually, the method to generate Slater determinants is the Thouless algorithm \parencite{thouless_stability_1960}. Starting from a Slater determinant $\ket{\psi_0}$ one can generate an ensemble of new determinants given by \parencite{google_ai_quantum_and_collaborators_hartree-fock_2020}:
\begin{equation}
    \ket{\psi(Z)} = e^{i\sum_{i,j} Z_{i,j} a^{\dagger}_i a_j }\ket{\psi_0}
\end{equation}
where Z is hermitian. In VQA algorithms \parencite{google_ai_quantum_and_collaborators_hartree-fock_2020}, this matrix is optimized with a PQC to minimize $\bra{\psi_Z}H_\mathrm{T}\ket{\psi_Z}$.

Thus far, we did not take into account correlations which are at the heart of the goal of quantum computing. Starting from the Thouless algorithm, one can extend it to gradually  insert excitation into the system. This is the Unitary Coupled Cluster (UCC) ansatz.

It is constructed from a parametrized cluster operator $T(\theta)$ which is a sum of particle-hole excitations of different orders:
\begin{align}
    \begin{split}
        T(\theta) &= T^1(\mathbf{\theta}) + T^2(\mathbf{\theta}) + \dots \\
        T^1(\theta) &= \sum_{i,j} \theta_{i,j} a^{\dagger}_i a_j\\
        T^2(\theta) &= \sum_{i,jkl} \theta_{i,jkl} a^{\dagger}_i a^{\dagger}_j a_k a_l.
    \end{split}
\end{align}
The sum is truncated due to the decrease impact of higher order terms.

The anstaz of the PQC is then:
\begin{equation}
    \ket{\psi(\theta)} = e^{(T(\mathbf{\theta})-T^{\dagger}(\mathbf{\theta}))}\ket{\psi_\mathrm{init}}.
\end{equation}

\textit{UCCSD} refers to the truncation after the double excitation and \textit{UCCSDT} to the truncation after the triple excitation. The next step is transforming the operators $T$ into qubits operators with methods describe in Sec.~\ref{sec:transform}. Finally, the Trotter decomposition can be used to decompose $e^{T(\theta)-T^{\dagger}(\theta)}$ into a sequence of gates. This method is very well suited for DQS. 

This method is quite old and has been proposed decades ago \parencite{helgaker_molecular_2014} and has been consequently widely applied. A good summary of implementations of the UCC ansatz can be found in \parencite{anand_quantum_2022}. These problems often concern small molecules, as this method is still benchmarked on NISQ computers.

Thus far, we only discussed the quantum part of the algorithm and how to leverage it for molecular and fermionic many-body problems but the problem can be directly simplified with chemistry and theoretical simplifications. For instance, molecular wave-functions in Eq.~ \ref{eq:2nd-quant-terms_main} have to be chosen. They often constructed from linear combinations of atomic orbitals, themselves considered as linear combination of Gaussian functions (STO-NG) basis. The more common ones are the STO-3G (for three gaussians) or STO-6G (for six gaussians) \parencite{hehre_selfconsistent_2003}. Therefore, choosing  3 or 6 functions can reduce or increase the complexity of the target Hamiltonian and thus, the efficiency of the VQA algorithm. But in the other hand, choosing a more accurate representation of orbitals leads to a better description of correlations in molecules or materials. The difficulty lies in finding the balance between chemical accuracy and VQA efficiency.
In addition, some low energy orbitals can be considered as "frozen" whereas high energy orbitals can be considered as unoccupied just as in \parencite{hempel_quantum_2018} where they only focused on a subspace of the orbitals in the lithium molecule. With all these approximations, they reduced the target Hamiltonian from 12 qubits and 200 Pauli strings to an effective Hamiltonian of only 3 qubits and 13 Pauli strings. 

In conclusion, efforts have to be done in optimizing the quantum algorithm but also in the chemistry side where approximations are really important to simplify the problem.

\section{Conclusion}
Simulating matters at atomic scale is at the center of much research since the advent of computing. It enables to anticipate the behavior of complex system and even creating new materials. Several methods (DFT, DMFT, QMC) have encountered great successes and are used all around the world for fundamental physics or applications. Recently, a new approach has emerged to help or even surpass these "classical" approaches: the quantum computing. 
Many methods theoretically exist to provide a quantum advantage on a quantum computer. Yet, noiseless computers with thousands of logical qubits and high fidelity gates do not exist and it is not even sure they will one day. What do exist today are NISQ device, with experimental proofs of their utility. Therefore, hybrid methods such as VQA have been developed to show a quantum advantage on these devices. There have been some successes but also failures with the discovery of Barren Plateau for instance. Today, a real quantum advantage on important problems for the society has not been found. The field of quantum computing is an experimental and theoretical exploration where industrial giants are involved with the promise to revolutionize the world and many architectures have their pros and cons. Among this emulation, one technology has proven a quantum advantage in theoretical physics problems: the neutral atoms (or Rydberg Quantum Processor RQP) \parencite{browaeys_many-body_2020}. Groundbreaking results have emerged from this technology, considering digital \parencite{evered_high-fidelity_2023,kalinowski_non-abelian_2023,cong_hardware-efficient_2022} or analog \parencite{scholl_quantum_2021,chen_continuous_2023,scholl_microwave_2022} quantum computing. RQP is one of the best candidates to simulate many-body physics. Regarding the goal of my PhD to perform quantum simulation of fermionic many-body physics, it has appeared to me that RQP is an excellent candidate to get closer to a quantum advantage in this field and, most importantly, being able to implement my algorithms on a real device soon.

\chapter{Quantum simulation with Rydberg atoms}\label{Rydberg}
\section{Forewords}
In this chapter, I give an overview of the experimental apparatus of the Rydberg Quantum Processor (RQP) (first section) and the protocol to perform digital and analog quantum computing (second section). Most of the contents of are adapted from \parencite{browaeys_many-body_2020,henriet_quantum_2020,scholl_simulation_2021}.
\section{Neutral atom arrays}

\begin{figure}[!h]
    \centering\includegraphics[width = 1 \linewidth]{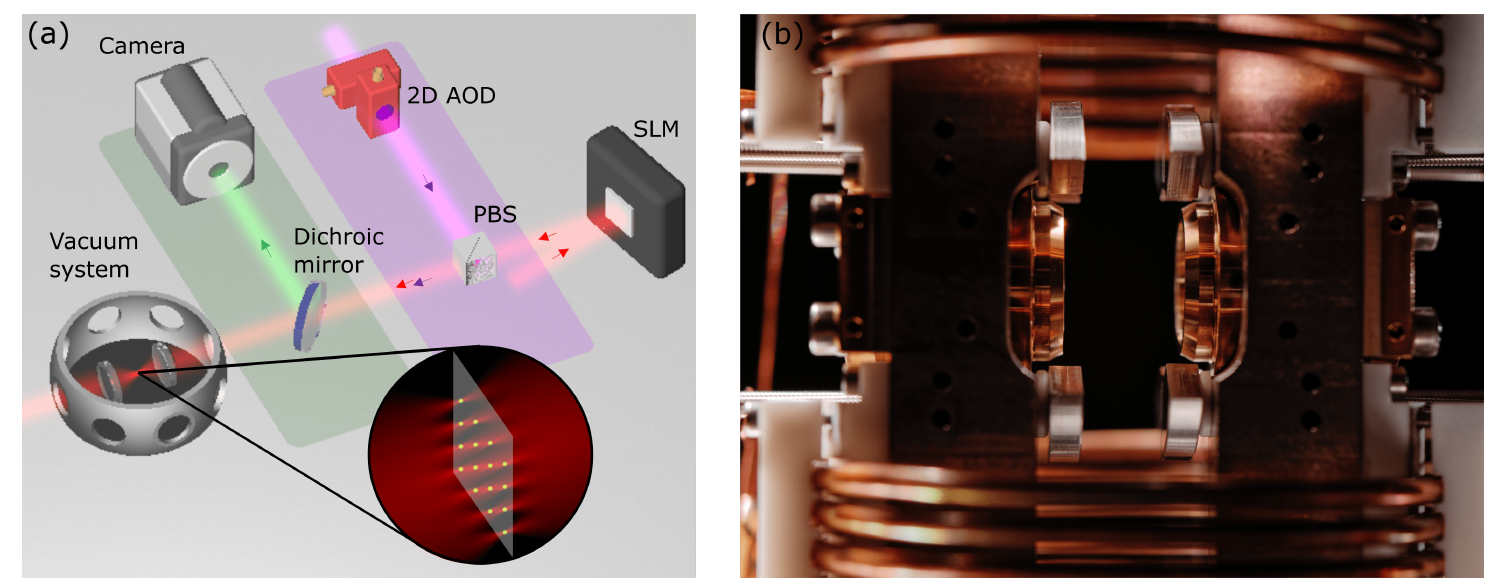}
    \caption{(a) Schematic representation of the main hardware components constituting the RQP. In red, the trapping laser is modulated by the Spatial Light Modulator (SLM) to produce microtraps \textit{i.e.} optical tweezers with the geometry desired (see inset). In purple, the moving tweezers allow reorganizing the atoms after the first fluorescence picture. They are controlled by a 2D acousto-optic laser beam deflector (AOD). In green, the fluorescence light is captured by a camera to obtain a bitstring. (b) Picture of the heart of experimental apparatus. The register is prepared at the center of the photography. Taken from \parencite{henriet_quantum_2020}.}
    \label{fig:schema_installation}
\end{figure}

\subsection{Initial state preparation}
\begin{figure}
    \centering
    \includegraphics[width = 1 \linewidth]{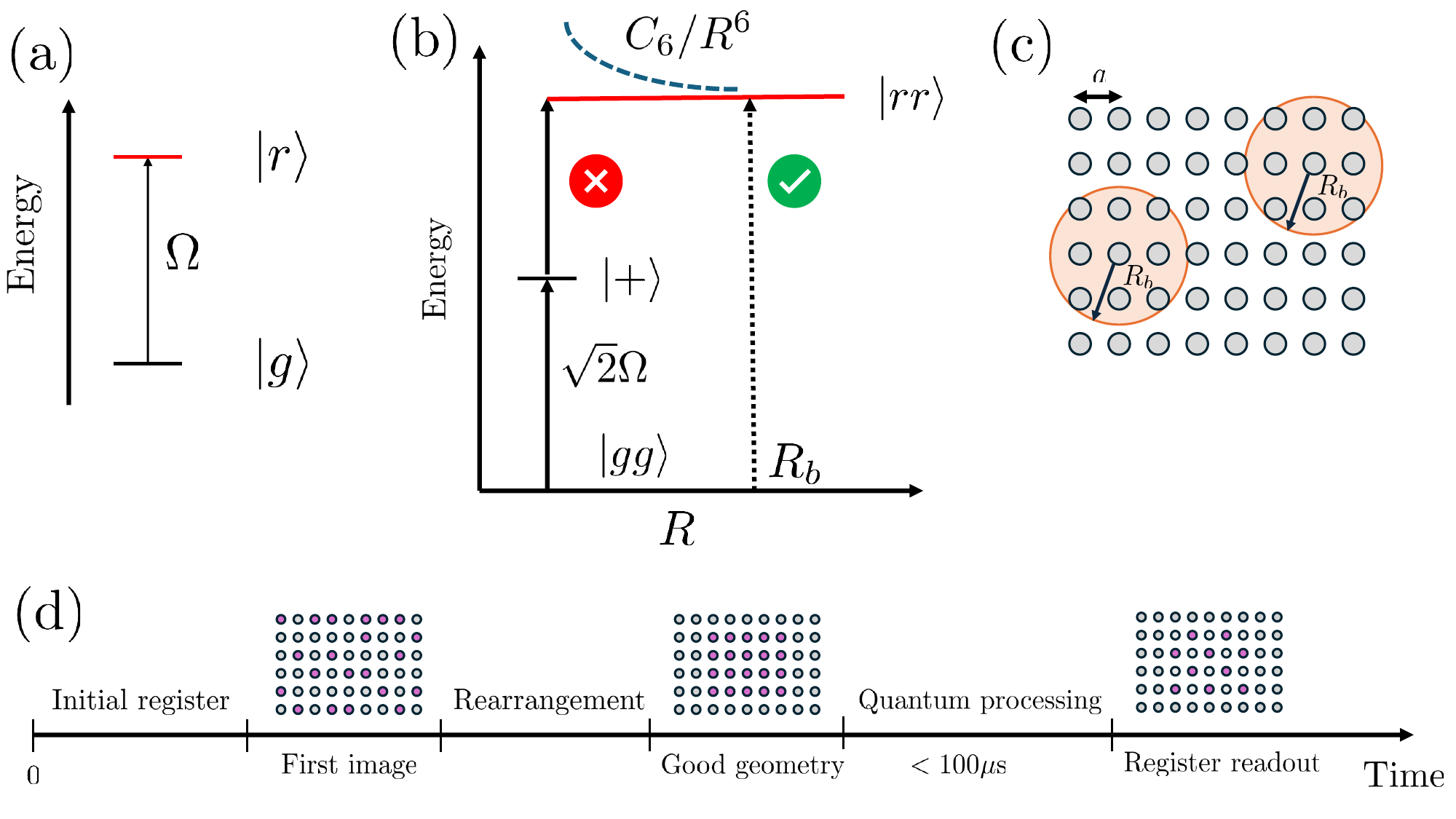}
    \caption{\textit{The Rydberg blockade and computation cycle} (a) The two chosen electronic states of atoms are coupled via the Rabi frequency $\Omega$. (b) Two atoms separated by a distance $R < R_b$ are only coupled to the state $\ket{+} = (\ket{rg} + \ket{gr})/\sqrt{2}$ and not $\ket{rr}$. If the distance is greater than $R_b$, the state $\ket{rr}$ is accessible. (c) Graphical representation of Rydberg blockade. A regular array with spacing $a$ is created with the SLM, each of the atoms prevent the atoms in the circle of radius $R_b$ from its center to be in the state $\ket{r}$ with it. (d) Temporal sequence of one computational cycle. The register is initialized with atoms trapped in a tweezer with a probability $1/2$. The atoms are then rearranged in the desired geometry. The quantum processing can now start and the cycle end with a picture of atoms: the register readout. All the process has to be done again for each measurement. The three first figures are inspired from Fig.~B1 of \parencite{browaeys_many-body_2020} and the last figure is inspired from Fig.~5B of \parencite{henriet_quantum_2020}.}
    \label{fig:rydberg_blockade}
\end{figure}

Quantum computing is by definition the manipulation of individual quantum objects (qubits) by experimental techniques. Quantum objects can take various forms such as gazes, dots, ions, photons, Josephson junctions and atoms. Until the end of the 1900s, controlling qubits with enough precision to entangle and apply logic operations on them was still a theoretical idea. Today, many platforms are able to control up to several hundreds of qubit at the same time and thus, give a hope to simulate quantum systems better than classical methods.
RQP is based on ensembles of individual atoms trapped in optical lattices. In this section, I give an overview of breakthrough methods which have led to being able to manipulates hundreds of neutral atoms individually and make them interact.

\begin{figure}[!h]
    \centering
    \includegraphics[width=1 \linewidth]{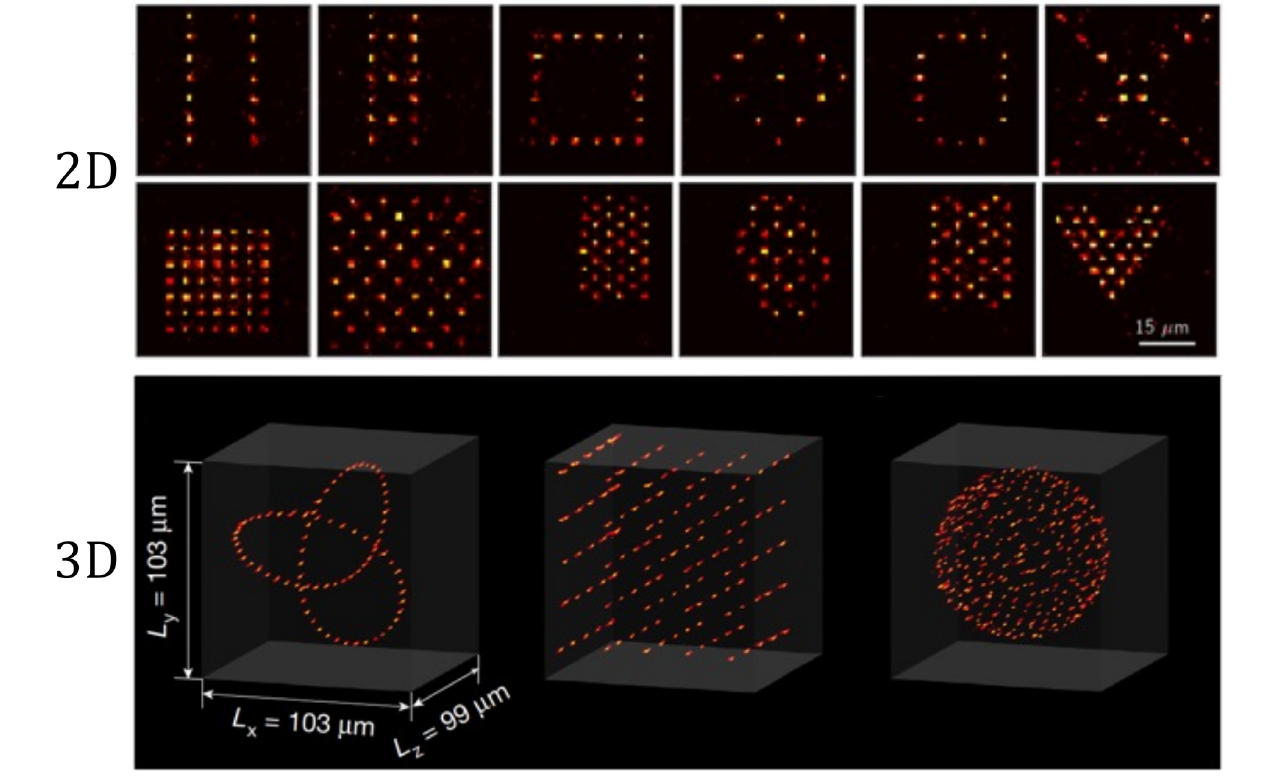}
    \caption{Examples of geometry in 2D and 3D experimentally feasible. Taken from \parencite{Barredo2018}.}
    \label{fig:rydberg_geo}
\end{figure}

RQP is fully controlled by light, from the state preparation and the set-up of the register to the quantum processing part and finally to readout the register. Each step needs different type of lights with different wavelengths and power. At the same time, electronic components and software stack monitor steps like rearranging atom tweezers or tune the magnitude of lasers achieving desired Hamiltonian.  
The first step to generate arrays of neutral atoms is to prepare a diluted atomic vapor inside an ultra-high vacuum system (to isolate atoms from the environment). To do so, one gram of solid rubidium (Rb) is heated up in an oven to transform it into a vapor. Contrary to other architectures such as superconductors, the device operates at room-temperature, leveraging the complications of a cooling system. At the output of the oven, atoms are slowed down to reach $\approx 10$ m/s.
The atoms of gas are then trapped inside a 3D magneto-optical trap (3D MOT). These atoms are the reservoir to a second trapping laser system which isolates each individual atoms, the so-called optical tweezers \parencite{schlosser_sub-poissonian_2001}. Two very useful characteristics of optical tweezer can be pointed out: at first, each tweezer contains at most 1 atom. This is ensured by the very fast light-assisted collision which happens if two atoms are in the trap. The number of tweezers and their arrangement in the 2D or 3D space is controlled by a Spatial Light Modulator (SLM) \parencite{nogrette_single-atom_2014} which is a "mask" that modulates the phase of the light into a desired intensity pattern (see Fig.~\ref{fig:rydberg_geo}). A strong asset of RQP is to fully control the geometry of the array and therefore connectivity between atoms (more details in the next section, Sec.~\ref{sec:spin_spin}). Nevertheless, the occupation of a trap is not deterministic: this means that a trap is filled with one atom with a probability of $p \sim 0.5$. To palliate this issue, a picture of atoms is taken by looking at their fluorescence and tweezers are moved one by one to arrange atoms in the desired way. This procedure has a success rate of $99\%$ and takes few tens of milliseconds. This is enabled by an algorithm calculating the optimal moves for each tweezer from the first picture. After all of this, the register is ready for the quantum processing. A schematic representation of the experimental apparatus is shown in Fig.~\ref{fig:schema_installation} with also a picture of the heart of the device.

\subsection{Generating spin-spin interactions}\label{sec:spin_spin}
In this subsection, I describe the physics behind the QRP and the tunable Hamiltonians that can be simulated thanks to this physics.
In order to perform quantum computing, atoms have to interact. To this aim, two electronic states are chosen to be the qubit $\ket{0}$ and $\ket{1}$: the groundstate and a high excited state of the atoms. A key asset of Rydberg atoms for quantum computing is the Rydberg blockade \parencite{saffman_quantum_2010}.The strong van der Waals interactions between two atoms in the Rydberg state can be used to prevent simultaneous excitations of two atoms. In an isolated atom, a laser field is coupling the ground state and the Rydberg state of an atom with a Rabi frequency, $\Omega$ Fig.~\ref{fig:rydberg_blockade}(a). In the case of two atoms separated by a distance $R$, they undergo a van der Waals interaction for the state $\ket{rr}$ and therefore there is a shift in energy by the quantity $C_6/R^6$ to reach this state. Thus, one can define the blockade radius defined by:
\begin{equation}
    R_b = (C_6/\hbar \Omega)^{1/6}.
\end{equation}
If the blockade condition is fulfilled $\hbar \Omega \ll C_6/R^6$ \textit{i.e.}  $R \ll R_b$, the state $\ket{rr}$ is non-resonant and the two atoms evolve from the state $\ket{gg}$ to the state $\ket{+} = (\ket{rg}+\ket{gr})/\sqrt{2}$ with a coupling $\sqrt{2} \Omega$ Fig.~\ref{fig:rydberg_blockade}(b). More precisely, it will oscillate between these two states at a frequency of $\sqrt{2}\Omega$, this is a Rabi oscillation. Depending on the geometry, it is possible to bring $N$ atoms to a fully entangled state $\sum_{i=1}^N \ket{g g \dots r_i \dots g}/\sqrt{N}$. If the system size is bigger than the Rydberg radius, some atoms can be excited and some not, leading to complex many-body states interesting for quantum simulation. Mathematically, the Rydberg blockade is written:
\begin{equation}
    \sum_{i,j} \frac{C_6}{R^6} n_i n_j
\end{equation}
where $n_i = (1+Z_i)/2$.
There is a second way to make atom interact. To do, two Rydberg states are chosen to be state $\ket{g}$ and $\ket{r}$. These two states are dipole-coupled and separated by a transition of typically 10 GHz. The dipole-dipole interaction leads to a coherent exchange  between the two Rydberg states. Contrary to van der Waals interaction, this scales like $C_3/R^3$ \parencite{barredo_coherent_2015,orioli_relaxation_2018}. In addition, $C_3 \propto 1-3\cos^2{\theta}$ where $\theta$ is the angle of the separation between the two atoms. Physically, the atoms reproduce a "XY" interaction, which is described by:
\begin{equation}
    \sum_{i,j}\frac{C_3}{R_{i,j}^3}(S^+_iS^-_j + S^-_i S^+_j) = 2\sum_{i,j} \frac{C_3}{R_{i,j}^3} (S^x_iS^x_j+S^y_iS^y_j)
\end{equation}

These two interactions (van der Waals and dipole-dipole) are the building blocks of quantum computing with RQP. Indeed, they depend on the distance between atoms, and they can bring entanglement in the system. As a result, depending on the geometry we choose, one can generate very complex many-body states. More details about Hamiltonian one can generate with this device will be given in Sec.~\ref{sec:analog_rydberg}.

The quantum processing is quite quick ($< 100$ $\mu$s) compared to the whole process which lasts around $200$ ms. After having done the quantum process, one needs to extract information on the system with a measurement.
\subsection{Register readout}
The last part of a quantum algorithm is the measurement of the system state. As we have seen before, it is crucial to optimize this step because it is the moment we extract the information from all the quantum processing sequence. In the RQP, it is quite "easy" because it only involves taking a final fluorescence image. Indeed, atoms in the state $\ket{g}$ will appear dark in the image, whereas atoms in the excited Rydberg state $\ket{r}$ will appear bright (see Fig.\ref{fig:rydberg_blockade}(d)). The efficiency of this method is of $98.6\%$ or more as reported in \parencite{fuhrmanek_free-space_2011}. One obtains a bit string from the readout register. After the measurement, all the process of register preparation, quantum processing and register readout has to be done again to obtain another bit string. As the total process takes $\approx 200-300$ ms, the device rate is of $2-3$ Hz and one can obtain several hundred measurements in a few minutes.
In the next section, we see how one can perform quantum simulation and computing with this device and the possible applications.
\section{Quantum simulation}
The quantum processing part is the one which changes depending on the problems and the algorithm we implement.  In this section, I will describe how to perform digital and analog quantum simulation on a RQP and the possibilities the platform can offer in these two paradigms.

\subsection{Digital quantum simulation}
DQS needs a great quantum coherence to work, \textit{i.e.} with the minimum of interaction with the environment. In addition, it has to be possible to perform at least single-qubit rotations and CNOT gates to be able to generate a universal gate set \parencite{chuang}. Hopefully, RQP is able to perform these gates. 

The first point is ensured by taken qubit $\ket{0}$ and $\ket{1}$ as the two hyperfine groundstates $F = 1$ and $F = 2$ of the rubidium atom. Indeed, they both have a very long lifetime (several years) and it therefore avoids quantum decoherence. This means that Rydberg states $\ket{r}$ are only intermediate states to perform gates, whereas they are used for the Rydberg blockade effect. 
Arbitrary rotation around the Bloch sphere can be performed by driving the qubit transition with a control field \parencite{yavuz_fast_2006, bluvstein_controlling_2021}. The laser-atoms interaction is characterized by the Rabi frequency $\Omega$, the detuning $\delta$ and its phase $\phi$. I will not go into experimental details, but these parameters are driven to perform rotations around $(x,y,z)$ axis with angles $(\tau\Omega \cos{\phi}, \tau \Omega \sin{\phi},\tau \delta)$ with $\tau$ the duration of the controlled sequence. Thus, any rotation around the Bloch sphere can be performed by controlling these four parameters. For instance, a pulse of area equal to $\pi$ where $\delta = 0$ will lead to a change $\ket{0} \rightarrow \ket{1}$. The same pulse on $\ket{1}$ will lead to $\ket{0}$ (the phase here is not important). This is the NOT gate:
\begin{equation}
    \mathrm{NOT} = \begin{pmatrix}
        0 & 1 \\
        1 & 0
    \end{pmatrix}
\end{equation}
It corresponds to half a Rabi oscillation and will also be called a $\pi$ pulse.
The Hadamard gate is performed by a $\pi$ pulse but where $\delta = \Omega$ and the phase $\phi = 0$. The fidelity of single-qubit gates is more than $99.5 \%$.

\begin{figure}[!h]
    \centering
   \begin{quantikz}
\lstick{$q_c$}&  &
\gategroup[2,steps=3, style={dashed, rounded corners, inner xsep=6pt, fill = blue!20}, background]{CNOT} & \gate[2]{CZ = \begin{pmatrix}
    1 & 0 & 0 & 0 \\
    0 & 1 & 0 & 0 \\
    0 & 0 & 1 & 0 \\
    0 & 0 & 0 & -1
\end{pmatrix}} &
 &  &\\
\lstick{$q_t$}&  & \gate{H} & & \gate{H} &  &
\end{quantikz}
    \caption{Decomposition of the CNOT gate into tensor product of Hadamard and CZ gates.}
    \label{fig:CZ_CNOT}
\end{figure}
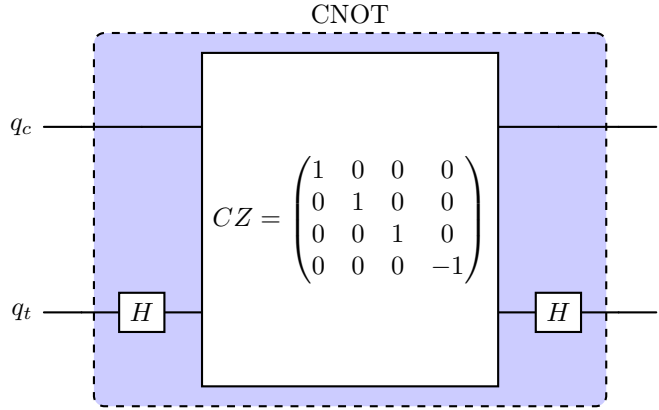

In order to perform a CNOT gate, one must decompose it into tensor products of Hadamard and CZ gates (Fig.~\ref{fig:CZ_CNOT}) because CZ gates are implementable with a specific sequence.

We consider two atoms in the state $\ket{0}$ or $\ket{1}$.
We consider that the state $\ket{1}$ is coupled with the state $\ket{r}$ whereas the state $\ket{0}$ is not. Finally, they are separated by a distance such that the Rydberg blockade condition is fulfilled. The protocol is then the following:

\begin{itemize}
    \item A $\pi$ pulse is applied to the control atom with a controlled phase.
    \item Then, a $2\pi$ pulse is applied to the target atom.
    \item Finally, a $\pi$ pule is applied on the control atom again.
\end{itemize}
Therefore, if the state is $\ket{00}$, these states are off-resonant and the state remains the same.

If the state is $\ket{10}$ or $\ket{01}$, only the qubit in the state $\ket{1}$ will undergo a $2\pi$ pulse with a phase $e^{i\pi} = -1$. If the state is $\ket{11}$, the first pulse will bring the control atom to the state $\ket{r}$ and because of the Rydberg blockade, the state $\ket{r}$ is off resonant for the target atom, and it stays in the state $\ket{1}$ after the $2\pi$ pulse. Finally, the last $\pi$ pulse swaps the state of the control atom to $\ket{0}$ again. The whole state picks up a factor $e^{i\pi}=-1$.

\begin{figure}[!h]
    \centering
\includegraphics[width = 0.7 \linewidth]{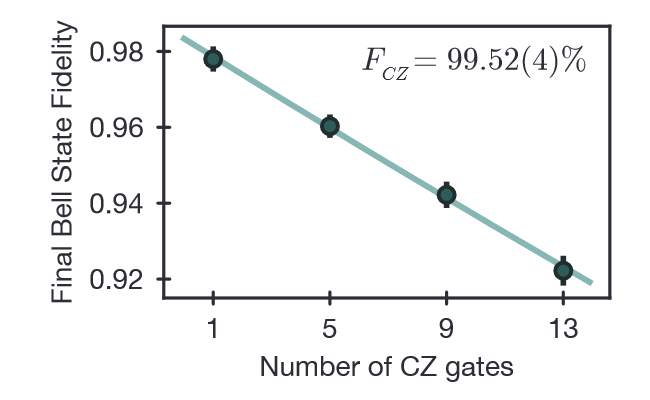}
    \caption{Fidelity of Bell's state as a function of number of CZ gates. Error bars represent 68$\%$ confidence intervals. Taken from \parencite{evered_high-fidelity_2023}.}
    \label{fig:CZ_lukhin}
\end{figure}

Finally, we have applied the gate:
\begin{equation}
    \begin{pmatrix}
        -1 & 0 & 0 & 0\\
         0 & -1 & 0 & 0 \\
         0 & 0 & -1 & 0 \\
         0 & 0 & 0 & 1
    \end{pmatrix} = e^{i\pi} \mathrm{CZ}
\end{equation}
which is the CZ gate up to a phase. We will call this method the $\mathrm{H-CZ CNOT}$ method following \parencite{isenhower_demonstration_2010}
Other protocols can be used to generate a CNOT gate \parencite{isenhower_demonstration_2010}. For a long time, the measured fidelity of the CNOT gate built this way was the major issue ($94.1 \%$ in 2018 \parencite{levine_high-fidelity_2018}) but recently, the Lukin's group has shown a fidelity of $99.5 \%$ on up to $60$ atoms in parallel, surpassing the surface code threshold for
error correction \parencite{evered_high-fidelity_2023} (see also Fig.~\ref{fig:CZ_lukhin}). Theoretically, achieving error correction is also at the center of interest for gates with RQP \parencite{jandura_optimizing_2023,ma2023highfidelity}. The high connectivity of Rydberg atoms and the growing fidelity of entangling gates pave the way toward digital quantum computing on a Rydberg device.

\subsection{Analog quantum simulation}\label{sec:analog_rydberg}
\begin{figure}[!h]
    \centering\includegraphics[width = 0.9 \linewidth]{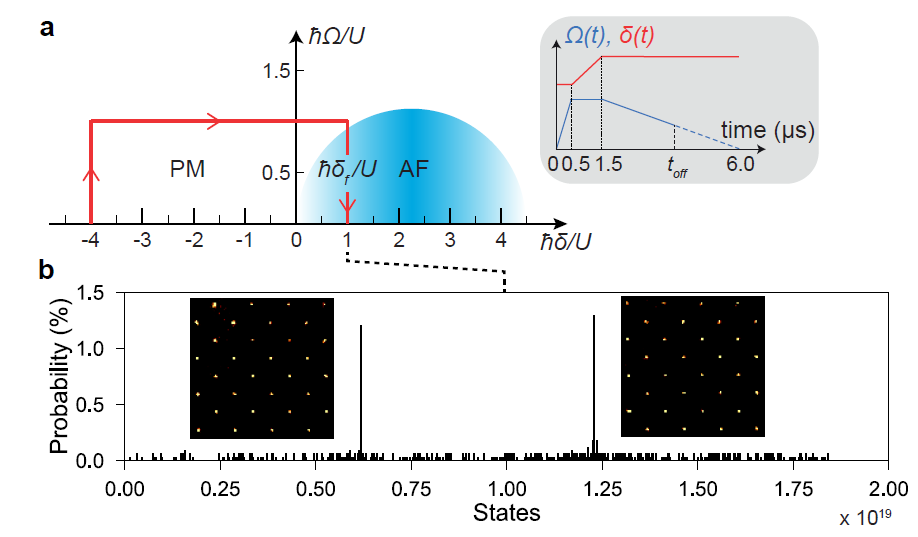}
    \caption{(a) Phase diagram of the Rabi frequency and the detuning to reach an antiferromagnetic state superposition. (b) State histogram for an $8 \times 8$ array at the end of the procedure. Antiferromagnetic states have a probability to be measured of $2.5 \%$ among more than $10^9$ states possible. Taken from \parencite{scholl_quantum_2021}.}
    \label{fig:Scholl_result}
\end{figure}

Neutral atoms arrays are mostly known for their results as analog quantum processor. Actually, it is possible to implement the Ising model by combining the van der Waals interaction and the tunable parameters: the Rabi frequency and the detuning. The resulting time-dependent Hamiltonian realized is (following the notation of \parencite{henriet_quantum_2020}):
\begin{equation}\label{eq:ising_ham}
    H_\mathrm{Rydberg}= \sum_{i\ne j}\frac{C_6}{|\textbf{r}_i-\textbf{r}_j|^{6}} n_i n_j +  \frac{\hbar\Omega (\tau)}{2}\sum_{i} S_i^x - \hbar \delta(\tau)\sum_{i} n_i
\end{equation}
with $\tau$ the time.
In practice, the Rabi frequency and the detuning are driven by a magnetic field ($\Omega$ is proportional to the transverse component whereas $\delta$ depends on the longitudinal component). The Ising Hamiltonian is the "toy" model to study magnetism and can help to tackle many problems from condensed matter, materials, chemistry to optimization problems for instance \parencite{lucas_ising_2014}. Recently, a "quantum advantage" of many-body problems has been demonstrated with the Ising model. An antiferromagnetic state is constructed from a parameters sequence with more than $200$ spins/qubits \parencite{scholl_quantum_2021}
(See Fig.~\ref{fig:Scholl_result}). Moreover, different phases of matter have been explored in \parencite{ebadi_quantum_2021} with this Hamiltonian on 256-atom
programmable quantum simulator.

In addition, one can use the dipole-dipole interaction to implement the "XY" Hamiltonian (following notation of \parencite{henriet_quantum_2020}):
\begin{equation}
      H=\frac{\hbar}{2}\Omega\sum_{j}S_{j}^{x} - \frac{\hbar}{2}\delta\sum_{j}S_{j}^{z}+2\sum_{i\ne j}\frac{C_{3}}{r^{3}_{i,j}}(S_{i}^{x}S_{j}^{x} + S_{i}^{y}S_{j}^{y}).
    \label{eq:xy_main}
\end{equation}

This Hamiltonian is another "simple" model to study magnetism in systems such that frustrated quantum magnets \parencite{balents_spin_2010} or topological materials \parencite{de_leseleuc_observation_2019}. In both cases, many applications are possible in the field of materials or chemistry but also in optimization, machine learning and so on. In addition, combining cleverly the capacity to drive the parameters and choose the geometry can lead to create effective XXZ models \parencite{signoles_glassy_2021} or non-spin conserving terms \parencite{whitlock_simulating_2017}. 

AQS is a strong asset of the RQP with results showing a "quantum advantage" over classical simulation for a specific theoretical physics problem \parencite{scholl_simulation_2021, chuang}. The Rabi frequency and the detuning being driven through time, it is possible to implement VQA with AQS where $\Omega, \delta \,\,\, \mathrm{and} \,\,\, \tau$ are the parameters (see Sec.~\ref{chemistry} for an example). Eventually, if the problem can be mapped into finding the groundstate of exactly Eq.~\ref{eq:ising_ham} or Eq.~\ref{eq:xy_main}, one can directly use the adiabatic theorem. Starting from a state which is the known ground state of a realizable Hamiltonian (for instance, the groundstate of Eq.~\ref{eq:ising_ham} for $\Omega = 0$ and $\delta \ll -|C_6|/R^6 $, the groundstate is $\ket{00 \dots 0}$ meaning that all atoms are in their electronic groundstate), one can tune the parameters slowly to reach (approximately) the many-body groundstate of the Ising Hamiltonian with the desired parameters. The same protocol applied for the XY Hamiltonian. This method is more efficient than VQA because measurements do not have to be repeated to feed a classical optimizer, the measurements are directly useful to get the groundstate.

\section{Conclusion}
RQP is based on a well established technology allowing to perform digital and analog quantum simulation. The whole process of quantum computing is enabled and fully optimized by light control, electronic components and software stacks. The Hamiltonians and gates implementable benefits from the good coherence time of Rydberg states (in the order of a hundred of microseconds) and the wide space of the parameters $\Omega$ and $\delta$ can explore. The device rate is between $2-3$ Hz (sometime it can go to 5 Hz): it means that one can obtain "only" $2-3$ bitstrings per second. The information is important if one wants to design an algorithm which can be experimentally implementable and only needs to run during a few days. RQP is a NISQ device and, therefore, is limited by several sources of noise. 
During my PhD, I tried to take into account all advantages but also limitations of the device to extract the full potential of it. For instance, in Chap.~\ref{chemistry}, the total number of measurements or shots is the criteria to stop the variational procedure whereas in Chap.~\ref{hubbard}, most source of noise are described and implemented to test the robustness of the algorithm.

\chapter{Digital-analog variational quantum eigensolver for chemistry}\label{chemistry}

\section{Forewords}
This chapter contains exactly my first publication available on arxiv and published in \textit{Physical Review A} \parencite{PhysRevA.107.042602}. It is the result of the first half of my PhD work. It has resulted from a strong collaboration with the start-up PASQAL and all simulations are performed with the python library pulser \parencite{silverio_pasqal-iopulser_2022}. 
In this chapter, I propose a new digital-analog variational algorithm to find groundstate of molecules on a Rydberg atom device and show numerically the result on $\mathrm{H}_2$, $\mathrm{LiH}$ and $\mathrm{BeH}_2$ molecules. To this aim, we take into account characteristics of the platform to optimize the VQA algorithm  and increase its efficiency. We propose a complete protocol from the register preparation to a measurement method. The stopping criteria of the algorithm is the number of measurements to fulfill feasible experiment today.
\section{Introduction}\label{sec:introduction}

Quantum simulation holds the promises to solve outstanding questions
in many-body physics, in particular finding the ground state of strongly interacting quantum systems \parencite{georgescu_quantum_2014, mcclean_theory_2016}. 
The determination of the ground state energies of complex molecules, one of the main tasks in quantum chemistry, is therefore an example of application where quantum simulation could be of interest. Along this line, proof-of-principle demonstrations were obtained using photons \parencite{Lanyon2010,Peruzzo2014}, ions \parencite{Shen2017,hempel_quantum_2018} or quantum circuits \parencite{kandala_hardware-efficient_2017}.   
The last two examples used a hybrid approach where a classical computer optimizes in an iterative way the results obtained by a quantum device that was operating in a digital mode, \textit{i.e.} as a series of one and two-qubit gates. 


Rydberg quantum simulators are another example of promising quantum simulation platforms, thanks to their potential for scaling the number of qubits and their programmability \parencite{browaeys_many-body_2020}. 
They rely on individual atoms 
trapped in arrays of optical tweezers that can interact when promoted to Rydberg states. The platform naturally implements spin Hamiltonians. Analog quantum simulation with hundreds of atoms has now been achieved \parencite{scholl_quantum_2021, ebadi_quantum_2022,chen_continuous_2023}. 

One appealing  feature of this platform is the ability to place the atoms in arbitrary position in two and three dimensions, thus allowing large flexibility in their connectivity. Another feature  is their ability to prepare different initial product states as heuristic trials before the unitary evolution (whether it is by a set of digital gates or the action of an analog Hamiltonian evolution). However, this freedom in register preparation has a significant time cost that adds to the repetition clock rate \parencite{henriet_quantum_2020}.

\begin{figure}[ht!]
    \centering
  \includegraphics[width=1\linewidth]{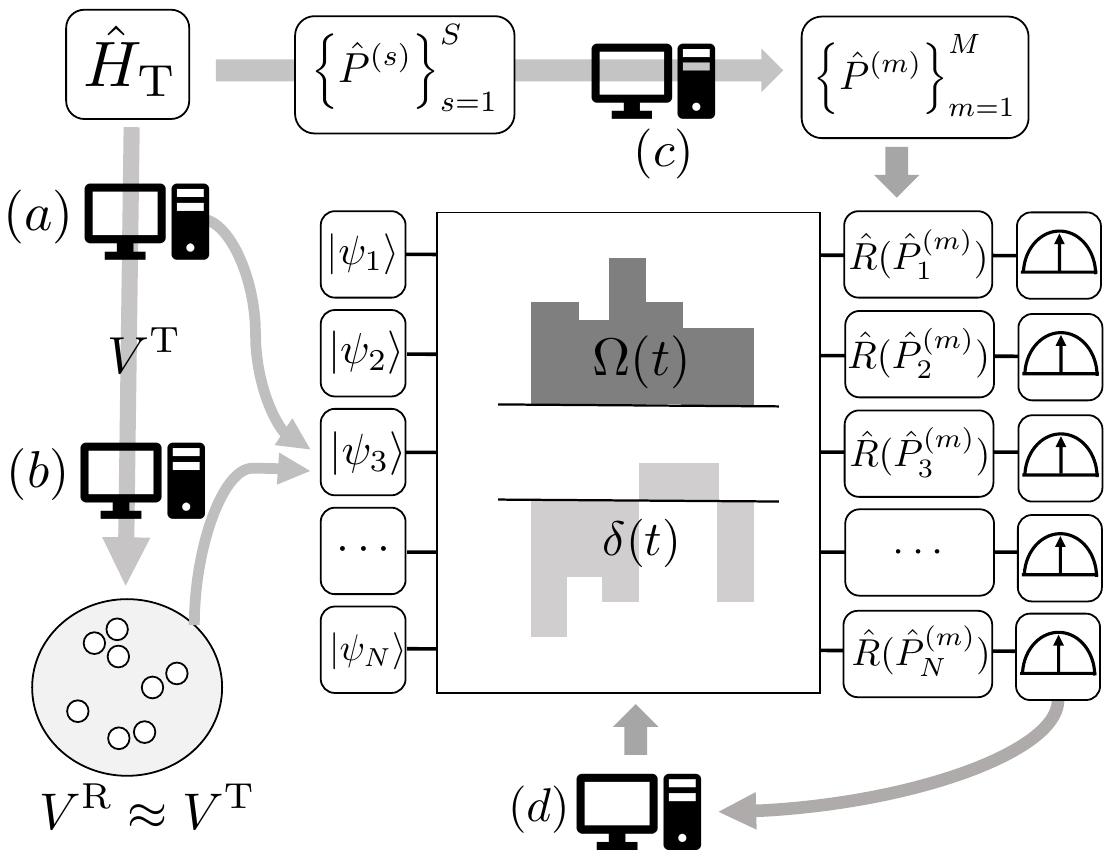}
\caption{Digital-Analog VQE on Rydberg atoms studied in this paper. The computer icons indicate where classical processing enhances and informs the cycle. First, a target molecular Hamiltonian $H_\text{T}$ is passed to computer \text{(a)}. This computer extracts a subset of terms that define a target interaction matrix $V^\text{T}$. Computer \text{(b)} then optimizes atom positions (with interaction matrix $V^\text{R}$) to approximate $V^\text{T}$. With a chosen register, computer \text{(a)} tests virtually some product states that can be experimentally implemented, warm-starting the algorithm. Meanwhile, computer \text{(c)} takes the Pauli strings $\{ \hat P ^{(s)}\}$ that constitute $ H_\text{T}$ and outputs a \emph{derandomized} set of Pauli measurement basis $\{ \hat R(\hat P ^{(m)})\}$ that will be used at the end of the circuit where the prepared state is measured in the $Z$-axis. A parameterized pulse acts on the initial state, and the readout data is used by computer \text{(d)} to estimate the expectation value of the target Hamiltonian $\langle  H_{\text{T}}\rangle$. Finally, computer \text{(d)} calculates new pulse parameters to update the quantum evolution. When a desired precision or some stopping criteria is reached, the best energy value is returned.}
    \label{fig:VQE_analog}
\end{figure}
Neutral atom devices are naturally suited for \emph{analog quantum algorithms}, where the analog blocks are represented by control pulses that drive the system (or subsets of it). Given a prepared state, the parameterized pulses can be adjusted to variationally improve on a given score of the state. Methods for the optimization of parameters have been the subject of intense exploration in recent years \parencite{progress_vqe, cerezo2020variational,cerezo_cost_2021,mcclean_theory_2016,barkoutsos_improving_2020,mcclean_barren_2018,gate-free-state, wakaura_evaluation_2021, banchi_measuring_2021, gacon_simultaneous_2021, piskor_using_2022}. Additionally, the information and ``cost functions'' from the prepared quantum system are obtained by repeatedly measuring the state in the computational basis, which constitutes an operational overhead. Recent results \parencite{huang_efficient_2021, elben_statistical_2019, nam_ground-state_2020,ebadi_quantum_2022, dalyac, kokail_self-verifying_2019} on protocols for the estimation of quantum observables are available and have helped to establish efficient measurement procedures based on generalized random measurements and a series of post-processing steps that are performed on a classical computer and that alleviate the measurement overhead. The types of randomized measurements that we shall describe in this paper require local rotations on the qubits of the register, thus constituting another ``digital'' layer, from a quantum circuit perspective. In fact, \emph{digital-analog} algorithms \parencite{parra-rodriguez_digital-analog_2020}, benefit from the fact that analog operations can be performed with much higher fidelities than when using digital gates, while local single-qubit gates can be added explicitly in crucial steps of the process (state preparation and measurement). 

In this paper, we explore the implementation of a digital-analog VQE algorithm in a Rydberg quantum simulator. We account for typical constraints of the platform: the local action is restrained to the initial state preparation and measurement, with Hamiltonian time-evolution acting on the entire system as a ``global'' gate. We study numerically this version of a VQE for the $\text{H}_{\text{2}}$ molecule using common ansatze, followed by a more efficient protocol for larger molecules. We discuss the embedding of the Hamiltonian in the atom register, the way in which the optimization of the pulse sequence can be performed and the necessity of including an efficient estimation of energies (namely, we explore the effect of a \emph{derandomization} estimation \parencite{huang_efficient_2021}) at each iteration step. We apply this numerically to the examples of LiH and BeH$_2$.
The manuscript is organized as follows: In section II we recall how the Variational Quantum Eigensolver (VQE) estimates the energy of the ground state of a molecular-based Hamiltonian. We then describe the basic ingredients of Rydberg Atom Quantum Processors and the Hamiltonians that they implement. We end the section by explaining the optimization cycle of variational quantum algorithms on these devices. In section III we describe the strategies for implementation of the VQE, going from a direct application of a Unitary Couple Cluster Ansatz, to the Quantum Alternating Operator ansatz and finally to a more hardware-oriented approach that combines elements of register preparation, pulse optimization and observable estimation. This is followed in section IV by numerical results of the error in energy obtained as a function of the number of repetitions of the experiment, an informative measure of the performance of hybrid classical-quantum implementations.

\section{Analog Variational Quantum Eigensolver with Rydberg atoms}
The Variational Quantum Eigensolver (or VQE) is a hybrid quantum-classical algorithm designed to find the lowest eigenvalue of a given Hamiltonian \parencite{fedorov_vqe_2022}. We describe below the origin of the Hamiltonians that we consider and how VQE can be studied with a Rydberg Quantum Processor.

\subsection{Hamiltonians from Quantum Computational Chemistry}\label{sec:chem}

We first recall the method used to express the electronic Hamiltonian 
as a spin model (see e.g.~\parencite{hempel_quantum_2018}). We start from the Born-Oppenheimer approximation of the Hamiltonian of the system, which considers the nuclei of the molecules as classical point charges:

\begin{equation}\label{elec-ham}
      H = -\sum_i \frac{\nabla_i^2}{2} - \sum_{i, I} \frac{\mathcal Z_I}{|\mathbf r_i - \mathbf R_I|} + \frac 1 2 \sum_{i\neq j} \frac{1}{|\mathbf r_i - \mathbf r_j|} 
\end{equation}
(in atomic units) where $\nabla_i$ is the kinetic energy term for the $i$-th electron, $\mathcal Z_I$ is the charge of the $I$-th nucleus, and $\mathbf r$, $\mathbf R$ denote the distance of the $i$-th electron and the $I$-th nucleus with respect to the center of mass, respectively. We aim to obtain the ground state energy of  \eqref{elec-ham}.

One needs to define a basis set in which to represent the electronic wavefunctions. We shall concentrate on the Slater-type orbital approximation for the basis set, with three Gaussian functions, STO-3G. This minimal basis set $\{\phi_i (\mathbf x_i)\}$ (where $\mathbf x_i=(\mathbf r_i, \sigma_i)$ encodes the $i$-th electron's spatial and spin coordinates) includes the necessary orbitals to represent the valence shell of an atom. Moreover, the wavefunctions need to be anti-symmetric under the exchange of electrons. This can be achieved through \emph{second quantization}, where one defines anticommuting fermionic creation/annihilation operators $\{a_p^\dagger\}, \{ a_p\}$ and rewrites the initial Slater determinant form of the wavefunction as $|\Psi \rangle = \prod_p ( a_p^\dagger)^{\phi_p} |\mathrm{vacuum}\rangle$, representing the occupation of each molecular orbital.

The fermionic operators are used to rewrite \eqref{elec-ham} as: 
\be
\label{eq:2nd-quant-ham}
  H = \sum_{p,q}h_{pq}{a}_p^{\dagger}{a}_q + \frac{1}{2}\sum_{p,q,r,s}h_{pqrs}{a}_p^{\dagger}{a}_q^{\dagger}{a}_s{a}_r.
\ee
where the coefficients $h_{pq}$ and $h_{pqrs}$ encode the spatial and spin configuration of each of the electrons and depend on the inter-nuclear and inter-electron distances $\mathbf R, \mathbf r$: 
\be\label{eq:2nd-quant-terms}
\begin{aligned}
    h_{pq}&=\int d \mathbf x \phi_p^* (\mathbf x)\left(-\frac{\nabla^2}{2}-\sum_i\frac{\mathcal Z_i}{|\mathbf R_i- \mathbf r|}\right)\phi_q(\mathbf x) \\
    h_{pqrs}&=\int d\mathbf x_1 d\mathbf x_2\frac{\phi_p^*(\mathbf x_1)\phi_q^*(\mathbf x_2)\phi_r(\mathbf x_1)\phi_s(\mathbf x_2)}{|\mathbf r_1- \mathbf r_2|}.
\end{aligned}
\ee
Next, we map the fermionic operators acting on Fock states of $n$ orbitals to a Hilbert space of operators acting on spin states of $N$ qubits. This corresponds to the quantum processors' effective interaction Hamiltonians, quantum gates and measurement basis. Useful maps of this kind include the Jordan-Wigner (JW) \parencite{JordanWigner1928} or the Bravyi-Kitaev (BK) \parencite{Bravyi2002} transformations. The obtained Hamiltonian is a sum of tensor products of single-qubit Pauli matrices:
\begin{align}\label{eq:pauli_ham}
    H_T = \sum_{s=1}^S {\bf c}_s \bigg( \bigotimes_{j=1}^N S_j^{(s)}\bigg)
\end{align}
where $ S_j \in \{\mathbb 1, S^z, S^y, S^z \}$,  $S$ is the number of Pauli strings in the Hamiltonian and $N$ the number of qubits.



\subsection{Rydberg Atom Quantum Processor}
Rydberg atom arrays are now well-established quantum simulation 
platforms \parencite{henriet_quantum_2020,browaeys_many-body_2020}. Briefly, atoms are trapped in optical tweezers, each
containing exactly one atom. The tweezers may be arranged in any 1D, 2D or 3D geometrical configurations. The register can be rebuilt after each computational cycle.
To perform quantum processing, we use the fact that the platform implements spin-like Hamiltonians, where the interactions originate from strong dipole-dipole couplings between atoms laser-excited to Rydberg states. 

Depending on the choice of atomic levels, the atoms experience different effective interactions. In the case of the Ising mode, $\ket{0}$ is a ``ground'' state prepared by optical pumping \parencite{browaeys_many-body_2020} and $\ket{1}$ is a Rydberg state of the atom. The Hamiltonian term for this interaction is:

\begin{align}\label{eq:ising}
    H_{\text{Ising}} &= \sum_{i > j}\frac{C_6}{r_{i,j}^{6}} n_i  n_j, 
   \end{align}
with $\hat n_i = |1\rangle_i \langle 1 | = (\mathbb 1_i + S^z_i)/2$ the projector on the Rydberg state and $r_{i,j}$ the distance between atoms. Here and below, $S^x_i,S^y_i$ and $S^z_i$ indicate the local Pauli operators.

If instead the two states chosen are two dipole-coupled Rydberg states (for example $\ket{0}=\ket{nS}$ and $\ket{1}=\ket{nP}$ for large $n$), the interaction is resonant and realizes 
a so-called ``XY'' or ``flip-flop'' term:

\begin{align}\label{eq:XY}
    H_{\text{XY}} &= \sum_{i \neq j}\frac{C_3}{r_{i,j}^{3}} (S^x_i S^x_j + S^y_i S^y_j),
\end{align}
where $C_3$ depends on the chosen Rydberg orbitals and their orientation with respect to the interatomic axis. It corresponds to a coherent exchange of neighboring spin states $\ket{10}$ to $\ket{01}$.

In addition, we can include time-dependent terms on the Hamiltonian, by means of a laser pulse (Ising mode) or a microwave field (XY mode) targeting the transition between the ground and excited states. This is represented by the following ``drive'' terms:

\begin{align}
H_{\text{drive}} = \frac{\hbar}{2}\sum_{i=1}^N \Omega_i(t) S^x_i - \hbar\sum_{i=1}^N \delta_i(t) n_i\ .
\end{align}
Here, $\Omega(t)$ is Rabi frequency and  $\delta(t)$ the detuning of the field  with respect to the resonant transition frequency. 
The addressing can be either global or local. In the procedure used in this work, the 
local addressing is restricted to the initial state preparation and the register readout stages. 

\subsection{Variational Algorithms on a Rydberg atoms device}

In the analog VQE algorithm, we seek to estimate the energy of the ground state of a qubit Hamiltonian called the \emph{target} Hamiltonian, $ H_{\text{T}}$,  by using an iterative method. 
The \emph{resource} Hamiltonian is the one realized by the hardware, 
and can be configured with different types of interactions ($ H_\mathrm{inter}$) (\ref{eq:ising}, \ref{eq:XY}) and driving fields ($H_\mathrm{drive}$): 
\begin{equation}
    H_{\text R} = H_{\rm inter} + H_{\rm drive}.
\end{equation}
Experimentally, the transition from the ground to the excited state is typically generated by a two-photon process, from which an approximate two level system is extracted, driven by an effective Rabi frequency $\Omega$ and detuning $\delta$ during the quantum processing stage. We use their values as parameters in our analog presentation of a VQE algorithm: The first step is to prepare the register of $N$ atoms with a geometry that determines the interaction terms $ H_{\rm inter}$ and then to initialize the system in a state $\ket{\psi_0}$. 
Then, a pulse sequence is applied to evolve the system under the resource Hamiltonian $H_{\text R}(\Omega(t), \delta(t))$ whose corresponding time-ordered unitary evolution operator is $U(t) = \mathcal T  \exp\big(-i \int_{0}^{t} H_{\text{R}}(\Omega(\tau), \delta(\tau) )d\tau\big)$. The final prepared state is:
\be
\label{eq:evolution_op}
\ket{\psi(\Omega, \delta,t)} = U(t)\ket{\psi_0}.
\ee
The energy of a prepared state will be calculated with respect to the target Hamiltonian: 
\be
\label{eq:energy_estimation}
E(\Omega, \delta, t) = \bra{\psi(\Omega, \delta, t)}H_{\text T}\ket{\psi(\Omega, \delta, t)}.
\ee
After each cycle, a classical optimizer adjusts the parameters $\Omega \rightarrow \Omega'$, $\delta \rightarrow \delta'$ and $t \rightarrow t'$ and we repeat the evolution of the initial quantum state $\ket{\psi_0}$ with the new parameter set $U(\Omega',\delta',t')$. We aim to obtain for each iteration $E(\Omega', \delta', t') \leq E(\Omega, \delta, t)$ \footnote{This classical optimization problem can be addressed for example by obtaining the gradient of the energy function.}. After several iterations of this loop, the variational scheme attempts to prepare a state whose energy is a good approximation of the ground state energy of $H_{\text T}$ \parencite{mcclean_theory_2016}.

\section{Description of the Protocols}

In this section, we describe two analog variational quantum algorithms for the estimation of the ground state energy and apply them to quantum chemistry problems. The protocols differ mainly by the choice of ansatz: one is the \emph{Unitary Coupled Cluster} (UCC) ansatz \parencite{bartlett_alternative_1989}, while the other is an adaptation of a \emph{hardware-efficient ansatz} \parencite{kandala_hardware-efficient_2017}, based on repeating alternating values of amplitude, frequency or phase of the applied pulses. We verify numerically the performance of these two types of ansatz in a Rydberg-based Quantum Processor (QP). Next we discuss a protocol for larger molecules tailored after the hardware capabilities. We begin by considering the prototypical example of the H$_2$ molecule.

\subsection{UCC ansatz on an analog quantum processor: application on H\pdfmath{\bf _2}}\label{sec:H2}

Numerous implementations of the VQE algorithm rely on the use of  digital gates. 
Recent experimental implementations for the H$_2$, LiH and BeH$_2$ molecules have been realized in \parencite{kandala_hardware-efficient_2017,hempel_quantum_2018}, with superconducting and trapped ions devices respectively. For the analog version of this algorithm on H$_2$, we consider the target Hamiltonian and the ansatz as in \parencite{hempel_quantum_2018}.
The Jordan-Wigner and Bravyi-Kitaev transformations lead to two different spin Hamiltonians of this molecule:

\begin{align}
\begin{split}
H_{\text{JW}} =& \, {\bf c_0} \mathbb{1} + {\bf c_1}(S^z_0 +  S^z_1) + {\bf c_2}( S^z_2 +  S^z_3) +\\
&{\bf c_3} S^z_3 S^z_2 +  {\bf c_4} S^z_2 S^z_0 + {\bf c_5}(S^z_2 S^z _0 + S^z_3 S^z_1) +\\
&{\bf c_6} (S^z_2 S^z_1 + S^z_3 S^z_0) + {\bf c_7} (S^x_3 S^y_2 S^y_1 S^x_0 +\\
& S^y_3 S^x_2 S^x_1 S^y_0 - S^x_3 S^x_2 S^y_1 S^y_0 + S^y_3 S^y_2 S^x_1 S^x_0)
\label{hamjw}
\end{split}
\end{align}    
and
\begin{align}
\begin{split}
H_{\text{BK}} =& \, {\bf f_0} \mathbb{1}  + {\bf f_1} S^z_0 + {\bf f_2} S^z_1 + {\bf f_3} S^z_2 + {\bf f_4}S^z_1 S^z_0 +\\ & {\bf f_5} S^z_2 S^z_0 + {\bf f_6} S^z_3 S^z_1 +  {\bf f_7} S^x_2 S^z_1 S^x_0 +
    {\bf f_8} S^y_2 S^z_1 S^y_0 + \\ 
    & {\bf f_9} S^z_2 S^z_1 S^z_0 + {\bf f_{10}} S^z_3 S^z_2 S^z_0 + {\bf f_{11}}S^z_3 S^z_2 S^z_1 + \\
    & {\bf f_{12}}S^z_3 S^x_2 S^z_1 S^x_0 + {\bf f_{13}} S^z_3 S^y_2 S^z_1 S^y_0 + {\bf f_{14}}S^z_3 S^z_2 S^z_1 S^z_0
\label{hambk}
\end{split}
\end{align}     
where the coefficients $\{\bf c_i\}$ and $\{\bf f_j\}$ are calculated from (\ref{eq:2nd-quant-terms}). Since in \eqref{hambk} qubits $1$ and $3$ are only affected by the operators $\mathbb{1}$ and $Z$ one can actually work with the following two-qubit effective Hamiltonian \parencite{omalley_scalable_2016}:
\begin{align}\begin{split}
H_{\text{BK}}^{\text{(eff)}}=& \, {\bf g_0} \mathbb{1} + {\bf g_1}S^z_0 + {\bf g_2}S^z_1 + {\bf g_3} S^z_0 S^z_1 + \\
& {\bf g_4}S^x_0 S^x_1 + {\bf g_5}S^y_0S^y_1.
\label{eqhambkef}
\end{split}\end{align}

Usually, a good ansatz $|\psi(\boldsymbol \theta)\rangle = U(\boldsymbol \theta) |\psi_0\rangle$ requires a balance between hardware constraints and symmetries in target Hamiltonian. However, using the `knobs' available on the hardware is often not efficient, and one thus needs additional guidance to reach the states we are looking for in a potentially very large Hilbert space. In this sense, the well-established Unitary Coupled Cluster (UCC) ansatz allows one to perform an unitary operation while keeping advantages of coupled cluster ansatz from chemistry \parencite{helgaker_molecular_2014}. 

In most cases, implementing the UCC ansatz in a quantum processor requires constructing a digital quantum circuit with full local addressing. An example where global addressing is sufficient is the H$_2$ molecule. The initial guess of the molecular wave function is a product state obtained from the classical Hartree-Fock calculation performed to determine the coefficients $\{\bf c_i\}$ and $\{\bf f_j\}$. Considering only relevant single and double excitations in the unitary coupled-cluster operator (UCC-SD) yields the following one-parameter unitary: 
\begin{equation}\label{UCCSD}
    U_{\text{UCC-SD}}(\theta) = e^{\theta (c_{2}^{\dagger}c_{3}^{\dagger}c_1c_0 - c_0^{\dagger}c_1^{\dagger}c_3c_2)}
\end{equation}
where the minimal set of orbitals are represented by the fermionic annihilation and creation operators  $c, c^{\dagger}$ \parencite{hempel_quantum_2018}. A Jordan-Wigner transformation on these operators leads to the UCC ansatz $\ket{\psi(\theta)} =\exp(-i\theta S^x_3 S^x_2 S^x_1 S^y_0) \ket{0011}$, where $\ket{0011}$ is the Hartree-Fock state. In the case of the (effective) Bravyi-Kitaev transform \eqref{eqhambkef}, we obtain the simpler UCC ansatz $\ket{\psi(\theta)}=\exp(-i\theta S^x_1 S^y_0)\ket{01}$.
 
Since the evolution Hamiltonian commutes with the XY Hamiltonian (\ref{eq:XY}), one can use the latter ansatz and attempt to drive the Rydberg QP in the XY mode, 
using $\Omega=0$ and non-zero local detunings, leaving the rest of the parameters to be set by variational optimization:
\begin{align}
\begin{split}
\label{eq:xyansatz}
\ket{\psi(\delta_0, \delta_1, t)} &= \exp \Big( -it(\delta_0 S^z_0 + \delta_1 S^z_1 + H_{\text{XY}}) \Big) \ket{01} \\ 
&= a(\delta_0,\delta_1,t) \ket{01} + b(\delta_0, \delta_1, t) \ket{10},
\end{split}
\end{align}
which coincides with the subspace reached with the UCC ansatz: 
\begin{equation}\label{eq:uccbk}
\exp(-i\theta S^x_1 S^y_0)\ket{01} = a(\theta) \ket{01} + b(\theta) \ket{10}.
\end{equation}

\begin{figure}
    \centering
    \includegraphics[width=0.7\linewidth]{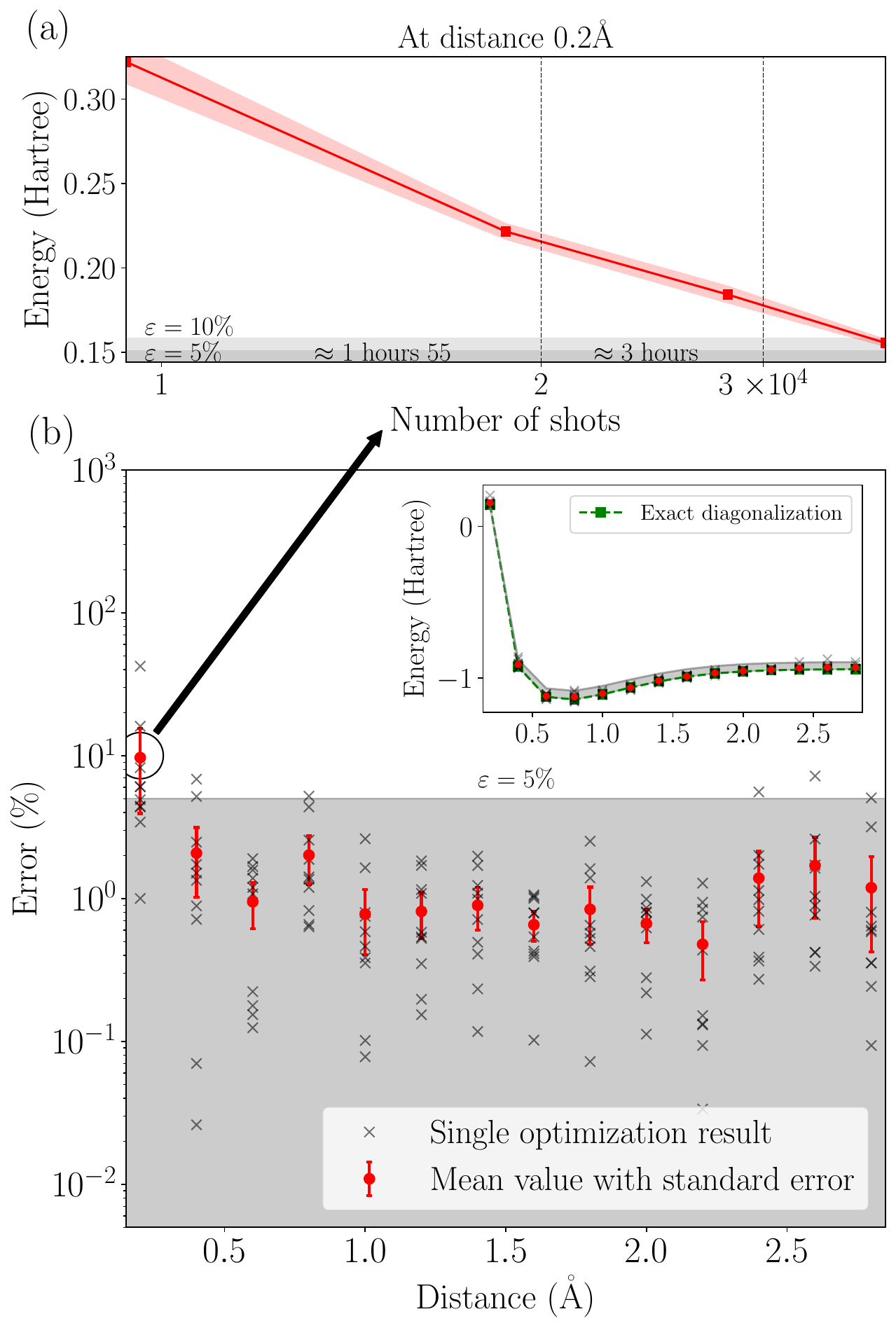}
    \caption{\textit{Numerical implementation of an analog VQE algorithm using a UCC ansatz.} 
    (a) A zoom on the smallest inter-atomic distance ($0.2$ \r{A}) shows the evolution of the optimization with respect to the number of shots. The differential evolution was set to perform at most 4 iterations (red squares). The red scale shows the errorbar over 20 realizations. It takes approximately $3.5$ hours of runtime for a QP operating at 3 Hz 
    to achieve $\epsilon = 10 \% $ (light grey scale) of error and $4$ hours to achieve $\epsilon = 5 \% $ (dark grey scale).
    (b) Relative error in percentage (red circles) between the mean VQE result and the numerically computed lowest eigenvalue of the target Hamiltonian (in STO-3G basis) over 10 realizations (gray crosses). The expected error is below $\varepsilon = 5 \%$ (grey area). The inset depicts the same result on an energy scale and compares it with the exact solution (green squares). The result is drawn as a function of hydrogen inter-atomic distance. 
    }
    \label{fig:H2_UCC_Figure_slave_spin}
\end{figure}

A numerical implementation of this protocol is shown in Fig. \ref{fig:H2_UCC_Figure_slave_spin}, where the classical optimization was performed with a differential evolution algorithm \parencite{storn_differential_1997}. We observe that the ground-state energy can be obtained with an error smaller than $5\%$ using less than $36500$ shots for each point. 

Such examples of a UCC ansatz implementable with an analog approach, often rely on finding symmetries between target and resource Hamiltonians \parencite{kokail_self-verifying_2019}. Nevertheless, this kind of protocol remains impractical for larger molecules because of the increasingly higher number of qubits and Pauli strings in the Hamiltonian. In order to use the analog approach for larger encodings, we explore other approaches below.

\subsection{Alternating pulses}

\begin{figure}[ht!]
    \centering
    \includegraphics[width=1.1\linewidth,trim={0cm 0cm -1cm 0.5cm},clip]{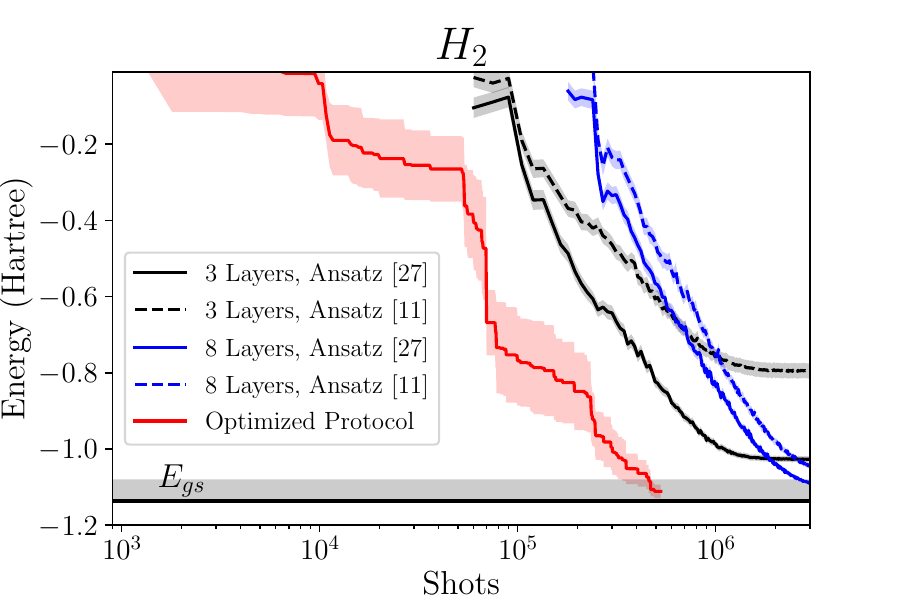}
    \caption{Evolution of the ground-state energy as a function of the accumulated number of shots for the H$_2$ molecule at a fixed inter-nuclear distance. We have averaged numerically 200 realizations of VQE with 3 and 8 layers using the two alternating pulse ansatz, \parencite{dalyac_qualifying_2021} (straight line), and \parencite{ebadi_quantum_2022} (dashed line). In the alternating operator approach, each Pauli string mean value is performed with 1000 shots, which for the H$_2$ Hamiltonian represents $1.5\times10^4$ shots before obtaining the first energy data point. Achieving energy errors below $5\%$ (gray area) requires at least $\mathcal O (10^6)$ shots in total. For the optimized procedure (adding more control over atom positions, pulse shaping and derandomization estimation), the same energy error typically requires $\mathcal O(10^5)$ shots.} \label{fig:QAOA_comp}
\end{figure}

We now describe an alternating operator approach, based on the QAOA algorithm \parencite{farhi_quantum_2014}. Let $|\psi_0\rangle$ be the state composed of all qubits in the ground state. The whole sequence is composed by alternating constant (global) pulses, corresponding to two non-commuting Hamiltonians $H_a, H_b$:
\begin{align}
    H_{a} &= \frac{\hbar}{2}\sum_{i=1}^N \Big( \Omega S^x_i - \delta S^z_i \Big) + H_{\text{inter}} \\
     H_{b} &= \frac{\hbar}{2}  \sum_{i=1}^N \Omega S^x_i  + H_{\text {inter}}.
\end{align}
These Hamiltonians define evolution operators $U_a(t)$ and $U_b(t)$, during a certain time $t$ (see \eqref{eq:evolution_op}). The ansatz of $L$ layers is written as: 
\be
\ket{\psi(\mathbf t_a, \mathbf t_b)}= \prod_{\ell=1}^L U_a(t^{\ell}_a) U_b(t^{\ell}_b) \ket{\psi_0},
\ee
where the arrays of parameters $\mathbf t_k = (t^{1}_k, \ldots, t^{\ell}_k, \ldots, t^{L}_k)$, $k \in \{a,b\}$, fix the duration of each pulse in the layer, as described in \parencite{dalyac_qualifying_2021}. 
As another example, a different choice of parameters was used in \parencite{ebadi_quantum_2022}, considering a single Hamiltonian:
\begin{align}
\begin{split}
    H = \frac{\hbar}{2}&\sum_{i=1}^N \Big( \Omega(t) e^{i\phi(t)} \ket{0}_i \bra{1} + \text{h.c.} \Big)  +  H_{\text{inter}}
\end{split}
\end{align}
with different time $\mathbf t = (t^{1}, \ldots, t^{\ell}, \ldots, t^{L})$ and phase $\boldsymbol \phi = (\phi^{1}, \ldots, \phi^{\ell}, \ldots, \phi^{L})$ arrays defining $L$ segments of the pulse. The corresponding ansatz is then:
%
\be
\ket{\psi(\mathbf t, \boldsymbol{\phi})}= \prod_{\ell=1}^L U(t^{\ell}, \phi^{\ell})  \ket{\psi_0}.
\ee
The two approaches can be implemented in existing experimental setups, especially when the target Hamiltonian is equal to the resource Hamiltonian (such as the case of the Maximal Independent Set problem with Unit Disks, which is native to the Rydberg atoms setting). However, these methods struggle to minimize the molecular target Hamiltonian energies within a limited number of iterations and measurement repetitions. The alternating pulse ansatz assumes an initial register configuration and initial guesses for the durations of the pulses in each layer, two tasks that are the subject of active research. The expectation is that a properly chosen register and an optimized pulse will drive the system to a low-energy state. In Fig. \ref{fig:QAOA_comp}, we compare numerically the  performance of the two alternating pulse ansatze discussed above (\parencite{dalyac_qualifying_2021}, \parencite{ebadi_quantum_2022}). We also included the procedure described in Sec. \ref{Param_pulse}, which addresses the embedding of the problem in the register and an estimation protocol for the observables. Comparing the required number of shots for these approaches highlights the necessity of including an efficient estimation protocol for the observables.

\subsection{Optimized Register and Iteratively Parameterized Pulses}\label{Param_pulse}

In this section, we present a more refined approach to deal with larger systems,  aiming at exploiting the capabilities already available in Rydberg simulators. To exemplify the procedure, we consider in the following  the Ising mode with the  resource Hamiltonian (\ref{eq:ising}).

\subsubsection{Atom register and initial state}\label{sec:embedding}

Even though we only consider global pulses for the processing stage, there still remains freedom in the choice of the positions of the atoms. This determines the strength of pairwise interactions and defines a connectivity graph whose edges correspond to the atoms that experience a blockade effect \parencite{henriet_quantum_2020} (a different graph structure can be defined for the XY mode \eqref{eq:XY}).

In order to find suitable atomic positions, the coordinates are optimized in the plane so that the associated interaction energy matrix resembles as much as possible the information contained in the target Hamiltonian. 
Since the latter contains  general Pauli strings, we consider a subset of terms whose coefficients can be expressed in terms of the coordinates of the atoms\footnote{A broader series of techniques for embedding the problem information into the atom register has been considered in \parencite{leclerc2022financial, coelho2022efficient}}. 
A simple choice consists in selecting the terms that can be directly compared with the Ising-like interaction of the atoms: Let the matrix $V^{\text{T}}$ be given by the positive coefficients of the terms with only two $Z$ operators in the molecular target Hamiltonian and $V^{\text{R}}$ (our ``register'' matrix) the resulting values of interaction strength $C_6/r_{i,j}^6$ for each pair $i,j$ of atom positions in the register. This defines a score function $\sum_{i,j}(V^{\text{T}}_{i,j} - V^{\text{R}}_{i,j})^2$ that we minimize numerically by varying the atom coordinates. 

\begin{figure}
    \centering
\includegraphics[width=0.7 \linewidth]{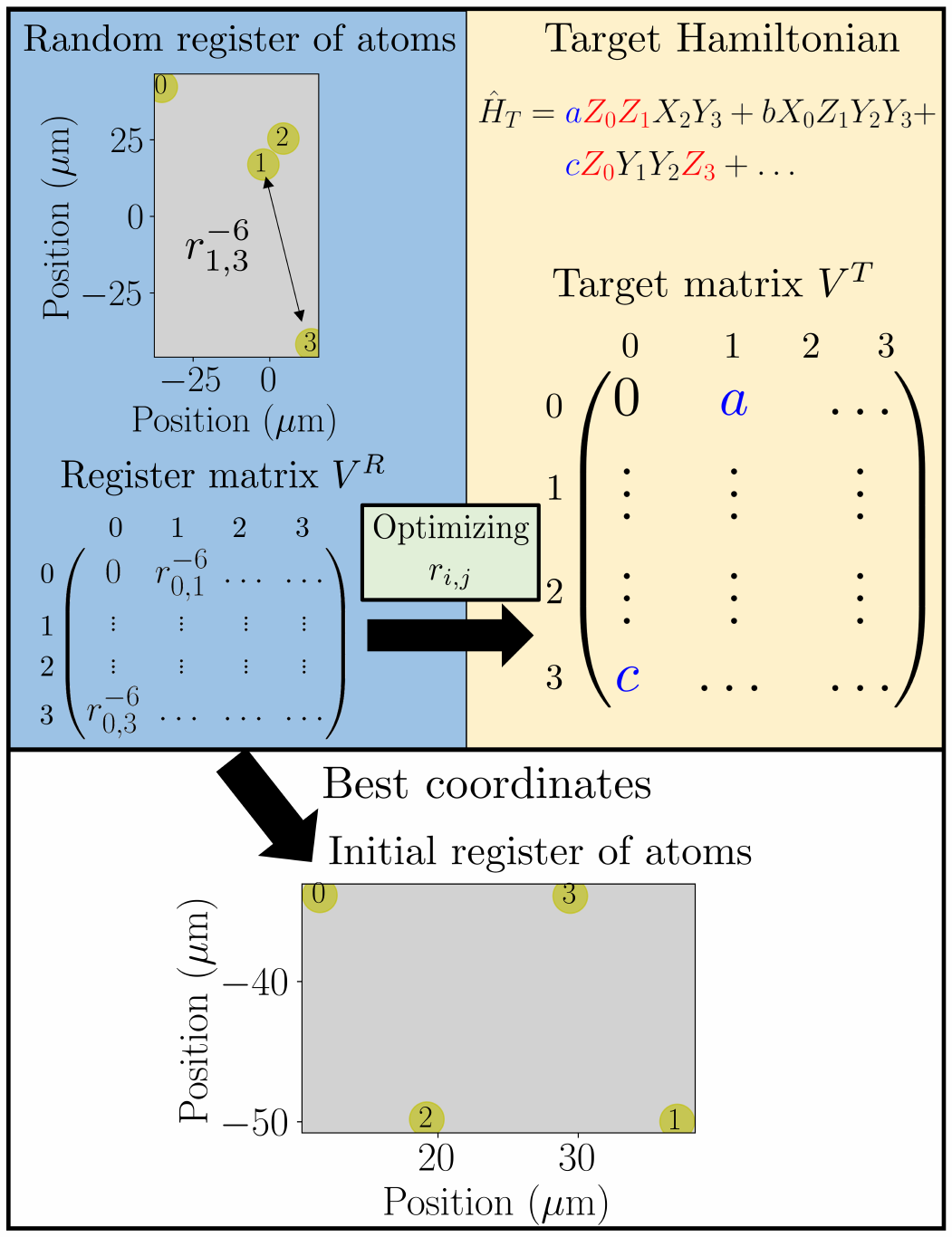}
    \caption{Protocol to optimize the positions of the atoms in a register based on a target Hamiltonian. Blue: we begin with a register of randomly placed atoms and all the Ising interaction terms are entered in a $N\times N$  matrix. Yellow: all positive coefficients before the Pauli strings with only two $Z$ operators in the target Hamiltonian are combined in another (target) matrix $N \times N$. The coordinates of the atoms are optimized to minimize the distance between the two matrices. We then obtain a new register on which we will apply the VQE sequence. }
    \label{fig:register_H2}
\end{figure}

The set of atomic positions that arises from this minimization will be our optimized register. Its geometry will be used to simulate the target Hamiltonian, but has no intrinsic chemical meaning. The information that is taken from the Hamiltonian can be chosen from other subsets of the Pauli strings (e.g. terms with 3 or more $Z$ operators) and different interpretations of how the coefficients constitute a register matrix. A different resource Hamiltonian, such as one with XY interactions, would imply a different choice of subset. In Fig. \ref{fig:register_H2}, we summarize graphically the procedure for the case of the $\text{H}_{\text{2}}$ molecule with the Jordan-Wigner transformation. It turns out that we obtain at a geometry very similar to the one heuristically picked for the alternating pulse ansatz.

\subsubsection{Optimization of the parameterized pulse sequence}\label{sec:pulse_opt_protocol}

\begin{figure}[h!]
    \centering
    \includegraphics[width=1\linewidth, trim={3.5cm 2cm 6.5cm 1cm},clip]{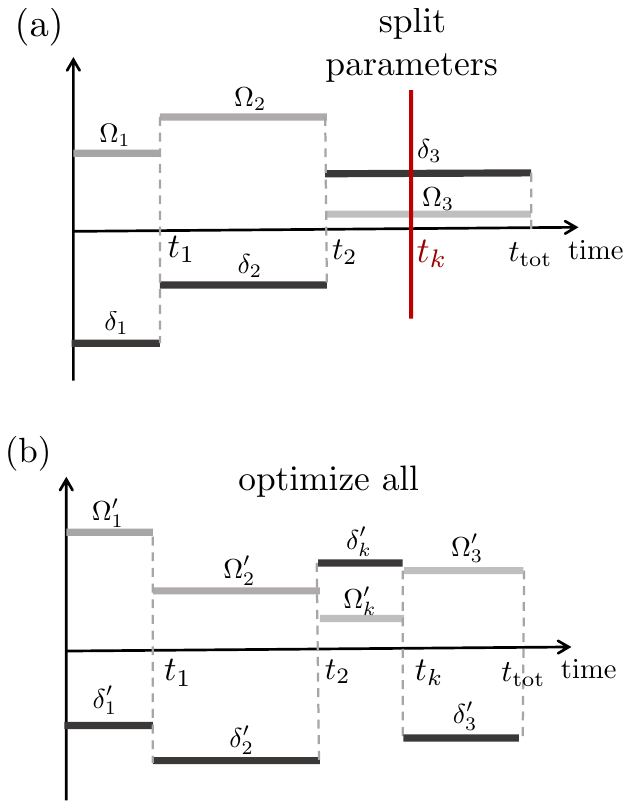}
    \caption{Iterative splitting and optimization of the pulse parameters: (a) Choose at random a time $t_{k}$, which will fall in the interval $(t_{i-1}, t_i)$, and accept it if $|t_{i-1}-t_{k}|$ and $|t_{k}-t_{i}|$ conform to the device response time. Split the corresponding $\Omega_{i}$ into two parameters $\Omega'_{i}, \Omega'_{k}$ with initial value equal to $\Omega_{i}$. Do the same to the set $\{ \delta_i \}$. (b) Optimize the new set of parameters to lower the energy of the prepared state (\ref{eq:energy_estimation}).}
    \label{fig:ctrl_pulse_sequence}
\end{figure}

We constructed a variation of the so-called \texttt{ctrl-VQE} protocol \parencite{gate-free-state} for the case of a global pulse on the register, in which the number of parameters increases at every optimization iteration, while the total time $t_\text{tot}$ remains fixed:

Consider a set of Rabi frequencies $\{\Omega_i\}_{i=1}^K$ and detunings $\{\delta_i\}_{i=1}^K$ defined discretely over a set of time labels $0 < t_1 < \ldots < t_K = t_\text{tot}$. 
Then, at iteration $k$, a new time label $0<t_{k}<t_\text{tot}$ is generated at random, lying between two previous time labels, $t_{i-1}< t_{k} < t_i$. To avoid labels too close to each other, we will accept $t_k$ if the intervals $|t_{i-1} - t_k|, |t_{k} - t_i|$ are large enough compared to the response time of the waveform generator of the machine (in the order of a few ns). The corresponding Rabi frequency $\Omega_i$ and detuning $\delta_i$ from the parent interval $[t_{i-1}, t_i]$ are then split into two independent parameters $\Omega_{i}', \Omega'_{k}$ and $\delta'_{i}, \delta'_{k}$ whose initial values are set equal to their parent parameters  (see Fig. \ref{fig:ctrl_pulse_sequence}). Finally, the new set of parameters $\{\Omega_1, \ldots, \Omega'_{i}, \Omega'_{k}, \ldots, \Omega_K \}$ (likewise for $\{\delta_i \}_{i=1}^K)$ is optimized starting from the previous iteration values. This algorithm acts therefore as a pulse shaping process. From time $t_{i-1}$ to $t_{i}$ the acting Hamiltonian is:
\begin{equation}
H_i = \frac{\hbar}{2} \Big( \Omega_i \sum_{j=1}^N S^x_j - \delta_i \sum_{j=1}^N S^z_j \Big) +  H_{\text{inter}}
\end{equation}
and our ansatz, for $K$ iterations, becomes:
\begin{equation}
    \ket{\psi(\boldsymbol \Omega,\boldsymbol \delta)} = \mathcal T \prod_{i=1}^K \exp\Big[-i \int_{t_i}^{t_{i+1}}  H_i(\tau) d\tau \Big ]\ket{\psi_0}
\end{equation}

Note that while the interval involves a constant Hamiltonian, we include a time-dependent integration at each interval, to indicate that the waveforms that compose the pulse can be adapted to hardware conditions (e.g. by being interpolated, or by adapting the shape with an envelope function).
\subsubsection{Energy estimation by derandomization}\label{derand}

In our algorithm implementation, we take as a figure of performance of the variational optimization run the \emph{total} number of shots required to achieve a given error threshold $\varepsilon$ for the energy. A bounded number of processing cycles is required to remain within a realistic time lapse for the entire implementation process. Rather than measuring several times each of the Pauli observables in the Hamiltonian, we use an estimation protocol (derandomization \parencite{huang_efficient_2021}) based on fixing local Pauli measurements from an originally random set. This allows to efficiently predict the energy of the prepared state $\ket{\psi}$,  $\langle H_\text{T}(\psi) \rangle$, at each loop of the optimization of the parameters. 

More specifically, the derandomization algorithm starts with an initial measurement set of $M$ random Pauli strings $\{ S^{(m)} \}_{m=1}^M$. A greedy algorithm improves the overall expected performance of the measurement set, effectively ``derandomizing'' the operators of each random Pauli string in sequence. The improvement is quantified by the average of the \emph{confidence bound}, which ensures that the empirical average \footnote{ 
A Pauli string $A$ \emph{hits} $B$, if by changing some operators in $A$ to $\mathbb 1$, we form $B$ (for example $ZX\mathbb{1}$ hits $\mathbb{1}X\mathbb{1}$ and $Z\mathbb{1}\mathbb{1}$). The empirical average is obtained using those Pauli measurement basis $\{S^{(m)}\}$ that hit an observable $S^{(p)}$, with the relevant measured bits expressed as $\pm$:
$$
\omega_s = \frac{1}{N_\mathrm{h}}\sum^M_{\substack{m: \\ S^{(m)} \text{hits} \: S^{(p)}}}  \Bigg(  \prod^N_{\substack{j : S^{(p)}_j \neq \mathbb 1}} \mathbf b^{(m)}_j \Bigg),
$$
where $N_{\text{h}}$ counts how many Pauli strings in the set $\{ S^{(m)}\}$ hit $S^{(p)}$, and $\mathbf b^{(m)} = \mathbf b^{(m)}_1 \cdots \mathbf b^{(m)}_N$ is the bitstring measured with the basis $S^{(m)}$

}  $\omega_p$ corresponding to the $p$-th term of $ H_\text{T}$ is within a desired accuracy $| \omega_s - \langle S^{(p)}\rangle |/|\langle S^{(p)}\rangle | < \epsilon$ and with a high probability. The total energy is finally estimated as $\langle H_\text{T}(\psi) \rangle\approx \sum_{p=1}^P \omega_p$.

While the pulses that prepare the state are global, the measurement itself requires the implementation of local rotations on the qubits. This can be achieved experimentally by using a toolbox such as the one described in \parencite{notarnicola_randomized_2021}, thus emphasizing the digital-analog interplay that is now within reach for next-generation neutral atom devices.

\section{Numerical Results} \label{numerical_result}

\subsection{Application on LiH and BeH\pdfmath{\bf _2} molecules}\label{sec:result_lih_beh2}

\begin{figure}[ht!]
    \centering
    \includegraphics[width=0.55\linewidth]{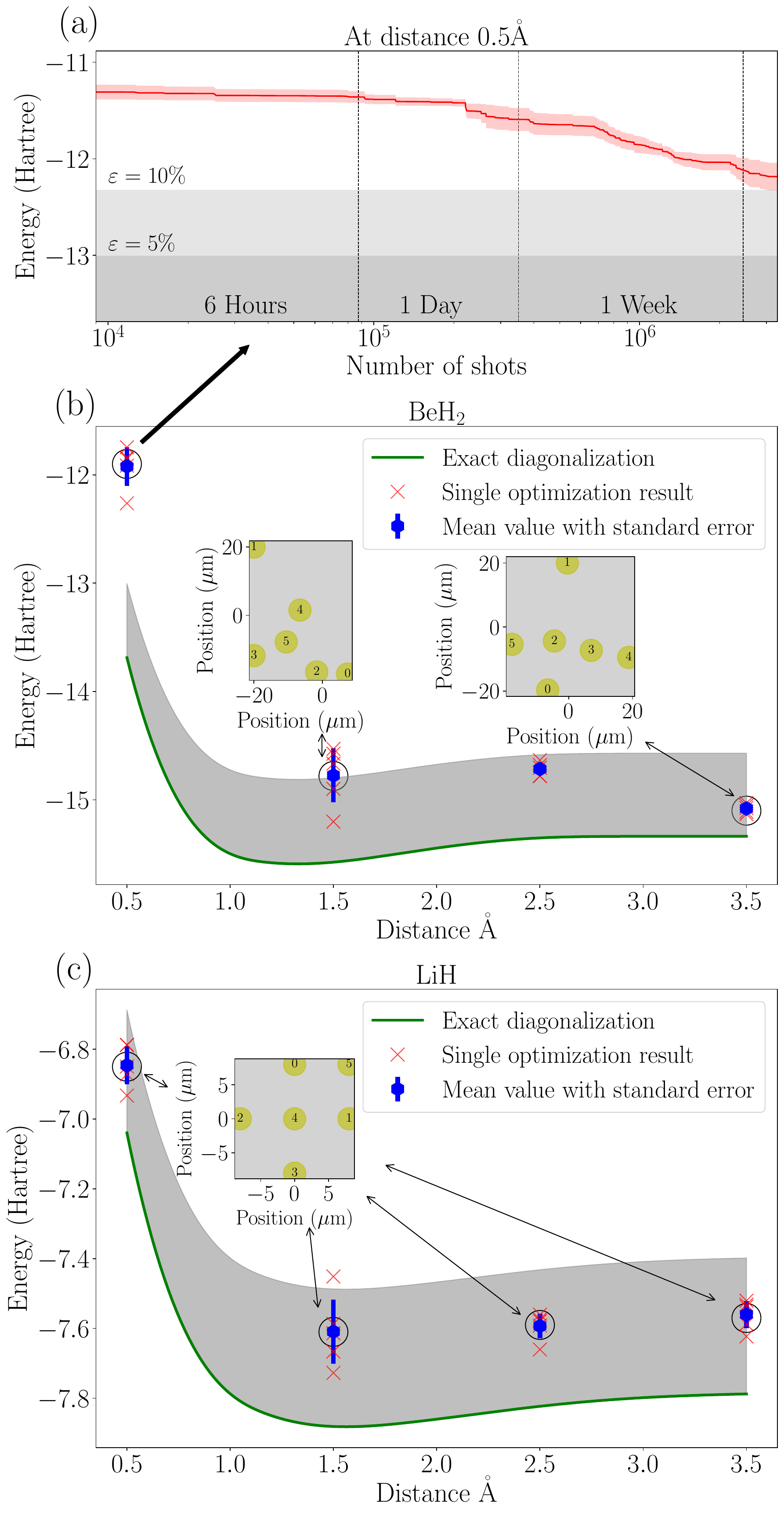} 
    \caption{\textit{Numerical results of the VQE algorithm with our digital-analog protocol} (a) for BeH$_2$ at an intermolecular distance of $0.5$ \r{A} where we increased the number of shots beyond 1 day of experiment and  observed the expected improvement over several days of calculations. The light grey shade and the dark grey shade indicate respectively $\varepsilon=10\%$ and $\varepsilon=5\%$ error benchmark. The red line shows the improvement mean value over 100 run.  
    (b) Result of BeH$_2$ molecule with an encoding of 6 qubits and 165 Pauli strings and (c) result of LiH molecule result with an encoding of 6 qubits and 118 Pauli strings. The insets show the register geometry at specific inter-nuclear distances. For the case of LiH a single heuristic choice was used, while for BeH$_2$ an optimized geometry was prepared at each inter-nuclear distance, minimizing the distance between selected terms of the target Hamiltonian and the interaction energies of the atoms in the register (see Sec. \ref{sec:embedding}). Blue squares:  mean value for several simulations.
    Green line: result from exact diagonalization. The gray shade indicates an $\varepsilon=5\%$ error benchmark. The total number of shots for each optimization result (red crosses) is set to 350000, corresponding roughly to a day of processing in a Rydberg QP.}
    \label{fig:results}
\end{figure}

We have applied the method described in section \ref{Param_pulse} to the LiH and BeH$_2$ molecules. Using the \texttt{Qiskit} \parencite{Qiskit} framework combined with \texttt{Pyquante} \parencite{muller_pyquante2_2022}, we calculate the one and two-body integrals of (\ref{eq:2nd-quant-terms}), encoding the problem into 6 qubits using the Bravyi-Kitaev method. The Hamiltonians contain 118 and 165 Pauli strings respectively. To design the pulse sequence and include realistic device constraints into the simulations we used the open source package \texttt{Pulser} \parencite{silverio_pasqal-iopulser_2022}. The Powell algorithm \parencite{powell_efficient_1964} was used for the classical optimization of the pulse values with 20 function evaluations for each iteration. The two initial Rabi frequency and detuning are chosen randomly in the interval $[0,2\times2\pi]$MHz for each optimization procedure. During the optimization, Rabi frequencies are bounded to this interval to remain within experimentally accessible values \parencite{scholl_quantum_2021}, while the interval for the detuning was taken as $[-2\times 2\pi,2\times2\pi]$MHz. We ran the algorithm five times for four different inter-nuclear distances $R$ yielding the results shown in Fig. \ref{fig:results}. The algorithm converges with small errors in most cases, but we notice the impact of the initial parameters on the obtained energies. For instance at $R=1.5$ {\AA}  for the BeH$_2$ molecule, the obtained energy values are up to $0.4$ Hartree apart. 

To optimize the register geometry, we took the coordinates as parameters, starting from random positions  and minimized the score function described in Sec. \ref{sec:embedding}. We optimized the atom register based on the Nelder-Mead method \parencite{nelder_simplex_1965}, with a few thousands function evaluations, and we also compared to several heuristic choices obtained by a term-by-term comparison with the interaction matrix (a well-performing choice of positions is shown for LiH in Fig. \ref{fig:results})

We define our error as
\be
\varepsilon = |{E_{\text{exact}} - E_{\text{estimated}}|/|E_{\text{exact}}}|,
\ee
where $E_{\text{exact}}=\langle  H_{\text{T}} \rangle$ is the exact diagonalization solution with respect to the target Hamiltonian and $E_{\text{estimated}}$ is the energy calculated with the optimized geometry, the optimized pulse sequence, and the derandomization estimation. The optimized  configurations, together with optimized pulse parameters and energies estimated at each iteration with derandomized measurements give rise to energy errors typically below the $\varepsilon = 5\%$ threshold in less than 350000 shots.

The implementation of the derandomization algorithm allows us to choose the number of measurements that we wish to take (our budget), for a given target accuracy of estimation, which we set to correspond to our $\varepsilon = 5\%$ benchmark. The resulting ``derandomized'' Pauli measurements included typically close to 20 different Pauli strings, calculated from the minimization of the average confidence bound that ensures an empirical average within the chosen $\varepsilon$. Since some derandomized Pauli strings have more operators in common with the terms in the target Hamiltonian (they ``hit'' more target observables), we adjusted the measurement repetitions to be spent proportionally more in them, which improved statistics. We also verified that the obtained accuracy improves upon increasing the allowed number of shots, although we don't expect a full convergence, given the incomplete information used to define $ H_{\text{T}}$.

\subsection{Roadmap for more complex molecules}

\begin{figure}[ht!]
    \centering
    \includegraphics[width=0.3\linewidth,trim={5.5cm 1.8cm 8.5cm 1cm},clip]{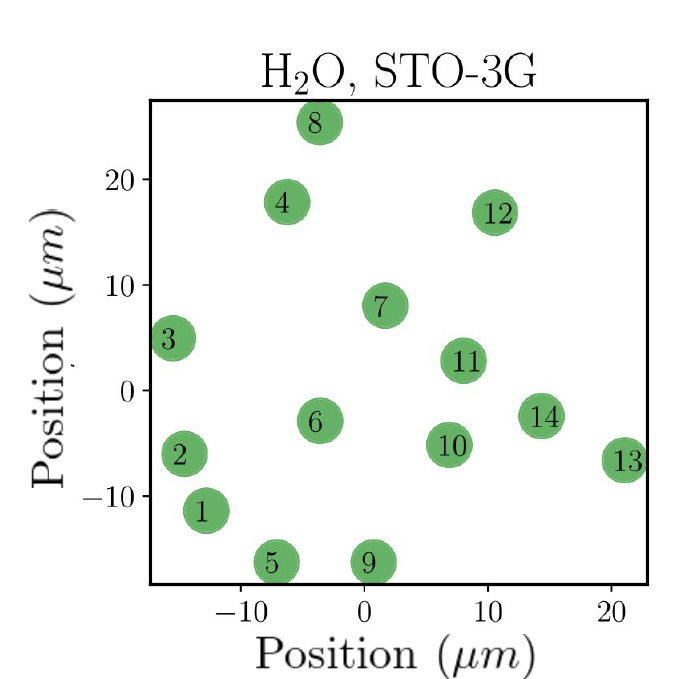}
    \includegraphics[width=0.3\linewidth,trim={5.5cm 2cm 8.5cm 1cm},clip]{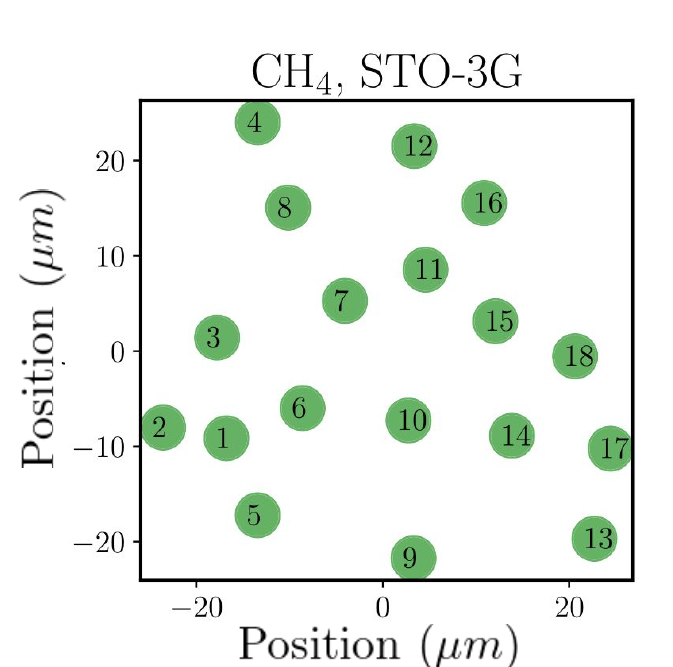}
    \includegraphics[width=0.3\linewidth,trim={5.5cm 1.8cm 8.5cm 1cm},clip]{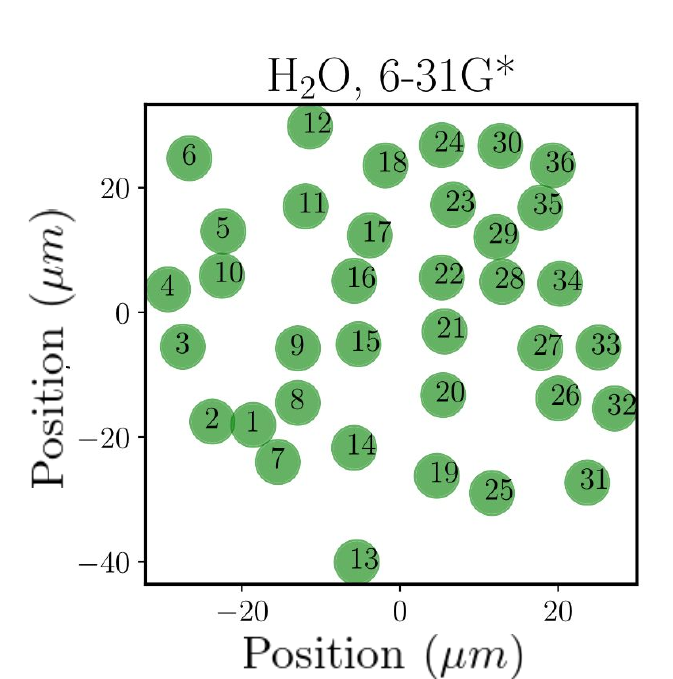}
    \includegraphics[width=0.3\linewidth,trim={5.5cm 2cm 8.5cm 1cm},clip]{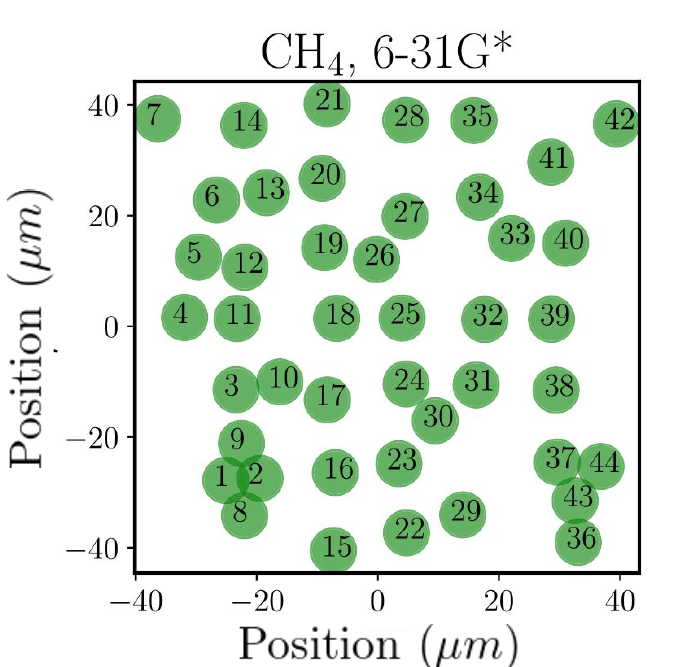}
    \caption{Example embeddings for the H$_2$O and CH$_4$ molecules. Using a Jordan-Wigner encoding, for two different basis, our H$_2$O Hamiltonian (left column) consisted of 14 qubits and 2110 terms (STO-3G basis), of which 595 were chosen as features for the reference Hamiltonian. Choosing a 6-31G* basis gives 36 qubits, with 83003 terms and 5594 of them relevant. For CH$_4$ (right column),  the STO-3G basis gives 18 atoms and 6892 Pauli terms, of which 1359 terms were used for the embedding. The 6-31G* basis requires 44 qubits, 297075 terms and we have picked an embedding with 11772 terms. }
    \label{fig:large_molecules}
\end{figure}

We discuss in this section some observations about the presented protocol for larger molecules and more complex basis sets, where the number of terms in the Hamiltonian and the required qubits to encode it grows quickly. Currently available neutral-atom devices can load hundreds of traps \parencite{schymik2022situ}, but the available space on the register will eventually become a resource limitation. In Fig. \ref{fig:large_molecules} we show the register optimization results for H$_2$O and CH$_4$ in different basis sets, where thousands of terms would need to be measured. Our simple restriction to $Z$-terms captures limited features of the Hamiltonian, mostly concentrating atoms where the largest values need to be reproduced. In fact, as the system size grows, we observe that most of the atoms in the register act as a ``background'' for these clusters. Choosing different terms from the Target Hamiltonian will bring forward other features, highlighting the opportunities of using learning methods to find more performing atom positions.  We have not addressed here the possibilities offered by  three-dimensional registers \parencite{Barredo2018}, which allow for more complex embeddings and have been already studied for graph-combinatorial problems \parencite{dalyac2022embedding}, although they can be straightforwardly included in the protocol.

\begin{figure}[ht!]
\centering
\includegraphics[width=0.8\linewidth]{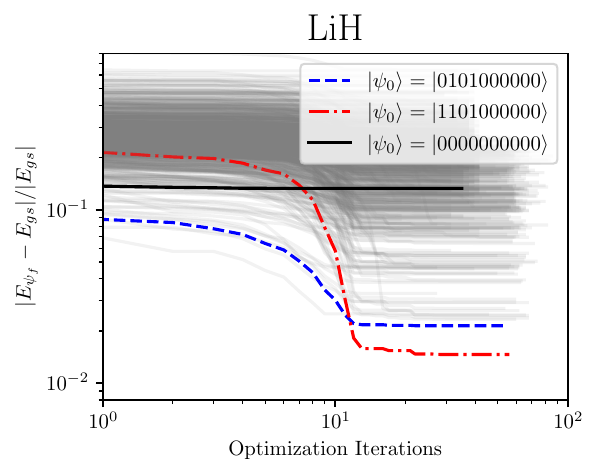}
\caption{Relative error $\varepsilon$ after the first step of the pulse optimization \ref{sec:pulse_opt_protocol}, using different initial product states. While the default $\ket{\psi_0}=\ket{0}^{\otimes N}$ (black line) barely improves its $\varepsilon$ by the optimization step, other product states (blue, dashed line and red, dashed-dotted line) achieve a very low error. We have used an 10-qubit Jordan-Wigner encoding of the LiH Molecular Hamiltonian under the STO-3G basis (276 terms), and averaged several instances for each product state.}
\label{fig:product_states}
\end{figure}

Two aspects of the optimization that rely on classical computation can be further refined: the choice of initial state and the selection of parameters: initializing the optimization with a product state from a Hartree-Fock state approximation can help exploring a lower energy set of output states. In Fig. \ref{fig:product_states} we have scanned through all product states of an 8-qubit system to select those who benefit the most from the first step of the pulse-optimization protocol presented above. The best choice of initial product state can then be used for the Rydberg QP implementation\footnote{Preparing the initial product state requires for example masking atoms with the help of a spatial light modulator.}. Comparing the energy $\langle \psi_0 | U(\theta)^\dagger H_{\text{T}} U(\theta) | \psi_0 \rangle$ of a candidate initial product state $|\psi_0\rangle$ evolving under a constant pulse parameterized by $\theta$ can be performed for example using tensor-network techniques over an HPC backend \parencite{bidzhiev2023emutn, rudolph_synergy_2022}. On the other hand, the number of parameters can be chosen at will, and do not depend on the number of qubits. More advanced control techniques can be applied here, and there is large choice of techniques and numerical tools in the subject of quantum optimal control. We recall that the optimization of the pulse sequence is not dependent on the embedding itself -- it is a global property of the system evolution, where  nearby atoms constitute blockade regions that characterize the final state. 

\section{Discussion}

In this work, we have numerically studied a digital-analog quantum algorithm in the context of quantum chemistry using an ideal Rydberg quantum processor as a hardware.  We have considered small molecules with resource Hamiltonians of two qubits for the H$_2$ molecule and six qubits for the LiH and BeH$_2$ molecules to demonstrate the applicability of our methods. Our purpose was to describe the construction of such an algorithm, discussing the cost  of each stage in terms of the number of measurement repetitions. Our numerical results should be viewed as a first benchmark and should trigger further explorations. Besides, it provides a roadmap for the improvement of Rydberg quantum processor, in particular in terms of cycle time. 

By considering the symmetries of H$_2$ Hamiltonian, we show how the UCC method efficiently and accurately approximates the ground state energy. However, finding a two-body Hamiltonian which commutes with a more general molecular target Hamiltonian is a hard problem. We therefore proposed another protocol for larger molecules: we optimized the geometries of the atomic array, pulse sequences and included an estimation method (derandomization) for the energy measurement. We targeted  $5\%$ of accuracy compared to the exact diagonalization method for Hamiltonian with 6 qubits and more than a hundred Pauli strings.

We observed that the geometry of the array has a significant impact on the result: In the case of LiH, where the target matrix $V^T$ does not provide much information due to the few terms with only two $Z$ operators, the optimized positions underperform with respect to a careful choice of positions, although scaling the heuristics that gives rise to such a geometry for molecules with large number of qubits in their encoding is impractical. This calls for the design of more advanced embedding algorithms. Indeed, for the case of BeH$_2$, the register optimization achieves rather small energy errors, especially for larger distances. 

Previous studies \parencite{progress_vqe} have quantified the demanding resource requirements for practical VQE applications. After several iterations of pulse optimization with energy estimation via derandomization, the error on the average energy of the final prepared state descends to $\varepsilon<5\%$, and expected to be obtained within a day of measurement in a typical current-day Rydberg QP.

In the numerical implementation we have used out-of-the-box optimizers with limitations for the numerical task at hand. Other possibilities include the use of an interpolated waveform for each set of parameters and shaping the pulse using bayesian optimization routines as  explored in \parencite{coelho2022efficient} for the study of combinatorial graph problems. Note that experimentally, one could clone several times the atom layout in spatially separated regions of the register (at least for a small number of qubits), multiplying the obtained number of bitstrings. To achieve close to $1\%$ relative error, we expect that at least a week of Rydberg QP runtime would be necessary (see extrapolation shown in Fig. \ref{fig:results}). The capacity of a circuit ansatz to construct a desired quantum state while keeping a small depth and number of parameters is studied by its \emph{expressibility} \parencite{sim2019expressibility, holmes2022connecting}. In the case of analog systems, this is an emerging topic of research \parencite{tangpanitanon2020expressibility}, with the goal of ensuring that a given ansatz could potentially lead to a good approximation of the ground state and achieve chemical accuracy.

The impact of experimental errors in a real-life implementation will also lead to performance reductions. SPAM (State Preparation And Measurement) errors are typically the largest source of discrepancy for the neutral atom devices \parencite{de_leseleuc_analysis_2018}, but the energy errors observed in numerical simulations remain low as long as the failure rates are small, given that the variational nature of the algorithm shows robustness to several types of errors \parencite{henriet_robustness_2020}. Recently \parencite{guo2022chemistry}, VQE was experimentally implemented in a superconducting quantum processor for H$_2$, LiH and F$_2$ with 4, 6 and 12 qubits, respectively, using the UCC ansatz and a different flavor of measurement protocol \parencite{wu_overlapped_2023}. The readouts went through an error mitigation post-processing routine, which showed that these techniques can greatly compensate the noise effects from their quantum processor, with an error reduction of up to two orders of magnitude and leading to chemical accuracy in some circumstances. We expect that such mitigation can be added to the protocols considered in this paper and will help experimental implementations in Rydberg QP.

To tackle quantum simulation algorithms for energy estimation, more developments in both quantum and classical parts of the hybrid algorithm are needed. Reaching chemical accuracy for molecules with a few tens of qubits remains an open challenge that can now begin to be explored in experimental devices. This will provide evidence to generate new and fundamental insights to understand under what conditions a computational advantage can be achieved. 

\chapter{Using Rydberg platform to simulate strong fermionic correlations in the 2D-Hubbard model}\label{hubbard}
\section{Forewords}
This chapter is a mix of the main text and the supplementary material of a preprint \parencite{michel2023hubbard} submitted to a peer-reviewed journal. It is the result of a collaboration with ATOS/EVIDEN.

\section{Introduction}
Decades of theoretical efforts have led to tremendous progress in the understanding of the exotic phases of strongly correlated electron systems. For instance, lots is known about the physics of their minimal model, the Hubbard model \parencite{Leblanc2015,Schafer2021}. Yet, the exponential difficulty of the underlying many-body problem still poses formidable challenges in low-temperature, doped phases relevant to cuprate superconductors, in multi-orbital settings relevant, for instance, to iron-based superconductors \parencite{Si2016} and the recent Moiré superconductors \parencite{Andrei2021}, or in out-of-equilibrium situations like sudden quenches that lead to a fast growth of entanglement.

\textit{Quantum processors}, i.e., controllable, synthetic quantum many-body systems \parencite{Ayral2023}, are promising  to solve these outstanding challenges \parencite{Feynman1982}.
Ultracold fermionic atoms trapped in optical lattices were already implemented more than a decade ago \parencite{schneider_metallic_2008,Jordens2008,Esslinger2010,schneider_fermionic_2012,schreiber_observation_2015,Hart2015,Cheuk2016,Boll2016,mazurenko_cold-atom_2017,Tarruell2018} as the most direct, or "analog", quantum processors of fermions. They allowed to observe signatures of, for instance, Mott physics, while operating---so far---at temperatures too high to gain insights into pseudogap or superconducting phases.

In contrast, universal "digital" quantum processors rely on  quantum bits encoded on two-level or "spin-1/2" systems, and operate logic gates on them. They in principle enable the simulation of the second-quantized fermionic problems explored in  materials science \parencite{Bauer2020} or chemistry \parencite{Cao2019}.

Yet,  early attempts are facing the physical limitations of these processors
in terms of the number of qubits and number of gates that can be reliably executed before decoherence sets in.
Fermionic systems are particularly demanding due to the loss of locality of the Hamiltonian \parencite{JordanWigner1928,Bravyi2002} or the need for auxiliary qubits \parencite{Verstraete2005,Setia2018,Derby2021} that come with translating to a qubit language. 
Both constraints generically lead to longer quantum programs, and hence an increased sensitivity to imperfections.
To alleviate those issues, hybrid quantum-classical methods \parencite{Bharti2021, Endo2021} such as the Variational Quantum Eigensolver (VQE, \parencite{Peruzzo2014}) were proposed, with many developments but without clear-cut advantage so far.

Despite remarkable recent progress towards large-size digital quantum processors, "analog" quantum processors remain a serious alternative to explore
fermionic problems. Beyond the aforementioned ultracold atoms, 
platforms include systems of trapped ions and cold Rydberg atoms.
The lesser degree of control of these machines---with a fixed, specific "resource" Hamiltonian that does not necessarily match the "target" Hamiltonian of interest --- is compensated by the large number of particles that can be controlled, with now up to a few hundreds of particles \parencite{scholl_quantum_2021,chen_continuous_2023, ebadi_quantum_2021}.
In addition, the parameters of the resource Hamiltonian are usually precisely controlled in time \parencite{glaetzle_designing_2015, bluvstein_controlling_2021, browaeys_many-body_2020, bloch2008many,scholl_quantum_2021,gonzalez-cuadra_fermionic_2023}.
This has enabled the use of analog quantum processors to study many-body problems in several recent works 
\parencite{kokail_self-verifying_2019, gonzalez-cuadra_fermionic_2023, arguello2019analogue, PhysRevA.107.042602, bloch2008many}. 
For instance, \parencite{kokail_self-verifying_2019} have investigated the physics of the Schwinger model---a toy problem for lattice quantum electrodynamics---by leveraging the similarity between the symmetries of a 20-ion quantum simulator and those of the Schwinger model. 
In general, however, such a similarity between target and resource 
Hamiltonians is rare. In particular, the question of how to tackle a fermionic many-body problem with a spin-based, analog simulator is an open problem. 

In this Letter, we propose a method to address this problem using a specific processor, namely an analogue Rydberg quantum processors \parencite{henriet_quantum_2020, browaeys_many-body_2020}.
By using a self-consistent mapping between the fermionic problem and a "slave-spin" model, we circumvent the nonlocality issues related to fermion-to-spin transformations.
We show that the method allows one to compute key properties of the Hubbard model in and out of equilibrium.
We show, through realistic numerical simulations, that it does so even in the presence of hardware imperfections like decoherence, readout error and shot noise coming from the finite number of repetitions of the experiment.

\section{Problem reduction through the slave-spin method.}
As a proof of concept, we tackle the single-band, half-filled Fermi-Hubbard model on a square lattice. Its Hamiltonian:
\begin{equation}
    H_\text{Hubbard} = \sum_{i,j,\sigma} t_{i,j} d_{i\sigma}^{\dagger} d_{j\sigma} + \frac{U }{2} \sum_{i} (n_{i}^d - 1)^2  
      \label{Hubbard_half}
\end{equation}
contains creation (resp. annihilation) operators $d^\dagger_{i\sigma}$ (resp. $d_{i\sigma}$) that create (resp. annihilate) an electron of spin $\sigma$ on lattice site $i$, with a hopping amplitude  $t_{i,j}$  between two sites (we will focus on nearest-neighbor hopping only, $t_{i,j} = - t \delta_{\langle i,j \rangle}$) and an on-site interaction $U$. (The chemical potential was set to $\, \mu=U/2$ to enforce half-filling.) 

This prototypical model of strongly-correlated electrons is generically hard to solve on classical computers, especially in out-of-equilibrium phases where the most advanced methods are usually limited to short-time dynamics. 
Instead of directly tackling this fermionic model, we resort to a kind of separation of variables that singles out two important degrees of freedom of the model, namely spin and charge.
Technically, this is achieved by resorting to a "slave-particle" method known as $Z_2$ slave-spin theory \parencite{ruegg_mathsfz_2-slave-spin_2010,PhysRevLett.112.177001}.     
We replace the fermionic operator $d^{\dagger}_{i\sigma}$ by the product of a pseudo-fermion operator $f^\dagger_{i\sigma}$ and an auxiliary spin field $S^z_i$ ($S^{a=x,y,z}_i$  denote the Pauli spin operators), namely $d_{i\sigma}^{\dagger} = S_i^z f_{i\sigma}^{\dagger}$.
The ensuing enlargement of the Hilbert space is compensated for by imposing constraints $\frac{S_i^x+1}{2} = (n^f_i-1)^2$ on each site. In the particle-hole symmetric case studied here, these constraints will be fulfilled automatically \parencite{schiro_quantum_2011}.

We then perform a mean-field decoupling of the pseudo-fermion and spin degrees of freedom $S^z_{i}S^z_{j}f_{i,\sigma}^{\dagger}f_{j,\sigma} \approx  \langle S^z_{i}S^z_{j}\rangle f_{i,\sigma}^{\dagger}f_{j,\sigma} + S^z_{i}S^z_{j} \langle f_{i,\sigma}^{\dagger}f_{j,\sigma} \rangle - \langle S^z_{i}S^z_{j} \rangle \langle f_{i,\sigma}^{\dagger}f_{j,\sigma} \rangle$. We obtain a sum of two self-consistent Hamiltonians $H \approx H_\mathrm{f} + H_\mathrm{s}$:
\begin{subequations}
\begin{alignat}{2}
    H_\mathrm{f} &= \sum_{i,j, \sigma} Q_{i,j} f_{i,\sigma}^{\dagger}f_{j,\sigma}
    \label{Ham_f}
    \\
    H_\mathrm{s} &=  \sum_{i,j} J_{i,j} S^z_{i}S^z_{j} + \frac{U}{4} \sum_{i} S_i^x \label{Ham_S},
\end{alignat}
\end{subequations}
with $Q_{i,j} = t_{i,j} \langle S_i^z S_j^z \rangle$ and $J_{i,j} = \sum_{\sigma} t_{i,j} \langle f_{i,\sigma}^{\dagger}f_{j,\sigma} \rangle$.

\begin{figure}
    \centering
    \includegraphics[width =  0.7 \linewidth]{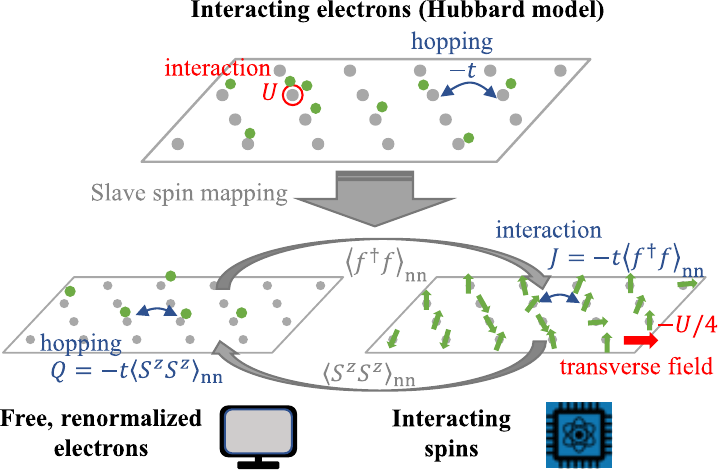}
    \caption{\textit{Slave spin mapping.} The Hubbard Hamiltonian (top) is mapped to two self-consistently determined simpler problems: an efficiently solvable free fermionic Hamiltonian with a renormalized hopping (bottom left, $H_\mathrm{f}(Q)$ in the text), and a transverse field Ising Hamiltonian (bottom right, $H_\mathrm{s}(J)$ in the text), which we tackle using a quantum Rydberg processor.}
    \label{fig:loop_SSMF_final}
\end{figure}

Solving the Hubbard model within slave spin theory amounts to solving these two self-consistently defined problems. This is done in an iterative fashion, as illustrated in Fig.~\ref{fig:loop_SSMF_final}: starting from an initial guess for the renormalized hopping $Q$ to initiate the self-consistent computation, 
(i) the correlation function $\langle f_{i,\sigma}^{\dagger}f_{j,\sigma} \rangle$ of the pseudo-fermion problem, needed to define the spin interaction $J_{i,j}$, can be computed efficiently on a classical computer using a Bogoliubov transformation (see Suppl. Mat \ref{subsec:bogoliubov} for more details),
(ii) the spin problem, on the other hand, is harder to tackle on classical computers.
We thus set out to compute its spin-spin correlation function with a quantum processor.
Since $H_\mathrm{s}$ is of infinite size, we first reduce it to a finite-size problem by using a cluster mean-field approximation, as done in e.g. \parencite{hassan_slave_2010}: 
we solve
\begin{equation}\label{Ham_C}
    H_\mathrm{s}^{\mathcal{C}} = \sum_{i,j \in \mathcal{C}} J_{i,j} S^z_{i}S^z_{j} + \frac{U}{4} \sum_{i\in \mathcal{C}} S_i^x + \sum_{i \in \mathcal{C}} h_i S_i^z,  
\end{equation}
where $\mathcal{C}$ denotes the set of $N$ cluster sites and $h_i  = 2 z_i \overline{J} \overline{m}$ is the self-consistent mean field that mimics the influence of the infinite lattice.
Here, $z_i$ is the number of neighbors of site $i$ outside the cluster, $\overline{J}=\frac{1}{N_p} \sum_{\langle i,j \rangle \in \mathcal{C}} J_{i,j}$ is the average nearest-neighbor coupling ($N_p$ is the number of nearest-neighbor links inside the cluster) and  $\overline{m} = \frac{1}{N}\sum_{i\in\mathcal{C}} \langle S^z_i \rangle$ is the average magnetization.
This model needs to be solved iteratively by starting from a guess for the mean field $\overline{m}$.
For a given value of this mean field, the finite spin problem defined by $H_\mathrm{s}^{\mathcal{C}}$ is solved using a quantum algorithm (described below). 
This yields the correlation function $\langle S^z_i S^z_j \rangle$ and closes the self-consistent loop, which runs until convergence.
At convergence, we can extract useful observables of the original Hubbard model. For instance, the quasiparticle weight $Z$ of the original model, which measures the quantum coherence of the fermionic excitations, can be obtained via the spin model's magnetization: $Z =  \overline{m}^2$ (we also have access to site-resolved magnetizations $\langle S_i^z \rangle$ and hence site-resolved quasiparticle weights).

\begin{figure}
    \centering
    \includegraphics[width =  0.7 \linewidth]{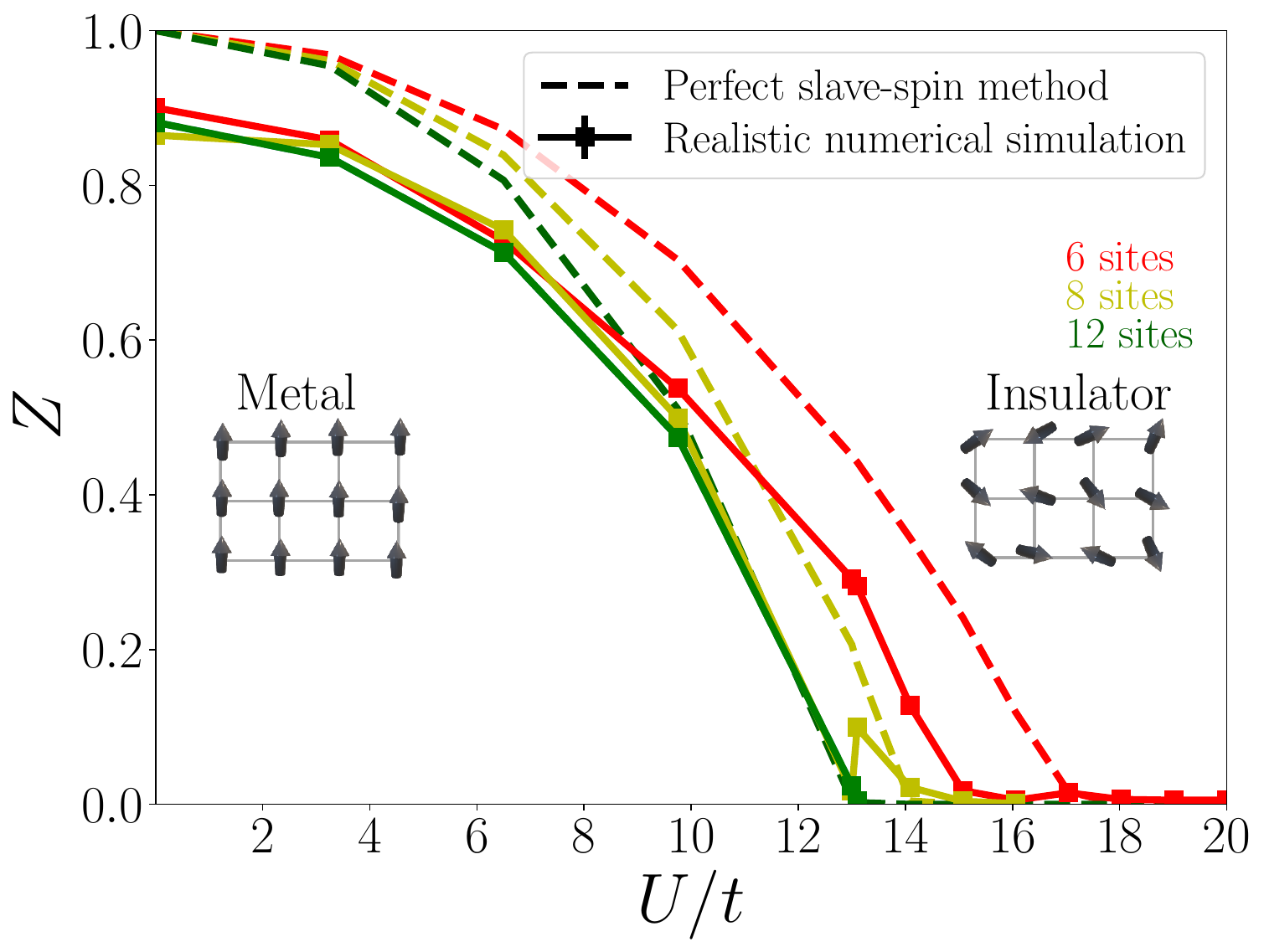}
    \caption{\textit{Mott transition observed with the slave-spin method on a realistic numerical simulation of Rydberg atoms device.} The characteristics of the device considered are  $t_\text{max} = 4 \, \mu$s, $\gamma = 0.02$ MHz, $N_\mathrm{s} = 150$, $\epsilon = \epsilon' = 3$\% and $k = 5$. The error bar is calculated from the errors of $\epsilon$ and $N_\mathrm{s}$.}
    \label{fig:particle_weights}
\end{figure}

\subsection{Details of the main equations}

We choose the most simple form of slave spins, introduced in \parencite{ruegg_mathsfz_2-slave-spin_2010}. We recall its main steps below.
We replace the fermionic operator $d^{\dagger}$ by the tensor product of a pseudo fermion operator (that follows the same anticommutation rules as $d^{\dagger}$) and an auxiliary spin field 
\begin{equation}
    d_{i\sigma}^{\dagger} = S_i^z f_{i\sigma}^{\dagger},
\end{equation}
where $S^z_i$ is the Pauli-$Z$ operator at site $i$ (later $S^a_i$, with $a \in \{x,y,z\}$, will denote the Pauli spin operators), and $f_{i\sigma}^\dagger$ and $f_{i\sigma}$ denote fermionic operators called pseudo-fermions.  
The $d$ and $f$ operators obey fermionic anticommutation relations due to the spin commutation relations.

By substituting, in $H_\mathrm{Hubbard}$, the original fermionic operators by new spin and pseudo fermion degrees of freedom, we effectively enlarge the Hilbert space where the new Hamiltonian, $H'_\mathrm{Hubbard}$, acts.
In practice, we want to map the original problem $H_\mathrm{Hubbard}$ to a Hilbert space of same size by looking at a restriction of the new Hamiltonian, $H'_\mathrm{Hubbard}$, on a restricted portion of the new Hilbert state, which is called the physical subspace. 
This is achieved by imposing a constraint: on each site $i$, we impose the relation \parencite{ruegg_mathsfz_2-slave-spin_2010}:
\begin{equation}\label{constraint_SSMF}
    \left(n_{i\uparrow}^f + n_{i\downarrow}^f - 1 \right)^2 = \frac{S^x_i+1}{2}
\end{equation}
to hold for the "physical states".
Among the eight possible local states, only four states (\textit{i.e.} the same number of original local states) verify this constraint:
\begin{subequations}\label{eq:constraint_states}
\begin{alignat}{3}
    \ket{n_{i}^d=0}&=\ket{S^x = 1, n_{i}^f = 0}, \\
    \ket{n_{i,\sigma}^d=1}&= \ket{S^x= -1, n_{i,\sigma}^f = 1},\,\sigma= \uparrow, \downarrow\\
    \ket{n_{i}^d=2} &= \ket{S^x= 1, n_{i}^f = 2}.
\end{alignat}
\end{subequations}
Assuming this constraint is satisfied in the physical subspace, we can transform the original Hubbard Hamiltonian, expressed as
\begin{align}
\begin{split}\label{Hubbard_half}
     H_\mathrm{Hubbard} =& \sum_{i,j,\sigma} t_{i,j} d_{i\sigma}^{\dagger} d_{j\sigma} + \frac{U}{2} \sum_{i} (n_{i}^d - 1)^2 \\
     \;\;+ & \left(\, \mu - \frac{U}{2} \right)\sum_{i} n_{i}^d,
\end{split}
\end{align}
to the following transformed Hamiltonian:
\begin{align}
\begin{split}\label{Ham_SS}
    H'_\mathrm{Hubbard} =& \sum_{i,j,\sigma} t_{i,j}  S^z_{i}S^z_{j}f_{i,\sigma}^{\dagger}f_{j,\sigma} + \frac{U}{2} \sum_i \left(\frac{S^x_i + 1}{2}\right) \\
     \;\;+ & \left(\, \mu - \frac{U}{2}\right)\sum_{i} (n_{i,\uparrow}^f + n_{i,\downarrow}^f),
\end{split}
\end{align}
via substitution of equality~\eqref{constraint_SSMF} in the interaction term of (\ref{Hubbard_half}). It is straightforward to see that $n_i^d = n_i^f$ considering (\ref{eq:constraint_states}).
At this point, no approximations have been made.

The next step is then to decouple fermions and spins with a mean-field approach
\begin{align}
    \begin{split}
    S^z_{i}S^z_{j}f_{i,\sigma}^{\dagger}f_{j,\sigma} \approx&  \langle S^z_{i}S^z_{j}\rangle f_{i,\sigma}^{\dagger}f_{j,\sigma} + S^z_{i}S^z_{j} \langle f_{i,\sigma}^{\dagger}f_{j,\sigma} \rangle -\\ &\langle S^z_{i}S^z_{j} \rangle \langle f_{i,\sigma}^{\dagger}f_{j,\sigma} \rangle.
    \end{split}
\end{align}
Therefore, Eq.~\eqref{Ham_SS} can be expressed as a sum of two Hamiltonians (neglecting constant terms and considering half-filling) $H'_\mathrm{Hubbard} = H_\mathrm{s} + H_\mathrm{f}$ with $H_\mathrm{s} = \sum_{i,j} t_{i,j}\langle f_{i,\sigma}^{\dagger}f_{j,\sigma} \rangle S_i^z S_j^z + \frac{U}{4}\sum_i S_i^x$, an Ising-like transverse-field Hamiltonian (TFIM) and $H_\mathrm{f} = \sum_{i,j,\sigma} t_{i,j} \langle S^z_{i}S^z_{j}\rangle  f_{i,\sigma}^{\dagger}f_{i,\sigma}$ describing the free renormalized electrons.
The correlators $\langle f_{i,\sigma}^{\dagger}f_{j,\sigma} \rangle$ and $\langle S^z_{i}S^z_{j}\rangle$ are obtained via auto-coherent loops until convergence is reached.

\subsection{Fulfillment of the constraint}
When performing loops described above, one must ensure that the constraint Eq.~\ref{constraint_SSMF} is imposed on each site. In practice, the mean-field simplification leads to 
\begin{equation}\label{eq:constraint_mean}
    \langle(n_{i,\uparrow}^f+ n_{i,\downarrow} - 1)^2\rangle_\mathrm{f} = \left\langle \frac{S_i^x + 1}{2}\right\rangle_\mathrm{s}.
\end{equation}

This equality can be enforced on all sites by using a Lagrange multiplier $\lambda_i$: one adds a term $H_{\lambda} = \sum_i \lambda_i ((n_i-1)^2-\frac{S_i^x+1}{2})$ to $H_\mathrm{s} + H_\mathrm{f}$ and optimizes the corresponding cost function.

In particle-hole symmetric cases (which includes our setting, namely the single-orbital, half-filled Hubbard model on a square---i.e bipartite---lattice), $\lambda_i$ should be zero to respect the symmetry of the energy spectrum around $0$ \parencite{yang_benchmarking_2019, schiro_quantum_2011}.

\subsection{Variants: Towards a multiorbital case}\label{sec:multiorbital}
The $Z_2$ slave spin theory used here is one among others.

Another related slave-spin approach \parencite{demedici_orbital-selective_2005, hassan_slave_2010} consists in enlarging the Hilbert space with the spin operator $S_{i,\sigma}^z$ such as 
\begin{equation}
    d_{i,\sigma}^{\dagger} = f_{i,\sigma}^{\dagger}S_{i,\sigma}^z 
\end{equation}

In this method, the physical states are $\ket{n_{i,\sigma}^d = 1} \Leftrightarrow \ket{n_{i,\sigma}^f = 1, S_{i,\sigma}^z = 1}$ and $ \ket{n_{i,\sigma}^d = 0} \Leftrightarrow \ket{n_{i,\sigma}^f = 0, S_{i,\sigma}^z = -1}$.
The constraint to be satisfied to only span physical states is then $n_{i,\sigma} = S_{i,\sigma}^z + \frac{1}{2}$.

While this method lends itself quite naturally to multiorbital models (see e.g \parencite{demedici_orbital-selective_2005}), the additional $\sigma$ dependency of the slave-spin operators (compared to the Z2 slave spins considered in our work) leads to an effective model which is more difficult to relate to existing experimental platforms.

\section{Solution of the two coupled subproblems}

In this section, we show how we solve numerically Eq. (\ref{Ham_f}) and Eq. (\ref{Ham_S}) to obtain matrices $J$ and $Q$. For $H_\text{s}$, we describe the embedding of the cluster mean-field theory.

\subsection{Solving the fermionic Hamiltonian $H_\mathrm{f}$ for $J$: Bogoliubov method} \label{subsec:bogoliubov}

To compute $J_{i,j}$, we need to compute the one-particle density matrix 
\begin{equation}
    G_{i,j}^\sigma = _{\mathrm{f}}\bra{\psi_0}f_{i,\sigma}^{\dagger}f_{j,\sigma}\ket{\psi_0}_{\mathrm{f}}.
\end{equation}

$H_\mathrm{f}$ can be rewritten as a matrix product:
\begin{equation}\label{eq:Hf_matrix_form}
H_\mathrm{f} = F^{\dagger}QF
\end{equation}
with $F^{\dagger} = (f_{1,\downarrow}^{\dagger},f_{1,\uparrow}^{\dagger},  f_{2,\downarrow}^{\dagger}, \dots)$ and $Q$ a Hermitian, $N\times N$, matrix.
$Q$ can be diagonalized numerically: $Q = LDL^{\dagger}$, with $D= \mathrm{diag} \lbrace \lambda_1, \lambda_2, \dots, \lambda_N  \rbrace$.
If the number of sites is even, the trace of $D$ vanishes and we obtain as many $\lambda_i <0$ as $\lambda_i > 0$.
It leads to define $C^{\dagger} = F^{\dagger}L \iff C = L^{\dagger}F$ and a diagonal form of $H_\mathrm{f}$ is obtained
\begin{equation}
    H_\mathrm{f} = \sum_{i,\sigma}\lambda_i c_{i,\sigma}^{\dagger}c_{i,\sigma}.
\end{equation}

The ground-state energy of this Hamiltonian is the sum of negative $\lambda_i$ and the groundstate is then a Slater determinant $\ket{0101\dots01}_\mathrm{C}$ in the $c$ basis with $1$ corresponding to negative energies and $0$ otherwise. To go back in the $f$ basis, one can use the $L$ matrices:
\begin{align}
G_{i,j}^\sigma
        &= \bra{\psi_0}\sum_{k,k',\sigma} L^{\dagger}_{k,i}L_{j,k'}c^{\dagger}_{k,\sigma}c_{k',\sigma}\ket{\psi_0}\\
        &= \sum_{k,k',\sigma}\delta_{k,k'} L^{\dagger}_{k,i}L_{j,k'}\langle c^{\dagger}_{k,\sigma}c_{k',\sigma}\rangle\\
        &= \sum_{k,\sigma} L^{*}_{i,k} n_{k,\sigma} L^{t}_{k,j}.
\end{align}
with $n_{k,\sigma}$ equal to $1$ for $k$ indices where $\lambda_k < 0$.
Numerically, it means that only matrices $L$ and eigenvalues $\lambda_i$ are needed to compute $J$.
This part can be dealt with a classical quantum computer 
as it has a polynomial complexity. Going into the thermodynamic limit makes it easier as the system is really translation invariant and the Hamiltonian is then diagonal in the Fourier space. In our work, we choose to solve $H_\mathrm{f}$ considering boundaries to show the effect of a finite-size system on the method. Further developments could be done to simplify this step and consider an infinite-size system.


\subsection{Solving the spin Hamiltonian via a cluster mean-field approach} \label{subsec:cluster_mean_field}

\begin{figure}
    \centering
    \includegraphics[width=1.0 \linewidth]{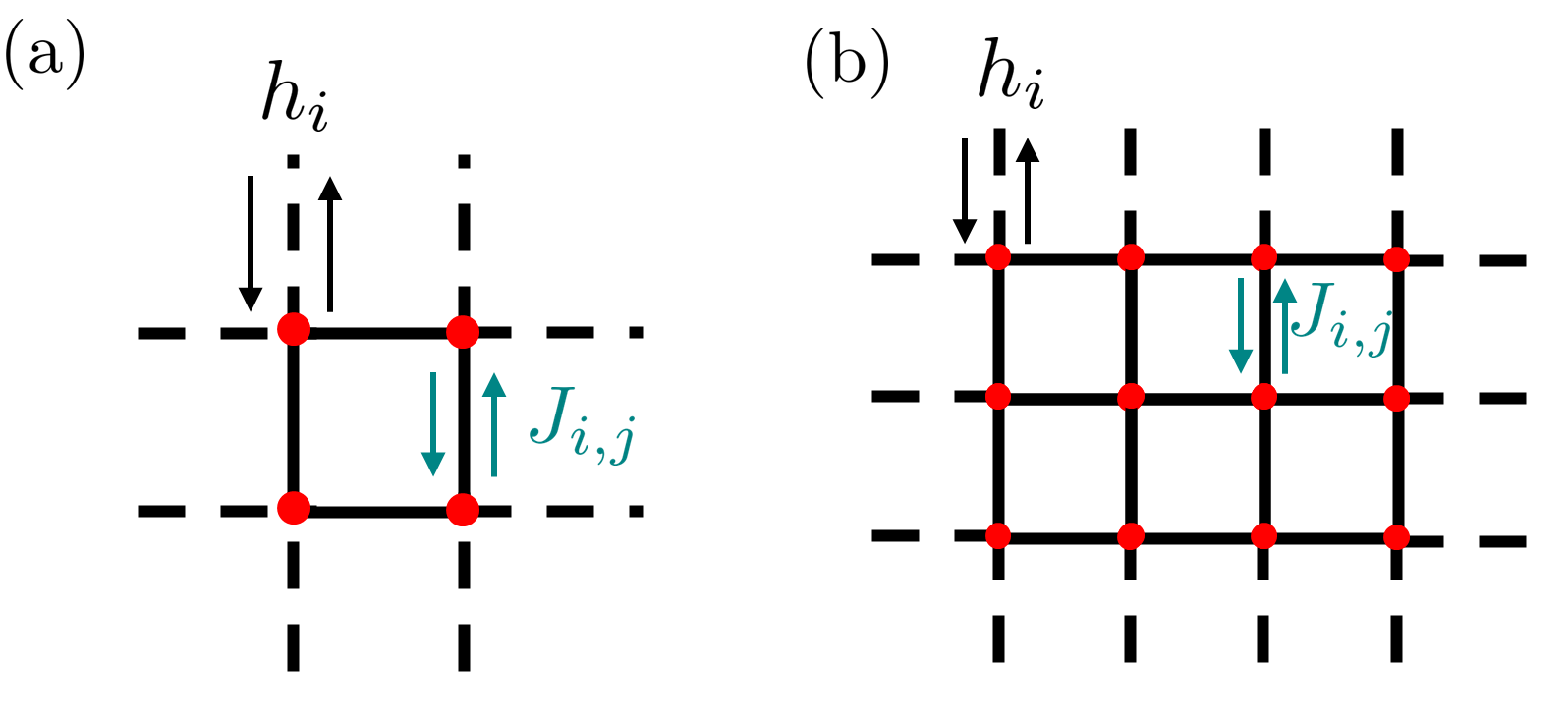}
    \caption{Schematic representation of the cluster geometry for (a) $N=4$ sites and (b) $12$ sites. The dashed black lines represent the interaction with the surrounding mean-field whereas the full black line show the interactions within the cluster. In the twelve sites lattice, the two sites inside the cluster do not interact with the mean-field. }
    \label{fig:CMFA}
\end{figure}

We now focus on the computation of $Q_{i,j}$. It requires the computations of the spin-spin correlation function $\langle S_i^z S_j^z \rangle$.

We consider a cluster of $N_\text{x}$ columns and $N_\text{y}$ rows (see Fig.~\ref{fig:CMFA} for an example) surrounded by a mean field. The number of sites in the cluster is then $N = N_\text{x} \times N_\text{y}$.


The cluster mean-field approximation leads  to
\begin{equation}
    S_i^z S_j^z \approx \langle S_i^z \rangle S_j^z + \langle S_j^z \rangle S_i^z - \langle S_i^z \rangle \langle S_j^z \rangle,
\end{equation}
where $i$ ($j$) is inside the cluster at the border of it and $j$ ($i$) is not. The mean-field parameter $\langle S^z_i \rangle$ is the same for all sites in the thermodynamic limit. As we consider finite-size systems,  we numerically compute 
\begin{equation}
    \overline{m} = \frac{1}{N} \sum_{i=1}^{N} \langle S_i^z \rangle.
\end{equation}
This mean magnetization will be the one outside the cluster following a self-consistent loop. 

Therefore, $\sum_{i,j} J_{i,j} S^z_{i}S^z_{j} = \sum_{i,j} J_{i,j} (m S_j^z + m S_i^z)  = m \sum_{i,j} J_{i,j} (S_j^z + S_i^z)$, neglecting constant terms. However, the matrix element $J_{i,j}$ is not known for a site $i$ inside the cluster and a site $j$ outside of it. In the thermodynamic limit, all $J_{i,j}$ are equals as it is the one-particle density matrix of a free fermionic system. We can thus take the mean value of all $J_{i,j}$ for nearest neighbors inside the cluster to guess the interaction between sites inside and outside the cluster. Let's then define 
\begin{equation}
\overline{J} = \frac{1}{N_\text{p}} \sum_{\langle i,j \rangle} J_{i,j}
\end{equation}
where the sum goes all over nearest neighbors in the cluster and $N_\text{p}$ is the number of such pairs. In the square lattice, each site has 4 nearest neighbors. We can define a number $z_i$ which is the number of neighbors outside the cluster for site $i$. For example, this number is equal to $0$ for a site which has 4 neighbors inside the cluster.

Finally we obtain a mean-field term

\begin{equation}
    \sum_{i \in \mathcal{C}} h_i S_i^z = 2 \overline{J} \overline{m} \sum_{i \in \mathcal{C}} z_i S_i^z
\end{equation}
and we obtain Eq.~\eqref{Ham_C}.



\subsection{Convergence of the self-consistent loop}
\begin{figure}
    \centering
    \includegraphics[width=0.7 \linewidth]{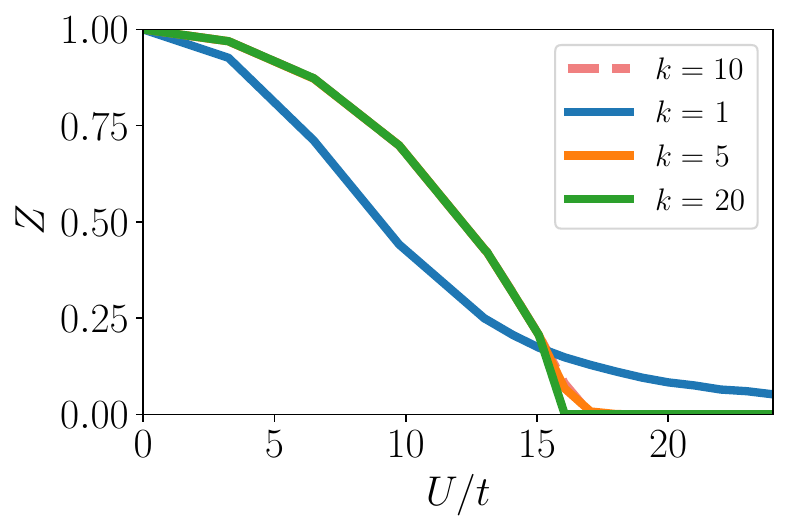}
    \caption{\textit{Impact of imposed number of loops $k$ in the slave-spin mean-field theory for a cluster of $N=4$ sites.} The resolution method is annealing and all sources of error are neglected.
    \label{fig:lmax}
    }
    
\end{figure}

\begin{figure}
    \centering
    \includegraphics[width = 0.8 \columnwidth]{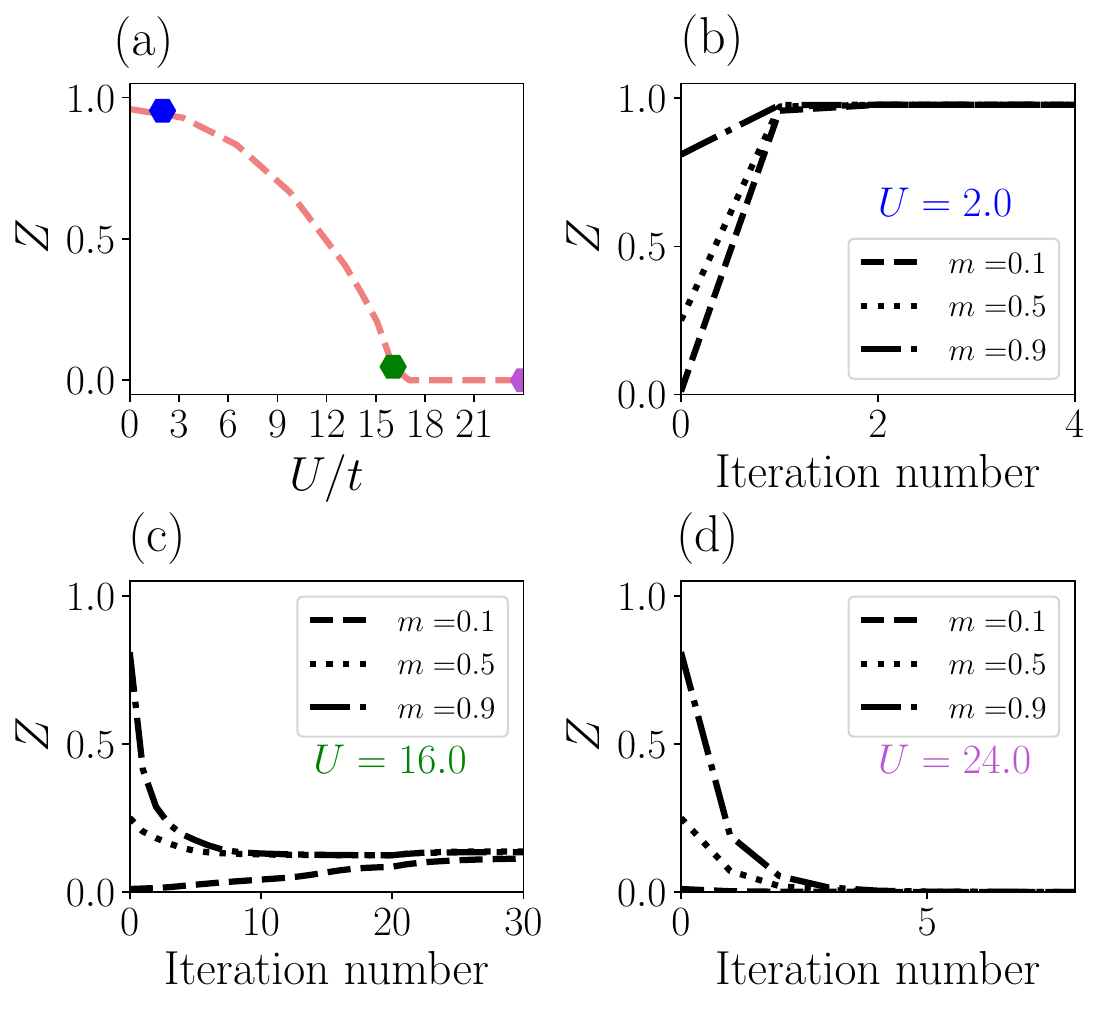}
    \caption{\textit{Evolution of $Z$ as a function of loop iterations for a 6 sites embedding}. (a) Mott transition for a $3 \times 3 $ cluster where three points are highlighted. The convergence of these points ((b) $U/t=2.0$, (b) $U/t=16.0$ and (d) $U/t=24.0$) during the slave-spin mean-field procedure is shown for different initial guess of the mean field $\overline{m}$ (0.1, 0.5 and 0.9). The solving method is annealing where all source of noise are neglected. The number of allowed iteration is increased to $100$ but the x-axis are limited to convergence in the three panels for sake of clarity and the error accepted is $\eta = 10^{-5}$.}
    \label{fig:h_CMFT_iterations}
\end{figure}

The self-consistent procedure to solve the inner loop is first to guess an initial value for the magnetization $m_0$, then to solve Eq.~(\ref{Ham_C}) and calculate $\overline{m} = \frac{1}{N} \sum_i^{N} \langle S_i^z \rangle$ in the groundstate obtained. The loop goes on until a convergence criteria is reached. In our simulation, two criteria are used: the number of inner loop and outer loop can be narrowed by a number $k$ (so the total number of loop allowed is $k \times k$). The second criteria is the norm of the difference between $Q$ at step $l-1$ and $Q$ at step $l$ for the outer loop and the norm of the difference between $\overline{m}$ at step $l-1$ and $\overline{m}$ at step $l$ for the inner loop. We choose a value $\eta$ such as the loop stop if one of the two norm is lower than $\eta$. in our simulation we choose $\eta = 0.01$.
The evolution of $Z$ as a function of iterations is shown in Fig.~\ref{fig:h_CMFT_iterations} for a cluster of $N = 6$ sites, $k=10$ and $\eta = 10^{-6}$. Different initial guess for $\overline{m}$ are tested and they all converge toward the same value which states for the robustness of the method. The convergence takes more time close to the transition value. The impact of the number of loops $k$ imposed is shown in Fig.~\ref{fig:lmax}.

\section{Quantum algorithm for the spin Hamiltonian.} 
Let us turn to the solution of the (cluster) spin problem, $H_\mathrm{s}^{\mathcal{C}}$.
It is nothing but the transverse-field Ising model, which has recently been claimed to be a potential candidate problem for reaching quantum advantage using gate-based quantum processors \parencite{Kim2023}.
As it turns out, its form is very similar to the Hamiltonian realized experimentally by Rydberg atoms trapped with optical tweezers~\parencite{browaeys_many-body_2020}:
\begin{equation}\label{eq:reshamglob}
H_\mathrm{Rydberg}= \sum_{i\ne j}\frac{C_6}{|\textbf{r}_i-\textbf{r}_j|^{6}} \hat{n}_i \hat{n}_j +  \frac{\hbar\Omega (\tau)}{2}\sum_{i} \hat{S}_i^x - \hbar \delta(\tau)\sum_{i} \hat{n}_i,
\end{equation}
where $\Omega(\tau)$ and $\delta(\tau)$ are the time-dependent Rabi and detuning drives, and $C_6$ the magnitude of the interatomic van der Waals interactions; $\hat{n}_i = (I + S^z_i)/2$.
The main difference between $H_\mathrm{Rydberg}$ and $H_\mathrm{s}^{\mathcal{C}}$ is the sign of the interaction: it is positive for Rydberg atoms, negative (since usually $t_{i,j}<0$) for the slave spin problem.
We can nevertheless make use of the Rydberg platform: instead of looking for its ground state, we are going to be looking for its most excited state, which we attain using an annealing procedure:
we start from drive parameters $\Omega(\tau=0) = 0$ and a large negative $\delta(\tau=0)$ so that the system's native initial state $\ket{\psi_\mathrm{start}}=\ket{g}^{\otimes N}$ is the most excited state of the initial Hamiltonian. We then, over a long enough annealing time, linearly ramp the Rabi and detuning drives to reach the final values $\hbar\Omega(\tau_\mathrm{max}) = \frac{U}{2}$, 
$\hbar\delta_i(\tau_\mathrm{max}) = \sum_{j \neq i} \frac{C_6}{r_{i,j}^6} + 4\overline{J}\overline{m}z_i$. 
Provided we can, in addition, optimize the atom positions in such a way that $\frac{C_6}{r_{i,j}^6} \approx -4J_{i,j}$ (details about this optimization are in Sec.~\ref{subsec:optim}), the final Hamiltonian will be $-H_\mathrm{s}^\mathcal{C}$.
Hence, following the adiabatic theorem (applied to the most excited state), the procedure should (approximately) bring the system to the most excited state of $-H_\mathrm{s}^\mathcal{C}$ and hence the ground state of $H_\mathrm{s}^\mathcal{C}$. We can finally measure the spin-spin correlation function on this state.

\section{Solving the spin model with a Rydberg platform: details}\label{subsec:rydberg}

In order to solve Eq.~\eqref{Ham_C}, we propose to use the Ising Hamiltonian Eq.~\eqref{eq:reshamglob} generated by Rydberg atoms device. 

As discussed in the main text, this is achieved via an annealing procedure whose final Hamiltonian is supposed to be as close as possible to the Hamiltonian whose ground state correlations functions we want to compute.

In this section, we discuss in more detail the annealing procedure and the deviations from the ideal case that we take into account.

\subsection{Optimization of the geometry} \label{subsec:optim}

The Hamiltonian we are considering, Eq.~\eqref{Ham_C}, displays a self-consistently determined spin coupling matrix $J_{i,j}$, while the Hamiltonian that is controlled in the experiment displays a van der Waals interaction term $\sum_{j \neq i} \frac{C_6}{r_{i,j}^6}$. This section explains how we optimize the atom positions to make both couplings match as much as possible.

\begin{figure}[!h]
    \centering
    \includegraphics[width = 0.6 \linewidth]{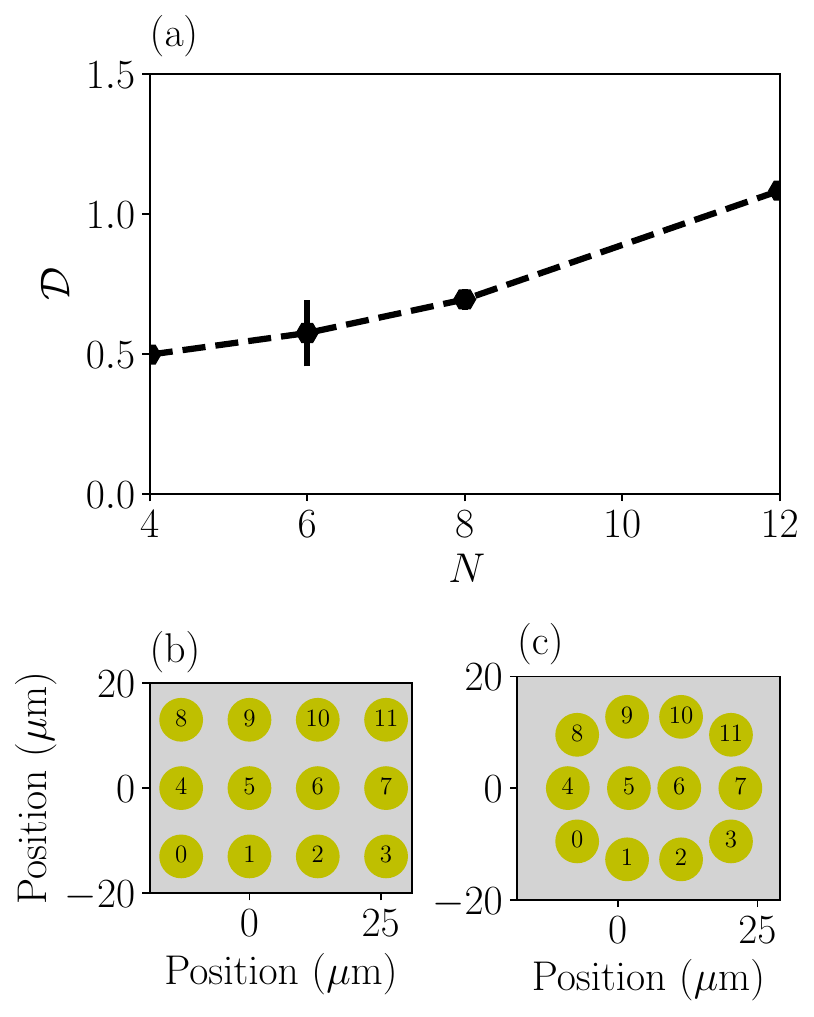}
    \caption{\textit{Optimization of geometry for an implementation on real device.}(a) Mean value of the cost function $\mathcal{D}$ (Eq.~\ref{eq:norm}) for different cluster size at $U = 13.1 $ MHz. The error bar shows the standard deviation over all $\mathcal{D}$ values encountered during loops. (b) Initial position of the atoms before optimization for a $N=12$ sites cluster in the last outer loop of the slave-spin mean-field method at $U=13.1$ MHz. (c) Position of atoms after the optimization of the geometry to minimize $\mathcal{D}$. }
    \label{fig:D_cost}
\end{figure}

Our goal is to minimize the cost function
\begin{equation}\label{eq:norm}
    \mathcal{D} = \sqrt{\sum_{i, j} \left ( \frac{C_6}{|r_{i,j}|^6} + 4J_{i,j} \right )^2 }.
\end{equation}
We use the conjugate gradient descent algorithm from the scipy library \parencite{2020SciPy-NMeth}\, with the following initial guess for the geometry: we place the atoms on a square lattice where the distance between nearest-neighbor atoms is $r_\text{init} = \text{max}_{i,j}(\frac{C_6}{|4J_{i,j}|})^{\frac{1}{6}}$.

The evolution of $\mathcal{D}$ as a function of the number of sites in the cluster is shown Fig.~\ref{fig:D_cost}. The optimization of the positions does not lead to a vanishing $\mathcal{D}$. In practice, the gradient descent algorithm can be trapped in numerous local minima, leading to a poor approximation of $-4 J_{i,j}$ by the interaction matrix element. In addition, difficulties can arise directly from the symmetries of the initial cluster guess. For instance, in the case of a $2 \times 2$ cluster, if distance between nearest-neighbors is called $a$, the distance between next nearest neighbors is always $a/\sqrt{2}$ whereas it should be $0$ for our model since $J_{i,j} = 0$ for next nearest neighbors. Therefore, in most cases, $\mathcal{D}$ is not exactly zero and finding the best geometry is not an easy task.

\begin{figure}
    \centering
    \includegraphics[width=1. \linewidth]{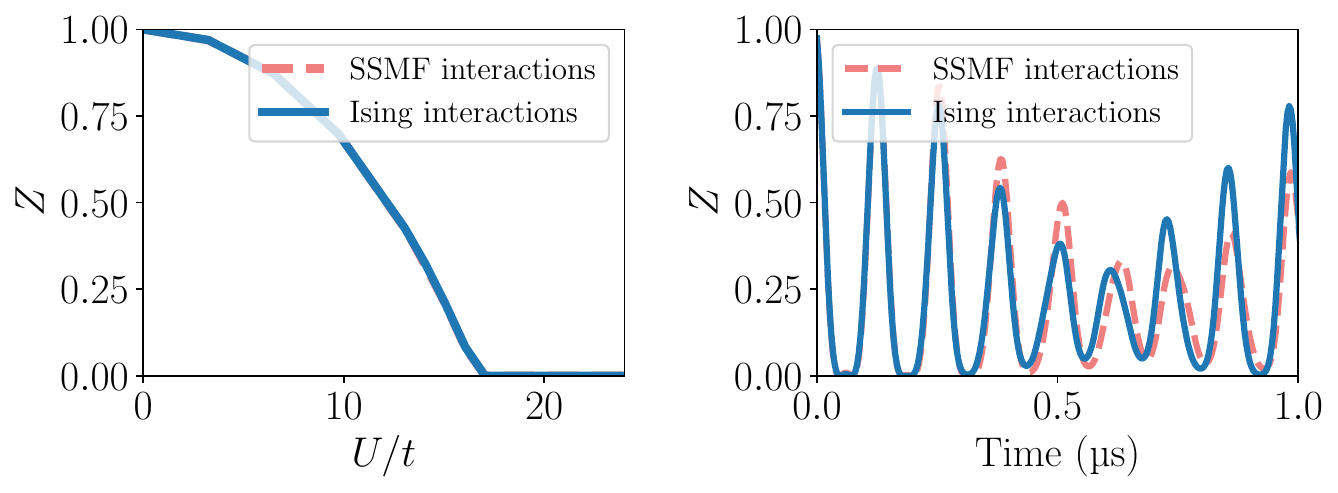}
    \caption{\textit{Impact of considering a realistic geometry on a cluster of $N=4$ sites}. \textit{Left:} Comparison of $Z$ values between method with the real matrix $J$ and the optimized one for 4 sites. \textit{Right:} $Z$ dynamics after a quench $U_\mathrm{f} = 13$ MHz with the same comparison. All other sources of noise are neglected. }
    \label{fig:optimized_geo}
\end{figure}

Despite these imperfections, it seems that the impact of considering an imperfect optimization of the geometry (leading to a nonzero $\mathcal{D}$) does not lead to significant changes. In Fig.~\ref{fig:optimized_geo}, we show the outcome of the equilibrium and out-of-equilibrium computations with a "perfect geometry" (assuming the coupling is actually $J_{i,j}$) and an imperfect geometry.
For the Mott transition, differences can be seen to be negligible for $N=4$ cluster sites. 
For the dynamical behavior, we can observe a change in amplitude  but the frequency remains the same as for the slave-spin mean-field interactions.
To illustrate the outcome of the optimization procedure, we also show an example of initial and optimized position for $N = 12$ cluster sites in Fig.~\ref{fig:D_cost}.
One can observe that the final pattern is slightly distorted compared to the translation-invariant initial pattern. This is due to the fact that the couplings at the edges of the cluster differ from the ones in the "bulk" of the cluster to account for the cluster's environment. As the cluster size grows, these edge effects will have less and less influence, and the optimization will become easier and easier.

Thus, we can conclude that the geometry optimization yields reasonably faithful interactions.

\subsection{Details of the annealing schedule}

Once the geometry is found, the atoms are prepared in the state $\ket{\psi_\mathrm{start}} = \ket{g}^{\bigotimes N}$.
The following Hamiltonian is the one applied at $\tau=0$
\begin{equation}\label{eq:Hstart}
    H_\mathrm{start} = \sum_{i\ne j}\frac{C_6}{|\textbf{r}_i-\textbf{r}_j|^{6}} \hat{n}_i \hat{n}_j -\hbar \delta_\mathrm{start} \sum_i n_i
\end{equation}
where $\delta_\mathrm{start}$ is set to $- 5$ MHz so that $\ket{\psi_\text{start}}$ is the most excited state of (\ref{eq:Hstart}).
The Rabi frequency and the detuning are then driven during a time $\tau_\text{max}$ to reach the Hamiltonian $-H_\mathrm{s}^{\mathcal{C}}$
A global addressing is performed for the Rabi frequency and for the detuning.

Following the procedure described in the main text, the Rabi frequency starts at $0$ MHz and is driven linearly to $\frac{U}{2}$ ($\hbar\Omega(\tau_\mathrm{max}) = \frac{U}{2}$). Similarly, the detunings are all prepared at a value $\delta_\mathrm{start}$ and are driven separately to values $\hbar\delta_i(\tau_\mathrm{max}) = \sum_{j \neq i} \frac{C_6}{r_{i,j}^6} + 4\overline{J}\overline{m}z_i$.

\begin{figure}
    \centering
    \includegraphics[width=0.7 \linewidth]{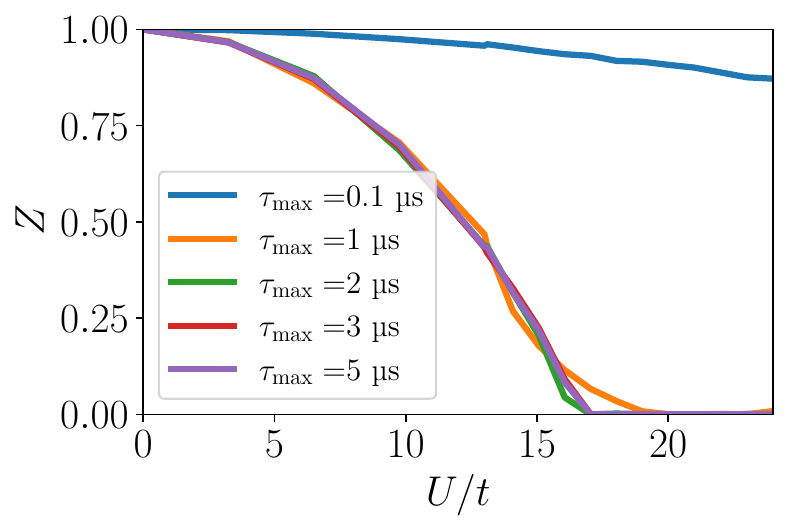}
    \caption{\textit{Impact of total annealing time for a cluster of $N = 4$ sites.} All other sources of error are neglected. 
    \label{fig:Figure_slave_spin_sev_tmax}
    }
    
\end{figure}

The effect of annealing time is shown in Fig.~\ref{fig:Figure_slave_spin_sev_tmax}. Starting from $\tau_\text{max} = 3 \, \, \mu$s, the impact seems to be insignificant. In our simulations, we choose  $\tau_\text{max} = 4 \, \mu$s to ensure a good convergence.

\begin{figure}
    \centering
    \includegraphics[width=0.7 \linewidth]{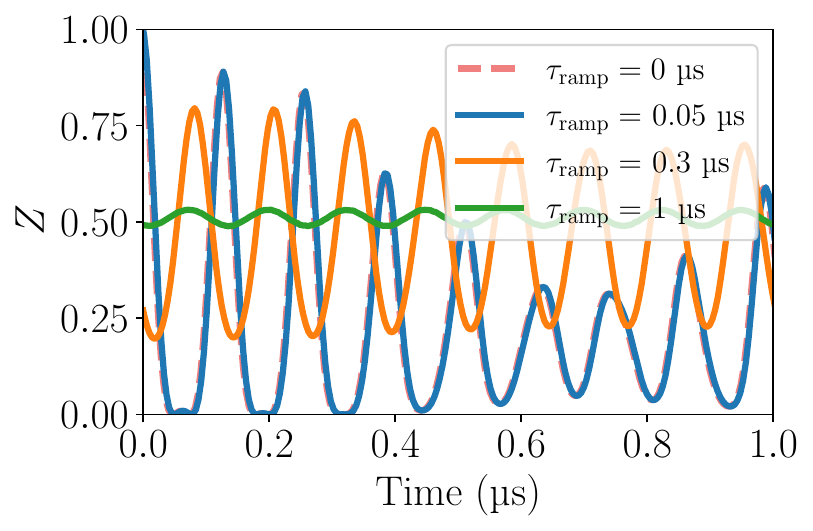}
    \caption{\textit{Impact of switch-on time $\tau_\mathrm{ramp}$ in the quench dynamic for a cluster of $N = 4$ sites.} We consider $U_\mathrm{f} = 13$ MHz. All other sources of error are neglected. 
    \label{fig:tramp}
    }
    
\end{figure}

To perform the study of the dynamics in the Hubbard model, we need to quench the value of the Rabi term. In practice, the quench is not instantaneous. In Fig.~\ref{fig:tramp}, we investigate the effect of the finite switch-on time $\tau_\mathrm{ramp}$ on the Rabi and detuning. We see that this time impacts the frequency of the signal for $\tau_\mathrm{ramp} \geq 0.3 \, \mu$s. At the actual device specifications $\tau_\mathrm{ramp} \approx 0.05\, \mu$s, the effects are negligible.  

\subsection{Experimental imperfections}

The algorithm described in the main text is designed to work on actual neutral atoms devices. As they are NISQ computer, it is necessary to evaluate effects of noise and limitations on the results of the method. In the following, methods to emulate noise are described and implemented in our code.

All numerical simulations are performed with the library QuTiP \parencite{johansson_qutip_2013} (exact diagonalization) and the Quantum Learning Machine. The SPAM error is implemented with a code from Pulser \parencite{silverio_pasqal-iopulser_2022}.

\subsubsection{Dephasing noise}
Decoherence during the annealing procedure is described via the Lindblad master equation (following \parencite{lienhard_observing_2018}):
\begin{equation}
    \frac{d \rho}{d\tau} = -i[H(\tau),\rho] - \frac{1}{2}\sum_{i =1}^{N} \gamma_i \Big [ \Big \{ L_i^{\dagger} L_i, \rho \Big \} -2L_i \rho L_i^{\dagger} \Big ] 
    \end{equation}
where $\rho$ is the density matrix describing the mixed state of the system and $H(\tau)$ is the resource Hamiltonian at a time $\tau$ during the annealing.
The jump operators $L_i$ corresponding to dephasing are equal to $n_i$. 
We choose to simplify the problem by taking a single dephasing parameter $\gamma_i = \gamma$.

\begin{figure}[!h]
    \centering
    \includegraphics[width=1.\linewidth]{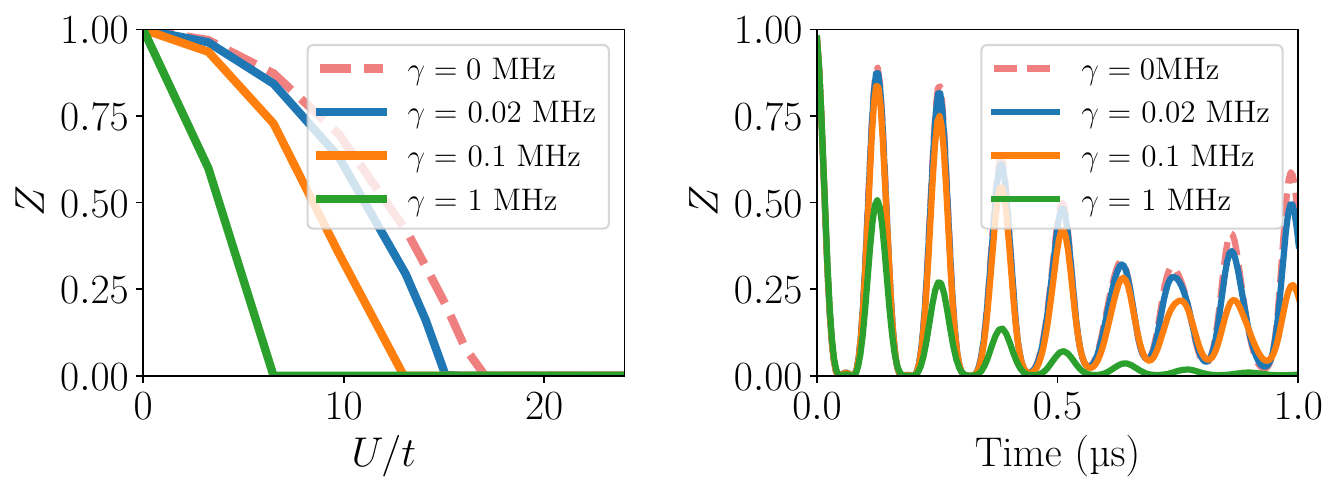}
    \caption{\textit{Effect of dephasing noise on the result of for a cluster of $N = 4$ sites.} \textit{Left:} Impact one the Mott phase transition. All other sources of noise are neglected. \textit{Right:} Impact on the quench dynamic for $U_\mathrm{f} = 13$ MHz. The color code is the same as for the top panel. }
    \label{fig:gamma}
\end{figure}

The effect of this noise is shown as a function of dephasing parameters in Fig.~\ref{fig:gamma}. For the quench dynamics, the dephasing damps the oscillations but do not change the frequency. We can see this behavior on the Mott transition where it is shifted for small values of $U/t$ for larger $\gamma$.

\subsubsection{Sampling and measurement error}

\begin{figure}[!h]
    \centering
    \includegraphics[width=1. \linewidth]{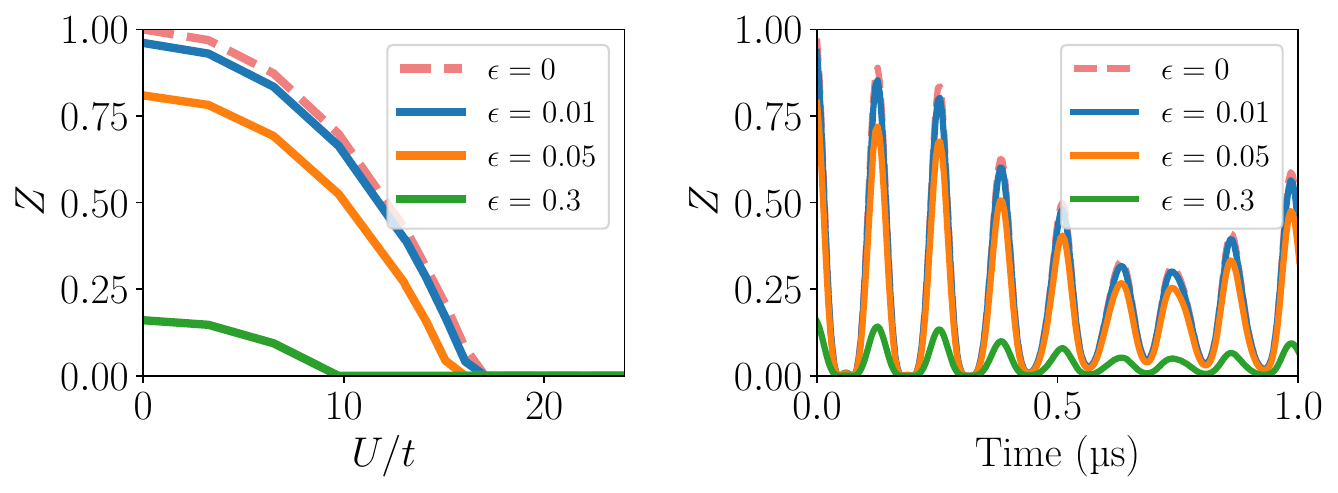}
    \caption{\textit{Impact of measurement error for a cluster of $N = 4$ sites}. \textit{Left:} shows the effect of $\epsilon = \epsilon'$ at equilibrium and \textit{right:} out of equilibrium for $U_\mathrm{f} = 13$ MHz. The number of shots considered for each measurement is $10^{6}$. All other sources of noise are neglected. }
    \label{fig:eps}
\end{figure}

\begin{figure}[!h]
    \centering
    \includegraphics[width=1.\linewidth]{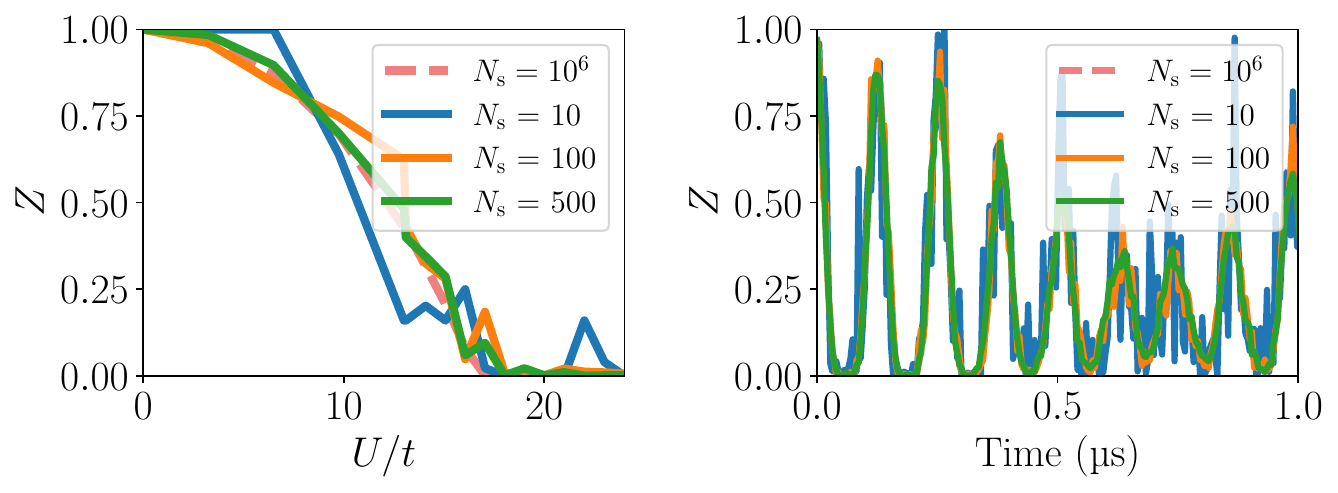}
    \caption{\textit{Impact of different sampling rate $N_\mathrm{s}$ on measured states for a cluster of $N = 4$ sites.} \textit{Left:} at equilibrium and \textit{right:} out of equilibrium ($U_\mathrm{f} = 13$ MHz). All other sources of error are neglected. }
    \label{fig:shotnoise}
\end{figure}

We simulate the sampling of states as in Rydberg atoms device by picking randomly $N_\text{s}$ times a bitstring with a probability equal to the probability to measure this bitstring in the z-basis on the device. The impact of such a procedure is shown Fig.~\ref{fig:shotnoise} at equilibrium and out of equilibrium. The impact is negligible starting from $N_\mathrm{s} \approx 100$. 
We model the readout error by a probability (in $\%$) of error $\epsilon$ of detecting an atom in a state $\ket{r}$ instead of its real state $\ket{g}$ and $\epsilon'$ of not detecting an excited atom. Experimentally, these values are around $2\sim 3\%$ \parencite{de_leseleuc_analysis_2018}. We choose $\epsilon = \epsilon' = 3\%$ for both values in the main text. The impact of such error for $\epsilon = \epsilon'$ is shown in Fig.~\ref{fig:eps}. Until $\epsilon \approx 5 \%$, the behavior of the system remains the same.

\section{Results at equilibrium}
We implemented this self-consistent procedure with a realistic numerical simulation of the annealing algorithm used to solve the spin problem with Rydberg atoms. We repeated the computation for several values of the local Hubbard interaction $U$ to obtain the evolution of the quasiparticle weight  $Z$ as a function of $U$, as shown in Fig.~\ref{fig:particle_weights}, for cluster sizes, and hence number of atoms, of 4, 6, 8 and 12.
The major experimental limitations were considered in order to account for the true potential of current devices: dephasing noise, shot noise, measurement error, global detuning, finite annealing times $\tau_\mathrm{max}$ and imperfect positioning of the atoms to reproduce the right magnetic coupling (see Sec.~\ref{subsec:rydberg} for more details).
Despite these limitations, leading to few points being far from the noiseless result due to error repetitions within all loops, the quasiparticle weight we obtain (solid lines) is fairly close to the one obtained with a perfect solution of the spin model (dashed lines). The Rydberg platform can thus be used to get a reasonable estimate of the Mott transition, i.e the value $U_c$ when $Z$ vanishes and the systems turns Mott insulating.
While for the half-filled, single-band model studied in this proof-of-concept example, classical methods can be implemented to efficiently solve the spin model (see e.g \parencite{Schuler2016}), other regimes are less readily amenable to a controlled classical computation: doped regimes, multi-orbital models, and dynamical regimes. We investigate the latter regime in the next paragraph.

\begin{figure}
    \centering
    \includegraphics[width = 0.8 \linewidth]{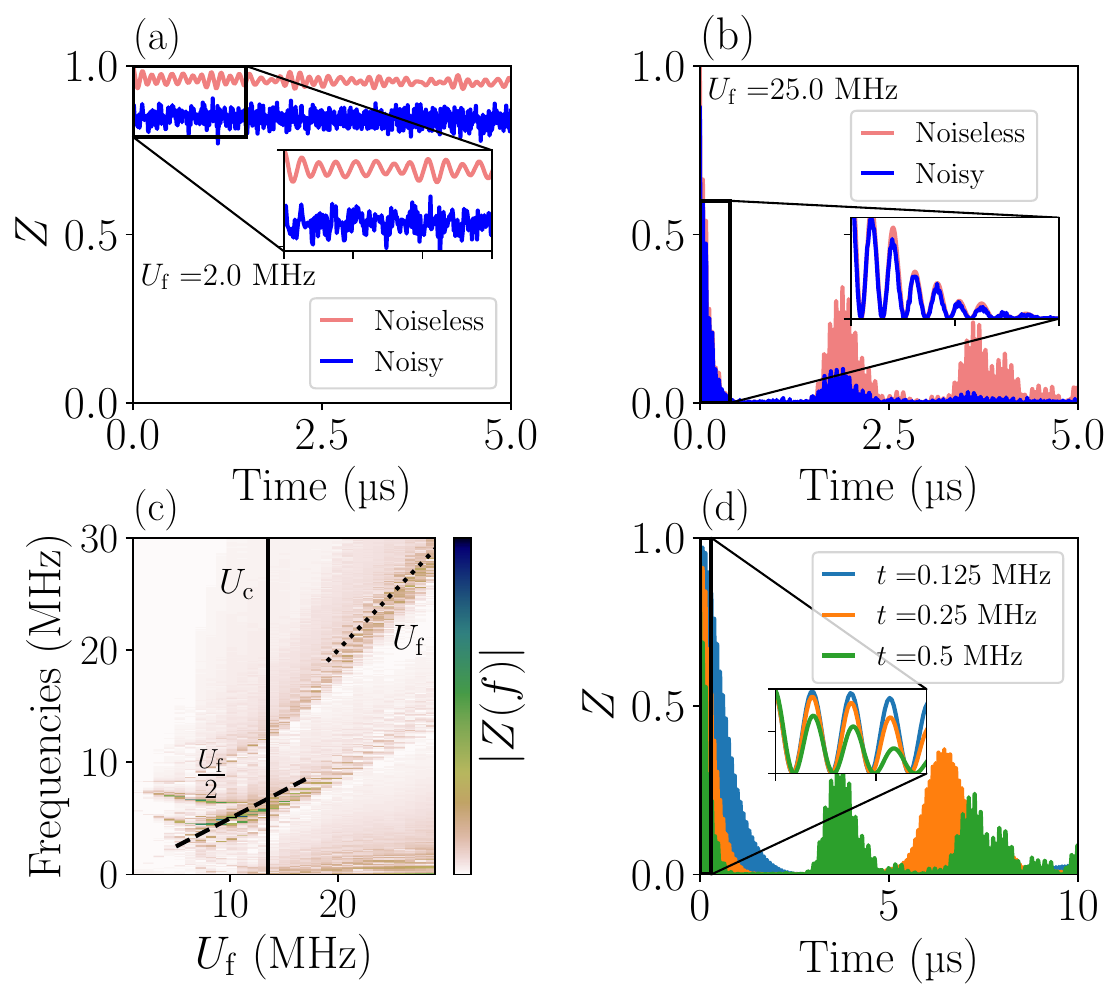}
    \caption{\textit{Dynamical response of the quasiparticle weight after an interaction quench.} $N=12$ spin cluster. Time evolution of $Z$ for (a) $U_\mathrm{f} = 2$ MHz and (b) $U_\mathrm{f} = 25$ MHz. The red line shows the noiseless annealing solution and the blue line a realistic numerical simulation on Rydberg atoms device ($\gamma = 0.02$ MHz, $\epsilon=\epsilon' = 3\%$, $N_\mathrm{s} = 150$ shots, realistic Ising interactions and a global detuning are imposed).
    (c) Fourier transform amplitude $|Z(f)|$ for several $U_\mathrm{f}$. 
     The vertical black line shows the equilibrium critical value $U_c$ as computed from Fig.~\ref{fig:particle_weights}.
     (d) Impact of the hopping terms $t$ on the damping of the response of $Z$ after a quench ($U_\mathrm{f} = 13 \approx U_c$). The blue, orange and green lines represent the result for $t = 0.125$, $0.25$ and $0.5$ MHz, respectively.  }
    \label{fig:quench}
\end{figure}

\section{Dynamics of the Hubbard model with the slave-spin method} 
We thus turn to a dynamical setting to emphasize the potential advantage brought by the use of quantum processors when used within this slave-spin framework.
Starting from a noninteracting ground state ($U=0$), we suddenly switch on the value of the local interaction to a final value $U_\mathrm{f}$.
Our goal is to validate that the method solved with a physically realizable quantum processor can recover the phenomenology observed in previous experimental and theoretical studies of quenched Hubbard systems \parencite{greiner_collapse_2002, kollath_quench_2007, schachenmayer_atomic_2011,lacki_dynamical_2019,
eckstein_thermalization_2009,iyer_coherent_2014,
schiro_quantum_2011,will_observation_2015,
riegger_interaction_2015, yang_benchmarking_2019}, to wit:
collapse and revival oscillations of various observables in the $U_\mathrm{f} \gg U_c$ regime, with a $2\pi/U_\mathrm{f}$ period, and a damping that increases with bandwidth. In the $U_\mathrm{f} \ll U_c$ regime, overdamped oscillations have been observed (see \parencite{eckstein_thermalization_2009} for instance).

\subsection{Dynamics in slave-spin theory} \label{subsec:dynamics}

In this subsection, we review how the slave-spin formalism extends to the time-dependent case.

After the introduction of the slave variables, we are considering the Hamiltonian:
\begin{equation}
H'_\text{Hubbard} = \sum_{i,j} S_i^z S_j^z f_i^{\dagger} f_j + \frac{U}{4} \sum_i S_i^x
\end{equation}
(Eq.~\eqref{Ham_SS} at half-filling and neglecting the constants).
At the mean-field level, the time-dependent solution of the Schrödinger equation
is of the form $ \ket{\Psi(\tau)} = \ket{\Phi_\text{f}(\tau)}\ket{\Psi_\text{s}(\tau)}$ with $\ket{\Phi_\text{f}(\tau)}$ (the time will be defined as $\tau$ to avoid confusion with the hopping) governed by a Schrödinger evolution with time-dependent Hamiltonians:
\begin{align}
    \begin{split}
H_\text{f}(t) &= \sum_{i,j} t_{i,j} \langle S_i^z S_j^z \rangle (\tau)f_i^{\dagger}f_j\\
H_\text{s}(\tau) &= \sum_{i,j} t_{i,j} S_i^z S_j^z \langle f_i^{\dagger}f_j \rangle (\tau) + \frac{U}{4} \sum_i S^x_i.
\end{split}
\end{align}
The initial state is of the form $\ket{\Psi(\tau = 0)} = \ket{\Phi_\text{f}(\tau=0)}\ket{\psi_\text{s}(\tau = 0)} $ with $\ket{\Phi_\text{f}(\tau=0)}$ (respectively $\ket{\Phi_\text{f}(\tau=0)}$) the ground states of $H_\text{f}(\tau < 0)$ (respectively $H_\text{s}(\tau < 0))$ found with the mean-field slave-spin procedure.
To solve these coupled equations, we a priori need to compute correlators $\langle S_i^z S_j^z \rangle (\tau)$ and $\langle f_i^{\dagger}f_j \rangle (\tau)$ and use them ton construct $H_\text{f}(\tau)$ and $H_\text{s}(\tau)$. We should then evolve the wavefunctions to obtain correlators for a time $\tau + d\tau$ and so on.

In fact, as stated in \parencite{schiro_quantum_2011}, the dynamics of the pseudo-fermions is trivial if our system is translation invariant (i.e $t_{i,j} = t_{i-j} $).
In this case, indeed, $\langle S_i^z S_j^z \rangle (\tau) = g_{i-j}(\tau)$. Thus, 
\begin{equation}
    H_\text{f}(\tau) = \sum_{i,j} t_{i-j}g_{i-j}(\tau) f_i^{\dagger}f_j.
\end{equation}
This Hamiltonian is then diagonal in the Fourier space
\begin{equation}
    H_\text{f}(\tau) = \sum_k \epsilon_k(\tau) f_k^{\dagger} f_k,
\end{equation}
with $\epsilon_k(\tau)$ the Fourier transform of $ t_{i-j}g_{i-j}(\tau)$, and $n_k = f_k^{\dagger}f_k$, with $f_k \propto \sum_i e^{i k R_i} f_i$. We denote as $\ket{\Phi_{\alpha}}$ the Fock states of the system associated with the transformed operators, $f_k$. They are the eigenstates of $H_\text{f}(\tau)$ at any time $\tau$.

The initial state $\ket{\Phi_{\alpha_0}}$ is the ground state of the system.
Let's consider the time evolution of an arbitrary state $\ket{\Phi_\text{f}(\tau)}$, we can decompose $\ket{\Phi_\text{f}(\tau)} = \sum_{\alpha} c_{\alpha}(\tau) \ket{\Phi_{\alpha}} $. Therefore,
\begin{align}
    \begin{split}
        \sum_{\alpha} i\partial_t c_{\alpha}(\tau) \ket{\Phi_{\alpha}} &= \sum_{\alpha} c_{\alpha}(\tau)H_\text{f}(\tau) \ket{\Phi_{\alpha}} \\
        &= \sum_{\alpha} c_{\alpha}(\tau)E_{\alpha}(\tau)\ket{\Phi_{\alpha}} 
    \end{split}
\end{align}
One can project onto $\bra{\Phi_{\alpha}(\tau)}$:
\begin{equation}
i \partial_t c_{\alpha}(\tau) = c_{\alpha}(\tau)E_{\alpha}(\tau).
\end{equation}
Thus, $c_{\alpha}(\tau) = c_{\alpha}(\tau = 0)e^{-i \int_0^t E_{\alpha}(\tau')d\tau'}$.

Therefore, starting from the groundstate, for $\alpha \neq \alpha_0$, $c_{\alpha}(\tau) = 0$ and $c_{\alpha_0}(\tau) = e^{-i\phi(t)}$. At the end of the day, 
\begin{equation}
    \ket{\Phi_\text{f}(\tau)} = e^{-i\phi(\tau)}\ket{\Phi_\text{f}(\tau = 0)}
\end{equation}

and the renormalized fermionic system remains in the groundstate up to global phase, meaning that $\langle f_i^{\dagger}f_j \rangle (\tau) = \langle f_i^{\dagger}f_j \rangle_0$ is independent of time. This enables us to only consider the correlator $\langle S_i^z S_j^z \rangle (\tau)$ during the quench.

\subsection{Frequency dependency on eigenenergies}

The link between eigenenergies of $H(U_\mathrm{f})$ and the frequency of oscillations can be derived: the initial state is $\ket{\psi_\mathrm{s}(\tau < 0)}$, the groundstate of $H_\mathrm{s}(U=0)$. We can decompose it in the basis of $H(U_\mathrm{f})$ eigenstates: $\ket{\psi_\mathrm{s}(\tau < 0)} = \sum_k a_k \ket{E_k}$ where $\ket{E_k}$ are eigenstates of $H(U_\mathrm{f})$ corresponding to an eigenenergy $E_k$. Let's now consider the value of an observable $\hat{O}$ through time. We obtain:
\begin{align}
    \begin{split}
        \langle \hat{O} \rangle (\tau) &= \bra{\psi_\mathrm{s}(\tau < 0)}e^{iH(U_\mathrm{f})\tau} \hat{O} e^{-iH(U_\mathrm{f})\tau} \ket{\psi_\mathrm{s}(\tau < 0)}\\
        &= \sum_{k,k'}a_k^{*}a_{k'}\bra{E_k} e^{iH(U_\mathrm{f})\tau} \hat{O} e^{-iH(U_\mathrm{f})\tau} \ket{E_{k'}}\\
        & = \sum_{k,k'}a_k^{*}a_{k'}e^{i(E_k -E_{k'})\tau} \bra{E_k}  \hat{O}  \ket{E_{k'}}
    \end{split}
\end{align} 
Therefore, frequencies of oscillations of any observable only depend on differences between eigenenergies of $H(U_\mathrm{f})$.

\subsection{Dynamics and constraint fulfillment}
In the $Z_2$ slave-spin theory, one can define the projector $Q_i = \Big ( \frac{S_i^x+1}{2}-(n_i-1)^2 \Big )^2 = \frac{1+S_i^x e^{i\pi n_i}}{2} $ such that $Q_i\ket{\Psi} = 0$ iff $\ket{\Psi}$ respects the constraint Eq.~\eqref{eq:constraint_states}.
Using the fact that $[H_\mathrm{Hubbard}',\prod_i Q_i ]=0$, the constraint is fulfilled during the quench dynamics because $\prod_i Q_i\ket{\Psi(\tau = 0)}=0$ (see \parencite{schiro_quantum_2011,ruegg_mathsfz_2-slave-spin_2010} for more details).

\subsection{Results out of equilibrium}

Here, we look for this phenomenology in the time evolution of the quasiparticle weight $Z$.
Within slave-spin applied to the single-site Hubbard model at half-filling, interaction quenches are particularly simple to implement: translation invariance on the lattice makes the dynamics of pseudo-fermions trivial when starting from an eigenstate \parencite{schiro_quantum_2011}. 
Thus, only the dynamics of the spin model are of interest: the procedure boils down to quenching the value of the transverse field in Eq.~\eqref{Ham_C} from $0$ to $U_\mathrm{f}/4$.
On our target Rydberg platform, this means switching the Rabi frequency from zero up to the desired value to obtain $U_\mathrm{f}$.
In practice, the switch-on time is not instantaneous but very fast (about $50$ ns to switch from $0$ MHz to $U_\mathrm{f} = 2$ MHz).
One can directly measure $\langle S_i^z \rangle$ on the device for different evolution times. 
The main limiting factor is thus the measurement rate, in addition to the aforementioned sources of noise.
In Fig.~\ref{fig:quench}, we show the oscillations we observe numerically for a cluster of $12$ sites, with and without noise.
The upper panels present the oscillation of $Z$ as function of time after a quench to $U_\mathrm{f} = 13$ MHz (a) and to $U_\mathrm{f}= 25$ MHz (b).
From Fig.~\ref{fig:particle_weights}, we know that the phase transition for such a cluster is $U_c \approx 13.5$.
In the case of $U_\mathrm{f} = 25$ MHz ($U_\mathrm{f} \gg U_c$), we clearly observe the damped oscillations, whether in the noiseless or the noisy setting. 
Because of dephasing noise of the experiment, the agreement between the noiseless and noisy curve becomes worse with time. However, during the first µs of observations, we recover the perfect signal and the estimation of the oscillation frequency is possible (insets in (a) and (b)).
For $U_\mathrm{f} = 13$ MHz ($U_\mathrm{f} \approx U_c$), we see that $Z$ quickly reaches a value $\approx 0.1$ (slightly higher than the $Z$ obtained for this value of $U$ at equilibrium), around which it oscillates. 
Panel (c) exhibits the Fourier transform of $Z(\tau)$ for various $U_\mathrm{f}$ for the exact slave-spin method (namely with an exact solution of the spin model).
For $U_\mathrm{f} < U_c $, components at $\omega = U_\mathrm{f}/2$ can be identified along with other contributions, while for $U_\mathrm{f}>U_c$, $\omega = U_\mathrm{f}$ end up dominating the spectrum.
This is expected from the physics of the Mott transition in the Hubbard model: above the transition, the single-particle spectrum displays a Mott gap of $U_\mathrm{f}$, while below it excitations between the quasiparticle band and the emerging Hubbard bands (with energy $U_\mathrm{f}/2$), and within the quasiparticle band, are possible.
Finally, panel (d) confirms the expected increase of the damping of oscillations with the hopping strength $t$.

\section{Conclusion} 
In this work, we introduced a hybrid quantum-classical method to study the equilibrium and dynamics of a prototypical model for strongly correlated fermions, the Hubbard model.
Our method makes use of a spin-based quantum processor but does not suffer from the usual overheads of translating fermionic problems to spin problems, namely long quantum evolutions (due to nonlocal spin terms) or auxiliary quantum degrees of freedom.
This is made possible by the use of an existing advanced mapping, the slave-spin method, that turns the difficult fermionic problem at hand into a free, and thus efficiently tractable, fermion problem that is self-consistently coupled to an interacting, yet {\it local} spin problem. 
This locality makes this spin problem well suited for current quantum processors based on spins (aka qubits).

Here, to solve the spin problem, we turned to an {\it analog} quantum processor made of Rydberg atoms, as opposed to a {\it gate-based} quantum processor.
Despite being a priori restricted---because of the limited number of knobs in the Hamiltonian---in the class of the problems that it can deal with, the Rydberg platform is particularly well suited for the spin problem at hand because its Hamiltonian can be made to almost exactly coincide with the effective spin Hamiltonian to be solved.
Moreover, its analog character allows one to circumvent the usual issues associated with gate-based algorithms, like trotterization when performing time evolution, or the variational aspects inherent to many NISQ algorithms like VQE or its temporal counterparts.
Finally, the number of Rydberg atoms that can be controlled in current experiments allow to tackle problem sizes that are very hard to reach using classical methods.

This proposal calls for many further investigations. An important step is an experimental validation with larger atom numbers than the 12 atoms we simulated here.
Other future improvements involve the slave-spin method itself: doped regimes (relevant to cuprate materials), multiorbital models (relevant to iron-based superconductors, where orbital-selective effects may appear \parencite{demedici_orbital-selective_2005}) pose various technical difficulties that warrant further theoretical developments. In particular, the fulfillment of the constraint to ensure the states remain in the physical subspace becomes more difficult in these regimes than in the half-filled, single-band case that we studied here. Going beyond the mean-field decoupling of the pseudo-fermion and spin variables is also another interesting avenue.


\chapter{Summary and outlook}\label{conclusion}
Quantum simulation is a potential very strong tool to solve plenty of problems. One of the most promising field is the simulation of materials and many-body systems. Indeed, quantum simulator are often using many-body system to perform computation and the first idea that comes to mind is to map real-life materials problem onto the many-body system of the simulator. The digital approach, despite being very flexible and promising for all types of problems, is still limited to small number of noisy qubit and the gates fidelity has not reached yet the famous three "9" ($99.9\%$) when several gates are piled up. For all these reasons, I choose to develop analog quantum algorithm during my thesis having in mind an experimental realization during or soon after my PhD.

The Rydberg platform offers a way to simulate many-body physics up to hundreds of spins with a great fidelity. The two Hamiltonian one can implement on it are at the center of the study of magnetism and spin dynamics. It is therefore a logical choice to develop quantum algorithm on this platform for simulating materials and electronic systems of interest. 

\paragraph{Quantum chemistry} During the course of my PhD, I have strongly collaborated with the company PASQAL. From this collaboration we have proposed a new hybrid algorithm which combines a digital and an analog approach. This algorithm has been numerically tested on the $\mathrm{H}_2$,LiH and $\mathrm{BeH}_2$ molecules. For the $\mathrm{H}_2$ molecule, we are using the "XY" interaction of the neutral atoms as it shares common symmetries with the molecular Hamiltonian. This leads to obtain the groundstate of the molecule with a great precision. For the two other molecules, the geometry of the atom array is optimized with respect to selected features of the molecular Hamiltonian transformed into a qubit Hamiltonian with a qubit-fermion mapping. The pulse sequence is designed to have a constant duration and to increase the number of parameters. This leads to enhance the expressibility of the quantum "circuit" during the whole procedure and at the same time, avoid the impact of the choice of the initial parameters. Each step of the variational procedure performs a derandomization energy estimation. This method allows to reduce drastically the number of measurement needed to estimate the energy of an Hamiltonian. We show that one can estimate the groundstate energy of the molecules for several molecular inter-atomic distances to a few percent points of error ($\approx 5 \%$). 

\paragraph{Simulating strong fermionic correlations in the 2D-Hubbard model}
The second half of my PhD was dedicated to develop an new hybrid algorithm to simulate a Fermi-Hubbard model at and out of equilibrium on a RQP. This work is the result of a strong collaboration with the company EVIDEN/ATOS and PASQAL.
The algorithm is based on the slave-spin approach: a condensed-matter method to map the complexity of the Fermi-Hubbard model into an Ising-like model. To this aim, the original problem Hilbert space is enlarged by adding a "slave" spin to each site of the model. In order to restrain the problem to the physical space, a constraint on the occupation and slave-spin values on each site has to be fulfilled. The crucial part is now to perform a mean-field approximation to decouple fermionic and slave spin degrees of freedom. This leads to approximate the original Hamiltonian as a sum of a free-fermions Hamiltonian and an Ising-like Hamiltonian. This two Hamiltonian are self-correlated. We choose to solve this model on a square bipartite lattice to elude the constraint.  The free-fermions system can be solved classically with a polynomial complexity. The Ising-like Hamiltonian is solved by a cluster mean-field approach. Spin-spin correlators are computed to obtain a new kinetic term for the fermionic Hamiltonian. Fermionic correlators are then computed to obtain the new spin interaction term of the Ising Hamiltonian. The self-correlated loop goes on until convergence.

This method is used to compute numerically the Mott transition of the Hubbard model at equilibrium. We show that we can simulate the Ising-like Hamiltonian on RQP to calculate the correlators and one can recover the Mott transition even if a noisy experimental implementation is considered. The dynamical behavior of the Hubbard model can also be studied in the slave-spin paradigm. The observation of out-of-equilibrium Fermi-Hubbard behavior is very difficult classically and being able to do it on a RQP could be a proof of a quantum advantage.

\paragraph{Perspectives}
Our work paves the way of quantum computing for electronic structure simulation with a quantum simulator. The two proposed hybrid algorithms are implementable on a NISQ computer, the RQP, and therefore can be run on real device really soon. This work is one of the first step toward simulation of fermionic many-body problems on analog quantum simulators made of arrays of neutral atoms. 

Experimentally, both algorithms developed in this manuscript could be implemented on real device today. The digital-analog eigensolver for chemistry is bounded by the number of measurements and therefore sticks to the constraint of a RQP. The next step could be to anticipate and mitigate well known errors \parencite{de_leseleuc_analysis_2018} of the architecture to ensure the success of an experimental implementation. In addition, more complex molecules could be treated with the hundreds of atoms available on today's devices. The slave-spin method should be run experimentally really soon and I look forward the result as it could bring a potential quantum advantage in this field. I show in appendix \ref{sec:tria} that this method can be implemented for a triangular lattice if the constraint is fulfilled and the half-filling condition is imposed. In addition, it is possible to add orbitals to site (see Sec.~\ref{sec:multiorbital}) and hence, distinguish sites by the number of orbitals and obtain "effective" atoms. One can therefore think of placing several multiorbitals 2D layers in parallel and as a result, modelling an effective 3D material. This could be a way to simulate paramagnetism in austenitic steel for instance.

On the theoretical side, improving the efficiency of the digital-analog variational eigensolver for chemistry could lead to being able to reach the ground-state energy of any Hamiltonian which can be described as sum of Pauli strings. Therefore, one can think of time-dependent Hamiltonian for instance and solve it with this method at each time step. This is a path toward simulating chemical reaction in oxides or in PV. A lot of work can still be done with the slave-spin approach in understanding more deeply the involvement of performing a mean-field approximation between electrons and slave spins degrees of freedom. The constraint is also a major point to ensure the theoretical validity of the method and further studies could be done to generalize its fulfillment on every lattices. 

Beyond that, it is far from a forgone conclusion that QC will replace actual "classical" simulation methods of electronic structures which benefit from dozen of years of studies and improvements. Therefore, a smart use of QC could be to improve this classical method in terms of precision and/or computing capacity. For instance, QC results for small or medium systems with only few dozens qubits could be use as input parameters of a DFT computation, provided that "good" physical quantities can be extracted from the quantum simulation. Another approach would be to compute a large part of the system's Hamiltonian with DFT and add corrections with QC which simulates a small part of the systems with more correlations. This embedding approach is an active field of research \parencite{wang_quantum_2011, rossmannek_quantum_2023, vorwerk_quantum_2022, li_toward_2022, huang_simulating_2022,
tilly_reduced_2021,
cao_ab_2023} and results of the work presented here could be a first step to improve ab initio simulations in chemistry and materials. Finally, QC could help directly classical simulations by computing directly Green's function \parencite{endo_calculation_2020} in DMFT or encoding the occupation number in orbitals in DFT \parencite{senjean_toward_2023}.

\addcontentsline{toc}{chapter}{Bibliography} 
\printbibliography

\appendix

\chapter{LiH and BeH$_\text{2}$ Hamiltonians}
In this section, examples of complete Hamiltonians of molecules LiH (for an inter-atomic distance of $1.5 $ $\mathrm{\mathring{A}}$) and BeH$_\text{2}$ (for an inter-atomic distance of $1.17$ $ \mathrm{\mathring{A}}$) obtained with the method described in Chap.~\ref{chemistry} Sec.~\ref{sec:chem} are shown.

\begin{equation}
\begin{split}
H_\mathrm{LiH} =& {\bf -0.19975} +{\bf 0.05393}Z_0 {\bf -0.12836}Z_1 {\bf -0.31773}Z_0 Z_1 {\bf -0.31773}Z_3 +{\bf 0.0605}Z_1 Z_3  \\&+{\bf 0.11409}Z_0 Z_1 Z_3 +{\bf 0.05362}Z_4 +{\bf 0.11434}Z_2 Z_4 {\bf -0.03787}Z_2 Z_3 Z_4 +{\bf 0.05362}Z_1 Z_2 Z_3 Z_4  \\&+{\bf 0.0836}Z_0 Z_1 Z_2 Z_3 Z_4 {\bf -0.03787}Z_5 +{\bf 0.05666}Z_1 Z_5 +{\bf 0.11434}Z_0 Z_1 Z_5 +{\bf 0.0836}Z_3 Z_5  \\&+{\bf 0.05666}Z_2 Z_3 Z_5 {\bf -0.12836}Z_4 Z_5 +{\bf 0.0847}Z_1 Z_4 Z_5 +{\bf 0.0605}Z_0 Z_1 Z_4 Z_5 +{\bf 0.05393}Z_3 Z_4 Z_5  \\&+{\bf 0.12357}Z_2 Z_3 Z_4 Z_5 +{\bf 0.01522}X_1 {\bf -0.01522}Z_0 X_1 +{\bf 0.01089}X_1 Z_3 {\bf -0.01089}Z_0 X_1 Z_3  \\&+{\bf 0.00436}X_1 Z_2 Z_3 Z_4 {\bf -0.00436}Z_0 X_1 Z_2 Z_3 Z_4 +{\bf 0.01273}X_1 Z_5 {\bf -0.01273}Z_0 X_1 Z_5 {\bf -0.00901}X_1 Z_4 Z_5  \\&+{\bf 0.00901}Z_0 X_1 Z_4 Z_5 +{\bf 0.00448}X_0 X_2 {\bf -0.00479}X_0 Z_1 X_2 {\bf -0.03512}X_0 Z_1 X_2 Z_3 {\bf -0.03512}Y_0 Y_2 Z_4  \\&{\bf -0.00479}Y_0 Y_2 Z_3 Z_4 +{\bf 0.00448}Y_0 Z_1 Y_2 Z_3 Z_4 {\bf -0.03306}X_0 Z_1 X_2 Z_5 +{\bf 0.00237}Y_0 Y_2 Z_3 Z_5 +{\bf 0.00237}X_0 Z_1 X_2 Z_4 Z_5  \\&{\bf -0.03306}Y_0 Y_2 Z_3 Z_4 Z_5 - {\bf (4\times 10^{-5})}X_0 X_1 X_2 {\bf -0.00277}Y_0 Y_1 X_2 +{\bf 0.01054}X_0 X_1 X_2 Z_3 +{\bf 0.01054}X_0 Y_1 Y_2 Z_4  \\&+{\bf 0.00277}Y_0 X_1 Y_2 Z_3 Z_4 - {\bf (4\times 10^{-5})}X_0 Y_1 Y_2 Z_3 Z_4 +{\bf 0.01173}X_0 X_1 X_2 Z_5 {\bf -0.00154}X_0 Y_1 Y_2 Z_3 Z_5 \\&{\bf -0.00154}X_0 X_1 X_2 Z_4 Z_5  +{\bf 0.01173}X_0 Y_1 Y_2 Z_3 Z_4 Z_5 +{\bf 0.01522}X_3 X_4 {\bf -0.00901}Z_1 X_3 X_4 +{\bf 0.01089}Z_0 Z_1 X_3 X_4 \\& +{\bf 0.00436}Y_3 Y_4{\bf -0.01273}Z_2 Y_3 Y_4 +{\bf 0.00436}X_3 X_4 Z_5 {\bf -0.01273}Z_2 X_3 X_4 Z_5 +{\bf 0.01522}Y_3 Y_4 Z_5 {\bf -0.00901}Z_1 Y_3 Y_4 Z_5  \\&+{\bf 0.01089}Z_0 Z_1 Y_3 Y_4 Z_5 +{\bf 0.00658}X_1 X_3 X_4 {\bf -0.00658}Z_0 X_1 X_3 X_4 +{\bf 0.00658}X_1 Y_3 Y_4 Z_5 {\bf -0.00658}Z_0 X_1 Y_3 Y_4 Z_5  \\&{\bf -0.00776}X_0 Z_1 X_2 X_3 X_4 +{\bf 0.00776}Y_0 Y_2 Y_3 Y_4 +{\bf 0.00776}Y_0 Y_2 X_3 X_4 Z_5 {\bf -0.00776}X_0 Z_1 X_2 Y_3 Y_4 Z_5 \\&+{\bf 0.00211}X_0 X_1 X_2 X_3 X_4  {\bf -0.00211}X_0 Y_1 Y_2 Y_3 Y_4 {\bf -0.00211}X_0 Y_1 Y_2 X_3 X_4 Z_5 +{\bf 0.00211}X_0 X_1 X_2 Y_3 Y_4 Z_5 \\& + {\bf 0.00004}X_5 + {\bf -0.00154}Z_1 X_5  +{\bf 0.01054}Z_0 Z_1 X_5 +{\bf 0.00277}Z_3 X_5 {\bf -0.01173}Z_2 Z_3 X_5 +{\bf 0.00004}Z_4 X_5  \\& +{\bf 0.00154}Z_1 Z_4 X_5  + {\bf -0.01054}Z_0 Z_1 Z_4 X_5 {\bf -0.00277}Z_3 Z_4 X_5 +{\bf 0.01173}Z_2 Z_3 Z_4 X_5 \\& +{\bf 0.00211}X_1 X_5 {\bf -0.00211}Z_0 X_1 X_5  + {\bf -0.00211}X_1 Z_4 X_5 +{\bf 0.00211}Z_0 X_1 Z_4 X_5 {\bf -0.00837}X_0 Z_1 X_2 X_5 \\& +{\bf 0.00837}Y_0 Y_2 Z_3 X_5 +{\bf 0.00837}X_0 Z_1 X_2 Z_4 X_5  \\&{\bf -0.00837}Y_0 Y_2 Z_3 Z_4 X_5 +{\bf 0.00303}X_0 X_1 X_2 X_5 {\bf -0.00303}X_0 Y_1 Y_2 Z_3 X_5 {\bf -0.00303}X_0 X_1 X_2 Z_4 X_5 \\&+{\bf 0.00303}X_0 Y_1 Y_2 Z_3 Z_4 X_5  +{\bf 0.00448}X_3 X_4 X_5 +{\bf 0.03306}Z_2 X_3 X_4 X_5 {\bf -0.00479}Y_3 Y_4 X_5 +{\bf 0.00237}Z_1 Y_3 Y_4 X_5 \\& {\bf -0.03512}Z_0 Z_1 Y_3 Y_4 X_5  +{\bf 0.00448}Y_3 X_4 Y_5 +{\bf 0.03306}Z_2 Y_3 X_4 Y_5 +{\bf 0.00479}X_3 Y_4 Y_5 {\bf -0.00237}Z_1 X_3 Y_4 Y_5 \\& +{\bf 0.03512}Z_0 Z_1 X_3 Y_4 Y_5  {\bf -0.00776}X_1 Y_3 Y_4 X_5 +{\bf 0.00776}Z_0 X_1 Y_3 Y_4 X_5 +{\bf 0.00776}X_1 X_3 Y_4 Y_5 {\bf -0.00776}Z_0 X_1 X_3 Y_4 Y_5 \\&{\bf -0.03074}Y_0 Y_2 X_3 X_4 X_5  +{\bf 0.03074}X_0 Z_1 X_2 Y_3 Y_4 X_5 {\bf -0.03074}Y_0 Y_2 Y_3 X_4 Y_5 {\bf -0.03074}X_0 Z_1 X_2 X_3 Y_4 Y_5 \\&+{\bf 0.00837}X_0 Y_1 Y_2 X_3 X_4 X_5  {\bf -0.00837}X_0 X_1 X_2 Y_3 Y_4 X_5  +{\bf 0.00837}X_0 Y_1 Y_2 Y_3 X_4 Y_5 +{\bf 0.00837}X_0 X_1 X_2 X_3 Y_4 Y_5   
\end{split}
\end{equation}

\begin{equation}
\begin{split}
H_{\text{BeH}_\text{2}} =& {\bf -1.90305} {\bf -0.48894}Z_0 +{\bf 0.14357}Z_1 {\bf -0.18803}Z_0 Z_1 +{\bf 0.12314}Z_2 +{\bf 0.18326}Z_0 Z_2  \\&+{\bf 0.10964}Z_1 Z_2 +{\bf 0.18222}Z_0 Z_1 Z_2 {\bf -0.48894}Z_3 +{\bf 0.1288}Z_0 Z_3 +{\bf 0.1136}Z_0 Z_1 Z_3  \\&+{\bf 0.11249}Z_2 Z_3 +{\bf 0.11746}Z_1 Z_2 Z_3 +{\bf 0.14357}Z_4 {\bf -0.18803}Z_3 Z_4 +{\bf 0.1136}Z_0 Z_3 Z_4  \\&+{\bf 0.10602}Z_0 Z_1 Z_3 Z_4 +{\bf 0.10306}Z_2 Z_3 Z_4 +{\bf 0.10577}Z_1 Z_2 Z_3 Z_4 +{\bf 0.12314}Z_5 +{\bf 0.11249}Z_0 Z_5  \\&+{\bf 0.10306}Z_0 Z_1 Z_5 +{\bf 0.10451}Z_2 Z_5 +{\bf 0.10785}Z_1 Z_2 Z_5 +{\bf 0.18326}Z_3 Z_5 +{\bf 0.10964}Z_4 Z_5  \\&+{\bf 0.11746}Z_0 Z_4 Z_5 +{\bf 0.10577}Z_0 Z_1 Z_4 Z_5 +{\bf 0.10785}Z_2 Z_4 Z_5 +{\bf 0.11352}Z_1 Z_2 Z_4 Z_5 +{\bf 0.18222}Z_3 Z_4 Z_5  \\&{\bf -0.00743}X_0 X_1 {\bf -0.00229}Y_0 Y_1 {\bf -0.00229}X_0 X_1 Z_2 {\bf -0.00743}Y_0 Y_1 Z_2 {\bf -0.00711}Y_0 Y_1 Z_3  \\&{\bf -0.00711}X_0 X_1 Z_2 Z_3 {\bf -0.00875}Y_0 Y_1 Z_3 Z_4 {\bf -0.00875}X_0 X_1 Z_2 Z_3 Z_4 {\bf -0.00352}Y_0 Y_1 Z_5 {\bf -0.00352}X_0 X_1 Z_2 Z_5  \\&{\bf -0.0072}Y_0 Y_1 Z_4 Z_5 {\bf -0.0072}X_0 X_1 Z_2 Z_4 Z_5 {\bf -0.04165}X_0 X_2 +{\bf 0.03769}X_0 Z_1 X_2 {\bf -0.00396}Y_0 Y_2  \\&+{\bf 0.00839}X_1 X_2 {\bf -0.01015}Z_0 X_1 X_2 {\bf -0.01015}Y_1 Y_2 +{\bf 0.00839}Z_0 Y_1 Y_2 +{\bf 0.01355}Z_0 X_1 X_2 Z_3  \\&+{\bf 0.01355}Y_1 Y_2 Z_3 +{\bf 0.01082}Z_0 X_1 X_2 Z_3 Z_4 +{\bf 0.01082}Y_1 Y_2 Z_3 Z_4 +{\bf 0.00854}Z_0 X_1 X_2 Z_5 +{\bf 0.00854}Y_1 Y_2 Z_5  \\&+{\bf 0.01408}Z_0 X_1 X_2 Z_4 Z_5 +{\bf 0.01408}Y_1 Y_2 Z_4 Z_5 +{\bf 0.03611}X_0 X_3 {\bf -0.03611}X_0 Z_1 X_3 {\bf -0.03611}X_0 X_3 Z_4  \\&+{\bf 0.03611}X_0 Z_1 X_3 Z_4 {\bf -0.02498}X_1 X_3 +{\bf 0.02498}Z_0 X_1 Z_2 X_3 +{\bf 0.02498}X_1 X_3 Z_4 {\bf -0.02498}Z_0 X_1 Z_2 X_3 Z_4  \\&{\bf -0.03615}X_2 X_3 +{\bf 0.03615}Z_1 X_2 X_3 +{\bf 0.03615}X_2 X_3 Z_4 {\bf -0.03615}Z_1 X_2 X_3 Z_4 {\bf -0.01573}X_0 X_1 X_2 X_3  \\&{\bf -0.01573}Y_0 X_1 Y_2 X_3 +{\bf 0.01573}X_0 X_1 X_2 X_3 Z_4 +{\bf 0.01573}Y_0 X_1 Y_2 X_3 Z_4 {\bf -0.02498}X_0 X_4 +{\bf 0.02498}X_0 Z_1 X_4  \\&+{\bf 0.02498}X_0 Z_3 X_4 Z_5 {\bf -0.02498}X_0 Z_1 Z_3 X_4 Z_5 +{\bf 0.02085}X_1 X_4 {\bf -0.02085}Z_0 X_1 Z_2 X_4 {\bf -0.02085}X_1 Z_3 X_4 Z_5  \\&+{\bf 0.02085}Z_0 X_1 Z_2 Z_3 X_4 Z_5 +{\bf 0.02464}X_2 X_4 {\bf -0.02464}Z_1 X_2 X_4 {\bf -0.02464}X_2 Z_3 X_4 Z_5 +{\bf 0.02464}Z_1 X_2 Z_3 X_4 Z_5  \\&+{\bf 0.01532}X_0 X_1 X_2 X_4 +{\bf 0.01532}Y_0 X_1 Y_2 X_4 {\bf -0.01532}X_0 X_1 X_2 Z_3 X_4 Z_5 {\bf -0.01532}Y_0 X_1 Y_2 Z_3 X_4 Z_5 \\& {\bf -0.00743}X_3 X_4  {\bf -0.00229}Y_3 Y_4 {\bf -0.00711}Z_0 Y_3 Y_4 {\bf -0.00875}Z_0 Z_1 Y_3 Y_4 {\bf -0.00352}Z_2 Y_3 Y_4 {\bf -0.0072}Z_1 Z_2 Y_3 Y_4  \\&{\bf -0.00229}X_3 X_4 Z_5 {\bf -0.00711}Z_0 X_3 X_4 Z_5 {\bf -0.00875}Z_0 Z_1 X_3 X_4 Z_5 {\bf -0.00352}Z_2 X_3 X_4 Z_5 {\bf -0.0072}Z_1 Z_2 X_3 X_4 Z_5  \\&{\bf -0.00743}Y_3 Y_4 Z_5 +{\bf 0.01972}Y_0 Y_1 Y_3 Y_4 +{\bf 0.01972}X_0 X_1 Z_2 Y_3 Y_4 +{\bf 0.01972}Y_0 Y_1 X_3 X_4 Z_5 +{\bf 0.01972}X_0 X_1 Z_2 X_3 X_4 Z_5  \\&{\bf -0.0173}Z_0 X_1 X_2 Y_3 Y_4 {\bf -0.0173}Y_1 Y_2 Y_3 Y_4 {\bf -0.0173}Z_0 X_1 X_2 X_3 X_4 Z_5 {\bf -0.0173}Y_1 Y_2 X_3 X_4 Z_5 {\bf -0.03615}X_0 X_5  \\&+{\bf 0.03615}X_0 Z_1 X_5 +{\bf 0.03615}X_0 Z_4 X_5 {\bf -0.03615}X_0 Z_1 Z_4 X_5 +{\bf 0.02464}X_1 X_5 {\bf -0.02464}Z_0 X_1 Z_2 X_5  \\&{\bf -0.02464}X_1 Z_4 X_5 +{\bf 0.02464}Z_0 X_1 Z_2 Z_4 X_5 +{\bf 0.04177}X_2 X_5 {\bf -0.04177}Z_1 X_2 X_5 {\bf -0.04177}X_2 Z_4 X_5  \\&+{\bf 0.04177}Z_1 X_2 Z_4 X_5 +{\bf 0.01232}X_0 X_1 X_2 X_5 +{\bf 0.01232}Y_0 X_1 Y_2 X_5 {\bf -0.01232}X_0 X_1 X_2 Z_4 X_5 {\bf -0.01232}Y_0 X_1 Y_2 Z_4 X_5  \\&{\bf -0.04165}X_3 X_5 +{\bf 0.03769}X_3 Z_4 X_5 {\bf -0.00396}Y_3 Y_5 +{\bf 0.00839}X_4 X_5 {\bf -0.01015}Z_3 X_4 X_5  \\&+{\bf 0.01355}Z_0 Z_3 X_4 X_5 +{\bf 0.01082}Z_0 Z_1 Z_3 X_4 X_5 +{\bf 0.00854}Z_2 Z_3 X_4 X_5 +{\bf 0.01408}Z_1 Z_2 Z_3 X_4 X_5 {\bf -0.01015}Y_4 Y_5  \\&+{\bf 0.01355}Z_0 Y_4 Y_5 +{\bf 0.01082}Z_0 Z_1 Y_4 Y_5 +{\bf 0.00854}Z_2 Y_4 Y_5 +{\bf 0.01408}Z_1 Z_2 Y_4 Y_5 +{\bf 0.00839}Z_3 Y_4 Y_5  \\&{\bf -0.0173}Y_0 Y_1 Z_3 X_4 X_5 {\bf -0.0173}X_0 X_1 Z_2 Z_3 X_4 X_5 {\bf -0.0173}Y_0 Y_1 Y_4 Y_5 {\bf -0.0173}X_0 X_1 Z_2 Y_4 Y_5 \\& +{\bf 0.01858}Z_0 X_1 X_2 Z_3 X_4 X_5  +{\bf 0.01858}Y_1 Y_2 Z_3 X_4 X_5 +{\bf 0.01858}Z_0 X_1 X_2 Y_4 Y_5 +{\bf 0.01858}Y_1 Y_2 Y_4 Y_5 \\& {\bf -0.01573}X_0 X_3 X_4 X_5 +{\bf 0.01573}X_0 Z_1 X_3 X_4 X_5  {\bf -0.01573}X_0 Y_3 X_4 Y_5 +{\bf 0.01573}X_0 Z_1 Y_3 X_4 Y_5 \\& +{\bf 0.01532}X_1 X_3 X_4 X_5 {\bf -0.01532}Z_0 X_1 Z_2 X_3 X_4 X_5 \\& +{\bf 0.01532}X_1 Y_3 X_4 Y_5  {\bf -0.01532}Z_0 X_1 Z_2 Y_3 X_4 Y_5   +{\bf 0.01232}X_2 X_3 X_4 X_5 {\bf -0.01232}Z_1 X_2 X_3 X_4 X_5 \\& +{\bf 0.01232}X_2 Y_3 X_4 Y_5  {\bf -0.01232}Z_1 X_2 Y_3 X_4 Y_5  +{\bf 0.01415}X_0 X_1 X_2 X_3 X_4 X_5 +{\bf 0.01415}Y_0 X_1 Y_2 X_3 X_4 X_5 \\& +{\bf 0.01415}X_0 X_1 X_2 Y_3 X_4 Y_5  +{\bf 0.01415}Y_0 X_1 Y_2 Y_3 X_4 Y_5 
\end{split}
\end{equation}

\chapter{Slave-spin theory for the Mott transition in the triangular lattice}\label{sec:tria}

\section{Forewords}

In the main chapter (Chap.~\ref{hubbard}), we focus on the square lattice because of it is a bipartite lattice and therefore the constraint of the slave-spin method is useless. In addition, it is the most studied lattice.
At first, we did not focus on this lattice but on the triangular lattice, (111) plane of the crystal structure of
face-centered cubic materials like iron.
The two main differences are the condition for half-filling and the constraint fulfillment. In order to keep the system at half-filling, we use an optimizer to find the good chemical potential for every values of $U/t$. Yet, the constraint in never satisfied in the following. Despite this strong approximation, I think it is interesting to see the result as the geometry is totally different and the cluster mean-field is not defined in the same way. The simulation are performed for cluster of 6 and 10 sites.

The out of equilibrium case is not studied here.
The content of this chapter is extracted from a first version of the article on the Slave-Spin theory.
\section{Solving the spin Hamiltonian $H_\mathrm{S}$ for $Q$: cluster mean-field}
 
We now focus on the computation of $Q_{i,j}$. It requires the computations the spin-spin correlation function $\langle S_i^z S_j^z \rangle$.
Many strategies are available. Most authors \parencite{ruegg_mathsfz_2-slave-spin_2010, demedici_orbital-selective_2005} handle $H_\mathrm{S}$ at the single-site mean-field level. This yields local observables, but neglects important spatial correlation effects, thereby limiting the scope of the method. 
In particular, spatial (or orbital) resolution in quasi-particle weights is a key factor to investigate e.g hot and cold spots observed in angle-resolved photoemission spectroscopy \parencite{damascelli_angle-resolved_2003, brown_angle-resolved_2020, hashimoto_energy_2014}.

One must thus go beyond single-site mean-field.
A straightforward approach would be to directly diagonalize a finite-site version of $H_\mathrm{S}$ and compute the finite-size ground-state.
However, finite-size effects would prevent us from observing the phase transition we are looking for, namely the transition from a Fermi liquid to a Mott insulator.

We thus tackle $H_\mathrm{S}$ (Eq.~\eqref{Ham_S}) with a cluster mean-field approach, as done in e.g \parencite{hassan_slave_2010}.  Finite-size effects are thus reduced thanks to the influence of a self-consistent mean field $m = \langle S^z \rangle $ at the boundary of the cluster, at the same numerical price as the exact diagonalization (except the procedure needs to be repeated in a self-consistent fashion until convergence of the mean field).

The cluster mean-field approximation leads  to
\begin{equation}
    S_i^z S_j^z \approx \langle S_i^z \rangle S_j^z + \langle S_j^z \rangle S_i^z - \langle S_i^z \rangle \langle S_j^z \rangle,
\end{equation}
where $i$ ($j$) is inside the cluster and $j$ ($i$) is not. The mean-field parameter $\langle S^z \rangle$ is the same for all sites in the thermodynamic limit. As we consider finite-size systems,  we numerically compute 
\begin{equation}
    m = \frac{1}{N_a} \sum_i^{N_a} \langle S_i^z \rangle
\end{equation}
With $N_a$ being the number of sites.
Eq.~(\ref{Ham_S}) can be rewritten as
\begin{equation}\label{tria:Ham_C}
    H_\mathrm{S}^{(C)} (J, m) = H_\mathrm{S} (J) + \sum_{i \in \mathrm{C}} h_i S_i^z  
\end{equation}
where $C$ is the cluster and $h_i =  \sum_j  mJ_{i,j} $

\begin{figure}[!h]
    \centering
    \includegraphics[width=\linewidth,
                   trim={3.0cm 0cm 3.0cm 0.1cm},
                   clip]{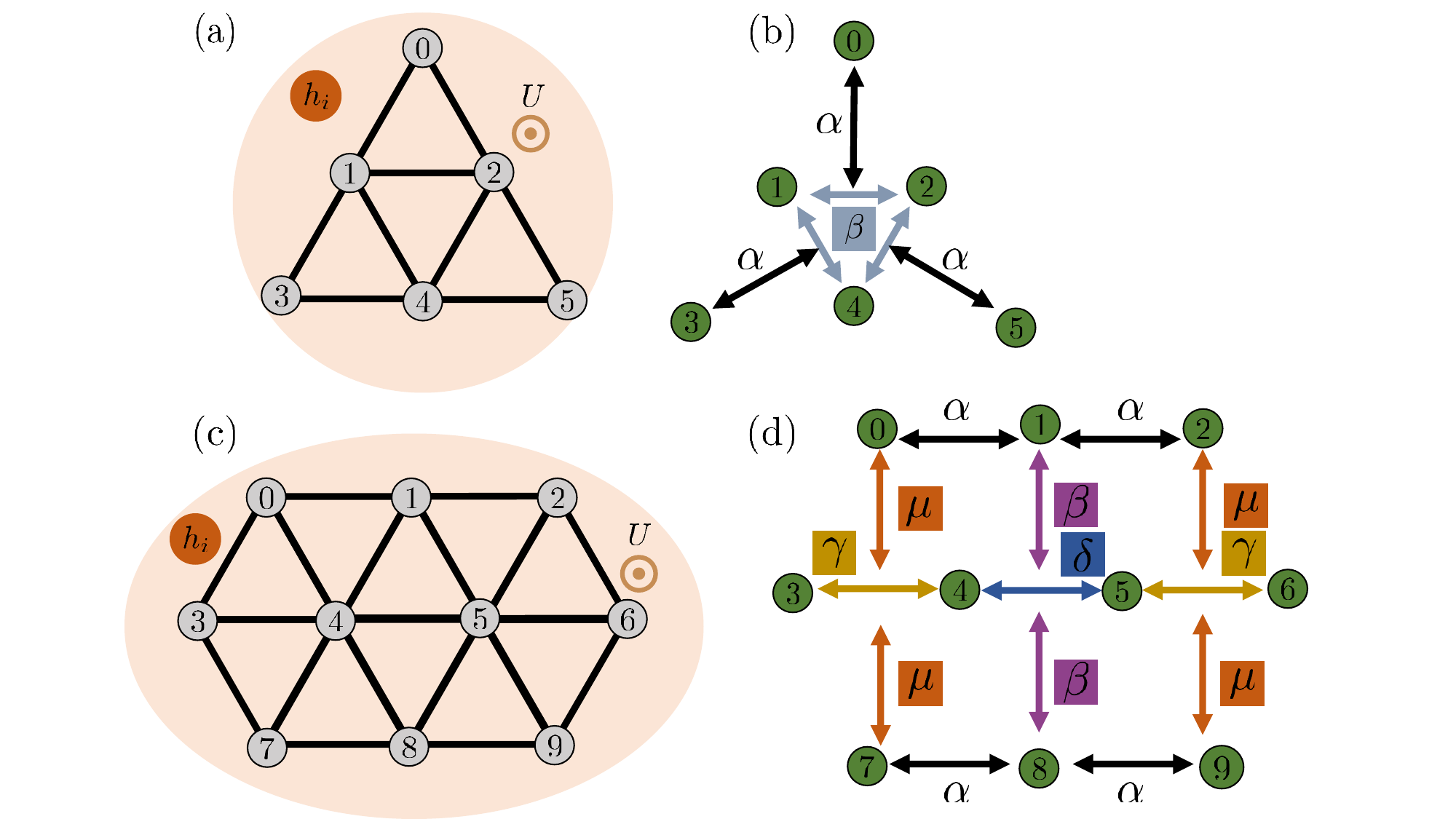}
    \caption{\textit{Cluster mean field and parameterized geometry of Rydberg atoms}.(a) Geometry of a 6-site cluster. After the slave spin mean field approach combined with the cluster mean field approach, all spins $S^z_i$ are immersed in a mean-field bath $h_i$ and undergo a transverse field  $\propto U$. (b) Optimization of the geometry with parameters $\alpha$ and $\beta$ to fulfill Eq.(\ref{eq:geometry}). (c-d) Same as (a-b), for a 10-site cluster. }
    \label{fig:SSMF_CMFT_embedding}
\end{figure}

The shape of the cluster and the mean field are illustrated in Fig.~\ref{fig:SSMF_CMFT_embedding} for clusters of size $6$ and $10$.

For instance, for the 6-site cluster, sites $0$, $3$ and $5$ have 2 nearest-neighbor whereas sites $1$, $2$ and $4$ are connected to 4 sites respectively. If we consider a hopping term of value $-1$ for connected sites and $0$ for the others, the matrix $t$ can be written:
\begin{equation}
    t = -\begin{pmatrix}
  0 & 1 & 1 & 0 & 0 & 0\\
 1 & 0 & 1 & 1 & 1 & 0\\
  1 & 1 & 0 & 0 & 1 & 1\\
  0 & 1 & 0 & 0 & 1 & 0\\
  0 & 1 & 1 & 1 & 0 & 1\\
  0 & 0 & 1 & 0 & 1 & 0\\
\end{pmatrix}
\end{equation}

This embedding will be the one studied in the rest of the article.

The self-consistent procedure to solve for the approximate ground state of $H_\mathrm{S}$, $|\psi_0\rangle_C$, and hence $Q_{i,j}$, is illustrated in Fig.~\ref{fig:loop_SSMF_final}.
The solution of the cluster Hamiltonian for a fixed mean field $h$, $H_\mathrm{S}^{(C)}$, is described in section~\ref{subsec:embedded}.


\section{Self-consistent outer cycle}

We are now ready to describe the outer self-consistent loop. It is illustrated in Fig.~\ref{fig:loop_SSMF_final}. 

In the regime of cluster mean-field theory, there is a difference between the quasi-particle weight $Z$ and the effective mass renormalization $\frac{m}{m*}$. across the Mott insulator phase, $Z$ vanishes to $0$ beside $g = \frac{m}{m*}$ stays finite \parencite{ruegg_mathsfz_2-slave-spin_2010,hassan_slave_2010,florens_slave-rotor_2004}. They are the order parameters of the Fermi-liquid/Mott insulator phase transition. 
The final output of our method is then the quasi-particle weight defined as 
\begin{equation}
    Z = h^2 =  \langle S^z \rangle^2.
\end{equation}
It is the order parameter for the phase transition : it is $\neq 0$ before transition and should be $= 0$ after the phase transition. The effective mass 
\begin{equation}
    g = \frac{m}{m^*} = \langle S_i^z S_{i+1}^z \rangle
\end{equation}
on the other hand, does not reach $0$ after the transition.
These definitions are true (i.e do not depend on the site we choose) if we consider a system with an infinite number of size. Numerically, these values are computed as mean values over all spins:
\begin{subequations}\label{eq:Zandg}
\begin{alignat}{2}
     &Z =   h^2 =  \Big (\frac{1}{N_a}\sum_{i}^{N_a} \langle S_i^z \rangle \Big )^2\\
    &g = \frac{1}{(N_a-1)}\sum_{i}^{N_a-1} \langle S_i^z S_{i+1}^z \rangle.
    \end{alignat}
\end{subequations}
In the following, the whole procedure to find these two values is named SSMF-CMFT for Slave Spin Mean Field - Cluster Mean Field Theory.

\subsection{Solving the embedded model}\label{subsec:embedded}

For a given value of the mean field $h$, we now need to compute $\langle S_i^z S_j^z \rangle$ and $\langle S_i^z \rangle$ with $\langle \dots \rangle = {}_{\mathrm{S}}\bra{\psi_0} \dots \ket{\psi_0}_\mathrm{S}$ to obtain $g$, $Q$ and $Z$ at the end of the loop. It is possible to compute these values without knowing the exact groundstate of the system but in our case, considering an implementation on a Rydberg quantum processor, it is simpler to get the groundstate of $H_\mathrm{S}^{(C)}(J,h)$.


\begin{figure}
    \centering
    \includegraphics[width=0.7\linewidth]{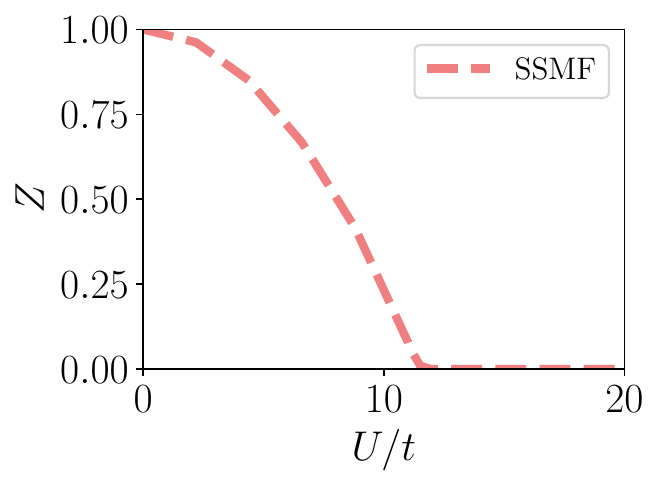}
    \caption{\textit{Quasiparticle weight and mass renormalization as a function of $U/t$ with the SSMF-CMFT method}. Systems plotted here 10 sites triangular lattices. $H_\mathrm{S}^{(C)}(J,h)$ is solved by exact diagonalization method.}
    \label{fig:tria_perfect_plot_Z_g}
\end{figure}


Exact diagonalization of the cluster Hamiltonian Eq.~\eqref{tria:Ham_C} is possible only for a few sites because of the exponential size of the corresponding Hilbert space.
We therefore turn to our Rydberg platform, characterized by the Hamiltonian given in Eq.~\eqref{eq:ising_ham}, to find the solution to our spin model Eq.~\eqref{tria:Ham_C}.

These two Hamiltonians have a major difference: while the Rydberg Hamiltonian has an antiferromagnetic interaction, the auxiliary spin Hamiltonian has a ferromagnetic interaction as long as $t_{i,j} > 0$.
We can nevertheless make use of the Rydberg platform: instead of looking for its ground state, we are going to be looking for its most excited state.

To this aim, we resort to an annealing procedure. The atomic register is prepared in a state 
\begin{equation}\label{eq:psi_start}
    \ket{\psi_\mathrm{start}}=\ket{g}^{\otimes N_\mathrm{a}}
\end{equation}
where $N_a$ is the number of atoms.

The adiabatic theorem states that a physical system remains in its instantaneous eigenstate if the Hamiltonian is driven slowly enough compare to the inverse square gap between the eigenvalue and the rest of the Hamiltonian's spectrum. In particular, this is true for the most excited state.
We can rewrite (\ref{tria:Ham_C}) as:
\begin{equation}\label{tria:Ham_target}
H_\mathrm{target} = \sum_{i,j,\sigma} (-J_{i,j}) S^z_{i}S^z_{j} - \sum_{i \in C} h_i S_i^z  -\frac{U}{4} \sum_{i} S_i^x
\end{equation}
that is the target Hamiltonian.
The first step is to calculate which experimental values should be obtained to go from Eq.~(\ref{eq:ising_ham}) to Eq.~(\ref{tria:Ham_target}).
The Rabi frequency, detuning and the interaction matrix of the atoms have to fulfil the following relations at the end of the annealing procedure:
\begin{align}
 \Omega(t_\mathrm{max}) &= \frac{U}{2}, \label{eq:omega_an}\\
  \delta_i(t_\mathrm{max}) &= \frac{1}{2} \sum_{j \neq i} \frac{C_6}{r_{i,j}^6} + 2h \sum_{j \neq i} J_{i,j}, \label{eq:delta_an}\\
\frac{C_6}{r_{i,j}^6} &= -4J_{i,j}. \label{eq:geometry}
\end{align}
In our embedding, sites can be described in two categories: those which have two nearest neighbours and those which have four nearest neighbours. It leads to a matrix $J_{i,j}$ of the form
\begin{equation}
    \begin{pmatrix}
  0. & a & a & 0. & 0. & 0.\\
  a & 0. & b & a & b & 0.\\
  a & b & 0. & 0. & b & a\\
  0. & a & 0. & 0. & a & 0.\\
  0. & b & b & a & 0. & a\\
  0. & 0. & a & 0. & a & 0.\\
\end{pmatrix}.
\end{equation}
It means that only two parameters are needed to satisfy Eq.~(\ref{eq:geometry}).
The parameters $\alpha$ and $\beta$ described in Fig. \ref{fig:SSMF_CMFT_embedding} are then optimized in each outer loop of the algorithm (Fig. \ref{fig:loop_SSMF_final}). The optimization of the geometry starts with an initial guess for parameters $\alpha$ and $\beta$. Then, a Nelder-Mead algorithm is used to minimize the value
\begin{equation}\label{eq:norm_geo}
    \sum_{i,j} \left|\frac{C_6}{r_{i,j}^6}+4J_{i,j}\right|
\end{equation}
It usually takes less than $50$ steps to find the final parameters with very good results (see Sec.\ref{sec:result_optimized_geo}).
In the following, we describe the experimental procedure to solve (\ref{tria:Ham_C}).
First, the atoms are prepared in the state (\ref{eq:psi_start}) and in the optimised geometry.
The following Hamiltonian is the one applied at $t=0$
\begin{equation}\label{eq:tria_Hstart}
    H_\mathrm{start} = \sum_{i\ne j}\frac{C_6}{|\textbf{r}_i-\textbf{r}_j|^{6}} \hat{n}_i \hat{n}_j -\hbar \delta_\mathrm{start} \sum_i n_i
\end{equation}
where $\delta_\mathrm{start}$ is set so that $\ket{\psi_\text{start}}$ is the most excited state of Eq.~(\ref{eq:tria_Hstart}). The Rabi frequency and the detuning are then driven during a time $t_{max}$ to reach the Hamiltonian Eq.~(\ref{tria:Ham_target}).
A global addressing is performed for the Rabi frequency whereas a local one is used for the detuning.  Following Eq.~(\ref{eq:omega_an}), the Rabi frequency starts at $0$ MHz and is driven linearly to $\frac{U}{2}$. Similarly, the detunings are all prepared at a value $\delta_\mathrm{start}$ and are driven separately to values Eq.~(\ref{eq:delta_an}). $h \in [0,1]$ and Eq.~(\ref{eq:geometry}) ensures that $\delta_i(t_\mathrm{max})$ is always positive.
Observables ($Q$, $Z$, $g$) are measured with a sample of shots obtained from the state of the atoms.

\section{Discussion on units}
Experimentally, the Rydberg states used impose a van der Waals coefficient $\frac{C_6}{h} = 1947.10^3$ MHz.$\mu$m$^6$ \parencite{scholl_quantum_2021}. This leads to fix all parameters in our protocol. The Rabi frequency (Eq.~(\ref{eq:omega_an})) can be driven up to a maximal value of $\frac{\Omega}{2\pi} \approx 2.5$ MHz. Therefore, considering $U \in [0,5]$ is tailored to our device. In addition, It is guessed that the phase transition is for $\frac{U}{t} \in [10,15]$ (see Sec.~\ref{sec:exact-diag-result}). We consider then $t=\frac{1}{3}$ such as all the interval is span.
This yields to value distance between atoms $\approx 11 \mu$m for the closest ones and $\approx 23 \mu$m for unintended interactions. Interaction values are then between $10^{-2}$ MHz and $\approx 0.8$ MHz.
Frequencies of local detuning are then imposed by Eq.~(\ref{eq:delta_an}). It is important to note that for $h$ values close to $1$ (i.e. $\frac{U}{t}\ll \frac{U_c}{t} $), all $\delta_i(t_\mathrm{max})$ are expected to reach $0$ MHz\footnote{We aim at $\frac{C_6}{r_{i,j}^6} = -4J_{i,j}$. If we consider this perfectly fulfilled, we can replace in Eq.~(\ref{eq:delta_an}) and we obtain $\delta_i(t_\mathrm{max}) = \frac{1}{2}(-4\sum_{i \neq j }J_{i,j})  + 2\times1\times \sum_{j \neq i} J_{i,j} = -2\sum_{j \neq i} J_{i,j} + 2\sum_{j \neq i} J_{i,j} = 0$}.
In our numerical simulation, optimization of geometry provides us $\delta_i(t_\mathrm{max}) \approx 10^{-2}$MHz. For values of $\frac{U}{t}$ at the phase transition and after, $h \approx 0$ and $\delta_i(t_\mathrm{max}) = -2\sum_{j \neq i} J_{i,j} \approx 0.8$ MHz.

\subsubsection{Numerical tools}
All numerical simulations are performed with the library qutip \parencite{johansson_qutip_2013} (exact diagonalization) and Quantum Learning Machine. The SPAM error is implemented with a code from Pulser \parencite{silverio_pasqal-iopulser_2022}.

During the SSMF-CMFT procedure, the vector state obtained for each diagonalization of Eq.~(\ref{tria:Ham_C}) is sampled to measure $Q$ and $Z$. Sampling the vector state above $100$ samples allows to see the Mott transition.
If the maximum number of iterations is set to $5$ for the SSMF loop and the CMFT loop, the maximal number of shots is $100 \times 25 = 2500$. The working rate of the device is $\approx 5$ Hz. It means it would take $500$ s to perform one complete run for a given $U$.


\section{Results }\label{sec:results}

\subsection{Exact diagonalization method}\label{sec:exact-diag-result}
This method is performed in Fig.~\ref{fig:tria_optimized_geo}.
The Mott transition appears for $U/t \approx 11.3$. The  Mott insulating phase occurs between $10.5$ and $16.2$ for other numerical methods like DMFT-exact diagonalization ($\frac{U}{t} = 15$, 8 sites), exact diagonalization on 12 sites ($\frac{U}{t} = 12$) and CDMFT ($\frac{U}{t} = 10.5$) \parencite{hassan_slave_2010}: the difference lies in the number of sites taken in the cluster and the choice of gauge considered in the slave-spin theory \parencite{demedici_orbital-selective_2005}.

\subsection{Impact of experimental parameters}

\begin{figure}[h!]
    \centering
    \includegraphics[width = 0.8 \linewidth]{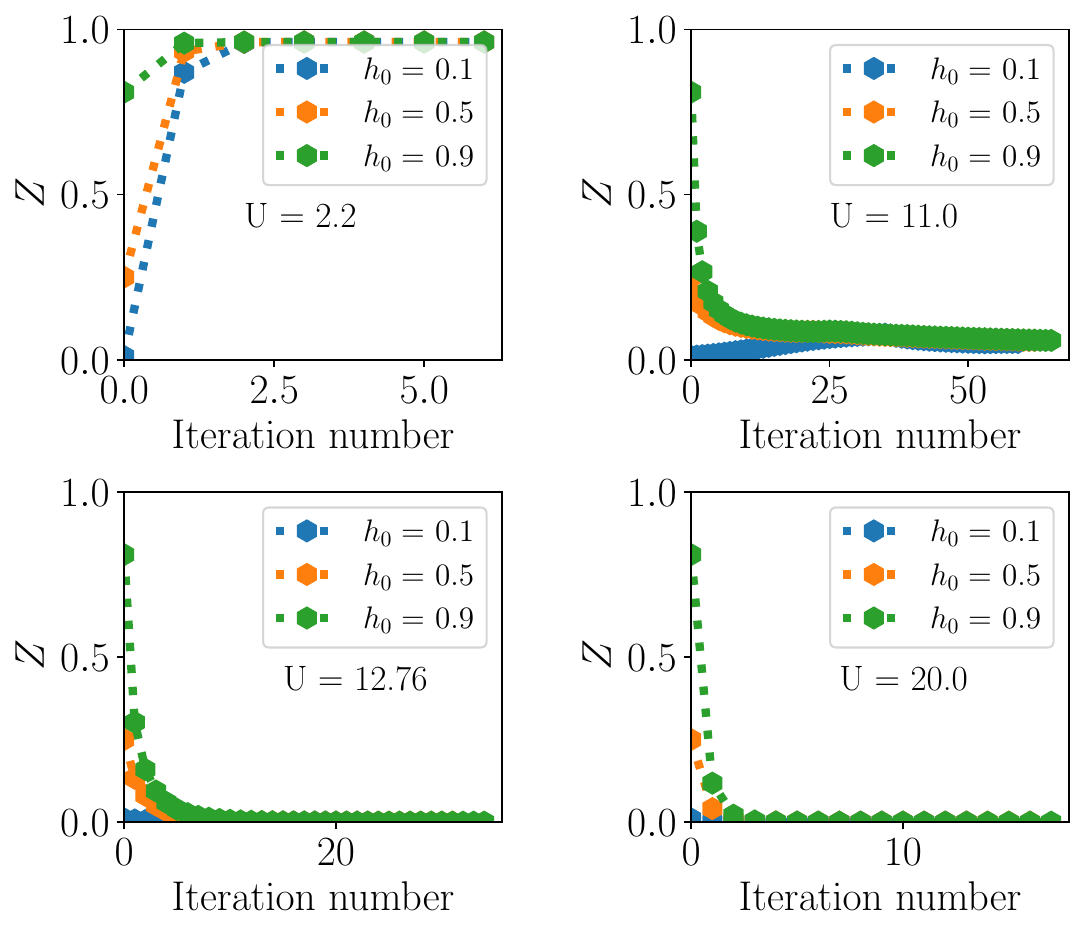}
    \caption{\textit{Evolution of $Z$ as a function of loop iterations for a 6 sites embedding}. Different starting point $h^0$ (0.1, 0.5 and 0.9) are considered for each value of $U/t$ (clockwise starting from upper left: $U/t=2.2$,  $U/t=11.0$, $U/t=20.0$ and $U/t=12.76$).  The solving method is exact diagonalization. The number of allowed iteration is increased to $2500$ and the error accepted is $\eta = 10^{-3}$.}
    \label{fig:tria_h_CMFT_iterations}
\end{figure}

The evolution of $h$ during the SSMF-CMFT procedure is shown Fig.~\ref{fig:tria_h_CMFT_iterations}. Different starting points $h^0$ are tested. If $h^0 = 0$ or $h^0 = 1$, the value of $h$ stays constant, it is therefore needed to choose $h^0 \in ]0,1[$. In this interval, $h$ always converges  towards the same value. Far from the phase transition, the convergence is relatively quick (less than $35$ iterations in total).
Near the phase transition, the convergence takes a lot more time as $h^2 = Z$ is an order parameter, but it finally reaches the same value regardless of initial guess for $h^0$. The same protocol can be applied to check the evolution of $Q$ during loops. The initial guess for $Q$ is always $t$ in all results.

\subsubsection{Optimized geometry \label{sec:result_optimized_geo}}
\begin{figure}
    \centering
    \includegraphics[width=0.7 \linewidth]{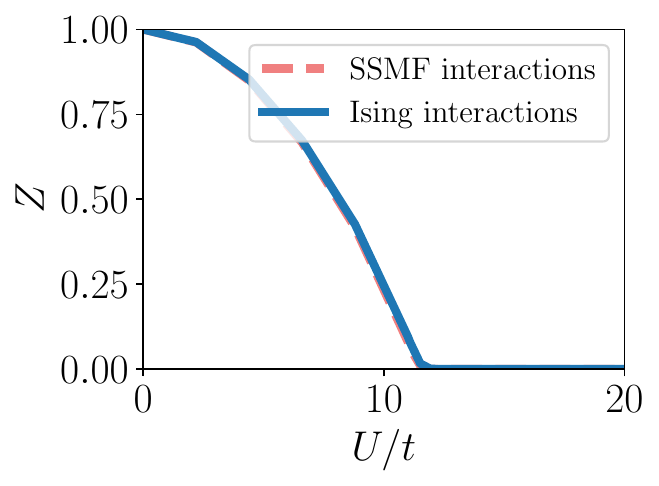}
    \caption{\textit{Impact of considering a realistic geometry}. Comparison of $Z$ values between method with the real matrix $J$ and the optimized one for 6 sites. In both cases, exact diagonalization method is performed to obtain the groundstate of $H_\mathrm{S}^{(C)}$}
    \label{fig:tria_optimized_geo}
\end{figure}

The result of the optimization of atomic positions is shown Fig.~\ref{fig:tria_optimized_geo} (6 sites). The difference in $Z$ values is under $0.1\%$. The norm of the difference Eq.~(\ref{eq:norm_geo}) after optimization is $\sim 0.18$ for each value of $\frac{U}{t}$.

\subsubsection{Dephasing noise}
 
\begin{figure}
    \centering
    \includegraphics[width = 0.7 \linewidth]{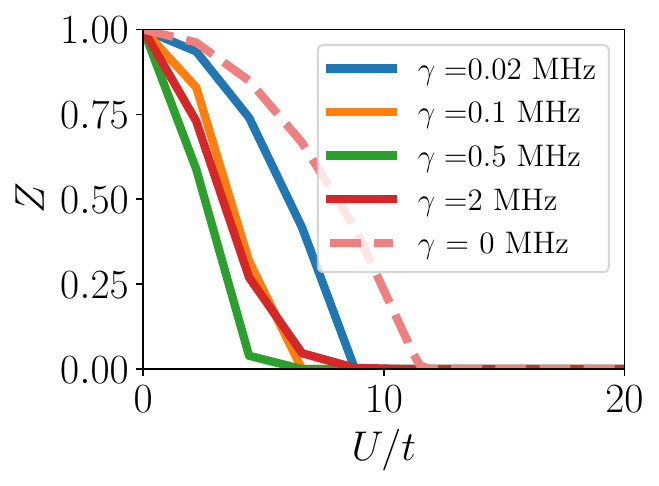}
    \caption{\textit{Impact of considering dephasing noise} A triangular lattice cluster of 6 sites is considered.}
    \label{fig:tria_gamma}
\end{figure}

The time of annealing is $t_{max}=$ 5 $\mu s$
 and the exact state vector is considered to measure $Z$.
The dephasing effect is shown in Fig. \ref{fig:tria_gamma}. We can notice that for small $\gamma$ ($0.01$ and $0.3$ MHz), the phase transition is still present but with smaller values of $\frac{U}{t}$. For larger values, the phase transition disappears.
Experimentally, $\gamma$ is measured around $0.02$ MHz. We will use this value in our simulations.

 \subsubsection{Measurement error}
 
\begin{figure}[!h]
    \centering
    \includegraphics[width=0.7 \linewidth]{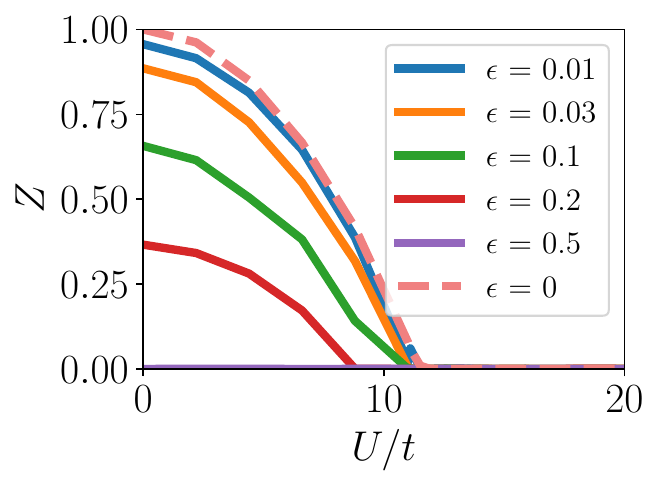}
    \caption{\textit{Impact of measurement error}. The top panel shows the effect of $\epsilon$ and the bottom panel shows the effect of $\epsilon'$. The number of shots considered for each measurement is $1000$. The time of annealing is 4 $\mu$s. }
    \label{fig:tria_epsprime}
\end{figure}

We show the effect of these errors in Fig. \ref{fig:tria_epsprime}. The trivial groundstate of Eq.~(\ref{Ham_C}) when $U=0$ is Eq.~(\ref{eq:psi_start}). Therefore,  a finite $\epsilon$ will have an impact on the measurement of this state.

 \subsubsection{Shot noise}
 
\begin{figure}
    \centering
    \includegraphics[width=0.6 \linewidth]{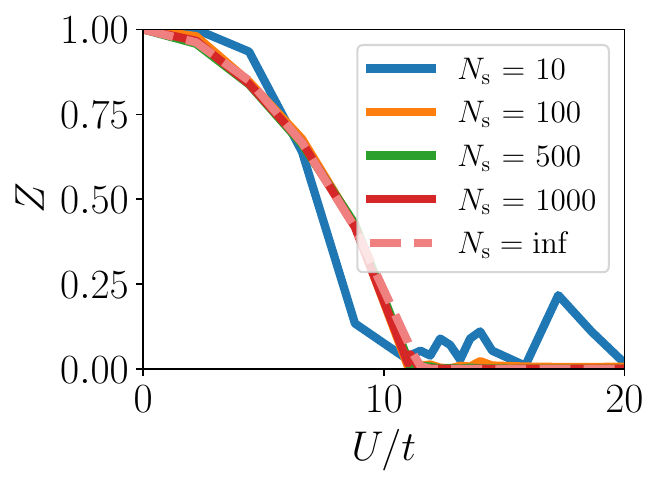}
    \caption{\textit{Impact of considering a realistic geometry}. Comparison of $Z$ values between method with the real matrix $J$ and the optimized one for 6 sites. In both cases, exact diagonalization method is performed to obtain the groundstate of $H_\mathrm{S}^{(C)}$}
    \label{fig:tria_nshot}
\end{figure}

 The effect of shot noise is shown in Fig.~\ref{fig:tria_nshot}. We clearly see that increasing the number of shot increases the precision of the result. In order to fit with experimental limitations and stay in the phase transition regime, we will impose $n_\mathrm{shot}=150$ for numerical simulations.

\begin{figure*}[t]
    \centering
    \includegraphics[width=\linewidth,
                     trim={0.1cm 0.1cm 0.1cm 0.1cm},
                     clip]{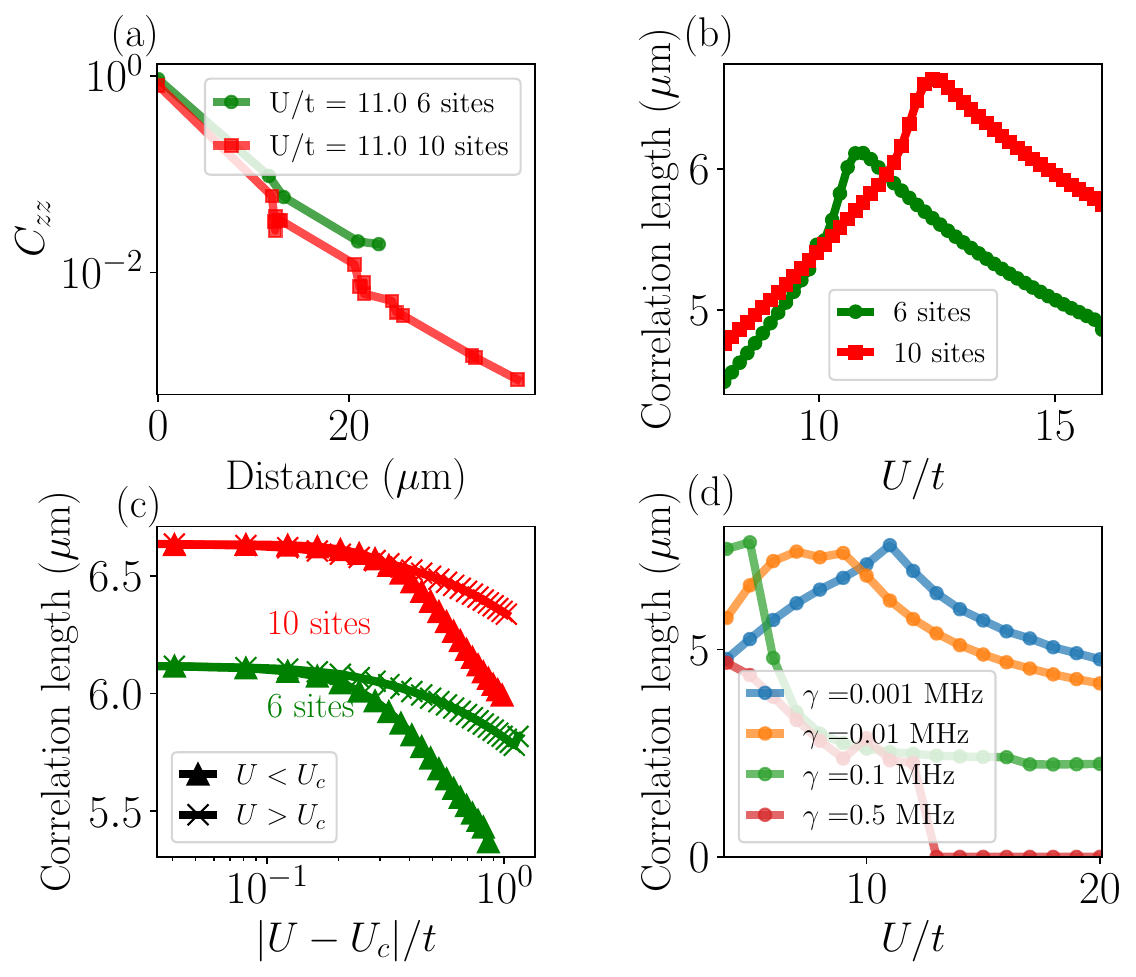}
    \caption{\textit{Correlations $C_{zz}(l)$ and correlation length of spins in the final state for a triangular lattice}. (a) correlations $C_{zz}$ vs. distance between atoms for an exact diagonalization method with an optimized geometry. values are shown for several values of $U/t$ and a system with 6 sites (circle dashed lines) and 10 sites (dotted square lines). (b) Correlation lenghts vs. $U/t$ for 6 sites (circle green line) and 10 sites (circle red line). The exact diagonalization method is used to solve the Ising Hamiltonian with an optimized geometry. (c) Correlation length for 6 sites vs. $U/t$ for different dephasing noise (blue line $\gamma = 0.001$ MHz, orange line $\gamma = 0.01$ MHz, green line $\gamma = 0.01$ MHz and red line $\gamma = 0.5$ MHz). The annealing procedure is performed to obtain correlations. the dashed black line depictes the result without dephasing ($\gamma = 0.001$ MHz). (d) Correlation length as function of $|U-U_c|/t$ on a log-scale for 6 and 10 sites.   }
    \label{fig:correl_length}
\end{figure*}

\section{Discussion}\label{sec:discussion}
\subsection{Convergence of the method}
Different initial guess of $h$ ($0.1$, $0.5$ and $0.9$) for this method are tested in Fig.~\ref{fig:tria_h_CMFT_iterations}. For $U \ll U_c$, $U \approx U_c$ and $U \gg U_c$, the convergent point is the same for all initial value, which confirms that this method is quite unrelated to the choice of the initial value of the mean field. In addition, when $U \approx U_c$, it takes more iteration steps to converge, which is normal close to the phase transition.

\subsection{Experimental feasibility}

The method proposed in this study offers several advantages with regard to experimental feasibility.
First, it avoids non-local terms that occurs when usual method of solving fermionic systems are used (Jordan-Wigner or Bravyi-Kitaev for instance). This leads to a relatively simple experimental setup where only annealing is performed. The scalability of annealing seems better compare to Variational Quantum Eigensolver algorithms for instance, where the number of parameters, the number of loop and the intractability of calculations increases with the system size \parencite{cerezo_variational_2021, fedorov_vqe_2022}. Finding the good optimizer in the classical part of this kind of algorithm is still an active field of research. In the work presented here, annealing only needs a global detuning and Rabi frequency, which suits well with experimental device. The low number of loop needed (set to a maximum value of $25$) leads to a short number of realizations. In addition, the classical part of the method only involves diagonalization of a free fermion system being scalable and easy to tract.
For the quantum part of the algorithm, it solves the two-dimensional transverse-field Ising model that exhibits several interesting properties as it is equivalent with the anisotropic limit of the three-dimensional lattice Ising model \parencite{du_croo_de_jongh_critical_1998,blote_cluster_2002}. In addition, The quantum phase transition of the 2D Ising transverse system is not quite well understood \parencite{schmitt_quantum_2022,hashizume_dynamical_2022,balducci_interface_2023} and some very recent works show a first-order phase transition for the first excited state of this model \parencite{yang_first-order_2023}. For all these reasons, Rydberg atoms can help to solve actual issues in the 2D ferromagnetic transverse Ising model and thus gained a good understanding of the 2D Hubbard model through SSMF.
Yet, the results above focus on the application of the slave-spin method applied to the half-filled, single-site Hubbard model and one should not expect them to prove any quantum advantage with respect to classical methods.
Indeed, in this parameters regime, classical methods like diagrammatic Monte-Carlo are efficient in the absence of a sign problem and can essentially reach the exact solution in the thermodynamic limit \parencite{Schuler2016}.
This therefore constitutes the short-term quantitative target for the Rydberg platform, provided a sufficient quality can be attained.

\chapter{Discussion on the correlation length of the system}
\section{Forewords}
Achieving a quantum advantage strongly relies being able to simulate strongly interacting state in the quantum simulator. This means that these states can not be reached by simulating small parts separately but the whole system is needed (all qubits are entangled). A good quantity to check if our system is interacting strongly is the correlation length. It describes the distance at which two spins interact in average. If the correlation length is small one spin "only sees" its nearest-neigbor and its state only depends of them. On the other hand, if the correlation length is very large, two very distant spins interact and it means that the state is very difficult to reach classically.

I propose therefore a little study of correlation length in the Ising with and without slave-spin theory and I focus on the difference between the ferrromagnetic model and the antiferromagnetic model.

\section{Correlation length with the slave-spin method implemented in a RQP}

A concrete quantitative yardstick for telling if our problem generates strongly interacting states, is the correlation length $\xi$ than can be achieved in the ground states prepared on the Rydberg platform.
Essentially, the size of the problem that is effectively handled by the quantum processor is of order $O(\xi^2)$ (for a 2D geometry), say $N_c^{(\mathrm{Q})} = \pi \xi^2 / 4$ for concreteness. If one denotes by $N_c^{(\mathrm{C})}$ the number of spins that can be successfully handled by a classical algorithm (for solving $H_\mathrm{S}$), it means one needs $\xi = (4 N_c^{(\mathrm{C})}/ \pi)^{1/2}$ to reach quantum advantage. Let us assess $\xi$ for the platform under consideration.
For this, we focus on the spin-spin correlation length, defined as 
\begin{equation}\label{eq:correl_length}
   C_{zz}(l) =  \langle S_0^x S_l^x \rangle - \langle S_0^x \rangle \langle S_l^x \rangle \propto e^{- l /\xi}.
\end{equation}
Correlation length and $C_{zz}(l)$ obtained with the SSMF-CMFT method for a triangular lattice ( the results shown in this appendix) are shown in Fig.~\ref{fig:correl_length}. A maximum value of the correlation length is observed for different values of $U/t$ which confirms the emergence of a phase transition. The dephasing noise has a strong impact on the value of the phase transition as notices in Fig. \ref{fig:tria_gamma}. The phase transition value is shifted to small $U/t$ values as the dephasing noise is increase. For very high values of $\gamma$, correlations become very close to $0$. It appears that the correlation length is smaller than the minimum distance between atoms ($ l \approx 7 $ $\mu$m$ < 11 $ $\mu$m).

This result seems to be well known for the 2D-ferromagnetic Ising system with a transverse field \parencite{rader_finite_2018}, especially when the longitudinal field is non null.

In order to confirm this result for a square lattice, I have run simulation on the Ising transverse field model on a square lattice with a simple cluster mean-field approach.

\begin{figure}
    \centering
    \includegraphics[width = 1 \linewidth]{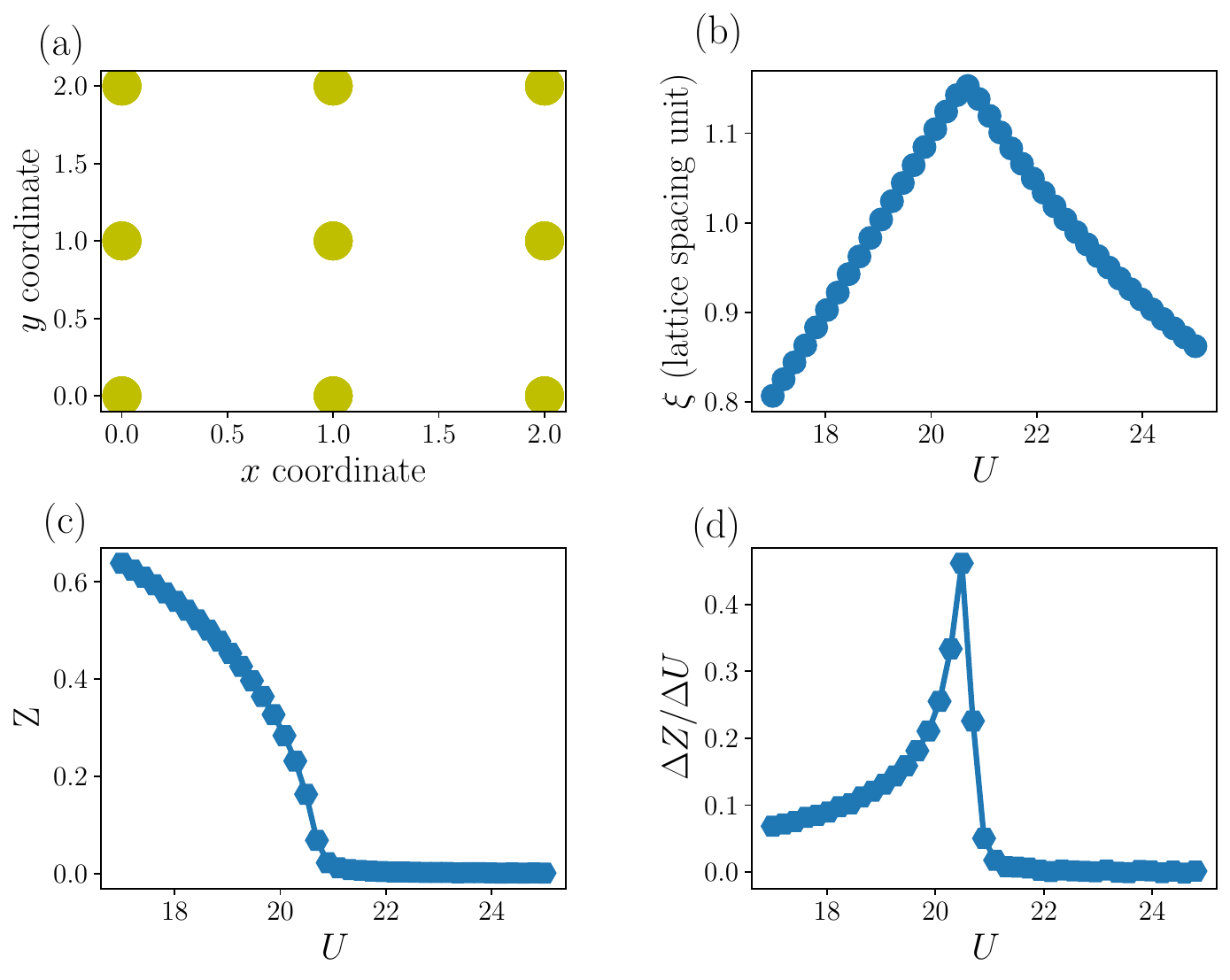}
    \caption{\textit{Correlations and phase transition for the ferromagnetic Ising transverse field model.}(a) A square lattice of $3 \times 3$ spins is used. (b) Correlation length as a function of $U$. The unit is the lattice spacing. The correlation length has a pic at $1.1$. (c) Evolution of $Z = \langle S^z \rangle^2$ as a function of $U$. (d) Variation of $Z$ with regard to variation $U$ as a function of $U$.}
    \label{fig:ferro_ising}
\end{figure}

\begin{figure}
    \centering
    \includegraphics[width = 1 \linewidth]{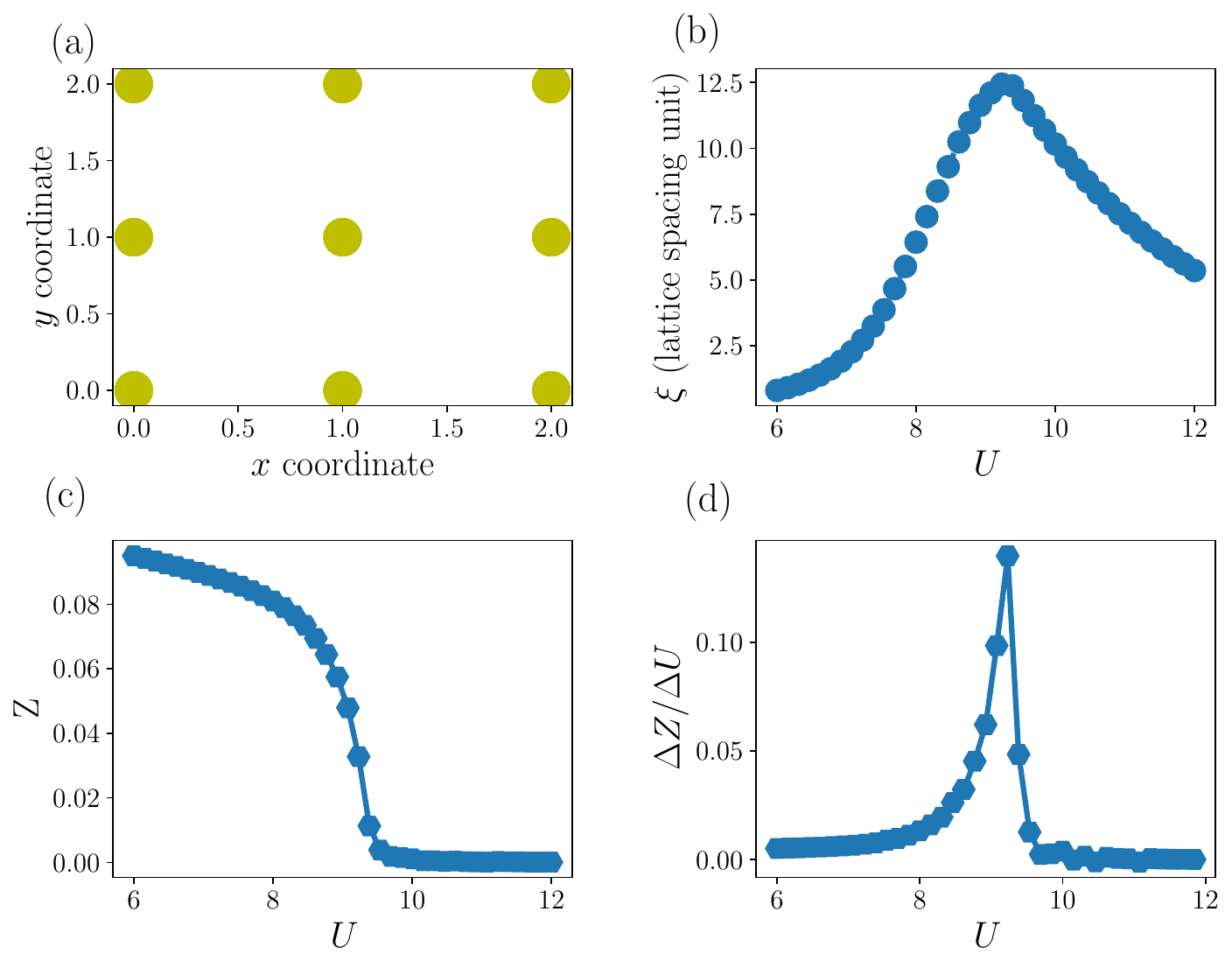}
    \caption{\textit{Correlations and phase transition for the antiferromagnetic Ising transverse field model.}(a) A square lattice of $3 \times 3$ spins is used. (b) Correlation length as a function of $U$. The unit is the lattice spacing. The correlation length has a pic at $12.5$. (c) Evolution of $Z = \langle S^z \rangle^2$ as a function of $U$. (d) Variation of $Z$ with regard to variation of $U$ as a function of $U$.}
    \label{fig:antiferro_ising}
\end{figure}

The considered Hamiltonian is:
\begin{equation}
    H = \sum_{\langle i,j \rangle} J_{i,j} S^z_iS^z_j + \overline{m}\sum_i S^z_i + U \sum_i S^z_i
\end{equation}
with $J_{i,j} =  -1$ for the ferromagnetic phase and $J_{i,j} =  1$ for the antiferromagnetic phase and $\overline{m} = 1/N \sum_i \langle S^z_i \rangle$. The lattice spacing (and the distance) is set to one between nearest neighbors.  Therefore, the distance between points is the Euclidean distance $\sqrt{|x_i-x_j|^2 + |y_i-y_j|^2}$ where $(x_i,y_i)$ and $(x_j,y_j)$ are the coordinates of the spins $i$ and $j$ respectively (see Fig.~\ref{fig:ferro_ising} and Fig.~\ref{fig:antiferro_ising}). The cluster mean-field resolution starts with a first guess of $\overline{m}$ and the groundstate is calculated with an exact diagonalization method. Then $\overline{m}$ is calculated with the system in this groundstate and the loop goes on until convergence. The last groundstate computed is the one used to calculate all correlators and then, $\xi$.

The computed values are shown in fig.~\ref{fig:ferro_ising} for the ferromagnetic phase and in  Fig.~\ref{fig:antiferro_ising} for the antiferromagnetic phase. In the case of ferromagnetic interactions, the correlation length goes to only one neighbor, in accordance with our result with the slave-spin theory whereas in the antiferromagnetic cases, the correlation length is more than $12$ times the distance between atoms in accordance with experimental results \parencite{scholl_quantum_2021}. In both cases, a phase transition is clearly seen as $\Delta Z / \Delta U$ diverges in both cases. The difference between this two transitions seems to not be clearly understood in the literature. The resolution of such a system with a cluster mean-field approach deserves more research for the slave-spin method to be more deeply understood.

\chapter{ Quasi-particle weight oscillations}
\section{Forewords}

In this appendix, I show the data I have obtained with the slave spin method for a quench $0 \rightarrow U_\mathrm{f}$ (see Chapter \ref{hubbard}) and how we can extract information from it even when a noisy numerical simulation is considered.

\section{Data of the signal}
\begin{figure}
    \centering   \includegraphics[width = 1\linewidth]{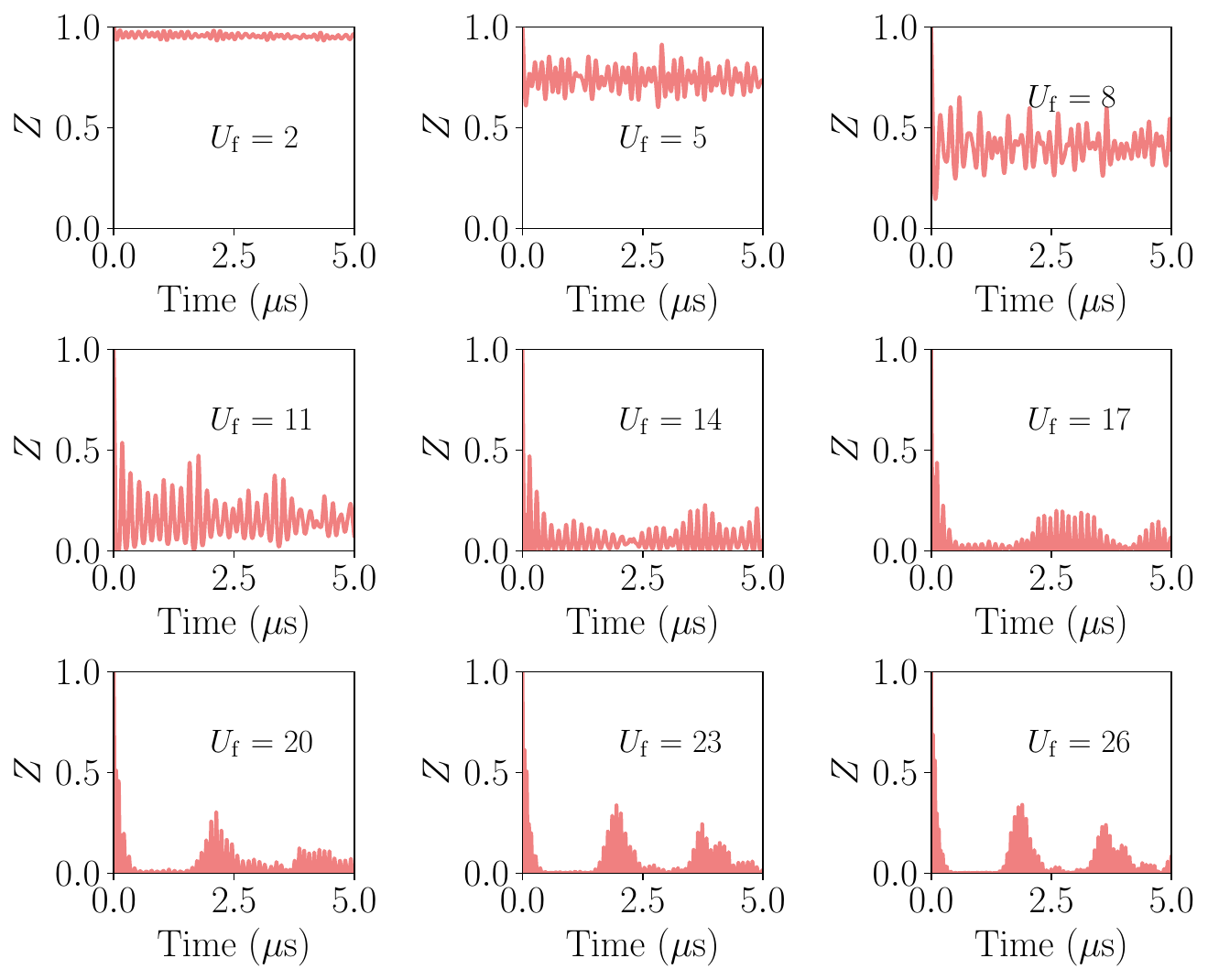}
    \caption{Response of the quasi-particle weight after a quench $0 \rightarrow U_\mathrm{f}$.}
    \label{fig:perfect_oscillations}
\end{figure}

Quasi-particle response for different $U_{f}$ is shown in Fig.~\ref{fig:perfect_oscillations}. The mean value of the oscillation converges to $0$ when $U_\mathrm{f}$ increases. 

\section{Fourier transform}
\begin{figure}
    \centering\includegraphics[width = 1 \linewidth]{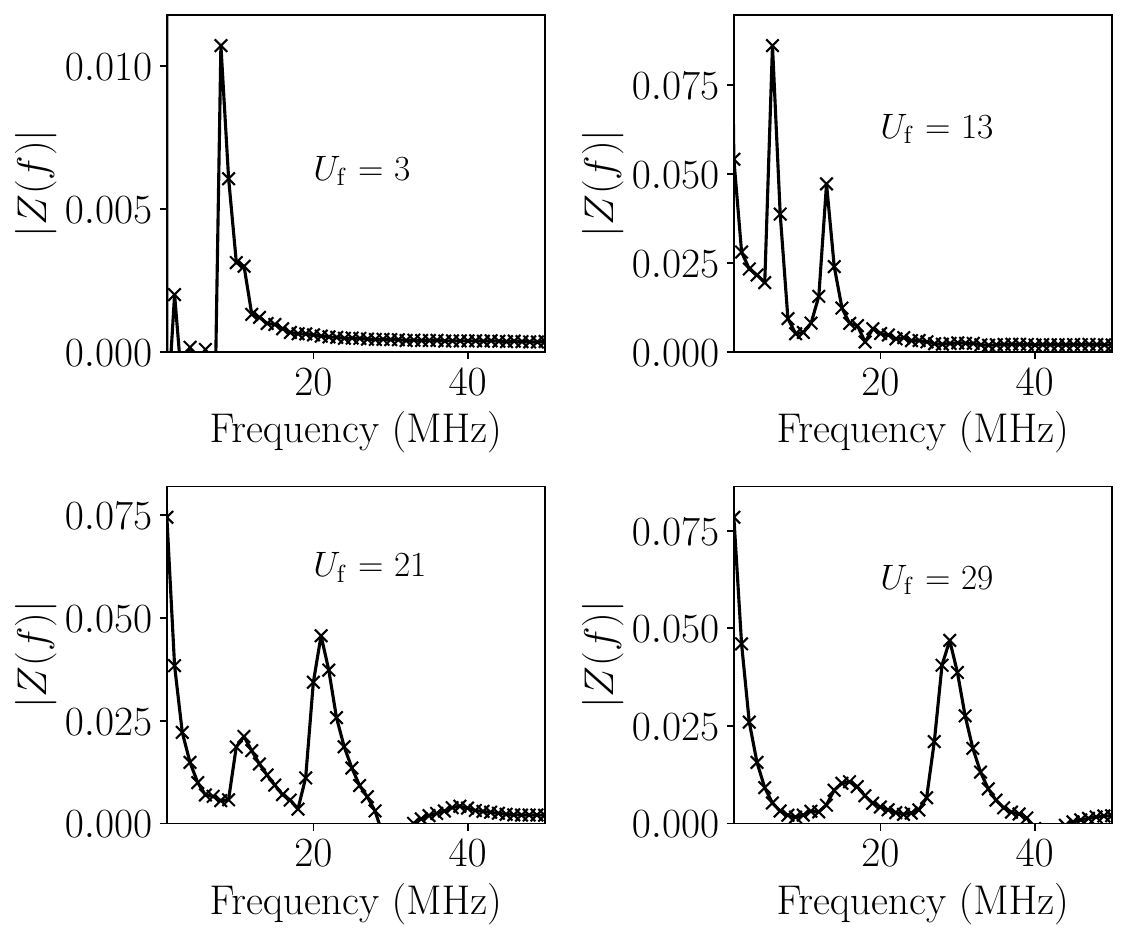}
    \caption{Fourier transform of the response of the quasi-particle weight after a quench.}
    \label{fig:FFT_perfect}
\end{figure}

The frequency of the oscillations of the quasi-particle weight obtained with the slave-spin method is computed from the Fourier transform of the signal. The signal is a list of points ($Z(\tau_k)$) taken during a time $\tau_\mathrm{max}$. The library NUMPY is used to perform the Fourier transform of the signal  version. The result is shown in Fig.~\ref{fig:FFT_perfect}. One can clearly the shift in frequency as $U_\mathrm{f}$ increases. In addition we see two pics at a frequency $U_\mathrm{f}$ and $U_\mathrm{f}/2$ as explained in chapter \ref{hubbard}.

\section{Effect of noise}
\begin{figure}
    \centering\includegraphics[width = 1 \linewidth]{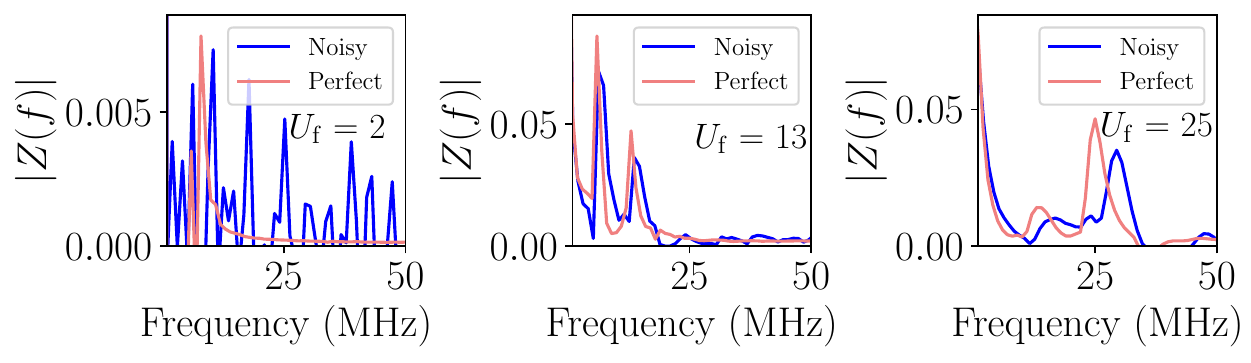}
    \caption{Fourier transform of the response of the quasi-particle weight after a quench considering a noiseless (pink) or a noisy numerical simulation (blue).}
    \label{fig:FFT_noise}
\end{figure}

The effect of noise is shown in Fig.~\ref{fig:FFT_noise}. When $U_\mathrm{f} \ll U_\mathrm{c}$, the oscillations are to small and the noise interferes with the signal. However, when $U_\mathrm{f} \geq U_\mathrm{c}$, the Fourier transform can clearly be recovered even in the presence of noise.

\chapter{Résumé en français}
EDF a pour objectif de délivrer une énergie électrique, stable et à bas prix quels que soient les défis sociétaux, politiques ou scientifiques rencontrés. Pour ce faire, il est apparu nécessaire, dans le projet de créer un réseau énergétique totalement décarboné d'ici 2050, d'augmenter la durée de vie des centrales nucléaires actuelles (à réacteur à eau pressurisée) au-delà de 60 ans. Une très bonne compréhension du vieillissement des matériaux (batteries, panneaux photovoltaïques, matériaux de structure dans les centrales nucléaires) devient alors primordiale pour anticiper l'apparition de défauts, de gonflement ou de corrosion sous contrainte par exemple dans les différents éléments de la centrale. Comme il est très difficile d'observer directement le vieillissement des matériaux dans un milieu fortement irradié, être capable de simuler numériquement ou expérimentalement correctement les matériaux d'intérêts dans les conditions réelles est la voie prise aujourd'hui à EDF. À cet effet, plusieurs méthodes de simulation numérique à l'échelle atomique sont utilisées et développées au sein d'EDF comme la théorie de la fonctionnelle de densité, la dynamique moléculaire, le Monte-Carlo cinétique ou la dynamique d'amas avec un certain succès. Ces simulations font partie d'un plus grand ensemble de simulations multi-échelles pour comprendre les mécanismes de vieillissement des matériaux. Néanmoins, certains états physiques de ces matériaux peuvent être très difficiles à simuler : paramagnétisme, électrons fortement corrélés, états excités. Des améliorations de modélisation et la levée de certains verrous sont par conséquents nécessaires.

Dans le même temps, la recherche sur le calcul et la simulation quantique a connu un très grand essor ces dernières années, que ce soit dans la recherche publique ou dans le cadre d'investissements privés. Le calcul quantique pourrait être une "révolution" dans les années à venir en surpassant largement les capacités de calcul des méthodes dites "classiques" pour des problèmes ciblés tels que : des problèmes d'optimisation, d'équation aux dérivées partielles, de cryptographie et évidemment de simulation de matériaux. Une technologie en particulier semble être adaptée à la simulation de corps en interaction : les atomes neutres (ou atomes de Rydberg). En effet, cette technologie permet d'implémenter nativement des Hamiltoniens de spins, générant ainsi des états fortement corrélés et exotiques de la matière. Ce concept de simulation quantique fut introduit par le physicien Richard Feynman et propose de simuler des systèmes d'intérêts, par exemple des électrons en interaction, à l'aide de systèmes artificiels contrôlables comme les atomes de Rydberg. Cela est maintenant permis grâce aux développements expérimentaux et théoriques importants de ces dernières années. Cependant, les algorithmes quantiques pour simuler la matière sont encore en développement et les simulateurs actuels, bien que très bien contrôlés, sont encore bruités et ne possèdent pas de qubits parfaits. Il est alors nécessaire de tester numériquement les nouveaux algorithmes sur un petit nombre de qubits dans les conditions les plus proches d'une expérience afin de les optimiser en vue d'une implémentation sur un simulateur réel.

Ma thèse s'inscrit à la fois dans la problématique de simulation de matériaux d'EDF et dans la recherche et l'algorithmie en simulation quantique.

Dans ce manuscrit, après l'introduction, je décris dans le chapitre 1 les méthodes numériques actuelles, dites de premier principe ou \textit{ab initio}, pour la simulation de matériaux à l'échelle atomique. Je présente leurs succès, mais aussi leurs limitations ainsi que les méthodes classiques qui proviennent de la théorie quantique. Ces méthodes sont très répandues dans le monde académique, mais aussi industriel, et ont permis beaucoup d'avancées dans la compréhension des phénomènes quantiques au sein des matériaux. Enfin, je donne un aperçu de la théorie sur laquelle est basée la simulation quantique et les méthodes qui permettent de passer d'un problème électronique à un problème décrit par des qubits. Dans le chapitre 3, je présente la plateforme de Rydberg et les applications possibles de cette plateforme pour la simulation et le calcul quantique ainsi que les récents résultats très encourageants obtenus. Dans le chapitre 4, je montre les résultats numériques d'un nouvel algorithme variationnel pour la chimie implémentable sur les atomes de Rydberg. Enfin, dans le chapitre 5, j'implémente numériquement un algorithme basé sur la méthode slave-spin pour simuler le comportement d'un modèle de Fermi-Hubbard à l'équilibre et hors équilibre. Cet algorithme est aussi conçu pour les atomes de Rydberg.

\textbf{Chapitre~\ref{Bibliography}  Simuler la matière corrélée : du classique au quantique }

Ce chapitre décrit les méthodes "classiques" actuelles pour simuler numériquement la matière ainsi que les bases théoriques pour le calcul quantique.

Tout d'abord, je présente la méthode Hartree-Fock puis la théorie de la fonctionnel de la densité (DFT) qui a pour but de résoudre l'équation de Schrödinger à l'aide de la densité d'état électronique du système plutôt que la fonction d'onde au prix d'approximations pour le terme d'échanges et de corrélations. Plusieurs approximations existent (approches locales, à gradients) et toutes possèdent leurs avantages et inconvénients. Je présente des développements récents permettant d'étendre la DFT, notamment sur la simulation du magnétisme. J'aborde ensuite les approches se basant sur le formalisme de la physique quantique avec notamment la théorie du champ moyen dynamique (DMFT) après avoir décrit la fonction de Green. Cette méthode permet de décrire un système de Fermi-Hubbard et notamment le phénomène de transition de Mott avec une bonne précision. Je décris aussi brièvement la méthode DFT+DMFT et la méthode de Monte-Carlo quantique. Je finis ce chapitre en donnant les bases théoriques au calcul quantique et aux différentes approches possibles (digitales ou analogues). J'expose la méthode d'estimation de phase et les méthodes variationnelles. Un approfondissement est proposé pour appliquer ces méthodes aux fermions en interaction.

\textbf{Chapitre~\ref{Rydberg} : Simulation quantique avec les atomes de Rydberg}

Ce chapitre me permet de décrire le fonctionnement expérimental de ce simulateur quantique. La première étape est la préparation des atomes froids sous forme de matrices d'atomes contrôlées par des lasers. La géométrie des atomes peut être totalement contrôlée pour réaliser n'importe quel type de réseaux 2D et 3D. Des états électroniques de Rydberg de ces atomes sont contrôlés à l'aide de laser, et permettent de considérer les atomes comme des qubits ou des spins. Ces spins effectifs interagissent soit par interaction d'Ising, soit par une interaction XY en fonction des états de Rydberg choisis.

Enfin, l'état du système est mesuré grâce à la fluorescence des atomes. Cette architecture permet de faire de la simulation digitale et analogue jusqu'à plus de 200 qubits avec des résultats récents très prometteurs.

\textbf{Chapitre~\ref{chemistry} : Algorithme quantique variationnel analogue-digital pour la chimie}

Dans ce chapitre, je décris, implémente et montre le résultat d'un algorithme que nous avons conçu pendant ma thèse, pour trouver l'énergie de l'état fondamental de molécules ($\mathrm{H}_2$, LiH et $\mathrm{BeH}_2$) avec les atomes neutres. Après avoir expliqué comment j'ai transformé les Hamiltoniens moléculaires en Hamiltonien de qubits, je présente comment optimiser la géométrie des atomes de Rydberg en fonction de l'Hamiltonien cible, les séquences de pulses optimisées qui sont implémentées ainsi que la méthode de mesure qui utilise la "derandomization". Je compare les résultats de ma méthode avec la méthode naïve d'opérateurs alternatifs et je montre que j'obtiens de meilleurs résultats plus rapidement (avec moins de mesures). Je montre que l'énergie fondamentale peut être retrouvée avec $5 \%$ d'erreur lorsque nous imposons un certain nombre de mesures (le critère d'arrêt de l'agorithme variationnel) et donnons des pistes pour améliorer la méthode. Ce travail a été publié dans \textit{Physical review A}.

\textbf{Chapitre~\ref{hubbard} : Utilisation de la plateforme de Rydberg pour simuler de la physique d'électrons fortement corrélés dans le modèle de Hubbard 2D}

Résoudre le modèle de Hubbard pourrait mener à l'explication de beaucoup de problématiques de matière condensée (par exemple, il devrait aider à décrire les supraconducteurs haute-température). Cependant, il est très difficile à résoudre classiquement avec certains types de dopage ou de modèles multi-orbitaux. Dans cette étude, nous proposons une méthode pour simuler le comportement d'un modèle de Hubbard 2D sur une plateforme de Rydberg. L'idée est d'utiliser la méthode des spins "esclaves" et d'un découplage des degrés de liberté avec une approche de champ moyen. En effet, les spins esclaves ajoutent artificiellement un degré de liberté aux fermions présents dans le système tout en restant dans un espace d'états physiques si une contrainte sur les opérateurs est appliquée. Un découplage en champ moyen est ensuite appliqué de façon à décorréler les degrés de liberté de charge (et de spin) de fermions et des spins esclaves. On obtient ainsi deux Hamiltoniens auto-cohérents :
\begin{itemize}
\item un Hamiltonien de fermions libres,
\item un Hamiltonien d'Ising en champ transverse.
\end{itemize}
La résolution de ces Hamiltoniens est réalisable avec une boucle auto-cohérente où résoudre l'un des deux permet de résoudre l'autre. L'idée majeure de cette approche est qu'il est possible de résoudre classiquement l'Hamiltonien de fermions libres alors que toute la complexité du modèle de départ à été transférée dans l'Hamiltonien d'Ising, qui est implémentable directement sur la plateforme de Rydberg. Il s'agit d'un algorithme hybride mais non variationnel. Je teste cette méthode sur un réseau carré à demi-remplissage et je montre que l'on peut retrouver une transition de Mott dans le cas théorique, mais aussi en considérant une émulation d'expérience sur une vraie architecture. Plusieurs sources de bruits possibles sont considérées (bruit de shot, déphasage, temps d'annealing, géométrie des atomes).
De plus, la dynamique des électrons est étudiée dans ce paradigme avec des résultats correspondant aux résultats théoriques et numériques de la littérature.

Cet algorithme peut être implémenté sur les simulateurs actuels avec un possible avantage quantique pour la dynamique des électrons, très difficile à simuler avec les méthodes numériques et théoriques actuelles.

\end{document}